%% file: main.tex
\documentclass[numberwithinsect]{no-lipics}
\DeclareTextAccentDefault{\k}{T1}

\usepackage{amssymb}
\usepackage{mathtools}
\usepackage{bm}
\usepackage{subcaption}
\usepackage{stmaryrd}
\usepackage[draft]{fixme}
\usepackage{booktabs}
\usepackage{todonotes}
\usepackage{accents}

\usetikzlibrary{calc, matrix, decorations.pathreplacing}

\input{macros}

\title{Tree Edit Distance: Near-Linear-Time Approximate, Sublinear, and Dynamic}

\author{Bingbing Hu}{University of California, San Diego\\La Jolla, United States}{bingbing@ucsd.edu}{https://orcid.org/0000-0001-8854-5969}{}
\author{Tomasz Kociumaka}{Max Planck Institute for Informatics, SIC,\\Saarbrücken, Germany}{tomasz.kociumaka@mpi-inf.mpg.de}{https://orcid.org/0000-0002-2477-1702}{}
\author{Jakob Nogler}{Massachusetts Institute of Technology\\Cambridge, United States}{jnogler@mit.edu}{https://orcid.org/0009-0002-7028-2595}{}
\author{Barna Saha}{University of California, San Diego\\La Jolla, United States}{barnas@ucsd.edu}{https://orcid.org/0000-0002-6494-3839}{}

\authorrunning{B. Hu, T. Kociumaka, J. Nogler, and B. Saha}

\begin{document}
\pagenumbering{roman}
\maketitle
\begin{abstract}
\input{abstract}
\end{abstract}
\clearpage
\newpage
\tableofcontents
\clearpage
\pagenumbering{arabic}

\input{s1_intro.tex}

\input{s2_overview.tex}
\input{s2_overview_dynamic}

\input{s3_prelims.tex}

\input{s4_decomp}

\input{s5_approx}

\input{s6_sublinear_revised}

\input{s7_dynamic_approximate}

\input{AI_acknowledgement}

\bibliographystyle{alphaurl}
\bibliography{main}
\newpage
\appendix
\input{appendix}

\end{document}

%% file: macros.tex
\usetikzlibrary{arrows,arrows.meta,shapes.misc, decorations.pathreplacing,decorations.pathmorphing,calligraphy, patterns.meta}
\usetikzlibrary{calc, matrix}
\tikzset{dotmark/.style={circle,fill,inner sep=1.5pt}}
\tikzset{emptymark/.style={circle,draw,fill=white,inner sep=1.5pt}}
\tikzset{crossmark/.style={thick,inner sep=1.5pt}}

\SetKwComment{Comment}{$\triangleright$\ \rm}{}
\SetKwInput{KwInput}{Input}
\SetKwInput{KwOutput}{Output}
\SetKwProg{Fn}{}{}{}

\allowdisplaybreaks

\def\twoheadleadsto{\tikz[baseline=(a.base)]{\draw[%
    decorate,decoration={zigzag,segment length=4, amplitude=.9},%
    ] (0,0) -- (.25, 0);%
    \draw[%
    -{Classical TikZ Rightarrow}.{Classical TikZ Rightarrow},%
    ] (.25, 0) -- (.4, 0);%
    \node (a) at (.4/2,-.03) {\phantom{\(\leadsto\)}};%
}}

\newcommand{\onto}{\twoheadleadsto}

\newcommand{\A}{\mathcal{A}}

\def\ShowAuthNotes{1}
\ifnum\ShowAuthNotes=1
\newcommand{\authnote}[3]{\textcolor{#3}{[{\bf #1:} { {#2}}]}}
\else
\newcommand{\authnote}[3]{}
\fi

\newcommand{\Oh}{\mathcal{O}}
\newcommand{\Ohtilde}{\tilde{\Oh}}
\newcommand{\tOh}{\Ohtilde}
\newcommand{\Ohhat}{\hat{\Oh}}
\DeclareMathOperator*{\poly}{poly}
\DeclareMathOperator*{\polylog}{polylog}
\DeclareMathOperator*{\argmin}{arg\,min}

\DeclareMathOperator{\E}{\mathbb{E}}

\newcommand{\sub}{\subseteq}
\newcommand{\preor}{\mathbin{\prec_{\mathrm{pre}}}}
\newcommand{\postor}{\mathbin{\prec_{\mathrm{post}}}}
\newcommand{\op}[1]{\texttt{(}_{#1}}
\newcommand{\cl}[1]{\texttt{)}_{#1}}

\DeclareMathOperator{\Prob}{\mathbf{Pr}}

\def\fragmentco#1#2{\bm{[}\,#1\,\bm{.\,.}\,#2\,\bm{)}}
\def\fragmentoc#1#2{\bm{(}\,#1\,\bm{.\,.}\,#2\,\bm{]}}
\def\fragmentoo#1#2{\bm{(}\,#1\,\bm{.\,.}\,#2\,\bm{)}}
\def\fragment#1#2{\bm{[}\,#1\,\bm{.\,.}\,#2\,\bm{]}}
\def\position#1{\bm{[}\,#1\,\bm{]}}

\newcommand{\ceil}[1]{\lceil #1 \rceil}
\newcommand{\floor}[1]{\lfloor #1 \rfloor}

\newcommand{\yes}{{\sc yes}\xspace}
\newcommand{\no}{{\sc no}\xspace}

\newcommand{\TD}{\operatorname{TD}}

\newcommand{\TDrec}{\ensuremath{\operatorname{TD}^{\textsf{rec}}}}
\newcommand{\Deltarec}{\ensuremath{\operatorname{\Delta}^{\textsf{rec}}}}
\newcommand{\TDsel}{\ensuremath{\operatorname{TD}^{\textsf{sel}}}}
\newcommand{\Deltasel}{\ensuremath{\operatorname{\Delta}^{\textsf{sel}}}}
\newcommand{\Ex}[2]{\E\!\left[#1 \;\middle|\; #2\right]}

\newcommand{\id}{\mathsf{id}}
\newcommand{\ids}{\mathsf{ids}}
\newcommand{\val}{\mathsf{val}}
\newcommand{\cost}{\mathsf{cost}}

\newcommand{\ed}{\mathsf{ed}}
\newcommand{\ted}{\mathsf{ted}}

\renewcommand{\Other}{\mathsf{tw}}
\newcommand{\Left}{\mathsf{l}}
\newcommand{\Right}{\mathsf{r}}
\newcommand{\LeftD}{\mathsf{ld}}
\newcommand{\ld}{\mathsf{ld}}
\newcommand{\RightD}{\mathsf{rd}}
\newcommand{\rd}{\mathsf{rd}}

\newcommand{\parent}{\mathsf{par}}

\newcommand{\dist}{\mathsf{dist}}
\newcommand{\penaltyFunc}{\mathsf{pen}}

\newcommand{\ul}[1]{\underaccent{\bar}{#1}}

\newcommand{\TED}{\textsf{TED}\xspace}
\newcommand{\GAPTED}{\textsf{Gap-TED}\xspace}

\newcommand{\ED}{\textsf{ED}\xspace}
\newcommand{\APSP}{\textsf{APSP}\xspace}

\newcommand{\mA}{\mathcal{A}}
\newcommand{\mB}{\mathcal{B}}

\newcommand{\mD}{\mathcal{D}}

\newcommand{\mI}{\mathcal{I}}

\newcommand{\mL}{\mathcal{L}}
\newcommand{\mS}{\mathcal{S}}
\newcommand{\mJ}{\mathcal{J}}

\newcommand{\hi}{\hat{\imath}}
\newcommand{\hj}{\hat{\jmath}}

\newcommand{\sT}{\mathsf{T}}

\newcommand{\sL}{\mathsf{L}}

\newcommand{\bF}{\mathbf{F}}

\newcommand{\bG}{\mathbf{G}}

\newcommand{\bT}{\mathbf{T}}

\renewcommand{\epsilon}{\varepsilon}
\newcommand{\biinterval}[2]{\langle #1, #2 \rangle}

\DeclarePairedDelimiter\abs{\lvert}{\rvert}
\DeclarePairedDelimiter\norm{\lVert}{\rVert}

\makeatletter
\let\oldabs\abs
\def\abs{\@ifstar{\oldabs}{\oldabs*}}
\let\oldnorm\norm
\def\norm{\@ifstar{\oldnorm}{\oldnorm*}}
\makeatother

\def\problembox#1{%
    \vspace{2mm}%
    \noindent\fbox{%
    \begin{minipage}{.985\linewidth}%
        #1
    \end{minipage}%
    }%
    \vspace{2mm}%
}

\newcommand{\defproblemwlist}[3]{%
    \problembox{%
        \textbf{#1}\\
        {\bf{Input:}} #2
        {\bf{Output:}} #3
    }%
}

\makeatletter
\renewenvironment{cases}{%
  \matrix@check\cases\env@cases
}{%
  \endarray\right.%
}
\def\env@cases{%
  \let\@ifnextchar\new@ifnextchar
  \left\lbrace
  \def\arraystretch{1.1}%
  \array{@{\;}c@{\quad}l@{}}%
}
\makeatother

\def\mid{\ensuremath :}

\def\emptyset{\varnothing}

\newcommand\thefont{\expandafter\string\the\font}

%% file: abstract.tex
Tree edit distance (\textsf{TED}), the analog of string edit distance (\textsf{ED}) for hierarchical data, asks for the minimum number of node insertions, deletions, and substitutions needed to transform one labeled rooted ordered tree into another.
At first glance, the tree structure makes the problem much harder: \textsf{ED} admits a classical $\mathcal{O}(n^2)$-time exact algorithm, whereas the fastest known exact algorithm for \textsf{TED} runs in $\tilde{\mathcal{O}}(n^{(3+\omega)/2})=\mathcal{O}(n^{2.686})$ time [Nogler, Polak, Saha, Vassilevska Williams, Xu, Ye; STOC'25].
Nevertheless, many state-of-the-art guarantees for \textsf{ED} have been recreated for \textsf{TED}.
These include linear-time exact algorithms for distances capped by a small power of $n$ [Das, Gilbert, Hajiaghayi, Kociumaka, Saha; STOC'23] and a truly subquadratic-time $(3+\varepsilon)$-approximation [Seddighin \& Seddighin; ITCS'22].
Further, the fastest $(1+\varepsilon)$-approximations take, up to sub-polynomial factors, $n^2$ time [Boroujeni, Ghodsi, Hajiaghayi, Seddighin; STOC'19].
Still, substantial polynomial gaps in running time or approximation ratio remain in several regimes, notably near-linear-time approximation, the sublinear query model, and the dynamic setting.

In this work, we significantly narrow these three gaps.
\begin{itemize}
    \item We give an $n^{o(1)}$-approximation algorithm running in $\mathcal{O}(n)$ time and, for every $0 < \delta < 1$, a $(\log n)^{\mathcal{O}(1/\delta)}$-approximation algorithm running in $\mathcal{O}(n^{1+\delta})$ time.
    This improves the previous near-linear-time $\mathcal{O}(\sqrt n)$-approximation [Boroujeni, Ghodsi, Hajiaghayi, Seddighin; STOC'19] and, up to polylogarithmic factors, brings the approximability of \textsf{TED} in this regime to parity with that of \textsf{ED}.
    \item Under query access to the balanced-parentheses (Euler-tour) representation of the input trees, we obtain the first sublinear-time algorithm for \textsf{TED}: for every $\tau \in [1,n]$, our procedure computes a $\tau n^{o(1)}$-approximation of the value $1+\textsf{ted}$ in $\mathcal{O}(n/\tau)$ time.
    This establishes the feasibility of sublinear-time \textsf{TED} approximation; in contrast, sublinear-time algorithms for \textsf{ED} approximation have been studied for more than two decades.
    \item In the dynamic setting, we present an exact algorithm for \textsf{TED} with amortized update time $\tilde{\mathcal{O}}(k^7)$, where $k$ is a cap beyond which the distance need not be reported, as well as a $(1+\varepsilon)$-approximate dynamic algorithm with amortized update time $\tilde{\mathcal{O}}_{\varepsilon}(k)$.
    The exact result is the first dynamic \textsf{TED} algorithm that beats recomputing from scratch in a non-trivial parameter regime, while the approximate result sharply strengthens the previously known dynamic guarantees [Das, Gilbert, Hajiaghayi, Kociumaka, Saha; FOCS'25] and, for any parameter regime, cannot be significantly improved without a breakthrough for static $(1+\varepsilon)$-approximation of \textsf{ED}.
\end{itemize}

All our results use dynamic programming on a new hierarchical tree decomposition, an analog of partitioning a string into blocks of geometrically decreasing sizes.
This decomposition combines structural properties that had previously been achieved only in isolation and provides a flexible framework for transferring algorithmic ideas from \textsf{ED} to \textsf{TED}, with many additional ingredients handling the obstacles that arise in each computational setting.

%% file: s1_intro.tex
\section{Introduction}

The \emph{Tree Edit Distance} (\TED), first introduced by Selkow in the late 1970s~\cite{Selkow77}, generalizes the classical \emph{String Edit Distance} (\ED).
Since its introduction, tree edit distance has attracted sustained attention due to its many applications, including computational biology~\cite{10.1093/bioinformatics/6.4.309,waterman1995introduction,gusfield_1997,HTGK03}, structured data and XML analysis~\cite{Chawathe99,KochBG03,FerraginaLMM09}, image processing~\cite{BK99,KTSK00,KSK01,SebastianKK04}, and compiler optimization~\cite{DMRW10}, among others; see~\cite{Bil05,Aku10} for surveys.

In the standard formulation, we are given two rooted, ordered, labeled forests $\bF$ and $\bG$, where each node carries a label from an alphabet~$\Sigma$.
The tree edit distance $\ted(\bF, \bG)$ is the minimum length of a sequence of node edits that transform $\bF$ into $\bG$.
The allowed edits are:
\begin{mylist}[label=(\roman*)]
\item substituting the label of a node;
\item deleting a node, with the children of the deleted node reattached to the node's parent in their original order;
\item inserting a new labeled node between a parent and a consecutive subset of its children.
\end{mylist}

Unlike \ED, where the $\Oh(n^2)$-time textbook algorithm is optimal \cite{BI18} under the Strong Exponential Time Hypothesis (SETH), algorithms for \TED have seen continuous improvements over the past five decades~\cite{Tai79,ZS89,Kle98,DMRW10,M22,Dur23}.
The current state-of-the-art algorithm for \TED, due to Nogler, Polak, Saha, Vassilevska Williams, Xu, and Ye~\cite{NPSVXY25}, runs in $\tOh(n^{(3+\omega)/2})$ time.\footnote{Here, $\omega \leq 2.371552$ denotes the matrix multiplication exponent \cite{VWXXZ24}.
Very recently, OpenAI~\cite{OAI26MM} has claimed $\omega \leq 9/4$.
The $\Ohtilde(\cdot)$ notation hides factors polylogarithmic in the input size. The $\Ohtilde_{\varepsilon}(\cdot)$ notation additionally hides factors depending on $\varepsilon$ only.}
This may not be the final word, as no lower bound better than the $n^{2-o(1)}$ bound inherited from \ED is known so far.
It is plausible, however, that \TED requires super-quadratic time, as suggested by gaps between \TED and \ED in their \emph{weighted} versions~\cite{BGMW20,WF74}, where operation costs depend on node labels (in contrast to the \emph{unweighted} \TED version considered here). Indeed, the weighted version of \TED has been shown to be equivalent to the \emph{All-Pairs Shortest Paths (\APSP)} problem~\cite{BGMW20, NPSVXY25}, which is conjectured to require almost cubic time.

Faster algorithms have been obtained when the tree edit distance is low, bounded by a (small) threshold~$k$.
The $\tOh(nk^2)$-time algorithm of Akmal and Jin~\cite{AJ21}, improving upon an earlier $\Oh(nk^3)$-time solution~\cite{10.1007/11496656_29}, is the fastest known for $k \gtrsim n^{1/4}$.
For smaller thresholds, the state of the art is the $\tOh(n + k^6)$-time algorithm of Kociumaka and Shahali~\cite{KS25}, which improves upon \cite{DGHKSS22,DGHKS23}.

The complexity of computing (unweighted) \TED drops substantially when approximation is allowed.
Boroujeni, Ghodsi, Hajiaghayi, and Seddighin~\cite{BGHS19} gave a $(1+\varepsilon)$-approximation algorithm in $\Ohtilde_{\varepsilon}(n^2)$ time, and more generally in $\Ohtilde_{\varepsilon}(nk)$ time, where $k = \ted(\bF, \bG)$, whereas Seddighin and Seddighin~\cite{SS22} presented an $\Oh(n^{1.99})$-time $(3+o(1))$-approximation.
For any parameter $\tau \le \sqrt{n}$, one can also achieve an $\Oh(\tau)$-approximation in $\Ohtilde(n^2 / \tau^2)$ time~\cite{BGHS19}; this yields an $\Oh(\sqrt{n})$-approximation in near-linear time.
A very recent work of Das, Gilbert, Hajiaghayi, Kociumaka, and Saha~\cite{DGHKS25} further improved upon this bound for small values of~$k$, with an $\Oh(k\log n)$-approximation in $\Ohtilde(n)$ time.

Despite this progress, \TED approximation remains poorly understood compared to \ED.
For the latter, a long line of research on near-linear-time approximation algorithms \cite{10.5555/279082.279125,10.1109/FOCS.2004.14,10.5555/1109557.1109644,AO09} culminated in a seminal result by Andoni, Krauthgamer, and Onak \cite{AKO10}, who achieved a $(\log n)^{\Oh(1/\delta)}$-approximation in $\Oh(n^{1+\delta})$ time, for any $0<\delta <1$.
After an eight-year hiatus, the study of this problem regained momentum with a breakthrough by Chakraborty, Das, Goldenberg, Koucký, and Saks~\cite{10.1145/3422823},
who obtained a constant-factor approximation for \ED in subquadratic time.
This led to a sequence of refinements~\cite{DBLP:journals/jacm/BoroujeniEGHS21,KS20,BR19,GRS:20},
ultimately yielding a near-linear-time constant-factor approximation by Andoni and Nosatzki~\cite{AN20}.
Specifically, for any constant $\delta > 0$, their algorithm achieves a $\Oh_\delta(1)$-factor approximation of \ED between strings of length~$n$ in $\Oh(n^{1+\delta})$ time.
In contrast, the best known near-linear-time algorithm for \TED achieves only an $\Oh(\min(k\log n, \sqrt{n}))$-factor approximation.
This gap highlights a glaring open question.

\pagebreak[4]
\begin{center}
    {\bf Question 1:} {\it What is the best approximation achievable for \TED in near-linear time?\\
    Can we achieve subpolynomial-factor approximation in near-linear or linear time?}
\end{center}

In a similar spirit, one may ask about the best possible \TED approximation achievable in \emph{sublinear} time.
Fueled by a growing interest in modern computational paradigms such as sampling and property testing, there is a substantial body of work on approximating \ED in sublinear time, beginning with the seminal paper of Batu, Ergün, Kilian, Magen, Raskhodnikova, Rubinfeld, and Sami~\cite{BEKMRRS03}, and followed by~\cite{AO12,GKS19,BCR20,KS20a,BCFN22,GKKS22,BCFK24}.
For any parameter $\tau \in \fragment{1}{n}$, Goldenberg, Kociumaka, Krauthgamer, and Saha~\cite{GKKS22} achieved an $\Oh(n / \tau)$-time $\Ohhat(\tau)$-factor approximation of the quantity $1 + \ed$, where the $\Ohhat$ notation hides subpolynomial factors.\footnote{Distinguishing $\ed = 0$ from $\ed > 0$ requires $\Omega(n)$ time, so multiplicative approximation of $\ed$ itself is infeasible.}
This trade-off can be improved in most parameter regimes and, if the edit distance is moderately small, even its $(\log n)^{\Oh(1)}$-factor approximation can be computed in truly sublinear time~\cite{BCFN22,BCFK24}.
In contrast, no sublinear-time \TED approximations are currently known, and it is not even clear how such an algorithm should access the two input forests in the first place.

\begin{center}
    {\bf Question 2:} {\it What are the best achievable approximation results for \TED under truly sublinear time constraints?
    Is there a natural query access model allowing for results similar to those known for \ED?}
\end{center}

Finally, we turn our attention to dynamic algorithms.
In this setting, the underlying forests undergo dynamic updates (node edits), and the goal is to maintain the tree edit distance efficiently.
Dynamic variants are particularly relevant for practical applications such as tracking updates to structured documents that evolve over time.
However, even for the simpler case of \ED, obtaining fast update times in the dynamic setting has proven to be challenging.
Only recently, Gorbachev and Kociumaka~\cite{GK25} presented an exact algorithm with $\tOh(k)$ update time, where $k$ denotes the current edit distance between the strings, improving upon an earlier $\tOh(n)$-update-time algorithm~\cite{CKM20}.
Obtaining $\tOh(k^{1-\varepsilon})$ update time for any polynomial value of $k$ would immediately yield an $\tOh(n^{2-\varepsilon})$-time static \ED algorithm and thus violate SETH.
For \TED, no non-trivial dynamic exact algorithm was previously known. Under the Combinatorial $K$-Clique Detection Conjecture, Hu, Nogler, and Saha~\cite[Theorem~19]{HNS26} rule out combinatorial exact algorithms with $p(n)+n(u(n)+q(n))=\Oh(n^{3-\varepsilon})$ for any constant $\varepsilon>0$, where $p(n)$, $u(n)$, and $q(n)$ denote preprocessing, update, and query time, respectively.
In particular, this excludes truly subquadratic update time in the regime $k=\Theta(n)$ when preprocessing is truly subcubic and query time is truly subquadratic.

For dynamic \ED approximation, Kociumaka, Mukherjee, and Saha~\cite{KMS23} achieved an $n^{o(1)}$-time update with an $n^{o(1)}$ approximation factor, and this remains the state of the art.
For dynamic \TED approximation, the only nontrivial algorithm, introduced very recently~\cite{DGHKS25}, achieves an $\Oh(\min(\sqrt{n \log n}, k \log n))$-factor approximation, where $\ted(\bF, \bG) \le k$, and supports updates in $\Oh(\polylog n)$ time.
To date, no known dynamic algorithm maintains a $(1+\varepsilon)$-approximation of \TED.

\begin{center}
    {\bf Question 3:} {\it What is the best update time needed to dynamically maintain \TED or its $(1+\varepsilon)$ approximation?\\
    Is there any approach significantly better than recomputing \TED from scratch after each update?}
\end{center}
In this paper, we make significant progress towards answering all the above questions.

\subparagraph*{Our Results.}
Our first main contribution is a near-linear-time polylogarithmic-factor approximation for \TED, which matches the near-linear-time bounds of \ED up to a subpolynomial factor.

\begin{restatable*}{mtheorem}{approxted}\label{thm:approxted}
There is a randomized algorithm that, given $n$-node forests $\bF$ and $\bG$ and a real parameter $0 < \delta < 1$, outputs an $(\log n)^{\Oh(1/\delta)}$-approximation of $\ted(\bF,\bG)$ in $\Oh(n^{1+\delta})$ time, with high probability.
\end{restatable*}

Our solution substantially generalizes the algorithm of Andoni, Krauthgamer, and Onak~\cite{AKO10} for \ED.
As discussed in \cref{sec:ov:static}, adapting their scheme to \TED requires many new notions, yet still falls short of our goal: it results in an $\Oh(n^{1+\delta}\cdot k)$-time algorithm for \TED approximation, which is slower than the known $(1+\varepsilon)$-approximation~\cite{BGHS19}.
However, with a novel insight reducing the number of dynamic-programming states, we recover the original trade-off known for \ED, even for $\delta = o(1)$. In particular, we achieve $2^{\Oh(\sqrt{\log n \log \log n})}$-approximation in
$n\cdot 2^{\Oh(\sqrt{\log n \log \log n})}$ time.

Our second result is a reduction that allows decreasing the running time at the cost of increasing the approximation ratio, with the product of the two essentially intact.
In order to allow sublinear time complexity, we consider a simple access model inspired by succinct (that is, $(2n+o(n))$-bit) forest representations~\cite{MR01}.
Namely, we model the forest as a balanced sequence of parentheses, where each node is represented by a pair of matching parentheses; see \cref{fig:forest_merged}.
We assume constant-time access to the following data: \begin{mylist}[label=(\arabic*)]
    \item What is the label of the node represented by the $i$th leftmost parenthesis?
    \item What is the position of the \emph{twin} $\Other(i)$ of the $i$th leftmost parenthesis? (In particular, the $i$th parenthesis is opening if and only if $i < \Other(i)$.)
\end{mylist}
We note that this is a much weaker model compared to the set of queries supported by static~\cite{MR01} and dynamic~\cite{NS14} succinct data structures.
In particular, accessing the parent of a given node requires time proportional to the parent's degree.
Nevertheless, we still achieve a result analogous to \cite[Theorem III.2]{GKKS22}: an $\Ohhat(\tau)$-approximation in $\Oh(n/\tau)$ time.

\begin{restatable*}{mtheorem}{sublinearted}\label{thm:sublinearted}
There is a randomized algorithm that, given $\tau \in \fragment{1}{2n}$ and access to $n$-node forests $\bF$ and $\bG$,
computes a $\tau\cdot 2^{\Oh(\sqrt{\log n \log \log n})}$-approximation of $1+\ted(\bF,\bG)$ in $\Oh(n/\tau)$ time correctly w.h.p.
\end{restatable*}

Regardless of the access model, \cref{thm:sublinearted} implies the first linear-time approximation of $\TED$. This is because $1+\ted(\bF,\bG)$ and $\ted(\bF,\bG)$ differ significantly only if $\bF$ is equal to $\bG$, which is easy to test.

\begin{corollary}\label{thm:linearted}
There is a randomized algorithm that, given $n$-node forests $\bF$ and $\bG$,
outputs a $2^{\Oh(\sqrt{\log n \log \log n})}$-approximation of $\ted(\bF, \bG)$ in time $\Oh(n)$ correctly with high probability.\lipicsEnd
\end{corollary}

Next, we move on to the dynamic setting, where our main outcome (and the most technically sophisticated result of this paper) is a dynamic algorithm that, for a fixed parameter $k$ and constant $\varepsilon>0$, supports $\Ohtilde_{\varepsilon}(k)$ amortized-time updates and maintains a $(1+\varepsilon)$-approximation of the \TED value capped at $k$.

\begin{restatable*}{mtheorem}{dynapproxted}\label{thm:dynapproxted}
There exists a randomized dynamic algorithm that, given a threshold $k$ and a parameter $0<\varepsilon\le 1$, maintains initially empty forests $\bF$ and $\bG$ subject to edits, and upon every update, in amortized time $\Ohtilde(k \varepsilon^{-\Oh(1)})$ outputs a $(1+\varepsilon)$-approximation of $\ted(\bF, \bG)$ or reports $\ted(\bF, \bG) > k$.
The algorithm is correct with high probability against an oblivious adversary.\footnote{Following standard practice for dynamic algorithms, we assume for \cref{thm:dynapproxted,thm:dynexactted} that an upper bound $N$ on the total number of nodes of $\bF$ and $\bG$ at all times is known in advance and that the number of updates is polynomial in $N$; the polylogarithmic factors hidden in $\Ohtilde(\cdot)$ are then in terms of $N$.}
\end{restatable*}

Any improvement in the dependence on $k$ would immediately yield a faster static $(1+\varepsilon)$-approximation of \TED, improving upon the $\Ohtilde_{\varepsilon}(nk)$-time algorithm of~\cite{BGHS19}.
More generally, faster updates even in a single polynomial low-distance regime would imply faster, albeit nonuniform, static approximation of \TED with no distance bound:
An $\Oh_{\varepsilon}(k^{0.99})$ amortized update time for $k = \Theta(N^{\kappa})$, where $N$ is the maximum total forest size and $0<\kappa \le 1$ is fixed, would give a randomized nonuniform $(1+\varepsilon)$-approximation of \TED for $m$-node forests in $\Ohtilde_{\varepsilon}(m^{1.99})$ time.\footnote{Set $k=2m\ge\ted(\bF,\bG)$ and add the same $\Theta(m^{1/\kappa})$ single-node trees with a fresh label to both forests, which preserves $\ted$ and makes $k=\Theta(N^{\kappa})$. Inserting the padding takes too long, so the advice holds polynomially many padded states of the dynamic algorithm: we insert the input into a random one and stop after $\Ohtilde_{\varepsilon}(m^{1.99})$ time. Suitable advice can be sampled (by a union bound over inputs) from a distribution of padded states from which inserting any input takes $\Ohtilde_{\varepsilon}(m^{1.99})$ expected time. It exists, as otherwise an oblivious adversary could keep inserting and deleting inputs that are slow for the current distribution, violating the amortized bound.}
Such a nonuniform bound would already constitute a major breakthrough for $(1+\varepsilon)$-approximation of both \TED and \ED.
For \ED, Mao and Rubinstein~\cite{MR26} give $\Oh_{\varepsilon}(n^2/2^{\log^c n})$ time for a constant $c>0$.\footnote{Very recently, OpenAI~\cite{OAI26ED} has claimed a randomized $(1+\varepsilon)$-approximation for \ED in expected $n^{1+o(1)}$ time for fixed $\varepsilon>0$.}

Since \cref{thm:dynapproxted} has a polynomial dependence on $\varepsilon^{-1}$, it immediately yields an exact algorithm with $\poly(k\log n)$ amortized update time\footnote{By setting $\varepsilon < 1/(2k)$, the additive approximation error becomes strictly less than $1/2$.}, which is already novel.
For weighted and unweighted \ED, such dynamic algorithms are derived by simply plugging in dynamic data structures for the so-called Longest Common Extension queries~\cite{GKKLS18,KK22} into $\Ohtilde(n+\poly(k))$-time static algorithms~\cite{LV88,CKW23,GK25}.
In contrast, their counterparts for \TED~\cite{DGHKSS22,DGHKS23,KS25} rely on highly non-trivial $\Oh(n)$-time preprocessing, which we avoid by using a dynamic program of a very different structure.

Stripped of approximation-specific techniques, the arguments behind \cref{thm:dynapproxted} yield a \emph{deterministic} exact algorithm with amortized update time $\Ohtilde(k^7)$, still far from the conditional $k^{2-o(1)}$ update-time lower bound for combinatorial algorithms when $k=\Theta(n)$, preprocessing is truly subcubic, and query time is truly subquadratic~\cite[Theorem~19]{HNS26}. We therefore view \cref{thm:dynexactted} as a warm-up for $(1+\varepsilon)$-approximation and a proof-of-concept showcasing a low-distance regime of $k \ll n^{1/7}$, in which our dynamic algorithm is faster than recomputing \TED from scratch after every update using state-of-the-art tools \cite{KS25,AJ21,NPSVXY25}.

\begin{restatable*}{mtheorem}{dynexactted}\label{thm:dynexactted}
There exists a deterministic dynamic algorithm that, given a threshold $k$, maintains initially empty forests $\bF, \bG$ subject to edits, and upon every update, in amortized time $\Ohtilde(k^{7})$ outputs $\ted(\bF, \bG)$ or reports $\ted(\bF, \bG) > k$.
\end{restatable*}

%% file: s2_overview.tex
\section{Overview}\label{sec:overview}

This section highlights the technical contributions of this work.
After introducing the basic notation (\Cref{sec:ov:prelim}), we focus on the static algorithms behind \cref{thm:approxted,thm:sublinearted} (\Cref{sec:ov:static}).
The tools and insights we develop also apply to the dynamic algorithms behind \cref{thm:dynapproxted,thm:dynexactted} (\Cref{sec:ov:dynamic}).

\subsection{Tree Edit Distance: Basic Notions}\label{sec:ov:prelim}
While tree edit distance is defined as the minimum number of node edits to transform one forest into another~\cite{Selkow77}, most algorithmic works use Tai's equivalent definition based on a matching between preserved nodes~\cite{Tai79}.
To build intuition, let us first interpret string edit distance through this lens.

\begin{definition}[String Edit Distance (\ED), String Alignment]\label{def:sed}
    Given two strings $X,Y\in \Sigma^*$, we say that a subset $\mA\sub \fragmentco{0}{\abs{X}}\times \fragmentco{0}{\abs{Y}}$ is an \emph{alignment} between $X$ and $Y$, also denoted $\mA : X\onto Y$, if every two elements $(x,y),(x',y')\in \mA$ satisfy $x<x'$ if and only if $y<y'$.
    The \emph{cost} of the alignment $\mA$ is defined as
   \[
    \cost(\mA) \coloneqq \abs{X}+\abs{Y} - 2\abs{\{(x,y)\in \mA : X\position{x}=Y\position{y}\}} - \abs{\{(x,y)\in \mA : X\position{x}\ne Y\position{y}\}}.
\]
    Finally, we denote by $\ed(X,Y)$ the minimum cost achieved by any alignment $X \onto Y$.
\end{definition}
Here, unmatched characters cost $1$, matched unequal pairs cost $1$,
and matched equal pairs cost~$0$.

The same matching view extends to rooted ordered forests with labels in
$\Sigma$. We identify a forest $\bF$ with its node set and write
$\ell:\bF\to\Sigma$ for its labeling function.
The forest alignments must be non-crossing with respect to both preorder $\preor$ and postorder $\postor$.

\begin{definition}[Tree Edit Distance (\TED), Forest Alignment]\label{def:ted}
    Given two forests $\bF$ and $\bG$, we say that a subset $\mA\sub \bF\times \bG$ is an \emph{alignment} between $\bF$ and $\bG$, also denoted $\mA : \bF\onto \bG$, if every two elements $(u,v),(u',v')\in \mA$ satisfy \begin{mylist}[label=(\arabic*)] \item\label{it:pre} $u \preor u'$ if and only if $v \preor v'$ and \item\label{it:post} $u \postor u'$ if and only if $v \postor v'$\end{mylist}.
    The \emph{cost} of the alignment $\mA$ is defined as
   \[
    \cost(\mA) \coloneqq \abs{\bF}+\abs{\bG} - 2\abs{\{(u,v)\in \mA : \ell(u)=\ell(v)\}} - \abs{\{(u,v)\in \mA : \ell(u)\ne \ell(v)\}}.
\]
    Finally, we denote by $\ted(\bF,\bG)$ the minimum cost achieved by any alignment $\bF \onto \bG$.
\end{definition}

One can equivalently replace either condition~\ref{it:pre} or
condition~\ref{it:post}, but not both, by the following ancestor-preservation
constraint:
\begin{mylist}[label=(\arabic*),start=3]
    \item $u$ is an ancestor of $u'$ if and only if $v$ is an ancestor of $v'$.
\end{mylist}

Modern \TED algorithms~\cite{Kle98,BGHS19,SS22,M22,NPSVXY25} combine the preorder and postorder traversals into a single bi-order traversal of the forest, also known as the Euler tour.

\begin{restatable}[Bi-order Traversal]{definition}{biorder}\label{def:bi-order}
The \emph{bi-order traversal} $\mS(\bF)$ of a size-$n$ forest $\bF$ is defined inductively:
\begin{itemize}
    \item The bi-order traversal of a forest $\bF$ is the concatenation of the bi-order traversals of the constituent trees $\bT_0,\ldots,\bT_{t-1}$ listed from left to right, that is, $\mS(\bF) \coloneqq \mS(\bT_0)\cdots \mS(\bT_{t-1})$.
    \item The bi-order traversal of a tree $\bT$ with root node $r$ consists of two copies of $r$ interleaved with the bi-order traversal of the forest obtained by removing $r$ from $\bT$, that is, $\mS(\bT) \coloneqq r \cdot \mS(\bT \setminus r) \cdot r$.
\end{itemize}
Each node $u\in \bF$ appears twice in $\mS(\bF)$; we denote by $\Left(u), \Right(u) \in \fragmentco{0}{2n}$ the positions of its left and right occurrences, and we introduce $\Other : \fragmentco{0}{2n}\to \fragmentco{0}{2n}$ so that $\Other(\Left(u)) = \Right(u)$ and $\Other(\Right(u)) = \Left(u)$.
\end{restatable}

\begin{figure}[t!]
    \centering
    \input{figures/pre/better_forest_merged_TO}
    \caption{Forests $\bF$ and $\bG$ with nodes $v_1,\ldots, v_9$ and $u_1, \ldots, u_8$, respectively. Their bi-order traversal strings and parenthesis representations are depicted below the forest illustrations.
    Consider an optimal edit sequence that transforms $\bF$ into $\bG$ by deleting $v_2,v_9\in \bF$ and inserting $u_4\in \bG$; the remaining nodes are matched, as indicated by blue dotted lines. The corresponding tree alignment is shown as blue lines in the parentheses notation. Parentheses of deleted or inserted nodes remain unaligned. Aligned parentheses come in pairs and do not cross.}
     \label{fig:forest_merged}
\end{figure}
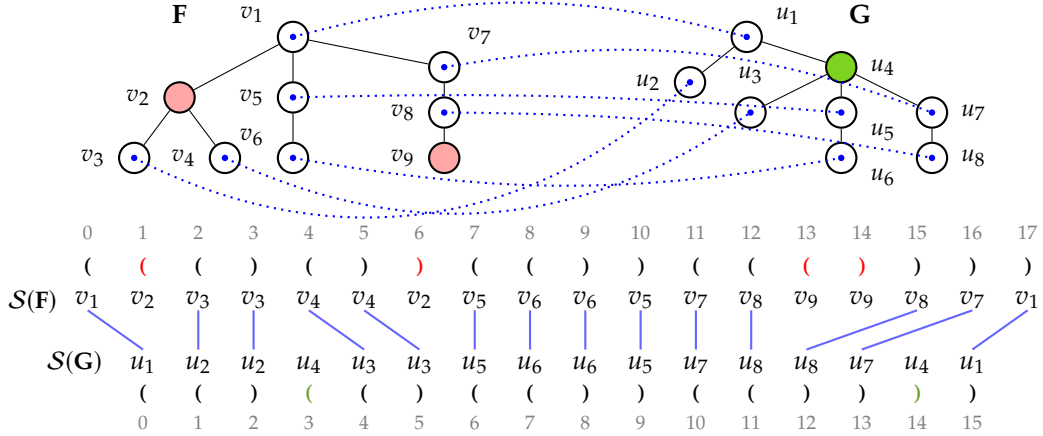

Equivalently, one can use the \emph{parentheses representation}~\cite{DGHKSS22,DGHKS23,KS25,DGHKS25}, where the two occurrences of a node $u\in \bF$ are replaced by labeled parentheses $\op{\ell(u)}$ and $\cl{\ell(u)}$.
We can obtain a string alignment $\mS(\mA):\mS(\bF)\onto \mS(\bG)$ from a tree alignment $\mA:\bF\onto \bG$ by matching $(\Left(u),\Left(v))$ and $(\Right(u),\Right(v))$ for each $(u,v)\in \mA$; see \cref{fig:forest_merged}.
For a string alignment $\mB : \mS(\bF)\onto \mS(\bG)$ to be obtainable from a tree alignment $\mA:\bF\onto \bG$, it suffices that $(\Other(x),\Other(y))\in \mB$ holds for every $(x,y)\in \mB$.
This view of tree alignments as constrained string alignments allows one to lift many \ED techniques to \TED; see~\cite{SS22,DGHKSS22,DGHKS23,DGHKS25}.

\subsection{Static Approximation Algorithms}\label{sec:ov:static}

Our static algorithm behind \cref{thm:approxted} is based on the framework developed by Andoni, Krauthgamer, and Onak~\cite{AKO10} for approximating \ED, and it achieves the same trade-off between approximation ratio $\Oh(c \log_c n)$ and running time $\Oh(nc \cdot (\log n)^{\Oh(\log_c n)})$ for a given parameter $c\in \fragment{2}{n}$.
However, our algorithm is not a simple adaptation. As we outline the approach of \cite{AKO10}, we point out the technical hurdles that need to be overcome.

Given two strings $X,Y$ of length $n$ and thresholds $k$ and $K$ such that $K/k$ is a sufficiently large multiple of $c\log_c n$, the algorithm of \cite{AKO10} distinguishes $\ed(X,Y)\le k$ from $\ed(X,Y)>K$; an $\Oh(c\log_c n)$-approximation then follows by trying geometrically increasing thresholds.
To this end, it decomposes the string
$X$ hierarchically into $\Oh(\log_c n)$ levels:
\begin{itemize}
    \item The topmost level consists of the entire interval $\fragmentco{0}{n}$.
    \item Each interval at a given level is partitioned into $c$ consecutive
    intervals of equal length, which form the next level.
    \item Each interval at the bottom level contains a single position of $X$.
\end{itemize}
This decomposition naturally induces a tree structure $T(X)$, where every node $v$ corresponds to an interval. We denote the interval represented by $v$ by $I_v$ and the substring of $X$ represented by $v$ by~$X_{I_v}$.

Andoni, Krauthgamer, and Onak~\cite{AKO10} define a quantity called \emph{Tree Distance} (not to be confused with Tree Edit Distance!) on top of the partition tree as follows.
\begin{definition}[$\TD$ for $\ED$ approximation \cite{AKO10}]
    For every $v \in T(X)$ and shift $s \in \fragment{-k}{k}$,
    \begin{enumerate}
        \item If $v$ is a leaf, then $\TD_{v,s}(X,Y)=\ed(X_{I_v},Y_{I_v+s})$.
        \item If $v$ has children $v_1,\ldots,v_c$, then
        \[
            \TD_{v,s}(X,Y) = \sum_{i=1}^c \min_{s_i\in \fragment{-k}{k}} \left(\TD_{v_i,s_i}(X,Y) + 2|s-s_i|\right)\qedhere
        \]
    \end{enumerate}
\end{definition}
Intuitively, each child $v_i$ independently chooses its best shift $s_i$ and pays the penalty of $2|s-s_i|$ for switching from the parent's shift $s$.

It turns out that $\TD_{r,0}(X,Y)$ is a good approximation of $\ed(X,Y)$, where $r$ is the root of $T(X)$. Specifically, an inductive argument implies that
\begin{equation}\label{eq:ako:approx} \ed(X,Y)\le \TD_{r,0}(X,Y) \le \ed_{\le k}(X,Y)\cdot \Oh(c\cdot \log_c n),\end{equation}
where $\ed_{\le k}(X,Y)$ is equal to $\ed(X,Y)$ when $\ed(X,Y)\le k$ and set to $\infty$ otherwise.

Even though computing $\TD_{r,0}(X,Y)$ would take $\Oh(nck)$ time, the extra $k$ factor can be removed if we only need to approximate the $\TD_{v,s}(X,Y)$ values. For this, every node $v$ is assigned two precisions:
\begin{itemize}
    \item A multiplicative accuracy $\alpha_v \geq 1$, which is 10 at the root and decays geometrically down the tree.
    \item An additive accuracy $\beta_v > 0$, which is $0.001K$ at the root and, for the children of each node $v$, chosen via a randomized process called \emph{precision sampling}, also introduced in \cite{AKO10}.
\end{itemize}
Then, the algorithm in \cite{AKO10} approximates $\TD_{v,s}(X,Y)$ by computing a value $\Delta_{v,s}(X,Y)$ such that
 \[
 \tfrac{1}{\alpha_v} \TD_{v,s}(X,Y) - \beta_v \leq \Delta_{v,s} \leq \alpha_v \TD_{v,s}(X,Y) + \beta_v.
 \]

Remarkably, when combining approximations $\tilde{\Delta}_{v_i,s}(X,Y)\coloneqq \min_{s_i\in \fragment{-k}{k}}(\Delta_{v_i,s_i}(X,Y) + 2|s-s_i|)$ with additive errors $\beta_{v_i}$ into an approximation $\Delta_{v,s}(X,Y)$, precision sampling guarantees that the additive error does not blow up: $\beta_v \ll \sum_{i=1}^c \beta_{v_i}$.

How does precision sampling help with the running time? If a node is ``short'', i.e., $|I_v|\le \beta_v$, then we can simply return $\Delta_{v,s}(X,Y) = 0$ for all $s$ without expanding its children. The expected number of nodes with $|I_v|>\beta_v$ is $\frac{n}{k}\cdot (\log n)^{\Oh(\log_c n)}$, by the properties of the precisions. Therefore, spending $\Ohtilde(kc)$ time on each such node gives total expected running time $nc\cdot (\log n)^{\Oh(\log_c n)}$. This concludes the description of \cite{AKO10}.

To use an approach in the spirit of \cite{AKO10}, we need to answer the following immediate questions:
\begin{mylist}
    \item What does a ``fragment'' of a tree look like?
    \item How do we decompose the input tree?
\end{mylist}
We describe our constructions in \Cref{sec:ov:subforests} and \Cref{sec:ov:decomp}, respectively.

\subsubsection{Generalizing Tree Edit Distance to Subforests}\label{sec:ov:subforests}
Almost all algorithms for computing or approximating \TED, including~\cite{Kle98,BGHS19,AJ21,M22,DGHKS23,NPSVXY25,KS25}, rely on dynamic programming whose states represent \emph{subforests} $\bF_{I}$ and $\bG_J$ defined by intervals $I=\fragmentco{i}{i'}$ and $J=\fragmentco{j}{j'}$ of positions in the bi-order traversals $\mS(\bF)$ and $\mS(\bG)$, respectively.
Each such subforest $\bF_{I}$ represents all nodes $u\in \bF$ both of whose parentheses $\Left(u)$ and $\Right(u)$ belong to $I$.
A subforest $\bF_{I}$ is called \emph{imbalanced} if there exists a node $u\in \bF$ that is \emph{partially contained} in $\bF_I$, that is, exactly one of $\Left(u)$ and $\Right(u)$ lies in $I$.
When considering $\ted(\bF_I, \bG_J)$, it is always assumed that only fully contained nodes can participate in the alignment.
Partially contained nodes must be deleted or inserted at the cost of $\frac{1}{2}$ per node~\cite{SS22,DGHKS23,KS25,DGHKS25}, so that the operation cost is distributed equally among the two parentheses affected.

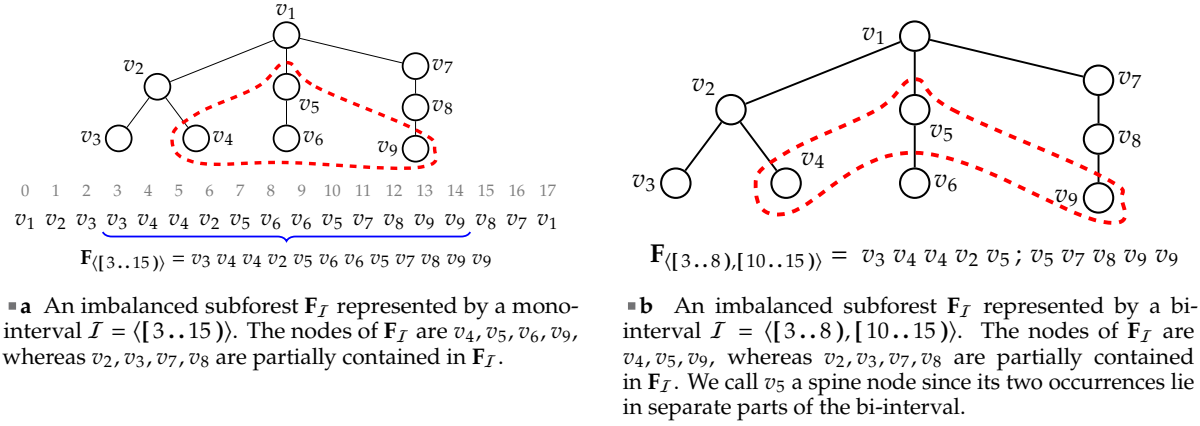
\begin{figure*}[t!]
    \centering
    \begin{subfigure}[t]{0.48\textwidth}
  \centering
  \resizebox{\linewidth}{!}{\input{figures/pre/subforest_mono}}
  \caption{An imbalanced subforest $\bF_{\mI}$ represented by a mono-interval $\mI=\langle \fragmentco{3}{15} \rangle$. The nodes of $\bF_{\mI}$ are $v_4,v_5,v_6,v_9$, whereas $v_2, v_3,v_7, v_8$ are partially contained in $\bF_{\mI}$.}
  \label{fig:subforest_to:a}
\end{subfigure}\hfill%
\begin{subfigure}[t]{0.48\textwidth}
  \centering
  \resizebox{\linewidth}{!}{\input{figures/pre/subforest_bi}}
  \caption{An imbalanced subforest $\bF_{\mI}$ represented by a bi-interval $\mI=\langle \fragmentco{3}{8}, \fragmentco{10}{15} \rangle$. The nodes of $\bF_{\mI}$ are $v_4,v_5,v_9$, whereas $v_2, v_3,v_7,v_8$ are partially contained in~$\bF_{\mI}$.
  We call $v_5$ a spine node since its two occurrences lie in separate parts of the bi-interval.}
  \label{fig:subforest_to:b}
\end{subfigure}

    \caption{Two imbalanced subforests. Partially contained nodes cannot be aligned in subforest alignments.}
    \label{fig:subforest_to}
\end{figure*}

Twinned parentheses can be far apart in the bi-order traversal string. Therefore, we also consider subforests $\bF_{\langle I_{\Left},I_{\Right}\rangle}$ defined by \emph{bi-intervals} $\langle I_{\Left},I_{\Right}\rangle$, that is, pairs of intervals,\footnote{For simplicity, this overview assumes that the two intervals of every bi-interval are disjoint and ordered from left to right, as the alignment definition below requires. In general, the two intervals on the target side ($\bG$) may overlap; see \cref{def:TED_on_biintervals}.} with fully and partially contained nodes defined in the natural way. As a compact notation, we denote a generic subforest by $\bF_{\mI}$, where $\mI$ can be a \emph{mono-interval} $\langle I \rangle$ or a \emph{bi-interval} $\langle I_{\Left},I_{\Right}\rangle$; see \cref{fig:subforest_to} for an illustration of the two types of subforests.

We emphasize that this notion generalizes both \emph{super-windows}
of~\cite{SS22} and \emph{contexts} of~\cite{DGHKSS22,DGHKS23,KS25,DGHKS25}, both of which are balanced by definition. See \cref{sec:comp_appendix} for an example explaining why the adaptation of \cite{AKO10} requires using imbalanced subforests $\bG_{\langle J_\Left, J_\Right \rangle}$ of $\bG$.

Still, for bi-intervals $\mI$ and $\mJ$, we can naturally define $\ted(\bF_\mI,\bG_\mJ)$ as follows (and analogously for mono-intervals by restricting $\mA$):

\begin{definition}[Tree Edit Distance (\TED), Alignment between Subforests]
    Given two forests $\bF,\bG$ and two bi-intervals $\mI = \langle I_1, I_2 \rangle, \mJ = \langle J_1, J_2 \rangle$, we say that a subset $\mA \sub (I_1 \times J_1) \cup (I_2 \times J_2)$ is an \emph{alignment} between $\bF_{\mI}$ and $\bG_{\mJ}$, also denoted $\mA : \bF_{\mI} \onto \bG_{\mJ}$, if every two elements $(x,y),(x',y')\in \mA$ satisfy \begin{mylist}[label=(\arabic*)] \item $x < x'$ if and only if $y < y'$ and \item $(x,y) \in \mA$ if and only if $(\Other(x), \Other(y)) \in \mA$\end{mylist}.
    The \emph{cost} of the alignment $\mA$ is defined as
    \begin{align*}
        \cost(\mA) \coloneqq
        & + \tfrac{1}{2}(|I_1|+|I_2| + |J_1|+|J_2|)
        -\abs{\{(x,y)\in \mA : \ell(\mS(\bF)\position{x})=\ell(\mS(\bG)\position{y})\}} \\
        &- \tfrac{1}{2}\abs{\{(x,y)\in \mA : \ell(\mS(\bF)\position{x}) \neq \ell(\mS(\bG)\position{y})\}}.
    \end{align*}
    Finally, we denote by $\ted(\bF_{\mI},\bG_{\mJ})$ the minimum cost achieved by any alignment $\bF_{\mI} \onto \bG_{\mJ}$.
\end{definition}
Intuitively, the left and right intervals would align independently were it not for the \emph{spine nodes}, whose two occurrences lie on different sides.

\subsubsection{Robust Hierarchical Decomposition}\label{sec:ov:decomp}
Our next challenge is to devise a hierarchical decomposition of a forest $\bF$ into subforests.
In $\ed(X,Y)$, for any decomposition of a string $X=X_1\cdot X_2$, one can compute $\ed(X,Y)$ as the minimum of $\ed(X_1,Y_1)+\ed(X_2,Y_2)$ across all decompositions $Y=Y_1\cdot Y_2$; in particular, $\ed(X_1,Y_1)$ and $\ed(X_2,Y_2)$ can be optimized independently.
For forests $\bF$ and $\bG$, this is true only if the decomposition of $\bF$ into two subforests $\bF_{\mI^1}$ and $\bF_{\mI^2}$ is such that no node $u\in \bF$ is partially contained in both $\bF_{\mI^1}$ and $\bF_{\mI^2}$; otherwise, we could not capture $u$ being aligned to some node $v\in \bG$.
Hence, we only decompose $\bF$ into balanced subforests.

Our forest decomposition has the following properties, comparable to those of the string decomposition in \cite{AKO10}:
\begin{enumerate}[(1)]
    \item\label{pr:depth} The hierarchical decomposition has $d = \Oh(\log_c n)$ levels.
    \item\label{pr:deg} For $t\in \fragmentoo{0}{d}$, each level-$t$ subforest is decomposed into $\Oh(c)$ level-$(t-1)$ subforests.
    \item\label{pr:size} Each level $t\in \fragmentco{0}{d}$ consists of $\Oh(n/c^t)$ subforests of size $\Oh(c^t)$ each.
\end{enumerate}

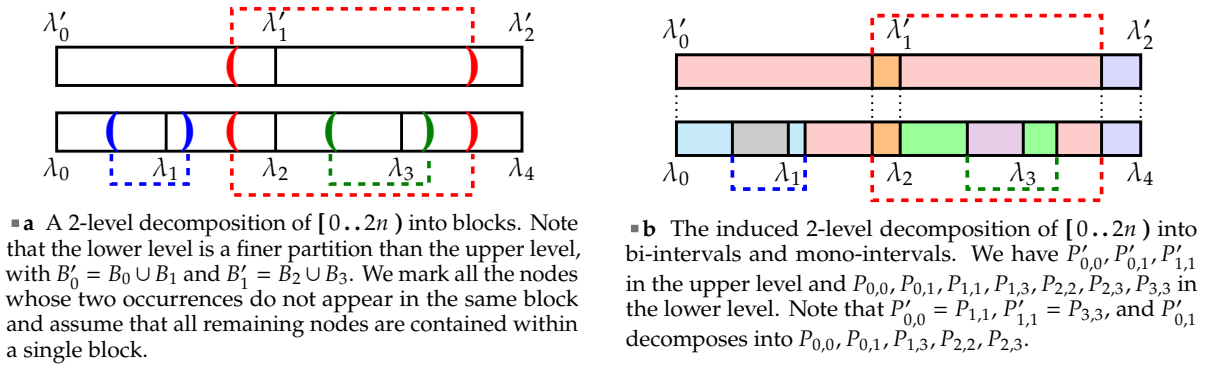
\begin{figure*}[t!]
    \centering

    \begin{subfigure}[t]{0.48\textwidth}
  \centering
  \resizebox{0.9\linewidth}{!}{\input{figures/decomp/block}}
  \caption{A 2-level decomposition of $\fragmentco{0}{2n}$ into blocks. Note that the lower level is a finer partition than the upper level, with $B'_0 = B_0\cup B_1$ and $B'_1 = B_2\cup B_3$.
   We mark all the nodes whose two occurrences do not appear in the same block and assume that all remaining nodes are contained within a single block.}
  \label{fig:decomp_to:a}
\end{subfigure}
\hfill
\begin{subfigure}[t]{0.48\textwidth}
  \centering
    \raisebox{-0.2em}{\resizebox{0.9\linewidth}{!}{\input{figures/decomp/partition}}}

  \caption{The induced 2-level decomposition of $\fragmentco{0}{2n}$ into bi-intervals and mono-intervals.
   We have $P_{0,0}', P_{0,1}', P_{1,1}'$ in the upper level and $P_{0,0}, P_{0,1}, P_{1,1}, P_{1,3}, P_{2,2}, P_{2,3}, P_{3,3}$ in the lower level. Note that $P_{0,0}' = P_{1,1}$, $P_{1,1}'=P_{3,3}$, and $P_{0,1}'$ decomposes into $P_{0,0}, P_{0,1}, P_{1,3}, P_{2,2}, P_{2,3}$.}
  \label{fig:decomp_to:b}
\end{subfigure}

    \caption{A 2-level decomposition of an $n$-node forest $\bF$ derived from a 2-level decomposition of $\fragmentco{0}{2n}$.}
    \label{fig:decomp_to}
\end{figure*}

To define the $t$-th level of our decomposition, consider the $t$-th level of the underlying decomposition of $\fragmentco{0}{2n}$, which partitions $\fragmentco{0}{2n}$ into $b$ blocks $B_0,\ldots,B_{b-1}$.
We define our decomposition by assigning each node $u \in \bF$ to one of the level-$t$ subforests $P_{i,j}$ indexed by integers $0\le i \le j < b$:
\begin{itemize}
    \item If $\Left(u)$ and $\Right(u)$ belong to distinct blocks $B_i$ and $B_j$, we assign $u$ to the subforest $P_{i,j}$.
    \item Otherwise, $\Left(u)$ and $\Right(u)$ belong to a single block $B_i$, and the same must be true for all descendants of $u$.
    In this case, we take the smallest position $x'\in B_i$ such that $x'\ge \Right(u)$ and $\Other(x')\notin B_i$.
    If such a position exists, the underlying node $u'$ has already been assigned by the previous rule (to $P_{i,j}$ or $P_{j,i}$ for some $j\in \fragmentco{0}{b}\setminus \{i\}$), and we assign $u$ to the same subforest.
    If $x'$ does not exist, we assign $u$ to $P_{i,i}$.
\end{itemize}
This rule guarantees that the non-empty subforests are defined by mono-intervals or bi-intervals: $P_{i,i}=\bF_{\langle I\rangle}$ for some suffix $I$ of $B_i$ and, if $i\ne j$, then $P_{i,j}=\bF_{\langle I, J\rangle}$ for some $I\sub B_i$ and $J\sub B_j$.
Moreover, in the latter case, $P_{i,j}$ has at least one \emph{spine node} for which $\Left(u)\in I$ and $\Right(u)\in J$; see \cref{fig:decomp_to} for a simple example.

\cref{sec:decomp} proves that this decomposition satisfies properties \ref{pr:depth}, \ref{pr:deg}, and \ref{pr:size}.

\subparagraph*{Extensions to the sublinear and dynamic settings.}
The decomposition also satisfies two additional useful properties. In particular, our sublinear-time algorithm behind \cref{thm:sublinearted} relies on the following property:
\begin{enumerate}[(1), resume]
    \item\label{pr:sub} For $t\in \fragmentco{0}{d}$, a uniformly random level-$t$ subforest can be sampled in $\Ohtilde(c^t)$ time w.h.p.
\end{enumerate}

Our dynamic algorithms instead generalize the construction by taking the underlying decomposition of $\fragmentco{0}{2n}$ from weight-balanced B-trees~\cite{AV96}, which, crucially for our reduction, evolve only through sibling merges and splits.
This yields the following property:
\begin{enumerate}[(1), resume]
     \item\label{pr:dyn} A node update (insertion or deletion) to $\bF$ can be implemented by modifying $\Oh(d)$ subforests from the decomposition, in the amortized sense.
\end{enumerate}

This property lets us design a dynamic-programming scheme that recovers a $(1+\varepsilon)$-approximation with the same running time as~\cite{BGHS19} while changing only a few states after an update.
Overall, as discussed at the beginning of \cref{sec:decomp}, we are not aware of any previous decomposition satisfying properties \ref{pr:depth}, \ref{pr:deg}, and \ref{pr:size}, let alone the extensions in \ref{pr:sub} and \ref{pr:dyn}.

\subsubsection{\boldmath A Static Algorithm in $\hat{\Oh}(nk)$ Time.}\label{sec:ov:slow-static}

With our definitions in place, we can finally generalize the Tree Distance framework. For a node $v$ in our hierarchical decomposition (associated with a mono- or bi-interval $\mI_v$) and shift $s \in \fragment{-2k}{2k}^{\ell_v}$, where $\ell_v \in \{1,2\}$ based on the interval type, we define:
\begin{enumerate}
    \item If $v$ is a leaf: $\TD_{v,s}(\bF,\bG)=\ted(\bF_{\mI_v},\bG_{\mI_v+s})$. (We note that in this base case, all $\mI_v$ are bi-intervals of the form $\langle \{\Left(u)\}, \{\Right(u)\}\rangle$ for some $u \in \bF$; thus, the shift $s = \langle s_1, s_2 \rangle$ consists of two components that act independently to shift $\mI_v$ to $\mI_v + s = \langle \{\Left(u)+s_1\}, \{\Right(u)+s_2\}\rangle$.)
    \item If $v$ has children $v_1,\ldots,v_c$:
    \begin{equation}\label{eq:ted_gen}
    \TD_{v,s}(\bF,\bG) = \sum_{i=1}^c \min_{s_i\in \fragment{-2k}{2k}^{\ell_{v_i}}} \left(\TD_{v_i,s_i}(\bF,\bG) + \mathsf{penalty}(s, s_i) \right),
    \end{equation}
    where the term $\mathsf{penalty}(s,s_i)$ generalizes the previous penalty, penalizing any interval in $\mI_{v_i}$ based on its relative shift from the containing interval in $\mI_v$.
\end{enumerate}

While our definitions preserve the required approximation guarantees, we still need to check whether the running-time guarantees are preserved. As before, precision sampling lets us evaluate \cref{eq:ted_gen} approximately on only $\frac{n}{k}\cdot (\log n)^{\Oh(\log_c n)}$ nodes instead of all $\Oh(n)$ nodes. Unfortunately, the framework breaks down here: the work per node increases from $\Oh(ck)$ to $\Oh(ck^2)$ because we must consider pairs of shifts. This extra $\Theta(k)$ factor causes our total running time to exceed the $\Ohtilde(nk)$ bound of the known $(1+\varepsilon)$-approximation for $\ted$~\cite{BGHS19}.

\subsubsection{State Sparsification}\label{sec:ov:sparse}

To recover the same running time as for strings, we must answer the following question: when $\mI_v=\langle I_{\Left}, I_{\Right}\rangle$ is a bi-interval ($\ell_v=2$), can we compute the approximation $\TD_{v,s}(\bF,\bG)$ using only a carefully selected subset of $\Ohtilde(k)$ shifts $S_v \subseteq \fragment{-2k}{2k}^{2}$?

Fix an optimal alignment $\mA : \bF \onto \bG$ of cost at most $k$.
A shift $\langle s_1, s_2\rangle$ is \emph{induced by} $\mA$ if $(i_1,i_1+s_1),\allowbreak(i_2,i_2+s_2)\in\mA$ for some $i_1\in I_{\Left}$ and $i_2\in I_{\Right}$.
The analog of \eqref{eq:ako:approx} still holds if every $S_v$ contains such a shift.
Every \emph{spine node} $u$ of $\bF_{\mI_v}$ aligned by $\mA$ induces such a shift (take $i_1=\Left(u)$ and $i_2=\Right(u)$).
Let $m_v$ be the number of spine nodes of $\bF_{\mI_v}$, and fix a threshold $\theta\in(0,\frac12]$ so that $1/\theta$ is polylogarithmic in $n$.
\begin{description}
    \item[Case 1: $\mA$ aligns at least $\theta m_v$ spine nodes of $\bF_{\mI_v}$.]
    Then, with high probability, polylogarithmically many ($\Oh(\theta^{-1}\log n)$) uniformly random spine nodes include an aligned one.
    Since $\mA$ has cost at most $k$, an aligned $u$ has only $\Oh(k)$ candidate partners $w\in\bG$, namely those with $|\Left(w)-\Left(u)|,|\Right(w)-\Right(u)|\le 2k$.
    Adding $\langle \Left(w)-\Left(u), \Right(w)-\Right(u)\rangle$ for all sampled $u$ and candidates $w$ yields $|S_v|=\Ohtilde(k)$ shifts, including one induced by $\mA$.

    \item[Case 2: $\mA$ aligns fewer than $\theta m_v$ spine nodes of $\bF_{\mI_v}$.]
    We decouple $v$: $\langle I_{\Left}\rangle$ and $\langle I_{\Right}\rangle$ become independent mono-intervals (each needing only $\Oh(k)$ shifts), and all spine nodes are deleted.
    Removing the fewer than $\theta m_v$ aligned spine pairs from $\mA$ costs less than $2\theta m_v$, whereas $\mA$ already pays more than $(1-\theta)m_v\ge m_v/2$ for deleting the remaining spine nodes; so the extra cost is at most $4\theta$ times the latter.
    Each spine node is charged at most once per level, so the extra cost is at most $4\theta\cdot\ted(\bF,\bG)$ per level.
    We choose $\theta$ to fit the sum over all levels into the approximation budget; here, $\theta=\frac12$ suffices for the $\Oh(c\log_c n)$ ratio.
\end{description}

\subparagraph*{Extension to the dynamic setting.} We mention here that this sparsification trick also enables $\Ohtilde_{\varepsilon}(k)$ rather than $\Ohtilde_{\varepsilon}(k^2)$ update time in our dynamic approximation algorithm. However, implementing it dynamically introduces two major challenges:
\begin{enumerate}[(a)]
    \item To achieve a $(1+\varepsilon)$-approximation, we must take $\theta$ of order $\varepsilon$ divided by the number of levels and consider finer $(1+\varepsilon)$-power length differences rather than just equal-sized bi-intervals like $\bG_{\mI + s}$. This is complicated because sampled spine nodes split each interval, creating up to four length differences. Furthermore, we lack the time to dynamically iterate through all $\Ohtilde(k)$ candidate matches for each sampled spine node, so we use nearest-neighbor search data structures.

    \item The samples and data structures must be maintained dynamically. For the specific nearest-neighbor search, we need to support a dynamic 2D point set under $(1+\varepsilon)$-approximate $L_1$ nearest-neighbor queries, for which we design new dynamic data structures.
\end{enumerate}

\subsubsection{Static Approximation in Almost-Linear Time}\label{sec:ov:ako}

Altogether, the sparsification from \cref{sec:ov:sparse}, along with some implementation details that we briefly sketch next, finally allows us to obtain the near-linear-time approximation algorithm.

Since our algorithm does not have access to the optimal alignment, it considers both cases for each node $v$. Thus, we approximate both the coupled tree-distance values $\TD_{v,s}(\bF,\bG)$ (computed among a smaller selection of $\Ohtilde(k)$ shifts) and their decoupled counterparts (with $\Oh(k)$ single shifts for the left and right parts of $\mI_v$).
The decoupled tree-distance values are defined using a formula analogous to \eqref{eq:ted_gen} in terms of both the coupled and decoupled tree-distance values for the children of $v$.

While the sparsified shift selection $S_v$ reduces the number of DP states, extra care is needed to implement the DP transitions efficiently.
For this, we design data structures that efficiently support queries on a set of weighted points: given a rectangle on the grid, report the contained point with the maximum weight (see \cite{ASR00,C88}). Moreover, since our tree-distance definition is now also defined on the left and right parts of bi-intervals, we carefully assign the precisions at each node so that the precision-sampling scheme still goes through.

\subsubsection{Static Approximation in Sublinear Time}

As a relatively straightforward consequence of the results of \cref{sec:decomp,sec:near-linear}, in \cref{sec:sublinear} we show how to improve the running time at the price of an increased approximation ratio.
The starting point of our result is a $2^{\Oh(\sqrt{\log n \log \log n})}$-factor approximation in $n\cdot 2^{\Oh(\sqrt{\log n \log \log n})}$ time.
For any parameter $\tau \ge 1$, we can decrease the running time by a factor of $\tau$ at the cost of increasing the approximation ratio by a factor of $\Ohtilde(\tau)$.
Technically, the quantity that we approximate multiplicatively is $1+\ted(\bF,\bG)$, and this is needed to overcome the $\Omega(n)$ lower bound for the time needed to test if $\bF=\bG$, that is, if $\ted(\bF,\bG)=0$.

We build upon a result for string edit distance by Goldenberg, Kociumaka, Krauthgamer, and Saha~\cite{GKKS22}.
Their reduction assumes access to an oracle that (for fixed parameters $k$ and $K$) distinguishes between $\ed(X,Y)\le k$ and $\ed(X,Y)>K$ in $n\cdot t(n)$ time for non-decreasing~$t$.
For any $K'\ge K$, it distinguishes between $\ed(X,Y)\le k$ and $\ed(X,Y)>K'$ in $\Oh(K/K'\cdot n\cdot t(n)\cdot \log^2 n)$ time, correctly w.h.p.

The algorithm of \cite{GKKS22} is very simple: decompose $\fragmentco{0}{n}$ into dyadic intervals, and, for each level $t$ of the decomposition, sample $\Theta(K/K'\cdot n\log n\,/\,2^t)$ uniformly random dyadic intervals of level $t$.
For each sampled interval $I$, the algorithm uses the oracle to distinguish $\ed(X_I,Y_I)\le k$ from $\ed(X_I,Y_I)>K$, and returns \yes if and only if all oracle calls returned \yes.
The monotonicity property implies that if $\ed(X,Y)\le k$, then $\ed(X_I,Y_I)\le k$ holds for every interval $I$ and thus the reduction returns \yes.
Conversely, a more insightful argument shows that, if $\ed(X,Y)>K'$, then there are $\Omega(K'/K)$ dyadic intervals $I$ for which $\ed(X_I,Y_I)>K$.
With high probability, the algorithm samples one and returns \no.

Thanks to our hierarchical forest decomposition, the same reduction can be used to distinguish $\ted(\bF,\bG)\le k$ from $\ted(\bF,\bG)>K'$.
Here, we rely on the ability to sample a level-$t$ subforest uniformly at random in $\Ohtilde(2^t)$ time w.h.p., without any preprocessing (just access to $\Other(x)$ and $\ell(\mS(\bF)\position{x})$ for any $x\in \fragmentco{0}{2\abs{\bF}}$).
Having sampled a uniformly random level-$t$ subforest $\bF_{\mI_v}$ of $\bF$, we generate the corresponding subforest $\bG_{\mI_v}$ of $\bG$ and use our almost-linear-time algorithm to distinguish $\ted(\bF_{\mI_v},\bG_{\mI_v})\le k$ from $\ted(\bF_{\mI_v},\bG_{\mI_v})>K$.
A caveat is that \cref{thm:approxted} supports complete forests only, so we need to observe that the underlying algorithm remains valid for two subforests, one of which is balanced.\footnote{Every such instance of the tree edit distance problem can be reduced to an instance with two complete forests, but doing so in a distance-preserving manner (that is, without any additive offset) is more difficult.}

%% file: figures/pre/better_forest_merged_TO.tex
\begin{tikzpicture}[
    treenode/.style={circle, draw=black, thick, minimum size=4mm, inner sep=0pt},
    rednode/.style={treenode, fill=red!35},
    greennode/.style={treenode, fill={rgb, 255:red, 126; green, 211; blue, 33}},
    every label/.style={font=\small, label distance=0.5mm},
    level 1/.style={sibling distance=15mm, level distance=8mm},
    level 2/.style={sibling distance=12mm, level distance=8mm},
    level 3/.style={sibling distance=12mm, level distance=8mm},
    corr/.style={draw=blue, thick, dotted}, 
    alignline/.style={draw=blue!60, thick}
]

\node[treenode, label=165:$v_1$] (v1) at (0,0) {}
    child { node[rednode, label=180:$v_2$] (v2) {}
        child { node[treenode, label=180:$v_3$] (v3) {} }
        child { node[treenode, label=180:$v_4$] (v4) {} }
    }
    child[sibling distance=20mm] { node[treenode, label=180:$v_5$] (v5) {}
        child { node[treenode, label=170:$v_6$] (v6) {} }
    }
    child[level distance=4mm, sibling distance=20mm] { node[treenode, label=45:$v_7$] (v7) {}
        child[level distance=6mm] { node[treenode, label=180:$v_8$] (v8) {}
            child[level distance=6mm] { node[rednode, label=180:$v_9$] (v9) {} }
        }
    };
\node[font=\bfseries] at (-1.5, 0.3) {F};

\begin{scope}[xshift=6cm]
    \node[treenode, label=10:$u_1$] (u1) at (0,0) {}
        child[level distance=6mm] { node[treenode, label=180:$u_2$] (u2) {} }
        child[level distance=4mm, sibling distance=25mm] { node[greennode, label=0:$u_4$] (u4) {}
            child[level distance=6mm] { node[treenode, label=90:$u_3$] (u3) {} }
            child[level distance=6mm] { node[treenode, label=-5:$u_5$] (u5) {}
                child[level distance=6mm] { node[treenode, label=-5:$u_6$] (u6) {} }
            }
            child[level distance=6mm] { node[treenode, label=0:$u_7$] (u7) {}
                child[level distance=6mm] { node[treenode, label=0:$u_8$] (u8) {} }
            }
        };
    \node[font=\bfseries] at (1.5, 0.3) {G};
\end{scope}

\draw[corr] (v1.center) to[bend left=15] (u1.center);
\draw[corr] (v3.center) to[bend right=35] (u2.center);
\draw[corr] (v4.center) to[bend right=30] (u3.center);
\draw[corr] (v5.center) to[bend left=3] (u5.center);
\draw[corr] (v7.center) to[bend left=15] (u7.center);
\draw[corr] (v6.center) to[bend right=10] (u6.center);
\draw[corr] (v8.center) to[bend left=5] (u8.center);

\foreach \n in {v1, v3, v4, v5, v6, v7, v8, u1, u2, u3, u5, u6, u7, u8} {
    \fill[blue] (\n.center) circle (1.2pt);
}

\begin{scope}[yshift=-3.0cm, xshift=3.5cm]
    \node[left, font=\small] at (-6.5, -0.5) {$\mS(\bF)$};
    \matrix (PF) [
        matrix of math nodes, ampersand replacement=\&,
        nodes={inner sep=1pt, text height=1.5ex, text depth=0.25ex, minimum width=4.5mm},
        column sep=8pt, row sep=2pt,
        row 1/.style={font=\scriptsize\color{gray}},
        row 2/.style={font=\scriptsize\boldmath}
    ] at (0,0) {
        0 \& 1 \& 2\& 3 \& 4 \& 5 \& 6 \& 7 \& 8 \& 9 \& 10 \& 11 \& 12 \& 13 \& 14 \& 15 \& 16 \& 17 \\
        ( \& \color{red}( \& ( \& ) \& ( \& ) \& \color{red}) \& ( \& ( \& ) \& ) \& ( \& ( \& \color{red}( \& \color{red}) \& ) \& ) \& ) \\
        v_1 \& v_2 \& v_3 \& v_3 \& v_4 \& v_4 \& v_2 \& v_5 \& v_6 \& v_6 \& v_5 \& v_7 \& v_8 \& v_9 \& v_9 \& v_8 \& v_7 \& v_1 \\
    };

    \node[left, font=\small] at (-5.9, -1.3) {$\mS(\bG)$};
    \matrix (PG) [
        matrix of math nodes, ampersand replacement=\&,
        nodes={inner sep=1pt, text height=1.3ex, text depth=0.25ex, minimum width=4.5mm},
        column sep=8pt, row sep=2pt,
        row 2/.style={font=\scriptsize\boldmath},
        row 3/.style={font=\scriptsize\color{gray}},
    ] at (0,-1.7) {
        u_1 \& u_2 \& u_2 \& u_4 \& u_3 \& u_3 \& u_5 \& u_6 \& u_6 \& u_5 \& u_7 \& u_8 \& u_8 \& u_7 \& u_4 \& u_1 \\
        ( \& ( \& ) \& \color{olive!50!green}( \& ( \& ) \& ( \& ( \& ) \& ) \& ( \& ( \& ) \& ) \& \color{olive!50!green}) \& ) \\
        0 \& 1 \& 2\& 3 \& 4 \& 5 \& 6 \& 7 \& 8 \& 9 \& 10 \& 11 \& 12 \& 13 \& 14 \& 15\\
    };

    \foreach \f/\g in {1/1, 3/2, 4/3, 5/5, 6/6, 8/7, 9/8, 10/9, 11/10, 12/11, 13/12, 16/13, 17/14, 18/16} {
        \draw[alignline] (PF-3-\f.south) -- (PG-1-\g.north);
    }
\end{scope}

\end{tikzpicture}

%% file: figures/pre/subforest_mono.tex
\begin{tikzpicture}[
  treenode/.style={circle, draw=black, thick, minimum size=4mm, inner sep=0pt},
  every label/.style={draw=none, rectangle, inner sep=1pt},
  level 1/.style={sibling distance=25mm, level distance=10mm},
  level 2/.style={sibling distance=15mm, level distance=10mm},
  scale=0.8
]

  \node[treenode, label=above:$v_1$] (v1) {}
    child { node[treenode, label=above left:$v_2$] (v2) {}
      child { node[treenode, label=left:$v_3$] (v3) {} }
      child { node[treenode, label=right:$v_4$] (v4) {} }
    }
    child { node[treenode, label=below right:$v_5$] (v5) {}
      child { node[treenode, label=right:$v_6$] (v6) {} }
    }
    child[level distance=6mm] { node[treenode, label=right:$v_7$] (v7) {}
      child[level distance=8mm] { node[treenode, label=right:$v_8$] (v8) {}
        child[level distance=8mm] { node[treenode, label=left:$v_9$] (v9) {} }
      }
    };

  \draw[red, dashed, line width=1.5pt] plot [smooth cycle, tension=0.7] coordinates {
    ([yshift=2mm]v5.north)              
    ([xshift=3mm]v5.east)                         
    ([yshift=1mm]v9.north)  
    ([xshift=1mm]v9.east)  
    ([yshift=-1mm]v9.south)             
    ([yshift=-2mm]v6.south)             
    ([yshift=-2mm]v4.south)             
    ([xshift=-2mm]v4.west)              
    ([yshift=2mm]v4.north) 
    ([xshift=-3mm]v5.west)  
  };
  
  \matrix (SeqL) [
    matrix of math nodes,
    nodes={inner sep=1.5pt, minimum width=3.5mm, text height=1.5ex, text depth=0.25ex},
    column sep=1pt, row sep=2pt,
    row 1/.style={font=\scriptsize\color{gray}} 
  ] at ([yshift=-1.0cm]v6.south) {
    0 & 1 & 2 & 3 & 4 & 5 & 6 & 7 & 8 & 9 & 10 & 11 & 12 & 13 & 14 & 15 & 16 & 17 \\
    v_1 & v_2 & v_3 & v_3 & v_4 & v_4 & v_2 & v_5 & v_6 & v_6 & v_5 & v_7 & v_8 & v_9 & v_9 & v_8 & v_7 & v_1 \\
  };   
  \draw[thick, blue, decorate, decoration={brace, amplitude=4pt, mirror}] 
    (SeqL-2-4.south west) -- (SeqL-2-15.south east) 
    node[midway, below, yshift=-4pt, font=\small, text=black] {$\bF_{\langle \fragmentco{3}{15}\rangle} = v_{3} \ v_{4} \ v_{4} \ v_{2} \ v_{5} \ v_{6} \ v_{6} \ v_{5} \ v_{7} \ v_{8} \ v_{9} \ v_{9}$};

\end{tikzpicture}

%% file: figures/pre/subforest_bi.tex
\tikzset{every picture/.style={line width=0.75pt}} 

\begin{tikzpicture}[
  treenode/.style={circle, draw=black, thick, minimum size=4mm, inner sep=0pt},
  every label/.style={draw=none, rectangle, inner sep=1pt},
  level 1/.style={sibling distance=25mm, level distance=10mm},
  level 2/.style={sibling distance=15mm, level distance=10mm}
]

  \node[treenode, label={[label distance=1.2mm]left:$v_1$}] (v1) {}
    child { node[treenode, label=above left:$v_2$] (v2) {}
      child { node[treenode, label=left:$v_3$] (v3) {} }
      child { node[treenode, label=above right:$v_4$] (v4) {} }
    }
    child { node[treenode, label=below right:$v_5$] (v5) {}
      child { node[treenode, label=right:$v_6$] (v6) {} }
    }
    child[level distance=6mm] { node[treenode, label=right:$v_7$] (v7) {}
      child[level distance=8mm] { node[treenode, label=right:$v_8$] (v8) {}
        child[level distance=8mm] { node[treenode, label=left:$v_9$] (v9) {} }
      }
    };

  \draw[red, dashed, line width=1.5pt] plot [smooth cycle, tension=0.7] coordinates {
    ([yshift=2mm]v5.north)              
    ([xshift=3mm]v5.east)                         
    ([yshift=1mm]v9.north)  
    ([xshift=1mm]v9.east)  
    ([yshift=-1mm]v9.south)             
    ([yshift=2mm]v6.north)             
    ([yshift=-2mm]v4.south)             
    ([xshift=-2mm]v4.west)              
    ([yshift=2mm]v4.north) 
    ([xshift=-3mm]v5.west)  
  };

  \node[yshift=-0.8cm] at (v6.south) 
    {$\bF_{\langle \fragmentco{3}{8},\fragmentco{10}{15}\rangle}=\ v_{3} \ v_{4} \ v_{4} \ v_{2} \ v_{5}\, ;\, v_{5} \ v_{7} \ v_{8} \ v_{9} \ v_{9}$};
\end{tikzpicture}

%% file: figures/decomp/block.tex
\begin{tikzpicture}[
    very thick, 
    every node/.style={font=\large}, 
    paren/.style={font=\huge\bfseries, inner sep=0pt}, 
    redbox/.style={red, dashed, line width=1.5pt},     
    bluebox/.style={blue, dashed, line width=1.5pt},
    greenbox/.style={green!50!black, dashed, line width=1.5pt},
    scale=0.8
]

\def\ytop{1.2}      
\def\ybot{0}      
\def\h{0.7}       
\def\ymidT{1.55}  
\def\ymidB{0.35}  

\def\xstart{0}    
\def\xend{8.5}     
\def\xlamA{2.0}   
\def\xlamB{4.0}   
\def\xlamC{6.3}   

\def\xBlueL{1.0}  
\def\xBlueR{2.4}  
\def\xRedL{3.2}   
\def\xRedR{7.6}   
\def\xGreenL{5.0} 
\def\xGreenR{6.8} 

\draw (\xstart, \ytop) rectangle (\xend, \ytop+\h);
\draw (\xlamB, \ytop) -- (\xlamB, \ytop+\h);

\node[above, yshift=-0.5mm] at (\xstart, \ytop+\h) {$\lambda'_0$};
\node[above, yshift=-0.5mm] at (\xlamB, \ytop+\h) {$\lambda'_1$};
\node[above, yshift=-0.5mm] at (\xend, \ytop+\h) {$\lambda'_2$};

\node[paren, red] (TopRedL) at (\xRedL, \ymidT) {(};
\node[paren, red] (TopRedR) at (\xRedR, \ymidT) {)};
\draw[redbox] (TopRedL.north) -- (TopRedL.north |- 0, \ytop+\h+0.7) -- (TopRedR.north |- 0, \ytop+\h+0.7) -- (TopRedR.north);

\draw (\xstart, \ybot) rectangle (\xend, \ybot+\h);
\draw (\xlamA, \ybot) -- (\xlamA, \ybot+\h);
\draw (\xlamB, \ybot) -- (\xlamB, \ybot+\h);
\draw (\xlamC, \ybot) -- (\xlamC, \ybot+\h);

\node[below, yshift=0.5mm] at (\xstart, \ybot) {$\lambda_0$};
\node[below, yshift=0.5mm] at (\xlamA, \ybot) {$\lambda_1$};
\node[below, yshift=0.5mm] at (\xlamB, \ybot) {$\lambda_2$};
\node[below, yshift=0.5mm] at (\xlamC, \ybot) {$\lambda_3$};
\node[below, yshift=0.5mm] at (\xend, \ybot) {$\lambda_4$};

\node[paren, blue] (BotBlueL) at (\xBlueL, \ymidB) {(};
\node[paren, blue] (BotBlueR) at (\xBlueR, \ymidB) {)};
\draw[bluebox] (BotBlueL.south) -- (BotBlueL.south |- 0, \ybot-0.6) -- (BotBlueR.south |- 0, \ybot-0.6) -- (BotBlueR.south);

\node[paren, green!50!black] (BotGreenL) at (\xGreenL, \ymidB) {(};
\node[paren, green!50!black] (BotGreenR) at (\xGreenR, \ymidB) {)};
\draw[greenbox] (BotGreenL.south) -- (BotGreenL.south |- 0, \ybot-0.6) -- (BotGreenR.south |- 0, \ybot-0.6) -- (BotGreenR.south);

\node[paren, red] (BotRedL) at (\xRedL, \ymidB) {(};
\node[paren, red] (BotRedR) at (\xRedR, \ymidB) {)};
\draw[redbox] (BotRedL.south) -- (BotRedL.south |- 0, \ybot-0.8) -- (BotRedR.south |- 0, \ybot-0.8) -- (BotRedR.south);

\end{tikzpicture}

%% file: figures/decomp/partition.tex
\begin{tikzpicture}[
    very thick, 
    every node/.style={font=\large}, 
    redbox/.style={red, dashed, line width=1.5pt},   
    bluebox/.style={blue, dashed, line width=1.5pt},
    greenbox/.style={green!50!black, dashed, line width=1.5pt},
    alignline/.style={thick, dotted},
    scale=0.8
]

\def\ytop{1.2}      
\def\ybot{0}      
\def\h{0.6}       

\def\xA{0}      
\def\xB{1.0}    
\def\xC{2.0}    
\def\xD{2.3}    
\def\xE{3.5}    
\def\xF{4.0}    
\def\xG{5.2}    
\def\xH{6.2}    
\def\xI{6.8}    
\def\xJ{7.6}    
\def\xK{8.3}    

\fill[red!20] (\xA, \ytop) rectangle (\xE, \ytop+\h);
\fill[orange!50] (\xE, \ytop) rectangle (\xF, \ytop+\h);
\fill[red!20] (\xF, \ytop) rectangle (\xJ, \ytop+\h);
\fill[blue!15] (\xJ, \ytop) rectangle (\xK, \ytop+\h);

\draw (\xA, \ytop) rectangle (\xK, \ytop+\h);
\draw (\xE, \ytop) -- (\xE, \ytop+\h);
\draw (\xF, \ytop) -- (\xF, \ytop+\h);
\draw (\xJ, \ytop) -- (\xJ, \ytop+\h);

\node[above, yshift=-0.5mm] at (\xA, \ytop+\h) {$\lambda'_0$};
\node[above, yshift=-0.5mm] at (\xF, \ytop+\h) {$\lambda'_1$};
\node[above, yshift=-0.5mm] at (\xK, \ytop+\h) {$\lambda'_2$};

\draw[redbox] (\xE, \ytop+\h) -- (\xE, \ytop+\h+0.7) -- (\xJ, \ytop+\h+0.7) -- (\xJ, \ytop+\h);

\fill[cyan!20] (\xA, \ybot) rectangle (\xB, \ybot+\h);
\fill[gray!40] (\xB, \ybot) rectangle (\xC, \ybot+\h);
\fill[cyan!20] (\xC, \ybot) rectangle (\xD, \ybot+\h);
\fill[red!20] (\xD, \ybot) rectangle (\xE, \ybot+\h);
\fill[orange!50] (\xE, \ybot) rectangle (\xF, \ybot+\h);
\fill[green!40] (\xF, \ybot) rectangle (\xG, \ybot+\h);
\fill[violet!20] (\xG, \ybot) rectangle (\xH, \ybot+\h);
\fill[green!40] (\xH, \ybot) rectangle (\xI, \ybot+\h);
\fill[red!20] (\xI, \ybot) rectangle (\xJ, \ybot+\h);
\fill[blue!15] (\xJ, \ybot) rectangle (\xK, \ybot+\h);

\draw (\xA, \ybot) rectangle (\xK, \ybot+\h);
\draw (\xB, \ybot) -- (\xB, \ybot+\h);
\draw (\xC, \ybot) -- (\xC, \ybot+\h);
\draw (\xD, \ybot) -- (\xD, \ybot+\h);
\draw (\xE, \ybot) -- (\xE, \ybot+\h);
\draw (\xF, \ybot) -- (\xF, \ybot+\h);
\draw (\xG, \ybot) -- (\xG, \ybot+\h);
\draw (\xH, \ybot) -- (\xH, \ybot+\h);
\draw (\xI, \ybot) -- (\xI, \ybot+\h);
\draw (\xJ, \ybot) -- (\xJ, \ybot+\h);

\node[below, yshift=0.5mm] at (\xA, \ybot) {$\lambda_0$};
\node[below, yshift=0.5mm] at (\xC, \ybot) {$\lambda_1$};
\node[below, yshift=0.5mm] at (\xF, \ybot) {$\lambda_2$};
\node[below, yshift=0.5mm] at (\xH, \ybot) {$\lambda_3$};
\node[below, yshift=0.5mm] at (\xK, \ybot) {$\lambda_4$};

\draw[bluebox] (\xB, \ybot) -- (\xB, \ybot-0.6) -- (\xD, \ybot-0.6) -- (\xD, \ybot);
\draw[greenbox] (\xG, \ybot) -- (\xG, \ybot-0.6) -- (\xI, \ybot-0.6) -- (\xI, \ybot);
\draw[redbox] (\xE, \ybot) -- (\xE, \ybot-0.8) -- (\xJ, \ybot-0.8) -- (\xJ, \ybot);

\draw[alignline] (\xA, \ytop) -- (\xA, \ybot+\h);
\draw[alignline] (\xE, \ytop) -- (\xE, \ybot+\h);
\draw[alignline] (\xF, \ytop) -- (\xF, \ybot+\h);
\draw[alignline] (\xJ, \ytop) -- (\xJ, \ybot+\h);
\draw[alignline] (\xK, \ytop) -- (\xK, \ybot+\h);

\end{tikzpicture}

%% file: s2_overview_dynamic.tex
\subsection{Dynamic Exact and Approximation Algorithms}\label{sec:ov:dynamic}

As the next application of our hierarchical forest decomposition, we discuss dynamic algorithms for maintaining $\ted(\bF,\bG)$ and its $(1+\varepsilon)$-approximation.

To discuss the technicalities in more detail, we step back and reintroduce some of the techniques mentioned earlier. The starting point for these dynamic algorithms is a new DP scheme for \TED based on our forest decomposition. At first glance, this DP scheme might appear much worse: it achieves a running time of $\Ohtilde(n^{7})$, and when parameterized by an upper bound $k$ on $\ted(\bF, \bG)$, the running time becomes $\Ohtilde(n k^{6})$. This is far from the $\Ohtilde(nk^2)$ state-of-the-art algorithm of Akmal and Jin \cite{AJ21}, and the $\Ohtilde(nk^3)$ time of Touzet \cite{10.1007/11496656_29} that the former algorithm improved upon.
So why even care?
Let us recall that our DP scheme has two key strengths:

\begin{itemize}
    \item Upon an update consisting of a label substitution, we only need to recompute up to $\Ohtilde(k^7)$ states of the DP, a quantity that does not polynomially depend on $n$. This stands in stark contrast to the algorithms from \cite{AJ21, 10.1007/11496656_29} and, generally, all classical DP algorithms that rely on the ``Decomposition Scheme'' framework \cite{DT03}. This includes most \TED algorithms, ranging from Zhang and Shasha \cite{ZS89} to Klein's algorithm \cite{Kle98} and Demaine et al. \cite{DMRW10}, where at least $\Omega(n)$ states need to be recomputed upon a single update, even when $k = \Ohtilde(1)$. Thus, there is hope to support deletions and insertions in time $\poly(k\log n)$.

    \item The second major advantage is that the DP scheme can be modified to recover a $(1+\varepsilon)$-approximation in $\Ohtilde_{\varepsilon}(nk)$ time, matching the state-of-the-art algorithm from \cite{BGHS19}, which is itself a result of independent interest.
\end{itemize}

Exploiting the first strength yields \cref{thm:dynexactted}, and combining it with the second yields \cref{thm:dynapproxted}. However, several technical steps that we describe in slightly more detail here are required to achieve these results. We start by introducing the basic static DP scheme and demonstrating how to obtain the $(1+\varepsilon)$-approximation. Finally, we explain how to make this scheme dynamic.

\newcommand{\hF}{T(\bF)}
\newcommand{\tedk}[1]{\ted^{(#1)}}
\newcommand{\tedko}[1]{\smash{\overline{\ted}}^{(#1)}}
\newcommand{\tedkt}[1]{\smash{\widetilde{\ted}}^{(#1)}}
\newcommand{\Wk}[1]{\smash{\overline{W}}^{(#1)}}
\newcommand{\Wt}[1]{\smash{\widetilde{W}}^{(#1)}}

\subsubsection{\boldmath An $\Ohtilde(nk^{6})$-time DP Algorithm for Exact Tree Edit Distance}

\begin{figure*}[!t]
     \centering
     \begin{subfigure}[b]{0.35\textwidth}
         \centering
         \scalebox{0.8}{\input{figures/dynamic/mono-to-mono-mono}}
         \caption{}
         \label{fig:relintro:a}
     \end{subfigure}
     \hfill
     \begin{subfigure}[b]{0.55\textwidth}
         \centering
          \scalebox{0.8}{\input{figures/dynamic/bi-to-bi-bi}}
         \caption{}
         \label{fig:relintro:b}
     \end{subfigure}
     \\[15pt]
     \begin{subfigure}[b]{0.35\textwidth}
         \centering
         \scalebox{0.8}{\input{figures/dynamic/mono-to-bi}}
         \caption{}
         \label{fig:relintro:c}
     \end{subfigure}
     \hfill
     \begin{subfigure}[b]{0.55\textwidth}
         \centering
           \scalebox{0.8}{\input{figures/dynamic/bi-to-bi-mono}}
         \caption{}
         \label{fig:relintro:d}
     \end{subfigure}
     \\[15pt]

    \caption{
        The four possible parent-child relations. The red and blue segments represent the intervals of the two children (thus, two segments constitute a bi-interval, whereas a single segment is a mono-interval). The union of the children's intervals forms the parent's interval. For instance, in \cref{fig:relintro:b}, the parent $v \in T(\bF)$ is associated with a bi-interval and has two children, $v_1$ and $v_2$, whose bi-intervals are contained inside the bi-interval of $v$ and one of them is nested around the other.
        In \cref{fig:relintro:c}, a node associated with a mono-interval has only one child that is associated with a bi-interval.
     }
     \label{fig:relintro}
\end{figure*}
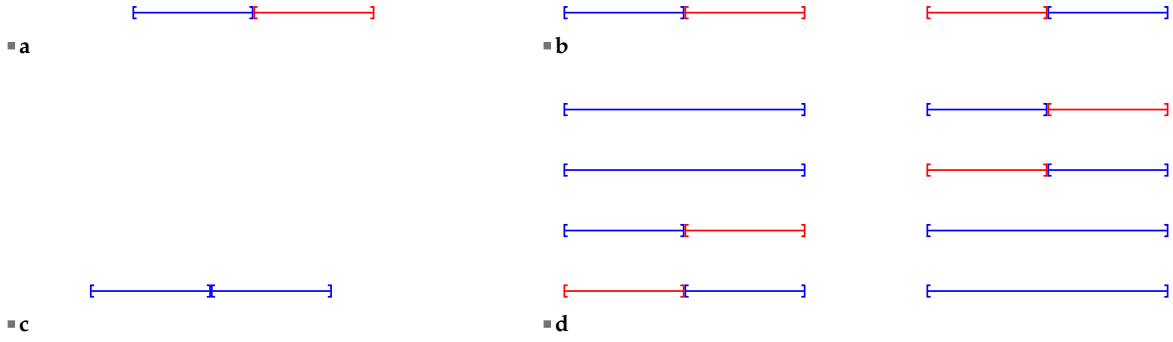

The starting point is once again our hierarchical decomposition, this time with $c=\Oh(1)$. Ideally, we want each node in the decomposition to have as few children as possible, preferably no more than two. Our hierarchical decomposition does not guarantee this directly, as it only bounds the number of children by a constant. To remedy this, we organize the children of each node in a binary tree, yielding a decomposition tree denoted by $T(\bF)$. \Cref{fig:relintro} illustrates the possible parent-child relations in $T(\bF)$.

In the remainder of the discussion, we assume that all parent-child relations are as depicted in \cref{fig:relintro:b}. Although this is clearly not possible for the root, this assumption simplifies the exposition and allows us to focus on the most difficult and interesting case. For a node $v \in T(\bF)$, we denote its corresponding bi-interval by $\mI_v = \langle I_{v,1}, I_{v,2} \rangle$, where $I_{v,i} = \fragmentco{a_{v,i}}{b_{v,i}}$ for $i \in \fragment{1}{2}$.

Our DP scheme is as follows:
for each node $v\in \hF$, we would like to determine $\ted(\bF_{\mI_v},\bG_{\mJ})$ for every (possibly imbalanced) subforest $\bG_{\mJ}$.
This allows us to recover $\ted(\bF_{\mI_v},\bG_{\mJ})$ from the two children $v_1,v_2$ as follows:

\begin{equation}\label{eq:rected}
\ted(\bF_{\mI_v},\bG_{\mJ}) = \min_{\text{$\mJ_1,\mJ_2$ partition $\mJ$ as in \cref{fig:relintro:b}}}\left(\ted(\bF_{\mI_{v_1}},\bG_{\mJ_1}) + \ted(\bF_{\mI_{v_2}},\bG_{\mJ_2})\right).
\end{equation}
Here, the bi-intervals $\mJ_1$ and $\mJ_2$ that we quantify over are required to partition $\mJ$ in a nested manner, exactly as $\mI_{v_1}$ and $\mI_{v_2}$ partition $\mI_v$. This means we are essentially quantifying over two positions within a bi-interval of $\mJ$, splitting it into two.

Since there are $\Ohtilde(n)$ nodes $v \in T(\bF)$ and $\Oh(n^4)$ possible choices for $\mJ$ (because a bi-interval has at most $4$ possible endpoints), and since we quantify over $\Oh(n^2)$ possible splitting points in \cref{eq:rected}, we obtain an overall running time of $\Ohtilde(n^7)$ by computing the values in a bottom-up fashion.

The next step is to modify the above scheme to obtain a running time parameterized by $k$. For this, we use the following insight: when considering an optimal alignment $\mA : \bF \onto \bG$, if $\cost(\mA) \leq k$, then $\mA$ must be \emph{$k$-bounded}, meaning that for every $(x,y) \in \mA$ we have $|x-y| \leq 2k$ (a similar observation already appears in \cite{KS25}). Since distant pairs $x$ and $y$ do not appear in $\mA$ anyway, we would ideally only need to compute $\ted(\bF_{\mI_v}, \bG_{\mJ})$ for $\mJ$ in the following set of size $\Oh(k^4)$:
\[
    W^{k}(\mI_v) = \{\mJ = \langle \fragmentco{a_1'}{b_1'}, \fragmentco{a_2'}{b_2'} \rangle \mid \forall_{i \in \fragment{1}{2}}\; |a_{v,i}-a_i'| \leq 2k\text{ and } |b_{v,i}-b_i'| \leq 2k\}
\]
Accordingly, the selections for $\mJ_1$ and $\mJ_2$ in \cref{eq:rected} have the additional restriction that they belong to $W^{k}(\mI_{v_1})$ and $W^{k}(\mI_{v_2})$, respectively, so we only consider $\Oh(k^2)$ pairs $(\mJ_1,\mJ_2)$ in \cref{eq:rected}.

The problem with this modification, however, is that even if the global alignment $\mA : \bF \onto \bG$ has a cost of at most $k$ (and is therefore $k$-bounded), it is not true that, for any $v \in T(\bF)$ and $\mJ \in W^{k}(\mI_v)$, the optimal alignment is $k$-bounded. Consequently, evaluating \cref{eq:rected} under this new restriction does not always compute the true $\ted(\bF_{\mI_v},\bG_{\mJ})$, but rather some other, potentially meaningless quantity.

Fortunately, it turns out that the recursion still yields a meaningful value; more specifically, it computes a value $t_v(\mJ)$ that satisfies $\ted(\bF_{\mI_v}, \bG_{\mJ}) \leq t_v(\mJ) \leq \ted^{(k)}(\bF_{\mI_v}, \bG_{\mJ})$, where the latter tree edit distance definition considers only $k$-bounded alignments. To be formally precise, we should adjust \cref{eq:rected} to $t_v(\mJ) = \min_{\mJ_1,\mJ_2} (t_{v_1}(\mJ_1) + t_{v_2}(\mJ_2))$ for each $v$ and $\mJ \in W^{k}(\mI_v)$. For the root $r$ of $T(\bF)$, the evaluation at $\mJ = \langle \fragmentco{0}{2|\bG|} \rangle$ is still useful because it satisfies $\ted(\bF, \bG) \leq t_r(\mJ) \leq \ted^{(k)}(\bF, \bG)$. This is sufficient to recover $\ted(\bF,\bG)$ whenever $\ted(\bF,\bG) \leq k$ because then $\ted(\bF,\bG) = \ted^{(k)}(\bF,\bG)$. The overall running time shrinks to $\Ohtilde(n \cdot k^4 \cdot k^2) = \Ohtilde(nk^{6})$.

\subsubsection{\boldmath Sparsifying the DP Algorithm to Achieve a $(1+\varepsilon)$-Approximation in Time $\Ohtilde_{\varepsilon}(nk)$}

As explained earlier, if a $(1+\varepsilon)$-approximation is sufficient, one can use two ideas to optimize the running time:
\begin{enumerate}[(1)]
    \item When implementing the recursion in \cref{eq:rected}, one might hope to restrict the evaluation to a subset of $\mJ_1$ and $\mJ_2$ instead of iterating over all possibilities.

    More specifically, in \cref{eq:rected}, it suffices to restrict $\mJ_1 = \langle J_1^{1}, J_2^1 \rangle$ such that $|J_1^1|=|I_{v_1,1}|+\delta_1$ and $|J_2^1|=|I_{v_1,2}|+\delta_2$ for $\delta_1, \delta_2 \in \Delta_{\varepsilon} \coloneqq \{0\} \cup \{\pm (1+\varepsilon)^p \}$. This now yields only $\Oh(\varepsilon^{-2} \log^2 n)$ possibilities for $\mJ_1$ (and consequently for $\mJ_2$, which is selected to form a partition). This approach works because the size difference $\big| |I_{v_1,1}| - |J_1^1| \big|$ provides a lower bound $\ted(\bF_{\mI_{v_1}}, \bG_{\mJ_1}) \geq \big| |I_{v_1,1}| - |J_1^1| \big| / 2$, and the distance $\ted(\bF_{\mI_v}, \bG_{\mJ})$ changes by at most two when moving the endpoints of the bi-intervals. Therefore, the larger the size difference, the more sparsely we can afford to sample the lengths. Overall, we incur a $(1+\Oh(\varepsilon))$-approximation overhead, but we only consider $\Oh(\varepsilon^{-2} \log^2 n) = \Ohtilde_\varepsilon(1)$ many pairs $(\mJ_1, \mJ_2)$.
    \label{it:sparse:2}

    \item Consider the set of values $\{t_v(\mJ) \mid \mJ \in W^{k}(\mI_v)\}$. These values are inherently related because they correspond to potentially overlapping bi-intervals. Thus, there is hope to select as small a subset $\widetilde{W}^{k}(\mI_v) \subseteq W^{k}(\mI_v)$ as possible such that the retained values $t_v(\mJ)$ still provide an efficiently computable $(1+\varepsilon)$-approximation for the omitted bi-intervals $\mJ$.

    To do this, we use the trick introduced earlier:
    \begin{itemize}
        \item For the dependent case, we aim to sample spine nodes within $\bF_{\mI_v}$ to form a set $Z_v$ of size $\Oh(\varepsilon^{-1} \log n)$. Then, for each $z \in Z_v$ and every possible $z' \in \bG$, we retain in $\widetilde{W}^{k}(\mI_v)$ only the bi-intervals $\mJ \in W^{k}(\mI_v)$ constructed using the pairs $(\Left(z), \Left(z'))$ and $(\Right(z), \Right(z'))$.

        The lengths we consider before and after $\Left(z')$ are $|\Left(z)-a_{v,1}|+\delta_1$ and $|b_{v,1}-\Left(z)| + \delta_1'$ for $\delta_1,\delta_1' \in \Delta_{\varepsilon}$, respectively. This allows us to charge any size difference to the size difference from the corresponding interval of $\mI_v$. Similarly, the lengths we consider before and after $\Right(z')$ are $|\Right(z)-a_{v,2}| + \delta_2$ and $|b_{v,2}-\Right(z)| + \delta_2'$ for $\delta_2,\delta_2' \in \Delta_{\varepsilon}$, respectively.

        \item For the decoupled case, we need to compute new values not present in the exact DP algorithm (because they are useless there). As with bi-intervals, for each $v$, each $i \in \fragment{1}{2}$, and each $\mJ = \langle J \rangle \in W^{k}(\langle I_{v,i} \rangle)$,\footnote{Similarly to bi-intervals, for a mono-interval we define $W^{k}(\langle I_{v,i} \rangle) = \{\mJ = \langle \fragmentco{a'}{b'} \rangle \mid |a_{v,i}-a'| \leq 2k, |b_{v,i} - b'| \leq 2k\}$.} where $|J| = |I_{v,i}| + \delta$ for $\delta \in \Delta_\varepsilon$, we want to approximate a value $t_{v,i}(\mJ)$ sandwiched between $\ted(\bF_{\langle I_{v,i} \rangle}, \bG_{\mJ})$ and $\ted^{(k)}(\bF_{\langle I_{v,i} \rangle}, \bG_{\mJ})$.
    \end{itemize}

    This now allows us to approximate each $\mJ \in W^{k}(\mI_v)$ using one of the stored values as follows: if an optimal alignment $\mA_v : \bF_{\mI_v} \onto \bG_{\mJ}$ maps more than an $\varepsilon$-fraction of the spine nodes, then our sampling guarantees that we hit at least one of these nodes. Consequently, we can approximate $\mA_v$ using one of the stored $\mJ' \in \widetilde{W}^{k}(\mI_v)$ constructed around this match. Otherwise, we can afford to simply ignore the spine nodes and treat the two sides independently.

    We remark that in the former case, we do not know \emph{a priori} which $z,z'$ to use in these four expressions. It turns out that here it suffices to try all $z \in Z_v$ along with all four possible lengths and to find the ``closest'' $z'$ that matches our description of $\mJ$ through these fixed quantities. At the data structure level, this reduces to $(1+\varepsilon)$-approximate nearest-neighbor queries w.r.t. the $L_1$ norm on a grid.
\end{enumerate}

Overall, our approach provides a method to compute a value $\tilde{t}_v(\mJ)$ for each $\mJ \in \widetilde{W}^{k}(\mI_v)$ (and similarly a value $\tilde{t}_{v,i}(\mJ)$ for the decoupled case) that satisfies $\ted(\bF_{\mI_v}, \bG_{\mJ}) \leq \tilde{t}_v(\mJ) \leq (1+\Oh(\varepsilon))^{h_v} \ted^{(k)}(\bF_{\mI_v}, \bG_{\mJ})$. Here, $h_v$ is the height of $v$ in $T(\bF)$, reflecting that the $(1+\varepsilon)$-errors accumulate multiplicatively. Since the root height is $h_r = \Oh(\log n)$, we can simply set $\varepsilon = \Theta(\varepsilon'/\log n)$ for our target precision $\varepsilon'$ and still achieve the desired approximation guarantee.
Since, across both the dependent and decoupled cases, we never store more than $\Ohtilde_{\varepsilon}(k)$ values for each $v \in T(\bF)$, the running time shrinks to $\Ohtilde_{\varepsilon}(n k)$.

\subsubsection{\boldmath An Exact Dynamic Algorithm with $\Ohtilde(k^8)$ Update Time}

Let us return to the exact case, but now consider the dynamic setting. Suppose we substitute the label of a single node $w \in \bG$,\footnote{In this presentation, we only consider updates in $\bG$. The case where the substitutions are in $\bF$ turns out to be easier.} and let $y \coloneqq \Left(w)$ and $y' \coloneqq \Right(w)$. How many values of $t_v(\mJ)$ for $v \in T(\bF)$ and $\mJ \in W^{k}(\mI_v)$ can potentially change? In other words, how many intervals $\mJ$ intersect the update positions $y$ and $y'$? A necessary condition is that $v \in A^{\circ}_k \coloneqq A^{\circ}_k(y) \cup A^{\circ}_k(y')$, where $A^{\circ}_k(x) \coloneqq \{v \in T(\bF) \mid (I_{v,1} \cup I_{v,2}) \cap \fragment{-2k+x}{x+2k} \neq \emptyset\}$, because the endpoints of any $\mJ \in W^{k}(\mI_v)$ are at most $2k$ further away than those of $\mI_v$.
At each level of $T(\bF)$, there are at most $\Oh(k)$ nodes $v \in A^{\circ}_k(x)$ for each position $x$. Thus, overall, $|A^{\circ}| = \Oh(k \log n)$. Consequently, the number of values $t_v(\mJ)$ to recompute (in a bottom-up fashion) amounts to $\Oh(|A^{\circ}| k^{4}) = \Oh(k^5 \log n)$, taking into account that the size of each $W^{k}(\mI_v)$ is bounded by $\Oh(k^4)$. Since each recomputation step takes $\Oh(k^2)$ time, this yields an overall recomputation time of $\Ohtilde(k^7)$, with a polylogarithmic dependence on $n$ only!

While for substitutions this looks promising, the update bounds become way too large if we consider the deletion (or insertion) of a node in $\bG$ at positions $y$ and $y'$. This happens because a deletion shifts the indices of all subsequent content in $\bG$. Consequently, even if the update positions $y$ and $y'$ are far away from the bi-interval $\mI_v$ of a node $v$, this index shift causes at least one bi-interval $\mJ$ to leave the set $W^{k}(\mI_v)$ while another enters it.\footnote{Here, we no longer view $\mI_v$ and $\mJ$ as being defined by positional indices. Instead, following a standard approach used in many dynamic string algorithms (e.g.,~\cite{KK22,KMS23}), we assign a persistent identifier to each character of $\mS(\bF)$ and $\mS(\bG)$, and represent each fragment by the identifiers of its leftmost and rightmost characters.} As a result, we would potentially need to recompute at least one value $t_v(\mJ)$ for every node $v \in T(\bF)$. This defeats our goal, as it yields an update time that grows linearly~in~$n$.

This problem, where the content of $\bG$ moves while the content of $\bF$ remains fixed, does not appear when $\bG = \bF$.
Indeed, in that case, as long as $v \notin A^{\circ}$, both $\mI_v$ and all $\mJ \in W^{k}(\mI_v)$ are sufficiently far from the update positions $y$ and $y'$. Because they ``shift together,'' no value $t_v(\mJ)$ needs to be recomputed for any $v \notin A^{\circ}$ and $\mJ \in W^{k}(\mI_v)$. Thus, in this special case, we can proceed just as we did for substitutions: we recompute $t_v(\mJ)$ bottom-up for all $v \in A^{\circ}$ and $\mJ \in W^{k}(\mI_v)$. As established previously, this recomputation takes $\Ohtilde(k^7)$ time. Note that we are temporarily ignoring the new nodes introduced into $T(\bF)$; here we assume that $T(\bF)$ is dynamic as well.\footnote{Technically, the decomposition tree prior to binarization is the dynamic tree we discussed previously. Whenever a node $v$ in this base tree undergoes a change, it correspondingly updates the $\Oh(1)$ binary structure involving $v$ in $T(\bF)$.} However, because these structural changes are on average only $\Oh(\log n)$, we can safely defer their analysis for now.

As noted in recent works on string algorithms \cite{KMS23,BGK25}, maintaining the $t_v(\mJ)$ values for distances within $\bF$ remains useful even when comparing against a different forest $\bG$. Indeed, assume that the optimal alignment $\mA : \bF \onto \bG$ has a cost of at most $k$ (in particular, it is $k$-bounded).
As long as $\mJ \in W^{k}(\mI_v)$ does not intersect any of the edits in $\mA$, the alignment implies the existence of some $\mL$ at a distance of at most $2k$ from $\mJ$ such that $\bF_{\mL}$ and $\bG_{\mJ}$ are isomorphic, denoted $\bF_{\mL} \equiv \bG_{\mJ}$.
Since $\mJ \in W^{k}(\mI_v)$, we conclude that $\mL \in W^{2k}(\mI_v)$. In other words, for $v \notin A^{\circ}(\mA)$, where we loosely define $A^{\circ}(\mA) \coloneqq \bigcup_{x \text{ is an update position in } \mA} A^{\circ}_k(x)$, we can recycle values from a data structure that maintains distances within $\bF$, using a threshold increased by a small multiplicative constant.\footnote{Note that employing a larger threshold only provides better values, as $\ted^{(\star)}(\bF_\mI, \bF_\mL)$ is monotonically non-increasing with respect to $\star$. At the same time, it is not necessarily true that $\ted^{(\star)}(\bF_\mI, \bF_\mL)=\ted^{(\star)}(\bF_\mI, \bG_\mJ)$ when $\bF_{\mL}\equiv \bG_{\mJ}$, but $\ted^{(\star')}(\bF_\mI, \bF_\mL)\le \ted^{(\star)}(\bF_\mI, \bG_\mJ)$ can be proved if $\star'-\star$ exceeds the total distance between the endpoints of $\mJ$ and $\mL$.} Consequently, we only need to explicitly compute $t_v(\mJ)$ for $v \in A^{\circ}(\mA)$.

Since we do not know $\mA$ a priori, we compute $t_v(\mJ)$ in a top-down fashion, recursively fetching the required values from the children to perform the $\Oh(k^2)$-time computation behind \cref{eq:rected}.
For $v \in T(\bF)$ and $\mJ \in W^{k}(\mI_v)$, whenever there exists an $\mL \in W^{2k}(\mI_v)$ such that $\bF_{\mL} \equiv \bG_{\mJ}$, we can simply copy the value from the other data structure. Based on the bounds presented above, we know that this copying step will fail for at most $\Oh(|A^{\circ}(\mA)| \cdot k^4) = \Ohtilde(k \cdot k \cdot k^4) = \Ohtilde(k^6)$ pairs $(v,\mJ)$. Consequently, the total running time is bounded by $\Ohtilde(k^8)$.

\subsubsection{Finding Nearby Isomorphic Subforests}

The final missing piece for the algorithm just described is a data structure that, given a node $v \in T(\bF)$ and a query $\mJ \in W^{k}(\mI_v)$, efficiently verifies whether such an $\mL \in W^{2k}(\mI_v)$ exists with $\bF_{\mL} \equiv \bG_{\mJ}$.

In the string setting, \cite{KMS23} used dynamic suffix arrays~\cite{KK22}, which allow one to test in $\Ohtilde(1)$ time whether the substring of $Y$ also occurs in some region of $X$.
For bi-intervals, that strategy becomes inefficient: although it allows detecting isomorphic occurrences of $\bG_{\langle J_{\Left} \rangle}$ near $\bF_{\langle I_{\Left} \rangle}$ and of $\bG_{\langle J_{\Right} \rangle}$ near $\bF_{\langle I_{\Right} \rangle}$, it does not help find occurrences that can be combined into an isomorphic occurrence of $\bG_\mJ$. This is particularly problematic if $|J_{\Left}|,|J_{\Right}|\ll k$, when many occurrence pairs often have to be verified.

Our solution is to first verify separately the equality of the corresponding fragments in the (labeled) parenthesis representations of $\mS(\bF)$ and $\mS(\bG)$. We maintain these representations using a standard dynamic string data structure~\cite{MSU97,ABR00,KK22}, in which strings (undergoing persistent splits and concatenations) are represented by abstract identifiers supporting $\Ohtilde(1)$-time lexicographic comparison. Furthermore, we must ensure that both trees share the exact same spine nodes. Since these nodes appear contiguously along a single root-to-leaf path, it suffices to locate the lowest and highest nodes of this path. We accomplish this using a dynamic 2D range emptiness query data structure from \cite{M03}.

Altogether, we maintain all identifiers for $\mL \in W^{2k}(\mI_v)$ in a data structure associated with $v \in T(\bF)$. Then, upon receiving a query $\mJ \in W^{k}(\mI_v)$, we compute its identifier in $\Ohtilde(1)$ time and search for a matching identifier among those stored at $v$. Updating this data structure involves, once again, recomputing the identifiers for all $\mL \in W^{2k}(\mI_v)$ that intersect the updates. By arguments similar to those presented above, this update can be performed in $\Ohtilde(k^5)$ time.

\subsubsection{\boldmath Improving the Update Time to $\Ohtilde(k^7)$}

The preceding algorithm achieves an update time of $\Ohtilde(k^8)$, which incurs an extra factor of $k$ compared to the $\Ohtilde(n k^7)$ running time of the static DP algorithm. While one might tolerate this increase in the exact setting (given that the polynomial degree is already quite high), there remains a compelling reason to eliminate this additional factor of $k$. Indeed, if the same extra factor were to propagate to an approximate dynamic counterpart, it would be undesirable, as an update time of $\Ohtilde_{\varepsilon}(k)$ is far better than $\Ohtilde_{\varepsilon}(k^2)$.

As previously mentioned, for a node $v \notin A^{\circ}$ located far from an update, maintaining the values $t_v(\mJ)$ for all $\mJ \in W^k(\mI_v)$ after the update is overly demanding. Because the content in $\bG$ shifts, newly entering bi-intervals necessitate recomputation. The first key observation is that if we only require maintaining $t_v(\mJ)$ for $\mJ \in W^{k-1}(\mI_v)$ following the update, we circumvent the need to process these newly entering bi-intervals.\footnote{An additional benefit of decrementing the threshold $k$ is that the upper bound $t_v(\mJ) \le \ted^{(k)}(\bF_\mI,\bG_\mJ)$ gets relaxed to $t_v(\mJ) \le \ted^{(k-1)}(\bF_\mI,\bG_\mJ)$; without this relaxation, many values $t_v(\mJ)$ for $\mJ\in W^{k-1}(\mI_v)$ would be invalidated.}
On the other hand, the algorithm we just described provides a method to initialize the values $t_v(\mJ)$ (if not explicitly, at least implicitly through copying) for all $v \in T(\bF)$ and $\mJ \in W^k(\mI_v)$ in $\Ohtilde(k^8)$ time.
Combining these two ideas, we can reinitialize the data structure every $k$ updates using the $\Ohtilde(k^8)$-time algorithm, but with a starting threshold of $2k$ instead of $k$. Then, over the course of the subsequent $k$ updates, we maintain the values $t_v(\mJ)$ such that during the $c$th update, we only track $\mJ \in W^{2k-c}(\mI_v)$. Although the range of $\mJ$ degrades over these $k$ updates, we only need to amortize the expensive $\Ohtilde(k^8)$ recomputation cost down to $\Ohtilde(k^7)$ per update.

To formalize this idea, we proceed in \emph{epochs} of $k$ updates. Let $\ul{\bG}$ and $\ul{\bF}$ be $\bG$ and $\bF$ at the beginning of the epoch.
Before the $c$th update of the epoch, we maintain the following invariant.
For each $v \in T(\bF)$ and $\mJ \in W^{2k-c}(\mI_v)$, at least one of these holds:
\begin{enumerate}[(a)]
\item We store a value $t_{v}(\mJ)$ such that $\ted(\bF_{\mI_v}, \bG_{\mJ}) \leq t_{v}(\mJ) \leq \ted^{(2k-c)}(\bF_{\mI_v}, \bG_{\mJ})$.
\label{it:intro:inv:a}
\item Node $v$ was also present in $T(\ul{\bF})$, and there is $\mL \in W^{4k}(\ul{\mI}_v)$ such that $\bG_{\mJ} \equiv \ul{\bF}_{\mL}$.
\label{it:intro:inv:b}
\item Neither of the above holds.
\label{it:intro:inv:c}
\end{enumerate}
The invariant also guarantees that the number of pairs $(v,\mJ)$ for which Invariant~\ref{it:intro:inv:c} holds is at most $\Phi_c$, where $\Phi_c$ is some potential that does not increase as $c$ grows.

At the beginning, by the previous discussion for $c = 0$, we can set $\Phi_0 = \Ohtilde(k^6)$ so that Invariant~\ref{it:intro:inv:c} holds for all $v \in A^{\circ}(\mA)$ (where this time $\mA$ has cost at most $2k$) and $\mJ \in W^{2k}(\mI_v)$; for the remaining $v$ and $\mJ$, we know that Invariant~\ref{it:intro:inv:b} holds.

If we explicitly require a value $t_v(\mJ)$ (thereby restoring Invariant~\ref{it:intro:inv:a} for $v$ and $\mJ$) and know that at least one invariant holds for its children, we can perform the $\Oh(k^2)$-time recomputation using those children. If Invariant~\ref{it:intro:inv:a} holds for them, the required values are already stored. Alternatively, if Invariant~\ref{it:intro:inv:b} holds, we can simply copy the values from a dynamic data structure that maintains distances within $\ul{\bF}$.\footnote{This data structure preserves $\ul{\bF}$ exactly as it was at the beginning of the epoch, catching up with updates to $\bF$ only at the epoch's conclusion. Additionally, we require a parallel data structure for checking identifiers as described; this also freezes $\ul{\bF}$ at its initial epoch state while keeping $\bG$ up to date.} Finally, if Invariant~\ref{it:intro:inv:c} holds, we can recursively recompute the values in $\Oh(k^2)$ time, thereby restoring Invariant~\ref{it:intro:inv:a}. Overall, any such recomputation takes $\Oh((1+\phi)k^2)$ time, where $\phi$ denotes the corresponding decrease in potential.

Upon an update, we now do the following: we restore Invariant~\ref{it:intro:inv:a} for all $v \in A^{\circ}$ (defined again with respect to the update positions $y, y'$) and $\mJ \in W^{2k-c}(\mI_v)$ by performing the aforementioned recomputation. This ensures that at least one of the invariants holds for all $v$ and $\mJ \in W^{2k-c-1}(\mI_v)$: either we explicitly restore Invariant~\ref{it:intro:inv:a}, or, because $v \notin A^{\circ}$, Invariant~\ref{it:intro:inv:b} or~\ref{it:intro:inv:c} remains valid.

Altogether, in each update, we spend $\Oh((|A^{\circ}|k^4 + \phi)k^2) = \Ohtilde(\phi k^2 + k^7)$ time, where $\phi$ is the potential decrease in the $c$th update. As $\Phi_0 = \Ohtilde(k^6)$, the latter quantity amortizes to $\Ohtilde(k^7)$ over $k$ updates.

\subsubsection{\boldmath Dynamic $(1+\varepsilon)$-Factor Approximation}

For the approximate setting, we would like to integrate the sparsified $(1+\varepsilon)$-approximate scheme into the dynamic algorithm above: instead of storing all $t_{v}(\mJ)$, we try to maintain, for each $v \in T(\bF)$, the values $\tilde{t}_v(\mJ)$ for a few bi-intervals $\mJ\in W^{2k}(\mI_v)$, as well as decoupled values $\tilde{t}_{v,1}(\mJ)$ and $\tilde{t}_{v,2}(\mJ)$ for a few mono-intervals $\mJ\in W^{2k}(\langle I_{v,1}\rangle)$ and $\mJ\in W^{2k}(\langle I_{v,2}\rangle)$, respectively.

However, even when translating the simplest case where we only have substitutions, we notice that we do not achieve the promised $\Ohtilde_{\varepsilon}(k)$ running time. Indeed, already for the first (left) of the two decoupled bi-interval sides, if we recompute $\tilde{t}_{v,1}(\mJ)$ for all $\mJ$ considered for $v \in A^{\circ}$, there are $\Ohtilde_{\varepsilon}(k)$ such $\mJ$ for each of the $\Oh(k \log n)$ nodes $v \in A^{\circ}$, leading to a quadratic dependence on $k$. Our solution is to impose an additional constraint on $\mJ$ and perform a finer analysis of the pairs $(v,\mJ)$ that we want to recompute, which at first glance are only those where $\mJ$ intersects the update. Let us explain our solution in relation to the left (first) side of the bi-interval.

We wrote ``at first glance'' because, when maintaining distances within $\bF$, we are interested in recomputing not only the pairs $(v,\mJ)$ that intersect the update, but also those where the distance from $\mJ$ to $I_{v,1}$ is already close to $2k$. In this latter scenario, a deletion between $\mJ$ and $I_{v,1}$ can improve the $k$-bounded distance $\ted^{(k)}(\bF_{\langle I_{v,1} \rangle}, \bG_{\mJ})$ by freeing up new alignment possibilities. For the remainder of the discussion, we ignore this second issue; instead, we present a solution and explain how it addresses the first issue (which ultimately resolves the second issue as well).

The additional constraint for the first issue is that we only consider mono-intervals $\mJ = \langle J \rangle$ such that $|J| \leq |I_{v,1}|\varepsilon^{-1}$. This restriction is harmless because if $|J| \geq |I_{v,1}|\varepsilon^{-1}$, then since $\ted(\bF_{\langle I_{v,1}\rangle}, \bG_{\mJ}) \geq \big| |J| - |I_{v,1}| \big| / 2$, one can infer $(|J| + |I_{v,1}|) / 2 \leq (1+\Oh(\varepsilon)) \ted(\bF_{\langle I_{v,1}\rangle}, \bG_{\mJ})$, meaning one might as well delete everything. Moreover, this restriction is helpful because, for a given $v$, if we consider all such $\mJ = \langle J \rangle$ and fix their length to $q$, then at most $q$ of them can intersect any update position $y, y'$.
Similar constraints are imposed for the other side and for the maintained bi-interval $\mJ$, where the bound is separate for either of the corresponding intervals $I_{v,1}$ and $I_{v,2}$ constituting $\mI_v$.

From the analysis point of view, this is sufficient, but there are still other details that we need to handle carefully. For instance, we now need to be able to support nearest-neighbor queries on a grid dynamically, for which we again employ dynamic 2D range emptiness queries \cite{M03}. Moreover, we must also be careful to use the same samples across the data structures that maintain distances within $\bF$ and those that maintain distances between $\bF$ and $\bG$.
All these challenges are carefully tackled in \cref{sec:dyn}, which is the most involved part of our manuscript.

%% file: figures/dynamic/mono-to-mono-mono.tex
\usetikzlibrary{arrows.meta, bending}

\begin{tikzpicture}
    \draw[Bracket-Bracket, thick, blue] (0,0) -- (2, 0);
    \draw[Bracket-Bracket, thick, red] (2,0) -- (4, 0);
\end{tikzpicture}

%% file: figures/dynamic/bi-to-bi-bi.tex
\usetikzlibrary{arrows.meta, bending}

\begin{tikzpicture}
    \draw[Bracket-Bracket, thick, blue] (0,0) -- (2, 0);
    \draw[Bracket-Bracket, thick, red] (2,0) -- (4, 0);
    
    \draw[Bracket-Bracket, thick, red] (6,0) -- (8, 0);
    \draw[Bracket-Bracket, thick, blue] (8,0) -- (10, 0);
\end{tikzpicture}

%% file: figures/dynamic/mono-to-bi.tex
\usetikzlibrary{arrows.meta, bending}

\begin{tikzpicture}
    \draw[Bracket-Bracket, thick, blue] (0,0) -- (2, 0);
    \draw[Bracket-Bracket, thick, blue] (2,0) -- (4, 0);
\end{tikzpicture}

%% file: figures/dynamic/bi-to-bi-mono.tex
\usetikzlibrary{arrows.meta, bending}

\begin{tikzpicture}
    \draw[Bracket-Bracket, thick, red] (0,0) -- (2, 0);
    \draw[Bracket-Bracket, thick, blue] (2,0) -- (4, 0);
    
    \draw[Bracket-Bracket, thick, blue] (6,0) -- (10, 0);


    \draw[Bracket-Bracket, thick, blue] (0,1) -- (2, 1);
    \draw[Bracket-Bracket, thick, red] (2,1) -- (4, 1);
    
    \draw[Bracket-Bracket, thick, blue] (6,1) -- (10, 1);


    \draw[Bracket-Bracket, thick, blue] (0,2) -- (4, 2);
    
    \draw[Bracket-Bracket, thick, red] (6,2) -- (8, 2);
    \draw[Bracket-Bracket, thick, blue] (8,2) -- (10, 2);


    \draw[Bracket-Bracket, thick, blue] (0,3) -- (4, 3);
    
    \draw[Bracket-Bracket, thick, blue] (6,3) -- (8, 3);
    \draw[Bracket-Bracket, thick, red] (8,3) -- (10, 3);
\end{tikzpicture}

%% file: s3_prelims.tex
\section{Preliminaries}
\label{sec:preliminaries}

\subparagraph*{Approximation.}
For two values $x$ and $y$, we say $x$ \emph{$(\alpha, \beta)$-approximates} $y$, or $x$ is an $(\alpha, \beta)$-approximation of $y$, if $\frac{y}{\alpha}-\beta \leq x \leq \alpha \cdot y + \beta$.

\subparagraph*{Sets.}
For integers \(i, j \in \mathbb{Z}\), we use the interval \(\fragment{i}{j}\) to represent the set \(\{i, \ldots, j\}\), and \(\fragmentco{i}{j}\) to represent the set \(\{i, \ldots, j - 1\}\).
We define \(\fragmentoc{i}{j}\) and \(\fragmentoo{i}{j}\) similarly.
Moreover, for a set $S \subseteq \mathbb{Z}$ and $x \in \mathbb{Z}$ we define $S + x \coloneqq \{s + x \mid s \in S\}$. We say that $\fragmentco{a}{b}$ \emph{precedes} $\fragmentco{c}{d}$ if $b \leq c$.

\subparagraph{Mono-intervals and Bi-intervals. }

Given intervals $I_1, \ldots, I_\ell \subseteq \mathbb{Z}$, we call the ordered sequence $\mI = \langle I_1, I_2, \ldots, I_\ell \rangle$ an \emph{interval collection}. In this work, we focus on two specific cases: when $\ell=1$, we say $\mI$ is a \emph{mono-interval}, and when $\ell=2$, we say $\mI$ is a \emph{bi-interval}. The \emph{size} of an interval collection is the sum of the lengths of its constituent intervals, defined as $\abs{\mI} = \sum_{i=1}^{\ell} \abs{I_i}$. Note that if intervals in the collection overlap, their overlapping portions are counted multiple times in the total size calculation. Lastly, with $U(\mI)$ we denote the union $\bigcup_{i=1}^{\ell} I_i$ and we say $\mI$ is \emph{disjoint} if $I_i$ and $I_j$ are pairwise disjoint for all distinct $i,j \in \fragment{1}{\ell}$. Sometimes, we slightly abuse notation and write $x \in \mI$ instead of $x \in U(\mI)$ for membership queries and other set operations such as taking the intersection.

\subparagraph*{Trees, Forests, and Subforests.}

In this paper, we consider ordered, rooted, labeled forests from an alphabet $\Sigma$.
For a forest, we consider it a tree with a ``virtual'' root. Given a node $v$ in a tree/forest, we use $\ell(v)$ to denote its label. We use $\parent(v)$ to denote its parent.
\biorder*

\begin{figure*}[htbp]
    \centering
    \input{figures/pre/forest_demo}
    \caption{A forest $\bF$, its traversal string, and parenthesis
    representation.}
     \label{fig:forest}
\end{figure*}

If we think of the first occurrence of a node as an opening parenthesis, and the last occurrence of the node as a closing parenthesis, then in the traversal string, pairs of opening-and-closing parentheses are either nested or separate. In fact, the traversal string can be viewed as a balanced string of parentheses. There cannot be the situation where two pairs of parentheses ``overlap'' each other. The following fact captures exactly this structural pattern. Also see \cref{fig:forest} for an illustration.
\begin{fact}\label{fct:non_crossing}
    Let $u, v$ be two different nodes in $\bF$ with $\Left(u) < \Left(v)$. Then we either have $\Left(u) < \Right(u) < \Left(v) < \Right(v)$, or $\Left(u) < \Left(v) < \Right(v) < \Right(u)$. Moreover, if $u$ appears before $v$ in the pre-order traversal but is not an ancestor of $v$, then we have the former case. If $u$ is an ancestor of $v$, then we have the latter case.
\end{fact}

Given an $n$-node forest $\bF$ and an interval collection $\mI$,
we denote with $\bF_\mI$ the pair $(\bF, \mI)$, which we refer to as the \emph{subforest} of $\bF$ indexed by $\mI$; informally, $\bF_\mI$ consists of the characters/parentheses of $\mS(\bF)$ indexed by the intervals of $\mI$, but all notions below are defined in terms of $\bF$ and the positions in $\mI$.
To accommodate intervals extending beyond $\fragmentco{0}{2n}$, we fix a symbol $\$ \notin \Sigma$ and set $\mS(\bF)\position{x} \coloneqq \$$ for every $x \in \mathbb{Z} \setminus \fragmentco{0}{2n}$; we call such $x$ a \emph{padding position}. A padding position is not an occurrence of any node, so $\Other(x)$ is undefined for it and, by \cref{def:TED_on_biintervals}\ref{def:TED_on_biintervals:ii} below, no alignment ever aligns it.

We further say $\bF_\mI$ is a \emph{balanced} subforest, if for each node $v \in \bF$ either both $\Left(v), \Right(v) \in U(\mI)$ or neither.
Given an ordered sequence of integers $s = \langle s_1, \ldots, s_\ell \rangle$,
also called \emph{shifts},
with the same number of integers as intervals in $\mI$,
we define $\bF_{\mI,s}$ to be the \emph{shifted} subforest $\bF_{\mJ}$,
where $\mJ = \langle I_1+s_1, \ldots, I_\ell +s_\ell \rangle$.
Lastly, if $\mI = \biinterval{\fragmentco{i}{j}}{\fragmentco{i'}{j'}}$ is a bi-interval, we define the set of \emph{spine nodes} of $\mI$ to be $\{v \in \bF \mid \Left(v) \in \fragmentco{i}{j}, \Right(v) \in \fragmentco{i'}{j'}\}$. See \cref{fig:subforest} for an example of subforests.

\begin{definition}[Equivalence of subforests]\label{def:equiv}
    Let $\bF$ and $\bG$ be forests, and let $\mI = \langle I_1, \ldots, I_\ell \rangle$ and $\mJ = \langle J_1, \ldots, J_\ell \rangle$ be interval collections with $I_i = \fragmentco{a_i}{b_i}$, $J_i = \fragmentco{a_i'}{b_i'}$, and $|I_i| = |J_i|$ for all $i \in \fragment{1}{\ell}$, and let $s_i \coloneqq a_i' - a_i$ denote the shift from $I_i$ to $J_i$. We write $\bF_{\mI} \equiv \bG_{\mJ}$ if all of the following hold:
    \begin{enumerate}[(i)]
        \item For all $i \in \fragment{1}{\ell}$ and $x \in I_i$, the characters $\mS(\bF)\position{x}$ and $\mS(\bG)\position{x + s_i}$ have the same label and are both first occurrences, both second occurrences, or both padding positions.
        \label{def:equiv:i}
        \item For all $i, j \in \fragment{1}{\ell}$, $x \in I_i$ that is not a padding position, and $x' \in I_j$, we have $x' = \Other(x)$ if and only if $x' + s_j = \Other(x + s_i)$.
        \label{def:equiv:ii}
        \item If $\ell = 2$ and both $I_1$ and $I_2$ are non-empty, then either $s_1 = s_2$, or the intervals $I_1$ and $I_2$ are disjoint, the intervals $J_1$ and $J_2$ are disjoint, and $I_1$ precedes $I_2$ if and only if $J_1$ precedes $J_2$. \qedhere
        \label{def:equiv:iii}
    \end{enumerate}
\end{definition}

\begin{figure*}[htbp]
    \centering

    \begin{subfigure}[t]{0.48\textwidth}
  \centering
  \resizebox{\linewidth}{!}{\input{figures/pre/subforest_mono}}
  \caption{The subforest indexed by the mono-interval $\langle \fragmentco{3}{15} \rangle$. It corresponds to nodes $v_4,v_5,v_6,v_9$. It is not a balanced subforest as $v_2, v_3, v_7, v_8$ appear only once in $\bF_{\langle \fragmentco{3}{15}}\rangle$.}
  \label{fig:subforest:a}
\end{subfigure}\hfill%
\begin{subfigure}[t]{0.48\textwidth}
  \centering
  \resizebox{\linewidth}{!}{\input{figures/pre/subforest_bi}}
  \caption{The subforest indexed by the bi-interval $\langle \fragmentco{3}{8}, \fragmentco{10}{15}\rangle$. It corresponds to nodes $v_4,v_5,v_9$. It is also not a balanced subforest. $v_5$ is a spine node as $\Left(v_5)$ and $\Right(v_5)$ appear in separate parts of the bi-interval.}
  \label{fig:subforest:b}
\end{subfigure}

    \caption{The above figures show instances of unbalanced subforests. When aligning a subforest with another subforest, the unbalanced parentheses automatically get deleted at cost $1/2$ each.}
    \label{fig:subforest}
\end{figure*}

\subparagraph*{Tree Alignments.}

In this paper, we try to visualize $\TED$ from a different perspective than the standard sequence-of-edit-operations perspective. Given a transformation from $\bF$ to $\bF'$, let $T \subseteq \bF \times \bF'$
be the set of all $(v,v')$ such that $v$ is matched or substituted with $v'$ in the least-cost transformation. Now, imagine taking $\mS(\bF)$ and $\mS(\bF')$, and connecting $\Left(v)$ with $\Left(v')$, and connecting $\Right(v)$ with $\Right(v')$ for every $(v,v') \in T$. If $\mA \subseteq \fragmentco{0}{2n} \times \fragmentco{0}{2n}$ is the set of connections that we draw, then it is not difficult to see that $\mA$ satisfies the following (Also see \cref{fig:alignment} for an illustration of a forest alignment.):
\begin{itemize}
    \item for any distinct $(x,y),(x',y')\in \mA$, we have that $x \neq x'$ and $y \neq y'$, i.e., no parenthesis can be matched more than once. Moreover, it cannot be the case where $x < x'$ and $y' < y$. We call this the \emph{non-crossing} property;
    \item if $(x,y) \in \mA$ then also $(\Other(x), \Other(y)) \in \mA$.
\end{itemize}

\begin{figure*}[htbp]
     \centering
    \input{figures/pre/ted_demo_better}
    \caption{An alignment perspective of tree edit operations. Suppose in an optimal edit sequence to transform $\bF$ into $\bG$, we delete the two red nodes, insert the green node, and the remaining nodes either remain unchanged or get substituted (which are indicated by blue dotted lines). The tree alignment corresponding to this edit sequence is shown below. Note that the parentheses corresponding to deleted or inserted nodes remain unaligned. The alignments appear in pairs, and they do not cross each other.}
     \label{fig:alignment}
\end{figure*}

In our paper, it will be important to examine the structure of $\mA$ from a local point of view. Instead of considering the entirety of $\fragmentco{0}{2n}$, we will focus on specific portions, namely, intervals contained within an interval collection. We capture this local view in the notion of \emph{alignment}.

\begin{definition}\label{def:TED_on_biintervals}
   Given two forests $\bF,\bG$, and two interval collections $\mI = \langle I_1, \ldots, I_\ell \rangle$ and $\mJ = \langle J_1, \ldots, J_\ell \rangle$ such that $\mI$ is disjoint, we say the subset $\mA \subseteq \bigcup_i I_i \times J_i$ is \emph{an alignment of $\bF_{\mI}$ onto $\bG_{\mJ}$},
   denoted also by $\mA : \bF_{\mI} \onto \bG_{\mJ}$, if all of the following hold:
   \begin{enumerate}[(i)]
       \item For every $i \in \fragment{1}{\ell}$ the elements in $\mA \cap (I_i \times J_i)$ are non-crossing;
       \label{def:TED_on_biintervals:i}
       \item For every $(x,y) \in \mA$ either $x = \Left(v), y = \Left(v')$ or $x = \Right(v), y = \Right(v')$ for some $v \in \bF, v' \in \bG$; 
       and\label{def:TED_on_biintervals:ii}
       \item  For every $(x, y) \in \mA$ also $(\Other(x), \Other(y)) \in \mA$.
       \label{def:TED_on_biintervals:iii} 
    \end{enumerate}
    We further say that $\mA$ is \emph{$k$-bounded} if also:
    \begin{enumerate}[(i), resume]
        \item For $(x,y) \in \mA$, we have $\abs{x-y} \leq 2k$.
        \label{def:TED_on_biintervals:iv} \qedhere
    \end{enumerate}
\end{definition}
Here are some important observations:
\begin{itemize}
    \item The assumption that $\mI$ is disjoint and \cref{def:TED_on_biintervals}\ref{def:TED_on_biintervals:i} imply that for every $x \in U(\mI)$ there is at most one $y$ such that $(x,y) \in \mA$. (The converse for $y \in U(\mJ)$ is not necessarily true because the $J_i$'s might overlap. A position $y \in J_i \cap J_j$ may be aligned once in $I_i \times J_i$ and once in $I_j \times J_j$.)
    This allows us to see $\mA$ as a (partial) function over $U(\mI)$ that maps an element $x$ to an element $y$ as long as $(x,y) \in \mA$ (unlike for $x$, notice that multiple $y$ might be mapped multiple times). Moreover, for a subset $X \subseteq X \subseteq U(\mI)$, we define $\mA(X) = \{\mA(x) \mid x \in X, \exists y \ (x,y) \in \mA\}$.
    \item \cref{def:TED_on_biintervals}\ref{def:TED_on_biintervals:ii} and \cref{def:TED_on_biintervals}\ref{def:TED_on_biintervals:iii} tell us that the elements of $\mA$ come in pairs of the form $(\Left(v),\Left(v'))$ and $(\Right(v),\Right(v'))$ for some $v \in \bF$ and $v' \in \bG$.
    \item The previous observation, together with \cref{def:TED_on_biintervals}\ref{def:TED_on_biintervals:i}, tells us that for any $i \in \fragment{1}{\ell}$, we have that $\mA \cap (I_i \times J_i)$ mimics a valid way how to draw connections between the substrings of $\mS(\bF)$ and $\mS(\bG)$ indexed by $I_i$ and $J_i$, respectively.
    Given different $i,j \in \fragment{1}{\ell}$,
    we have that $\mA \cap (I_i \times J_i)$ and $\mA \cap (I_j \times J_j)$ behave independently if not for the vertices $v \in \bF$ such that $\Left(v)$ and $\Right(v)$ are contained in different $I_i$'s.

    \item Padding positions in $U(\mI)$ or $U(\mJ)$ are not occurrences of nodes, so by \cref{def:TED_on_biintervals}\ref{def:TED_on_biintervals:ii} they are never aligned; that is, they are always deleted.
\end{itemize}
Although in the above definition we define $\mI$ and $\mJ$ to be interval collections of arbitrary length, we emphasize that in our paper, we will only consider \TED on mono-intervals and bi-intervals.

\subparagraph*{Tree Edit Distance Defined over Tree Alignments.}
Next, we aim to assign \emph{costs} to alignments,
which, for forests $\bF$ and $\bG$ with $n$ and $n'$ nodes, in the specific case $\mI = \langle \fragmentco{0}{2n} \rangle$ and $\mJ = \langle \fragmentco{0}{2n'} \rangle$ allow us to recover $\ted(\bF, \bG) = \ted(\bF_{\langle \fragmentco{0}{2n} \rangle}, \bG_{\langle \fragmentco{0}{2n'} \rangle})$.

\begin{definition}\label{def:cost_on_biintervals}
    Given two forests $\bF,\bG$ with $n$ and $n'$ nodes, respectively, and two interval collections $\mI = \langle I_1, \ldots, I_\ell \rangle$ and $\mJ = \langle J_1, \ldots, J_\ell \rangle$ such that $\mI$ is disjoint, consider an alignment $\mA : \bF_{\mI} \onto \bG_{\mJ}$. Then:
    \begin{itemize}
        \item For $(x,y) \in \mA$ (or alternatively $(\Other(x), \Other(y)) \in \mA$) where $x$ indexes the node $v \in \bF$ and $y$ indexes the node $v' \in \bG$,
        we say that $\mA$ \emph{aligns $x$ with $y$} or \emph{aligns $v$ with $v'$}. Furthermore, if $\ell(v) = \ell(v')$, we say that $\mA$ \emph{matches $x$ and $y$} or \emph{matches $v$ and $v'$}; otherwise, we say \emph{$\mA$ substitutes $x$ with $y$} or \emph{$\mA$ substitutes $v$ with $v'$}.
        \item We say $x \in U(\mI)$ is deleted if there is no $y$ such that $(x,y) \in \mA$.
        \item We say that $y \in J_i$ is deleted w.r.t. $I_i \times J_i$ if there is no $x$ such that $(x,y) \in \mA \cap (I_i \times J_i)$.
    \end{itemize}
    We define the \emph{cost of $\mA$}, denoted by $\cost(\mA)$, to be
    \begin{align*}
    \cost(\mA) \coloneqq
    \tfrac{1}{2} (\abs{\mI} + \abs{\mJ})
    -\sum_{i=1}^{\ell} \left(  \abs{\{x \mid x \in I_i, x \text{ matched by }\mA\}} + \tfrac{1}{2}\abs{\{x \mid x \in I_i, x \text{ substituted by }\mA\}}\right).
\end{align*}
    Lastly, we denote with $\ted(\bF_{\mI}, \bG_{\mJ})$ the minimum cost achieved by any alignment of $\bF_{\mI} \onto \bG_{\mJ}$, and with $\ted^{(k)}(\bF_{\mI}, \bG_{\mJ})$ the minimum cost achieved by any $k$-bounded alignment of $\bF_{\mI} \onto \bG_{\mJ}$.
\end{definition}

In other words, the cost of an alignment can be broken down as follows.
For every deleted $x \in U(\mI)$ and every deleted occurrence $y \in J_i$ (counted once for each $i \in \fragment{1}{\ell}$ with $y \in J_i$, as the $J_i$ may overlap) we charge $1/2$ to $\cost(\mA)$, for every matched $x \in U(\mI)$ and $y \in J_i$ we charge nothing, and for every substituted $x \in U(\mI)$ and $y \in J_i$ we charge $1/4$.

The relation $\equiv$ (\cref{def:equiv}) is designed so that local distances are invariant under it: for $\bG'$ and $\mL = \langle L_1, \ldots, L_{\ell} \rangle$ such that $\bG_{\mL}' \equiv \bG_{\mJ}$, we have $\ted(\bF_{\mI}, \bG_{\mJ}) = \ted(\bF_{\mI}, \bG_{\mL}')$, because replacing every $(x, y) \in \mA \cap (I_i \times J_i)$ by $(x, y + s_i)$, where $s_i$ is the shift from $J_i$ to $L_i$, transforms every alignment $\mA : \bF_\mI \onto \bG_\mJ$ into an alignment $\bF_\mI \onto \bG'_\mL$ of the same cost, and vice versa. Symmetrically, if $\mI$ and $\mI'$ are disjoint and $\bF'_{\mI'} \equiv \bF_{\mI}$, then $\ted(\bF_{\mI}, \bG_{\mJ}) = \ted(\bF'_{\mI'}, \bG_{\mJ})$ by replacing $(x,y)$ with $(x + s_i, y)$.

Further, we remark that when our interval collections $\mI$ and $\mJ$ are mono-intervals spanning entire forests $\bF$ and $\bG$, an optimal alignment $\mA : \bF \onto \bG$ corresponds exactly to a set of connections that we draw for an optimal transformation of $\bF$ into $\bG$. Thus, we obtain:

\begin{remark}
    For forests $\bF$ and $\bG$ with $n$ and $n'$ nodes, respectively, we have 
    \[\ted(\bF_{\langle \fragmentco{0}{2n} \rangle}, \bG_{\langle \fragmentco{0}{2n'} \rangle}) = \ted(\bF, \bG)\].
\end{remark}

Next, we prove some useful properties of tree edit distance on interval collections that we will use in later sections:

\begin{proposition}\label{prop:size_lb}
    Let $\mI = \langle I_1, \ldots, I_\ell \rangle$ and $\mJ = \langle J_1, \ldots, J_\ell \rangle$ be interval collections such that $\mI$ is disjoint. Then, $\frac{1}{2}\abs{\abs{\mI} - \abs{\mJ}} \leq \frac{1}{2}\sum_{i = 1}^{\ell}\abs{\abs{I_i}-\abs{J_i}} \leq \ted(\bF_\mI, \bG_{\mJ}) \leq \ted^{(k)}(\bF_\mI, \bG_{\mJ})$.
\end{proposition}

\begin{proof}
   By the triangle inequality, we have $\abs{\abs{\mI} - \abs{\mJ}}  = \abs{\sum_{i=1 \ldots \ell}\abs{I_i} - \sum_{i = 1}^{\ell}\abs{J_i}}\leq \sum_{i = 1 \ldots \ell}\abs{\abs{I_i}-\abs{J_i}}$. Note that each alignment $\mA$ generates at least $\abs{\abs{I_i} - \abs{J_i}}$ deletions in $\mA \cap (I_i \times J_i)$ (because we can align at most $\min(|I_i|,|J_i|)$ pairs $(x,y) \in I_i \times J_i$). Thus, $\frac{1}{2}\sum_{i = 1}^{\ell}\abs{\abs{I_i}-\abs{J_i}} \leq \ted(\bF_\mI, \bG_{\mJ})$.
\end{proof}

\begin{proposition}\label{prop:triangle_inequality}
    Let $\mI = \langle I_1, \ldots, I_\ell \rangle$, $\mJ = \langle J_1, \ldots, J_\ell \rangle$ and $\mL = \langle L_1, \ldots, L_\ell \rangle$ be interval collections such that $\mI$ is disjoint. Then,
    \begin{align*}
        \ted(\bF_{\mI}, \bG_{\mL}) &\leq \ted(\bF_\mI, \bG_{\mJ}) + \frac{1}{2}(\abs{\mL} - \abs{\mJ}) + 2\sum_{i=1}^{\ell} \min (\abs{J_i \setminus L_{i}}, \abs{I_i}) \\
        &\leq \ted(\bF_\mI, \bG_{\mJ}) + \frac{1}{2}(\abs{\mL} - \abs{\mJ}) + 2\sum_{i=1}^{\ell} \abs{J_i \setminus L_{i}}.
    \end{align*}
    In particular, when $\mJ=\mI+s$ and $\mL=\mI+r$ for two shifts $s = \langle s_1, \ldots s_\ell \rangle$, $r = \langle r_1, \ldots, r_\ell \rangle$, then $\ted(\bF_{\mI}, \bG_{\mI, s}) \leq \ted(\bF_\mI, \bG_{\mI,r}) + 2\sum_{i=1}^{\ell}\abs{s_i - r_i}$, because $|\mL| = |\mJ|$ and $|J_i \setminus L_i| \leq |s_i - r_i|$. Moreover, the inequality also holds when we replace $\ted$ with $\ted^{(k)}$.
\end{proposition}

\begin{proof}
    Consider an optimal alignment $\mA : \bF_\mI \onto \bG_{\mJ}$. To prove \cref{prop:triangle_inequality}, it suffices to construct an alignment $\mA' : \bF_\mI \onto \bG_{\mL}$ such that $\cost(\mA') - \cost(\mA) \leq \frac{1}{2}(\abs{\mL} - \abs{\mJ}) + 2\sum_{i=1}^{\ell} \abs{J_i \setminus L_{i}}$. We obtain $\mA'$ by removing elements from $\mA$: for every $i \in \fragment{1}{\ell}$ and $(x,y) \in \mA$, if $x \in I_i$ but $y \notin L_i$, then we remove $(x,y)$ together with $(\Other(x), \Other(y))$. In short, we remove the pairs that no longer fit the definition of an alignment $\bF_\mI$ onto $\bG_{\mL}$. By the definitions of $\cost(\mA)$ and $\cost(\mA')$, we can bound the change in cost as follows:
    \begin{align*}
    \cost(\mA') - \cost(\mA)
    &\leq \frac{1}{2}(\abs{\mL} - \abs{\mJ}) + \sum_{i=1}^{\ell} \abs{(\mA \setminus \mA') \cap (I_i \times J_i)}  \\
     &\leq \frac{1}{2}(\abs{\mL} - \abs{\mJ}) + 2\sum_{i=1}^{\ell} \min (\abs{J_i \setminus L_{i}}, \abs{I_i}).
    \end{align*}
    Each $(x,y)$ that does not fit into the definition anymore is such that $x \in I_i$ and $y \in J_i \setminus L_i$ for some $i$ (doubling takes into account the associated $(\Other(x), \Other(y))$).
    To see that this last inequality also holds for $\ted^{(k)}$, it suffices to notice that if $\mA$ is an optimal $k$-bounded alignment, then also $\mA'$ is $k$-bounded.
\end{proof}

\subparagraph*{Decompositions of Interval Collections.}
In this work, we estimate $\ted(\bF_{\mI}, \bG_{\mI,s})$ by decomposing a single interval collection $\mI$ into multiple interval collections. In what follows, we define this notion of decomposition.

\begin{definition} \label{def:interval_partitions}
    Let $\mI = \langle I_1, \ldots, I_\ell \rangle$ be an interval collection,
    and let $\mI^1, \ldots, \mI^c$ be $c$ interval collections such that $\mI^i = \langle I^i_1, \ldots, I^{i}_{\ell_i} \rangle$ contains $\ell_i$ intervals.
    We say that $\{\mI^i\}_{i=1}^{c}$ together with a map $\mathsf{pi}$ that assigns to every part $I^i_j$ an index $\mathsf{pi}(i,j) \in \fragment{1}{\ell}$ is a \emph{partition} of $\mI$ if all of the following hold:
    \begin{enumerate}[(i)]
        \item $I^i_j \subseteq I_{\mathsf{pi}(i,j)}$ for all $i$ and $j$;
        \label{def:interval_partitions:iii}
        \item For every $k \in \fragment{1}{\ell}$, the parts $I^i_j$ with $\mathsf{pi}(i,j) = k$ form a partition of $I_k$ in the normal interval sense.
        \label{def:interval_partitions:i}
    \end{enumerate}
    If $\mI$ is disjoint, then $\mathsf{pi}$ is determined by the nonempty parts and we omit it.
\end{definition}

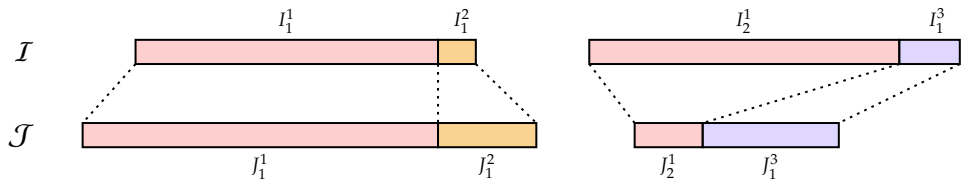
\begin{figure*}[htbp]
     \centering
    \input{figures/pre/ted_decomp_ints}
    \caption{An illustration of the input partitions in \cref{prop:ted_decomp}. Note that the two sides of $\mJ$ do not need to be necessarily disjoint, but the sides of $\mI$ do.
    \cref{prop:ted_decomp} establishes that $\ted(\bF_{\mI}, \bG_{\mJ})$ is upper-bounded by the sum of the tree edit distances of the individual partitions.}
     \label{fig:ted_decomp}
\end{figure*}

\begin{proposition}\label{prop:ted_decomp}
    Let $\bF, \bG$ be two forests, and let $\mI = \langle I_1, \ldots, I_\ell \rangle$ and $\mJ = \langle J_1, \ldots, J_{\ell} \rangle$ such that $\ell \in \{1,2\}$ and $\mI$ is disjoint. Consider partitions $\{\mI^{i} = \langle I^i_1, \ldots, I^{i}_{\ell_i} \rangle \}_{i=1}^c$ and $\{\mJ^{i} = \langle J^i_1, \ldots, J^{i}_{\ell_i} \rangle \}_{i=1}^c$ of $\mI$ and $\mJ$ with the same map $\mathsf{pi}$, such that $\ell_i \in \{1,2\}$ and such that, for every $k \in \fragment{1}{\ell}$, $I^i_j$ is the $q$th part (from the left) assigned to $I_k$ if and only if $J^i_j$ is the $q$th part assigned to $J_k$ (see also \cref{fig:ted_decomp}).
    Then, we have that
    \[
    \ted(\bF_{\mI}, \bG_{\mJ}) \leq \textstyle \sum_{\hi \in \fragment{1}{c}}\ted(\bF_{\mI^{\hat{\imath}}}, \bG_{\mJ^{\hat{\imath}}}) \quad \text{ and } \quad 
    \ted^{(k)}(\bF_{\mI}, \bG_{\mJ}) \leq \textstyle \sum_{\hi \in \fragment{1}{c}}\ted^{(k)}(\bF_{\mI^{\hat{\imath}}}, \bG_{\mJ^{\hat{\imath}}}).
    \]
\end{proposition}

\begin{proof}
    For every $i \in \fragment{1}{c}$, let $\mA^{i} : \bF_{\mI^{i}} \onto \bG_{\mJ^{i}}$ be an optimal alignment. We want to argue that $\mA \coloneqq \bigcup_{i=1}^c \mA^{i}$ satisfies the definitions of an alignment from $\bF_{\mI}$ to $\bG_{\mJ}$. 
    Clearly, Properties~\ref{def:TED_on_biintervals:ii} and~\ref{def:TED_on_biintervals:iii} hold.
    For Property~\ref{def:TED_on_biintervals:i}, we need to combine Property~\ref{def:TED_on_biintervals:i} of the $\mA^i$s and the fact that the two partitions $\{\mI^{i}\}_{i=1}^c$ and $\{\mJ^{i}\}_{i=1}^c$ preserve orderings.
    Moreover, since $\{\mI^{i}\}_{i=1}^c$ and $\{\mJ^{i}\}_{i=1}^c$ are both partitions and $\mI$ is disjoint, we have that $\cost(\mA) = \sum_{i=1}^c \cost(\mA^{i})$. Thus, $\ted(\bF_\mI, \bG_{\mJ}) \leq \cost(\mA) = \sum_{i=1}^c \cost(\mA^{i}) = \sum_{i=1}^{c} \ted(\bF_{\mI^{i}}, \bG_{\mJ^{i}})$, which proves the first inequality.
    For the second we observe that if all $\mA^{i}$s are $k$-bounded, then so is $\mA$.
\end{proof}

\subparagraph*{Augmented Alignments.}

It is often useful to decompose an optimal alignment $\mA$ from $\bF_{\mI}$ to $\bG_{\mJ}$ into multiple components so that we can study the behavior of the optimal alignment locally. For a position $x \in U(\mI)$ that is matched or substituted by $\mA$, it is clear that $\mA(x)$ is defined. However, if $x$ is deleted by $\mA$, then $\mA(x)$ is no longer well-defined. To remedy this issue, we introduce an augmentation $\pi_{\mA}$ of $\mA$.

\begin{definition}[Augmented alignment]\label{def:ghost_mappings}
    Consider two $n$-node forests $\bF, \bG$ and two interval collections $\mI = \langle I_1, \ldots, I_\ell \rangle$, $\mJ = \langle J_1, \ldots, J_\ell \rangle$, where $I_i = \fragmentco{a_i}{b_i}$ with $a_i < b_i$ (that is, all intervals of $\mI$ are non-empty) and $J_i = \fragmentco{a_i'}{b_i'}$.
    Moreover, let us fix an arbitrary optimal alignment $\mA : \bF_{\mI} \onto \bG_{\mJ}$ (which is possibly also $k$-bounded).

    Based on $\mA$, for each $i \in \fragment{1}{\ell}$, we define a mapping $\pi_{\mA,i}: \fragment{a_i}{b_i} \rightarrow \fragment{a_i'}{b_i'}$ as follows.
    For $x \in \fragment{a_i}{b_i}$,
    \begin{itemize}
        \item If $x = a_i$, then set $x_{\mathrm{pre}} = x_{\mathrm{suc}} = a_i$ and $y_{\mathrm{pre}} = y_{\mathrm{suc}} = a_i'$.
        \item If $x = b_i$, then set $x_{\mathrm{pre}} = x_{\mathrm{suc}} = b_i$ and $y_{\mathrm{pre}} = y_{\mathrm{suc}} = b_i'$.
        \item Otherwise, let $x_{\mathrm{suc}}$ be the smallest position $x' \in \fragmentco{x}{b_i}$ for which there is $y'$ such that $(x',y') \in \mA$ and set $y_{\mathrm{suc}} = y'$.
    If no such $x'$ exists, set $x_{\mathrm{suc}} = b_i$ and $y_{\mathrm{suc}} = b_i'$.
    Similarly, let $x_{\mathrm{pre}}$ be the largest position $x' \in \fragmentco{a_i}{x}$ for which there is $y'$ such that $(x',y') \in \mA$ and set $x_{\mathrm{pre}} = x'+1$ and $y_{\mathrm{pre}} = y' + 1$.
    If no such $x'$ exists, set $x_{\mathrm{pre}} = a_i$ and $y_{\mathrm{pre}} = a_i'$.
    \end{itemize}

    Now, define $\pi_{\mA,i}(x)$ as
    \[
        \pi_{\mA,i}(x) \coloneqq \max(y_{\mathrm{pre}}, y_{\mathrm{suc}} - (x_{\mathrm{suc}} - x)).
    \]

\end{definition}

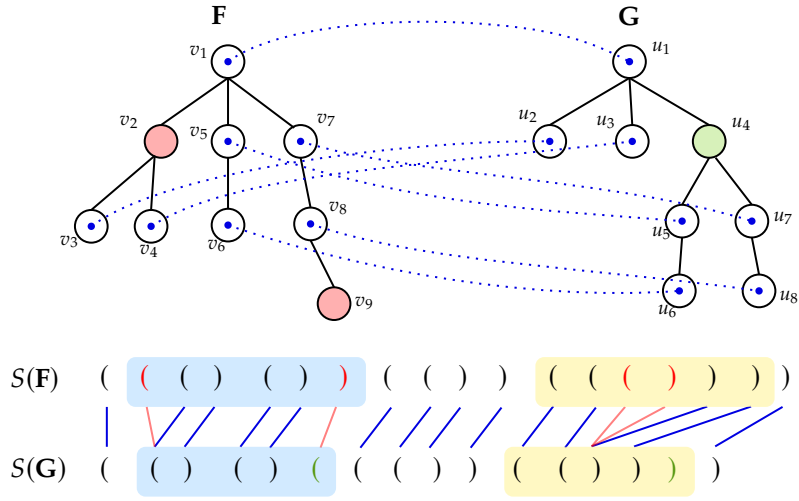
\begin{figure*}[htbp]
     \centering
    \input{figures/pre/ghost_mappings}
    \caption{The augmentation $\pi_{\mA}$ of an alignment $\mA : \bF \onto \bG$. The blue lines indicate existing alignments in $\mA$. Note that $v_2, v_9,$ and $u_4$ are unaligned in $\mA$, but in the augmented alignment, we align them according to $\pi_{\mA}(\cdot)$, as indicated by red lines. The blue blocks show the balanced subforest $\bF_{\langle \fragmentco{1}{7}\rangle}$ and its image $\bG_{\langle\fragmentco{1}{6}\rangle}$ under $\mA$, which is not balanced. Note that the restriction of $\mA$ to $\bF_{\langle \fragmentco{1}{7}\rangle}$ and $\bG_{\langle\fragmentco{1}{6}\rangle}$ aligns $v_3$ to $u_2$, and $v_4$ to $u_3$, which is indeed a locally optimal alignment. The yellow blocks illustrate another such instance.}
     \label{fig:ghost}
\end{figure*}

Observe that the augmented alignment $\pi_{\mA,i}$ always maps $a_i$ to $a_i'$ and $b_i$ to $b_i'$. For every $(x,y) \in \mA$ with $x \in \fragmentoo{a_i}{b_i}$ (equivalently $x \neq a_i$), we have $\pi_{\mA, i}(x) =y$.

The following proposition states that the range of the augmented alignment always bounds the range of the actual alignment.
\begin{proposition}\label{clm:TD_ub_subs_ted}
    Let $\bF, \bG, \mI, \mJ, \mA$ be as in \cref{def:ghost_mappings}.
    For any $i \in \fragment{1}{\ell}$, consider two positions $x,x' \in \fragment{a_i}{b_i}$, and take the set of indices $S \subseteq \fragmentco{x}{x'}$ that are either substituted or matched by $\mA$ . Then, $\{\mA(s) \mid s \in S\} \subseteq \fragmentco{\pi_{\mA, i}(x)}{\pi_{\mA,i}(x')}$. Moreover, $\pi_{\mA,i}$ is nondecreasing on $\fragment{a_i}{b_i}$.
\end{proposition}
\begin{proof}
    If $S$ is empty, then the statement is trivial. Now, consider $\hat{x} \in \fragmentco{x}{x'}$ such that there exists a $\hat{y}$ with $(\hat{x},\hat{y}) \in \mA$. We show $\pi_{\mA,i}(x) \leq \hat{y} < \pi_{\mA,i}(x')$. If $x = a_i$, then $\pi_{\mA,i}(x) = a_i' \leq \hat{y}$. Otherwise, let $x_{\mathrm{pre}}, y_{\mathrm{pre}}, x_{\mathrm{suc}}, y_{\mathrm{suc}}$ be defined for $x$ as in \cref{def:ghost_mappings}; since $\hat{x} \in \fragmentco{x}{b_i}$ is aligned, we have $x_{\mathrm{suc}} \leq \hat{x}$, and the non-crossing property gives $y_{\mathrm{pre}} \leq y_{\mathrm{suc}} = \mA(x_{\mathrm{suc}}) \leq \hat{y}$, so $\pi_{\mA,i}(x) \leq \max(y_{\mathrm{pre}}, y_{\mathrm{suc}}) \leq \hat{y}$. If $x' = b_i$, then $\pi_{\mA,i}(x') = b_i' > \hat{y}$. Otherwise, the largest aligned position in $\fragmentco{a_i}{x'}$ is some $x'' \geq \hat{x}$, so $y'_{\mathrm{pre}} = \mA(x'') + 1 \geq \hat{y} + 1$ by the non-crossing property, and $\pi_{\mA,i}(x') \geq y'_{\mathrm{pre}} > \hat{y}$.

    For monotonicity, let $x < x'$. If some $\hat{x} \in \fragmentco{x}{x'}$ is aligned, then $\pi_{\mA,i}(x) \leq \mA(\hat{x}) < \pi_{\mA,i}(x')$ by the above. Otherwise, $x$ and $x'$ share the same $x_{\mathrm{pre}}, y_{\mathrm{pre}}, x_{\mathrm{suc}}, y_{\mathrm{suc}}$, and $\max(y_{\mathrm{pre}}, y_{\mathrm{suc}} - (x_{\mathrm{suc}} - x))$ is nondecreasing in $x$; the cases $x = a_i$ and $x' = b_i$ follow from $a_i' \leq \pi_{\mA,i} \leq b_i'$.
\end{proof}
In particular, $\pi_{\mA}(\mL)$ in \cref{def:align_interval} below is an interval collection, and the images of disjoint intervals of $\mL$ inside the same $I_i$ are disjoint and appear in the same order.

\begin{proposition}\label{clm:ghost_kbounded}
    Let $\bF, \bG, \mI, \mJ, \mA$ be as in \cref{def:ghost_mappings}.
    Suppose $\mA$ is a $k$-bounded alignment.
    For any $i \in \fragment{1}{\ell}$, consider a position $x \in \fragment{a_i}{b_i}$. Then,
    we have $|x - \pi_{\mA,i}(x)| \leq \max \{ \abs{a_i - a_i'}, \abs{b_i - b_i'}, 2k\}$.
\end{proposition}
\begin{proof}
    First, we notice that $|x_{\mathrm{suc}} - y_{\mathrm{suc}}| \leq \max \{ |b_i - b_i'|, 2k\}$ and $|x_{\mathrm{pre}} - y_{\mathrm{pre}}| \leq \max \{ |a_i - a_i'|, 2k\}$ because $\mA$ is $k$-bounded.
    This means that
    \[
        \pi_{\mA,i}(x) - x = \max(y_{\mathrm{pre}}, y_{\mathrm{suc}} - (x_{\mathrm{suc}} - x)) - x
        \geq y_{\mathrm{suc}} - x_{\mathrm{suc}} \geq -\max \{ |b_i - b_i'|, 2k\}.
    \]
    Moreover, from $x \geq x_{\mathrm{pre}}$ follows
    \[
        \pi_{\mA,i}(x) - x = \max(y_{\mathrm{pre}}, y_{\mathrm{suc}} - (x_{\mathrm{suc}} - x)) - x
        \leq \max(y_{\mathrm{pre}} - x_{\mathrm{pre}}, y_{\mathrm{suc}} - x_{\mathrm{suc}})
        \leq \max \{ |a_i - a_i'|, |b_i - b_i'|, 2k\}.
    \]
    We conclude $|x - \pi_{\mA,i}(x)| \leq \max \{ |a_i - a_i'|, |b_i - b_i'|, 2k\}$.
\end{proof}

\begin{definition}\label{def:align_interval}
    Let $\bF, \bG, \mI, \mJ, \mA$ be as in \cref{def:ghost_mappings}.
    We are given a disjoint interval collection $\mL = \langle \fragmentco{c_1}{d_1}, \ldots, \fragmentco{c_{\ell'}}{d_{\ell'}} \rangle$
    such that for every non-padding position $x \in U(\mL)$, if $\Other(x) \in U(\mI)$, then $\Other(x) \in U(\mL)$, together with a specified parent-index map $i(j)$ satisfying 
    \[
    a_{i(j)} \leq c_j \leq d_j \leq b_{i(j)}.
    \]
    The map is also defined for empty intervals.
    We define the \emph{image} of $\mL$ under $\mA$ as
    \[
        \pi_{\mA}(\mL) \coloneqq \langle \fragmentco{\pi_{\mA, i(1)}(c_1)}{\pi_{\mA, i(1)}(d_1)}, \ldots, \fragmentco{\pi_{\mA, i(\ell')}(c_{\ell'})}{\pi_{\mA, i(\ell')}(d_{\ell'})} \rangle. \qedhere
    \]
\end{definition}

\begin{proposition}\label{clm:sub_opt}
    Let $\bF, \bG, \mI, \mJ, \mA$ be as in \cref{def:ghost_mappings}.
    Moreover, let $\mL$ be as in \cref{def:align_interval},
    then $\mA_{\mL} \coloneqq \mA \cap (U(\mL) \times U(\pi_{\mA}(\mL)))$ that aligns $\bF_{\mL}$ onto $\bG_{\pi_{\mA}(\mL)}$ 
    is optimal. For the $k$-bounded variant, choose $\mathcal A$
optimal among $k$-bounded alignments instead; then
$\mathcal A_{\mathcal L}$ is optimal among $k$-bounded
alignments as well.
\end{proposition}
\begin{proof}
    For the sake of brevity, set $\mJ' \coloneqq \pi_{\mA}(\mL)$.
    For the sake of contradiction, assume there is $\bar{\mA}_{\mL} : \bF_{\mL} \onto \bG_{\mJ'}$
    such that $\cost(\bar{\mA}_{\mL}) < \cost(\mA_{\mL})$.
    Let $\bar{\mA} \coloneqq (\mA \setminus \mA_{\mL}) \cup \bar{\mA}_{\mL}$.
    It suffices to show that $\bar{\mA}$ is a valid alignment of $\bF_{\mI}$ onto $\bG_{\mJ}$
    of cost $\cost(\mA) - \cost(\mA_{\mL}) + \cost(\bar{\mA}_{\mL})$, as we would obtain a contradiction with the optimality of $\mA$.
    To see why $\bar{\mA}$ is a valid alignment, observe that \cref{def:TED_on_biintervals}\ref{def:TED_on_biintervals:i} holds because both $\bar{\mA_{\mL}}$ and $\mA \setminus \mA_{\mL}$ are non-crossing within themselves by definition, and are non-crossing with each other due to \cref{clm:TD_ub_subs_ted}. \cref{def:TED_on_biintervals}\ref{def:TED_on_biintervals:ii} and \ref{def:TED_on_biintervals:iii} still hold by definition of alignments. Thus, $\cost(\mA_{\mL}) = \ted(\bF_{\mL}, \bG_{\mJ'})$. Note that the above exchange argument also holds when $\mA$ is chosen to be optimal among $k$-bounded alignments.
\end{proof}

\begin{proposition}\label{clm:TD_ub_sum_ted}
    Let $\bF, \bG, \mI, \mJ,$ and $\mA$ be as in \cref{def:ghost_mappings}, and let $\{\mI^{\hat{\imath}}\}_{\hat{\imath}=1}^c$ be a partition of $\mI$ such that 
    whenever both occurrences of a node $v$ belong to $U(\mI)$, they both belong to the same piece $U(\mI^{\hi})$ (note that a sufficient condition for this is that $\bF_{\mI^{\hi}}$ is a balanced forest for every $\hi \in \fragment{1}{c}$).

    Let $\mJ^{\hi}$ be the image of $\mI^{\hi}$ under $\mA$, i.e., $\mJ^{\hi} \coloneqq \pi_{\mA}(\mI^{\hi})$. Then, $\sum_{\hi=1}^{c} \ted(\bF_{\mI^{\hi}}, \bG_{\mJ^{\hi}}) = \ted(\bF_{\mI}, \bG_{\mJ})$. Furthermore, if $\mA$ is optimal among $k$-bounded alignments, then $\sum_{\hi=1}^{c} \ted^{(k)}(\bF_{\mI^{\hi}}, \bG_{\mJ^{\hi}}) = \ted^{(k)}(\bF_{\mI}, \bG_{\mJ})$.
\end{proposition}
\begin{proof}
    We observe that if $\{\mI^{\hi}\}_{\hi=1}^c$ is a partition of $\mI$, then $\{\mJ^{\hi}\}_{\hi=1}^c$ is also a partition of $\mJ$. Let $\mA_{\mI^{\hi}}$ be defined as $\mA \cap (U(\mI^{\hi}) \times U(\mJ^{\hi}))$. By our assumption that there is no node $v \in \bF_{\mI}$ with $\Left(v)$ and $\Right(v)$ contained in different parts $\mI^{\hi}$, it follows that the sets $\{\mA_{\mI^{\hi}}\}_{\hi=1}^c$ partition $\mA$. Therefore, from \cref{clm:sub_opt}, we conclude that $\ted(\bF_{\mI}, \bG_{\mJ}) = \cost(\mA) = \sum_{\hi=1}^c \cost(\mA_{\mI^{\hi}}) = \sum_{\hi=1}^c \ted(\bF_{\mI^{\hi}}, \bG_{\mJ^{\hi}})$.
    The same proof works for $k$-bounded alignments as well.
\end{proof}

\begin{proposition}\label{clm:TD_ub_diff_ted}
    Consider any disjoint interval collection $\mI$ over $\fragmentco{0}{2n}$ such that $\bF_{\mI}$ is a balanced subforest. Take an optimal alignment $\mA : \bF \onto \bG$. Let $\fragmentco{a_i}{b_i}$ be an interval of $\mI$, and let $a \leq b$ be two positions in $\fragment{a_i}{b_i}$. 
    Then we have $\abs{(a - \pi'(a)) - (b - \pi'(b))} \leq 2 \cdot \ted(\bF_{\mI}, \bG_{\mJ'})$, where $\mJ'$ is the image of $\mI$ under $\mA$, and $\pi'(c) \in \{\pi_{\mA}(c), \mA(c)\}$ is chosen independently for $c \in \{a, b\}$, with $\mA(c)$ allowed only if $c$ is aligned by $\mA$. (Since $\mA(c) = \pi_{\mA}(c)$ for aligned $c \neq 0$, the choice only matters for $c = 0$, where $\pi_{\mA}(0) = 0$.)
    
    In particular, by choosing $\mI = \langle \fragmentco{0}{2n}\rangle$ and $a = 0$ (which implies $\pi_{\mA}(a) = 0$), we have that $\abs{b - \pi'(b)} \leq 2 \cdot \ted(\bF, \bG) = 2 \cdot \cost(\mA)$ for any $b \in \fragment{0}{2n}$. Crucially, this implies that any alignment $\mA : \bF \onto \bG$ of cost at most $k$ is $k$-bounded. 
    In other words, $\ted(\bF,\bG) = \ted^{(k)}(\bF,\bG)$, provided that $\ted(\bF, \bG) \leq k$.
\end{proposition}
\begin{proof}
    Let $\mA_{\mI} \coloneqq \mA \cap (U(\mI) \times U(\mJ'))$. By \cref{clm:sub_opt} with $\mL = \mI$, $\mA_{\mI}$ is an optimal alignment of $\bF_{\mI}$ onto $\bG_{\mJ'}$, so $\cost(\mA_{\mI}) = \ted(\bF_{\mI}, \bG_{\mJ'})$.
    We may assume $\pi'(a) \leq \pi'(b)$: for $a < b$ this follows, depending on the choices of $\pi'$, from \cref{clm:TD_ub_subs_ted}, the monotonicity of $\pi_{\mA}$, or the non-crossing property; for $a = b$ the bound is symmetric in $\pi'(a)$ and $\pi'(b)$.
    By the same facts, $\mA_{\mI}$ aligns the positions in $\fragmentco{a}{b}$ only into $\fragmentco{\pi'(a)}{\pi'(b)} \subseteq J_i'$, and no other position of $U(\mI)$ into $\fragmentco{\pi'(a)}{\pi'(b)}$.
    Hence, by the pigeonhole principle, at least $\abs{(b - a) - (\pi'(b) - \pi'(a))} = \abs{(a - \pi'(a)) - (b - \pi'(b))}$ parentheses of $\bF_{\langle \fragmentco{a}{b} \rangle}$ or $\bG_{\langle \fragmentco{\pi'(a)}{\pi'(b)} \rangle}$ are deleted by $\mA_{\mI}$, which implies $\ted(\bF_{\mI}, \bG_{\mJ'}) = \cost(\mA_{\mI}) \geq \frac{1}{2} \abs{(a - \pi'(a)) - (b - \pi'(b))}$.
\end{proof}

%% file: figures/pre/forest_demo.tex
\tikzset{every picture/.style={line width=0.75pt}} 

\begin{tikzpicture}[x=0.75pt,y=0.75pt,yscale=-1,xscale=1]

\draw   (193.6,41.8) .. controls (193.6,37.27) and (197.27,33.6) .. (201.8,33.6) .. controls (206.33,33.6) and (210,37.27) .. (210,41.8) .. controls (210,46.33) and (206.33,50) .. (201.8,50) .. controls (197.27,50) and (193.6,46.33) .. (193.6,41.8) -- cycle ;
\draw   (160,81.8) .. controls (160,77.27) and (163.67,73.6) .. (168.2,73.6) .. controls (172.73,73.6) and (176.4,77.27) .. (176.4,81.8) .. controls (176.4,86.33) and (172.73,90) .. (168.2,90) .. controls (163.67,90) and (160,86.33) .. (160,81.8) -- cycle ;
\draw   (193.6,81.8) .. controls (193.6,77.27) and (197.27,73.6) .. (201.8,73.6) .. controls (206.33,73.6) and (210,77.27) .. (210,81.8) .. controls (210,86.33) and (206.33,90) .. (201.8,90) .. controls (197.27,90) and (193.6,86.33) .. (193.6,81.8) -- cycle ;
\draw   (230,81.8) .. controls (230,77.27) and (233.67,73.6) .. (238.2,73.6) .. controls (242.73,73.6) and (246.4,77.27) .. (246.4,81.8) .. controls (246.4,86.33) and (242.73,90) .. (238.2,90) .. controls (233.67,90) and (230,86.33) .. (230,81.8) -- cycle ;
\draw   (155,124.4) .. controls (155,119.87) and (158.67,116.2) .. (163.2,116.2) .. controls (167.73,116.2) and (171.4,119.87) .. (171.4,124.4) .. controls (171.4,128.93) and (167.73,132.6) .. (163.2,132.6) .. controls (158.67,132.6) and (155,128.93) .. (155,124.4) -- cycle ;
\draw   (125,124.4) .. controls (125,119.87) and (128.67,116.2) .. (133.2,116.2) .. controls (137.73,116.2) and (141.4,119.87) .. (141.4,124.4) .. controls (141.4,128.93) and (137.73,132.6) .. (133.2,132.6) .. controls (128.67,132.6) and (125,128.93) .. (125,124.4) -- cycle ;
\draw   (193.6,123.8) .. controls (193.6,119.27) and (197.27,115.6) .. (201.8,115.6) .. controls (206.33,115.6) and (210,119.27) .. (210,123.8) .. controls (210,128.33) and (206.33,132) .. (201.8,132) .. controls (197.27,132) and (193.6,128.33) .. (193.6,123.8) -- cycle ;
\draw   (235,123.2) .. controls (235,118.67) and (238.67,115) .. (243.2,115) .. controls (247.73,115) and (251.4,118.67) .. (251.4,123.2) .. controls (251.4,127.73) and (247.73,131.4) .. (243.2,131.4) .. controls (238.67,131.4) and (235,127.73) .. (235,123.2) -- cycle ;
\draw   (246.8,163.2) .. controls (246.8,158.67) and (250.47,155) .. (255,155) .. controls (259.53,155) and (263.2,158.67) .. (263.2,163.2) .. controls (263.2,167.73) and (259.53,171.4) .. (255,171.4) .. controls (250.47,171.4) and (246.8,167.73) .. (246.8,163.2) -- cycle ;
\draw    (201.8,50) -- (201.8,73.6) ;
\draw    (201.8,90) -- (201.8,115.6) ;
\draw    (201.8,50) -- (168.2,73.6) ;
\draw    (165,89) -- (163.2,116.2) ;
\draw    (165,89) -- (133.2,116.2) ;
\draw    (201.8,50) -- (235,75) ;
\draw    (238.2,90) -- (243.2,115) ;
\draw    (243.2,131.4) -- (255,155) ;

\draw (191,12.4) node [anchor=north west][inner sep=0.75pt]    {$v_{1}$};
\draw (146,60.4) node [anchor=north west][inner sep=0.75pt]    {$v_{2}$};
\draw (111,125.4) node [anchor=north west][inner sep=0.75pt]    {$v_{3}$};
\draw (156,132.4) node [anchor=north west][inner sep=0.75pt]    {$v_{4}$};
\draw (212,85.2) node [anchor=north west][inner sep=0.75pt]    {$v_{5}$};
\draw (203.8,135.4) node [anchor=north west][inner sep=0.75pt]    {$v_{6}$};
\draw (246,72.4) node [anchor=north west][inner sep=0.75pt]    {$v_{7}$};
\draw (251,112.4) node [anchor=north west][inner sep=0.75pt]    {$v_{8}$};
\draw (266,157.4) node [anchor=north west][inner sep=0.75pt]    {$v_{9}$};
\draw (26,190.4) node [anchor=north west][inner sep=0.75pt]    {$\mathcal{S}(\mathbf{F}) =v_{1} \ v_{2} \ v_{3} \ v_{3} \ v_{4} \ v_{4} \ v_{2} \ v_{5} \ v_{6} \ v_{6} \ v_{5} \ v_{7} \ v_{8} \ v_{9} \ v_{9} \ v_{8} \ v_{7} \ v_{1}$};
\draw (75,222.4) node [anchor=north west][inner sep=0.75pt]    {$\ ( \ \ \ ( \ \ \ \ ( \ \ \ \ ) \ \ ( \ \ \ \ \ ) \ \ \ ) \ \ ( \ \ \ \ ( \ \ \ \ \ ) \ \ \ ) \ \ \ ( \ \ \ ( \ \ \ ( \ \ \ \ \ ) \ \ \ ) \ \ \ ) \ \ \ \ )$};

\end{tikzpicture}

%% file: figures/pre/ted_demo_better.tex
\tikzset{every picture/.style={line width=0.75pt}} 

\begin{tikzpicture}[x=0.75pt,y=0.75pt,yscale=-1,xscale=1]

\draw   (193.6,41.8) .. controls (193.6,37.27) and (197.27,33.6) .. (201.8,33.6) .. controls (206.33,33.6) and (210,37.27) .. (210,41.8) .. controls (210,46.33) and (206.33,50) .. (201.8,50) .. controls (197.27,50) and (193.6,46.33) .. (193.6,41.8) -- cycle ;
\draw  [fill={rgb, 255:red, 255; green, 0; blue, 0 }  ,fill opacity=0.31 ] (160,81.8) .. controls (160,77.27) and (163.67,73.6) .. (168.2,73.6) .. controls (172.73,73.6) and (176.4,77.27) .. (176.4,81.8) .. controls (176.4,86.33) and (172.73,90) .. (168.2,90) .. controls (163.67,90) and (160,86.33) .. (160,81.8) -- cycle ;
\draw   (193.6,81.8) .. controls (193.6,77.27) and (197.27,73.6) .. (201.8,73.6) .. controls (206.33,73.6) and (210,77.27) .. (210,81.8) .. controls (210,86.33) and (206.33,90) .. (201.8,90) .. controls (197.27,90) and (193.6,86.33) .. (193.6,81.8) -- cycle ;
\draw   (230,81.8) .. controls (230,77.27) and (233.67,73.6) .. (238.2,73.6) .. controls (242.73,73.6) and (246.4,77.27) .. (246.4,81.8) .. controls (246.4,86.33) and (242.73,90) .. (238.2,90) .. controls (233.67,90) and (230,86.33) .. (230,81.8) -- cycle ;
\draw   (155,124.4) .. controls (155,119.87) and (158.67,116.2) .. (163.2,116.2) .. controls (167.73,116.2) and (171.4,119.87) .. (171.4,124.4) .. controls (171.4,128.93) and (167.73,132.6) .. (163.2,132.6) .. controls (158.67,132.6) and (155,128.93) .. (155,124.4) -- cycle ;
\draw   (125,124.4) .. controls (125,119.87) and (128.67,116.2) .. (133.2,116.2) .. controls (137.73,116.2) and (141.4,119.87) .. (141.4,124.4) .. controls (141.4,128.93) and (137.73,132.6) .. (133.2,132.6) .. controls (128.67,132.6) and (125,128.93) .. (125,124.4) -- cycle ;
\draw   (193.6,123.8) .. controls (193.6,119.27) and (197.27,115.6) .. (201.8,115.6) .. controls (206.33,115.6) and (210,119.27) .. (210,123.8) .. controls (210,128.33) and (206.33,132) .. (201.8,132) .. controls (197.27,132) and (193.6,128.33) .. (193.6,123.8) -- cycle ;
\draw   (235,123.2) .. controls (235,118.67) and (238.67,115) .. (243.2,115) .. controls (247.73,115) and (251.4,118.67) .. (251.4,123.2) .. controls (251.4,127.73) and (247.73,131.4) .. (243.2,131.4) .. controls (238.67,131.4) and (235,127.73) .. (235,123.2) -- cycle ;
\draw  [fill={rgb, 255:red, 255; green, 0; blue, 0 }  ,fill opacity=0.31 ] (246.8,163.2) .. controls (246.8,158.67) and (250.47,155) .. (255,155) .. controls (259.53,155) and (263.2,158.67) .. (263.2,163.2) .. controls (263.2,167.73) and (259.53,171.4) .. (255,171.4) .. controls (250.47,171.4) and (246.8,167.73) .. (246.8,163.2) -- cycle ;
\draw    (201.8,50) -- (201.8,73.6) ;
\draw    (201.8,90) -- (201.8,115.6) ;
\draw    (201.8,50) -- (168.2,73.6) ;
\draw    (165,89) -- (163.2,116.2) ;
\draw    (165,89) -- (133.2,116.2) ;
\draw    (201.8,50) -- (235,75) ;
\draw    (238.2,90) -- (243.2,115) ;
\draw    (243.2,131.4) -- (255,155) ;
\draw   (395,41.8) .. controls (395,37.27) and (398.67,33.6) .. (403.2,33.6) .. controls (407.73,33.6) and (411.4,37.27) .. (411.4,41.8) .. controls (411.4,46.33) and (407.73,50) .. (403.2,50) .. controls (398.67,50) and (395,46.33) .. (395,41.8) -- cycle ;
\draw   (355,81.8) .. controls (355,77.27) and (358.67,73.6) .. (363.2,73.6) .. controls (367.73,73.6) and (371.4,77.27) .. (371.4,81.8) .. controls (371.4,86.33) and (367.73,90) .. (363.2,90) .. controls (358.67,90) and (355,86.33) .. (355,81.8) -- cycle ;
\draw   (396.4,81.8) .. controls (396.4,77.27) and (400.07,73.6) .. (404.6,73.6) .. controls (409.13,73.6) and (412.8,77.27) .. (412.8,81.8) .. controls (412.8,86.33) and (409.13,90) .. (404.6,90) .. controls (400.07,90) and (396.4,86.33) .. (396.4,81.8) -- cycle ;
\draw  [fill={rgb, 255:red, 126; green, 211; blue, 33 }  ,fill opacity=0.31 ] (435,81.8) .. controls (435,77.27) and (438.67,73.6) .. (443.2,73.6) .. controls (447.73,73.6) and (451.4,77.27) .. (451.4,81.8) .. controls (451.4,86.33) and (447.73,90) .. (443.2,90) .. controls (438.67,90) and (435,86.33) .. (435,81.8) -- cycle ;
\draw   (421.4,121.8) .. controls (421.4,117.27) and (425.07,113.6) .. (429.6,113.6) .. controls (434.13,113.6) and (437.8,117.27) .. (437.8,121.8) .. controls (437.8,126.33) and (434.13,130) .. (429.6,130) .. controls (425.07,130) and (421.4,126.33) .. (421.4,121.8) -- cycle ;
\draw   (420,156.8) .. controls (420,152.27) and (423.67,148.6) .. (428.2,148.6) .. controls (432.73,148.6) and (436.4,152.27) .. (436.4,156.8) .. controls (436.4,161.33) and (432.73,165) .. (428.2,165) .. controls (423.67,165) and (420,161.33) .. (420,156.8) -- cycle ;
\draw   (456.4,121.8) .. controls (456.4,117.27) and (460.07,113.6) .. (464.6,113.6) .. controls (469.13,113.6) and (472.8,117.27) .. (472.8,121.8) .. controls (472.8,126.33) and (469.13,130) .. (464.6,130) .. controls (460.07,130) and (456.4,126.33) .. (456.4,121.8) -- cycle ;
\draw   (460,156.8) .. controls (460,152.27) and (463.67,148.6) .. (468.2,148.6) .. controls (472.73,148.6) and (476.4,152.27) .. (476.4,156.8) .. controls (476.4,161.33) and (472.73,165) .. (468.2,165) .. controls (463.67,165) and (460,161.33) .. (460,156.8) -- cycle ;
\draw    (403.2,50) -- (363.2,73.6) ;
\draw    (403.2,50) -- (404.6,73.6) ;
\draw    (403.2,50) -- (443.2,73.6) ;
\draw    (446.4,90) -- (464.6,113.6) ;
\draw    (443.2,90) -- (429.6,113.6) ;
\draw    (429.6,130) -- (428.2,148.6) ;
\draw    (464.6,130) -- (468.2,148.6) ;
\draw [color={rgb, 255:red, 0; green, 0; blue, 225 }  ,draw opacity=1 ] [dash pattern={on 0.84pt off 2.51pt}]  (201.8,41.8) .. controls (223.82,25.29) and (277.48,19.92) .. (321.25,22.31) .. controls (365.02,24.69) and (383.42,31.37) .. (403.2,41.8) ;
\draw [shift={(403.2,41.8)}, rotate = 27.8] [color={rgb, 255:red, 0; green, 0; blue, 225 }  ,draw opacity=1 ][fill={rgb, 255:red, 0; green, 0; blue, 225 }  ,fill opacity=1 ][line width=0.75]      (0, 0) circle [x radius= 1.34, y radius= 1.34]   ;
\draw [shift={(201.8,41.8)}, rotate = 323.13] [color={rgb, 255:red, 0; green, 0; blue, 225 }  ,draw opacity=1 ][fill={rgb, 255:red, 0; green, 0; blue, 225 }  ,fill opacity=1 ][line width=0.75]      (0, 0) circle [x radius= 1.34, y radius= 1.34]   ;
\draw [color={rgb, 255:red, 0; green, 0; blue, 225 }  ,draw opacity=1 ] [dash pattern={on 0.84pt off 2.51pt}]  (133.2,124.4) .. controls (173.2,94.4) and (303.2,80.6) .. (363.2,81.8) ;
\draw [shift={(363.2,81.8)}, rotate = 1.15] [color={rgb, 255:red, 0; green, 0; blue, 225 }  ,draw opacity=1 ][fill={rgb, 255:red, 0; green, 0; blue, 225 }  ,fill opacity=1 ][line width=0.75]      (0, 0) circle [x radius= 1.34, y radius= 1.34]   ;
\draw [shift={(133.2,124.4)}, rotate = 323.13] [color={rgb, 255:red, 0; green, 0; blue, 225 }  ,draw opacity=1 ][fill={rgb, 255:red, 0; green, 0; blue, 225 }  ,fill opacity=1 ][line width=0.75]      (0, 0) circle [x radius= 1.34, y radius= 1.34]   ;
\draw [color={rgb, 255:red, 0; green, 0; blue, 225 }  ,draw opacity=1 ] [dash pattern={on 0.84pt off 2.51pt}]  (163.2,124.4) .. controls (204.2,97.6) and (295.2,94.6) .. (404.6,81.8) ;
\draw [shift={(404.6,81.8)}, rotate = 353.33] [color={rgb, 255:red, 0; green, 0; blue, 225 }  ,draw opacity=1 ][fill={rgb, 255:red, 0; green, 0; blue, 225 }  ,fill opacity=1 ][line width=0.75]      (0, 0) circle [x radius= 1.34, y radius= 1.34]   ;
\draw [shift={(163.2,124.4)}, rotate = 326.83] [color={rgb, 255:red, 0; green, 0; blue, 225 }  ,draw opacity=1 ][fill={rgb, 255:red, 0; green, 0; blue, 225 }  ,fill opacity=1 ][line width=0.75]      (0, 0) circle [x radius= 1.34, y radius= 1.34]   ;
\draw [color={rgb, 255:red, 0; green, 0; blue, 225 }  ,draw opacity=1 ] [dash pattern={on 0.84pt off 2.51pt}]  (201.8,81.8) .. controls (287.2,114.6) and (366.2,116.6) .. (429.6,121.8) ;
\draw [shift={(429.6,121.8)}, rotate = 4.69] [color={rgb, 255:red, 0; green, 0; blue, 225 }  ,draw opacity=1 ][fill={rgb, 255:red, 0; green, 0; blue, 225 }  ,fill opacity=1 ][line width=0.75]      (0, 0) circle [x radius= 1.34, y radius= 1.34]   ;
\draw [shift={(201.8,81.8)}, rotate = 21.01] [color={rgb, 255:red, 0; green, 0; blue, 225 }  ,draw opacity=1 ][fill={rgb, 255:red, 0; green, 0; blue, 225 }  ,fill opacity=1 ][line width=0.75]      (0, 0) circle [x radius= 1.34, y radius= 1.34]   ;
\draw [color={rgb, 255:red, 0; green, 0; blue, 225 }  ,draw opacity=1 ] [dash pattern={on 0.84pt off 2.51pt}]  (201.8,123.8) .. controls (259.2,143.6) and (347.2,162.6) .. (428.2,156.8) ;
\draw [shift={(428.2,156.8)}, rotate = 355.9] [color={rgb, 255:red, 0; green, 0; blue, 225 }  ,draw opacity=1 ][fill={rgb, 255:red, 0; green, 0; blue, 225 }  ,fill opacity=1 ][line width=0.75]      (0, 0) circle [x radius= 1.34, y radius= 1.34]   ;
\draw [shift={(201.8,123.8)}, rotate = 19.03] [color={rgb, 255:red, 0; green, 0; blue, 225 }  ,draw opacity=1 ][fill={rgb, 255:red, 0; green, 0; blue, 225 }  ,fill opacity=1 ][line width=0.75]      (0, 0) circle [x radius= 1.34, y radius= 1.34]   ;
\draw [color={rgb, 255:red, 0; green, 0; blue, 225 }  ,draw opacity=1 ] [dash pattern={on 0.84pt off 2.51pt}]  (238.2,81.8) .. controls (295.2,99.6) and (393.2,99.6) .. (464.6,121.8) ;
\draw [shift={(464.6,121.8)}, rotate = 17.27] [color={rgb, 255:red, 0; green, 0; blue, 225 }  ,draw opacity=1 ][fill={rgb, 255:red, 0; green, 0; blue, 225 }  ,fill opacity=1 ][line width=0.75]      (0, 0) circle [x radius= 1.34, y radius= 1.34]   ;
\draw [shift={(238.2,81.8)}, rotate = 17.34] [color={rgb, 255:red, 0; green, 0; blue, 225 }  ,draw opacity=1 ][fill={rgb, 255:red, 0; green, 0; blue, 225 }  ,fill opacity=1 ][line width=0.75]      (0, 0) circle [x radius= 1.34, y radius= 1.34]   ;
\draw [color={rgb, 255:red, 0; green, 0; blue, 225 }  ,draw opacity=1 ] [dash pattern={on 0.84pt off 2.51pt}]  (243.2,123.2) .. controls (300.6,143) and (382.2,144.6) .. (468.2,156.8) ;
\draw [shift={(468.2,156.8)}, rotate = 8.07] [color={rgb, 255:red, 0; green, 0; blue, 225 }  ,draw opacity=1 ][fill={rgb, 255:red, 0; green, 0; blue, 225 }  ,fill opacity=1 ][line width=0.75]      (0, 0) circle [x radius= 1.34, y radius= 1.34]   ;
\draw [shift={(243.2,123.2)}, rotate = 19.03] [color={rgb, 255:red, 0; green, 0; blue, 225 }  ,draw opacity=1 ][fill={rgb, 255:red, 0; green, 0; blue, 225 }  ,fill opacity=1 ][line width=0.75]      (0, 0) circle [x radius= 1.34, y radius= 1.34]   ;
\draw [color={rgb, 255:red, 0; green, 0; blue, 225 }  ,draw opacity=1 ][fill={rgb, 255:red, 0; green, 0; blue, 225 }  ,fill opacity=1 ][line width=0.75]    (155,215) -- (155,235) ;
\draw [color={rgb, 255:red, 0; green, 0; blue, 225 }  ,draw opacity=1 ][fill={rgb, 255:red, 0; green, 0; blue, 225 }  ,fill opacity=1 ][line width=0.75]    (195,215) -- (180,235) ;
\draw [color={rgb, 255:red, 0; green, 0; blue, 225 }  ,draw opacity=1 ][fill={rgb, 255:red, 0; green, 0; blue, 225 }  ,fill opacity=1 ][line width=0.75]    (210,215) -- (195,235) ;
\draw [color={rgb, 255:red, 0; green, 0; blue, 225 }  ,draw opacity=1 ][fill={rgb, 255:red, 0; green, 0; blue, 225 }  ,fill opacity=1 ][line width=0.75]    (238,215) -- (223,235) ;
\draw [color={rgb, 255:red, 0; green, 0; blue, 225 }  ,draw opacity=1 ][fill={rgb, 255:red, 0; green, 0; blue, 225 }  ,fill opacity=1 ][line width=0.75]    (253.33,215.33) -- (237.87,235.47) ;
\draw [color={rgb, 255:red, 0; green, 0; blue, 225 }  ,draw opacity=1 ][fill={rgb, 255:red, 0; green, 0; blue, 225 }  ,fill opacity=1 ][line width=0.75]    (298.67,215.33) -- (283.2,235.47) ;
\draw [color={rgb, 255:red, 0; green, 0; blue, 225 }  ,draw opacity=1 ][fill={rgb, 255:red, 0; green, 0; blue, 225 }  ,fill opacity=1 ][line width=0.75]    (318.5,215) -- (303.03,235.13) ;
\draw [color={rgb, 255:red, 0; green, 0; blue, 225 }  ,draw opacity=1 ][fill={rgb, 255:red, 0; green, 0; blue, 225 }  ,fill opacity=1 ][line width=0.75]    (333.33,215.33) -- (317.87,235.47) ;
\draw [color={rgb, 255:red, 0; green, 0; blue, 225 }  ,draw opacity=1 ][fill={rgb, 255:red, 0; green, 0; blue, 225 }  ,fill opacity=1 ][line width=0.75]    (354,215.33) -- (338.53,235.47) ;
\draw [color={rgb, 255:red, 0; green, 0; blue, 225 }  ,draw opacity=1 ][fill={rgb, 255:red, 0; green, 0; blue, 225 }  ,fill opacity=1 ][line width=0.75]    (380,214.67) -- (364.53,234.8) ;
\draw [color={rgb, 255:red, 0; green, 0; blue, 225 }  ,draw opacity=1 ][fill={rgb, 255:red, 0; green, 0; blue, 225 }  ,fill opacity=1 ][line width=0.75]    (401.33,215.33) -- (385.87,235.47) ;
\draw [color={rgb, 255:red, 0; green, 0; blue, 225 }  ,draw opacity=1 ][fill={rgb, 255:red, 0; green, 0; blue, 225 }  ,fill opacity=1 ][line width=0.75]    (457.2,214.8) -- (399.2,234.8) ;
\draw [color={rgb, 255:red, 0; green, 0; blue, 225 }  ,draw opacity=1 ][fill={rgb, 255:red, 0; green, 0; blue, 225 }  ,fill opacity=1 ][line width=0.75]    (479.2,214.8) -- (420.47,234.87) ;
\draw [color={rgb, 255:red, 0; green, 0; blue, 225 }  ,draw opacity=1 ][fill={rgb, 255:red, 0; green, 0; blue, 225 }  ,fill opacity=1 ][line width=0.75]    (495,215) -- (461.2,234.8) ;

\draw (192,12.4) node [anchor=north west][inner sep=0.75pt]    {$\mathbf{F}$};
\draw (396,12.4) node [anchor=north west][inner sep=0.75pt]    {$\mathbf{G}$};
\draw (106,193.4) node [anchor=north west][inner sep=0.75pt]    {$S(\mathbf{F})$};
\draw (106,238.4) node [anchor=north west][inner sep=0.75pt]    {$S(\mathbf{G})$};
\draw (151,192.4) node [anchor=north west][inner sep=0.75pt]    {$($};
\draw (151,238.4) node [anchor=north west][inner sep=0.75pt]    {$($};
\draw (191,192.4) node [anchor=north west][inner sep=0.75pt]    {$($};
\draw (176,238.4) node [anchor=north west][inner sep=0.75pt]    {$($};
\draw (208,192.4) node [anchor=north west][inner sep=0.75pt]    {$)$};
\draw (193,238.4) node [anchor=north west][inner sep=0.75pt]    {$)$};
\draw (233,192.4) node [anchor=north west][inner sep=0.75pt]    {$($};
\draw (218,238.4) node [anchor=north west][inner sep=0.75pt]    {$($};
\draw (251,192.4) node [anchor=north west][inner sep=0.75pt]    {$)$};
\draw (236,238.4) node [anchor=north west][inner sep=0.75pt]    {$)$};
\draw (293,192.4) node [anchor=north west][inner sep=0.75pt]    {$($};
\draw (313,192.4) node [anchor=north west][inner sep=0.75pt]    {$($};
\draw (278,238.4) node [anchor=north west][inner sep=0.75pt]    {$($};
\draw (298,238.4) node [anchor=north west][inner sep=0.75pt]    {$($};
\draw (331,192.4) node [anchor=north west][inner sep=0.75pt]    {$)$};
\draw (318,238.4) node [anchor=north west][inner sep=0.75pt]    {$)$};
\draw (351,192.4) node [anchor=north west][inner sep=0.75pt]    {$)$};
\draw (336,238.4) node [anchor=north west][inner sep=0.75pt]    {$)$};
\draw (376,192.4) node [anchor=north west][inner sep=0.75pt]    {$($};
\draw (396,192.4) node [anchor=north west][inner sep=0.75pt]    {$($};
\draw (358,238.4) node [anchor=north west][inner sep=0.75pt]    {$($};
\draw (381,238.4) node [anchor=north west][inner sep=0.75pt]    {$($};
\draw (398,238.4) node [anchor=north west][inner sep=0.75pt]    {$)$};
\draw (456,192.4) node [anchor=north west][inner sep=0.75pt]    {$)$};
\draw (478,192.4) node [anchor=north west][inner sep=0.75pt]    {$)$};
\draw (418,238.4) node [anchor=north west][inner sep=0.75pt]    {$)$};
\draw (493,192.4) node [anchor=north west][inner sep=0.75pt]    {$)$};
\draw (458,238.4) node [anchor=north west][inner sep=0.75pt]    {$)$};
\draw (171,192.4) node [anchor=north west][inner sep=0.75pt]    {$\textcolor[rgb]{1,0,0}{(}$};
\draw (271,192.4) node [anchor=north west][inner sep=0.75pt]    {$\textcolor[rgb]{1,0,0}{)}$};
\draw (256,237.4) node [anchor=north west][inner sep=0.75pt]    {$\textcolor[rgb]{0.35,0.6,0.07}{(}$};
\draw (436,238.4) node [anchor=north west][inner sep=0.75pt]    {$\textcolor[rgb]{0.35,0.6,0.07}{)}$};
\draw (413,192.4) node [anchor=north west][inner sep=0.75pt]    {$\textcolor[rgb]{1,0,0}{(}$};
\draw (436,192.4) node [anchor=north west][inner sep=0.75pt]    {$\textcolor[rgb]{1,0,0}{)}$};

\end{tikzpicture}

%% file: figures/pre/ted_decomp_ints.tex
\definecolor{blockred}{RGB}{251,160,160}
\definecolor{blockorange}{RGB}{245,166,35}
\definecolor{blockviolet}{RGB}{191,176,255}

\begin{tikzpicture}[
    line width=0.75pt,
    block/.style={draw=black, fill opacity=0.5},
    link/.style ={dash pattern=on 1pt off 2pt},
    lab/.style  ={font=\scriptsize, inner sep=2pt},
  ]

  \def\h{0.32}      
  \def\rowI{0}      
  \def\rowJ{-1.1}   

  \foreach \i/\x in {0/1.2, 1/5.2, 2/5.7, 3/7.2, 4/11.3, 5/12.1}
    \coordinate (I\i) at (\x,\rowI);
  \foreach \i/\x in {0/0.5, 1/5.2, 2/6.5, 3/7.8, 4/8.7, 5/10.5}
    \coordinate (J\i) at (\x,\rowJ);

  \foreach \l/\r/\c/\name/\dx in {%
      I0/I1/blockred/{I^1_1}/0,
      I1/I2/blockorange/{I^2_1}/2,
      I3/I4/blockred/{I^1_2}/0,
      I4/I5/blockviolet/{I^3_1}/3}
  {
    \draw[block, fill=\c] (\l) rectangle ([yshift=-\h cm]\r);
    \node[lab, above, xshift=\dx pt] at ($(\l)!0.5!(\r)$) {$\name$};
  }

  \foreach \l/\r/\c/\name/\dx in {%
      J0/J1/blockred/{J^1_1}/0,
      J1/J2/blockorange/{J^2_1}/0,
      J3/J4/blockred/{J^1_2}/0,
      J4/J5/blockviolet/{J^3_1}/0}
  {
    \draw[block, fill=\c] (\l) rectangle ([yshift=-\h cm]\r);
    \node[lab, below, xshift=\dx pt]
      at ($([yshift=-\h cm]\l)!0.5!([yshift=-\h cm]\r)$) {$\name$};
  }

  \foreach \i in {0,...,5}
    \draw[link] ([yshift=-\h cm]I\i) -- (J\i);

  \node[left] at (0,-0.16) {$\mathcal{I}$};
  \node[left] at (0,-1.26) {$\mathcal{J}$};

\end{tikzpicture}

%% file: figures/pre/ghost_mappings.tex
\tikzset{every picture/.style={line width=0.75pt}} 

\begin{tikzpicture}[x=0.75pt,y=0.75pt,yscale=-1,xscale=1]

\draw [color={rgb, 255:red, 0; green, 0; blue, 225 }  ,draw opacity=1 ][fill={rgb, 255:red, 0; green, 0; blue, 225 }  ,fill opacity=1 ][line width=0.75]    (151,220) -- (151,240) ;
\draw [color={rgb, 255:red, 0; green, 0; blue, 225 }  ,draw opacity=1 ][fill={rgb, 255:red, 0; green, 0; blue, 225 }  ,fill opacity=1 ][line width=0.75]    (190,220) -- (175,240) ;
\draw [color={rgb, 255:red, 0; green, 0; blue, 225 }  ,draw opacity=1 ][fill={rgb, 255:red, 0; green, 0; blue, 225 }  ,fill opacity=1 ][line width=0.75]    (205,220) -- (190,240) ;
\draw [color={rgb, 255:red, 0; green, 0; blue, 225 }  ,draw opacity=1 ][fill={rgb, 255:red, 0; green, 0; blue, 225 }  ,fill opacity=1 ][line width=0.75]    (233,220) -- (218,240) ;
\draw [color={rgb, 255:red, 0; green, 0; blue, 225 }  ,draw opacity=1 ][fill={rgb, 255:red, 0; green, 0; blue, 225 }  ,fill opacity=1 ][line width=0.75]    (248.33,220.33) -- (232.87,240.47) ;
\draw [color={rgb, 255:red, 0; green, 0; blue, 225 }  ,draw opacity=1 ][fill={rgb, 255:red, 0; green, 0; blue, 225 }  ,fill opacity=1 ][line width=0.75]    (293.67,220.33) -- (278.2,240.47) ;
\draw [color={rgb, 255:red, 0; green, 0; blue, 225 }  ,draw opacity=1 ][fill={rgb, 255:red, 0; green, 0; blue, 225 }  ,fill opacity=1 ][line width=0.75]    (313.5,220) -- (298.03,240.13) ;
\draw [color={rgb, 255:red, 0; green, 0; blue, 225 }  ,draw opacity=1 ][fill={rgb, 255:red, 0; green, 0; blue, 225 }  ,fill opacity=1 ][line width=0.75]    (328.33,220.33) -- (312.87,240.47) ;
\draw [color={rgb, 255:red, 0; green, 0; blue, 225 }  ,draw opacity=1 ][fill={rgb, 255:red, 0; green, 0; blue, 225 }  ,fill opacity=1 ][line width=0.75]    (349,220.33) -- (333.53,240.47) ;
\draw [color={rgb, 255:red, 0; green, 0; blue, 225 }  ,draw opacity=1 ][fill={rgb, 255:red, 0; green, 0; blue, 225 }  ,fill opacity=1 ][line width=0.75]    (375,219.67) -- (359.53,239.8) ;
\draw [color={rgb, 255:red, 0; green, 0; blue, 225 }  ,draw opacity=1 ][fill={rgb, 255:red, 0; green, 0; blue, 225 }  ,fill opacity=1 ][line width=0.75]    (396.33,220.33) -- (380.87,240.47) ;
\draw [color={rgb, 255:red, 0; green, 0; blue, 225 }  ,draw opacity=1 ][fill={rgb, 255:red, 0; green, 0; blue, 225 }  ,fill opacity=1 ][line width=0.75]    (452.2,219.8) -- (394.2,239.8) ;
\draw [color={rgb, 255:red, 0; green, 0; blue, 225 }  ,draw opacity=1 ][fill={rgb, 255:red, 0; green, 0; blue, 225 }  ,fill opacity=1 ][line width=0.75]    (474.2,219.8) -- (415.47,239.87) ;
\draw [color={rgb, 255:red, 0; green, 0; blue, 225 }  ,draw opacity=1 ][fill={rgb, 255:red, 0; green, 0; blue, 225 }  ,fill opacity=1 ][line width=0.75]    (490,220) -- (456.2,239.8) ;
\draw [color={rgb, 255:red, 255; green, 0; blue, 0 }  ,draw opacity=0.5 ][fill={rgb, 255:red, 0; green, 0; blue, 225 }  ,fill opacity=1 ][line width=0.75]    (171,220) -- (175,240) ;
\draw [color={rgb, 255:red, 255; green, 0; blue, 0 }  ,draw opacity=0.5 ][fill={rgb, 255:red, 0; green, 0; blue, 225 }  ,fill opacity=1 ][line width=0.75]    (266,220) -- (258.2,240.47) ;
\draw [color={rgb, 255:red, 255; green, 0; blue, 0 }  ,draw opacity=0.5 ][fill={rgb, 255:red, 0; green, 0; blue, 225 }  ,fill opacity=1 ][line width=0.75]    (411,220) -- (394.2,239.8) ;
\draw [color={rgb, 255:red, 255; green, 0; blue, 0 }  ,draw opacity=0.5 ][fill={rgb, 255:red, 0; green, 0; blue, 225 }  ,fill opacity=1 ][line width=0.75]    (431,220) -- (394.2,239.8) ;
\draw  [draw opacity=0][fill={rgb, 255:red, 152; green, 208; blue, 255 }  ,fill opacity=0.43 ] (161,199.38) .. controls (161,196.96) and (162.96,195) .. (165.38,195) -- (276.63,195) .. controls (279.04,195) and (281,196.96) .. (281,199.38) -- (281,215.63) .. controls (281,218.04) and (279.04,220) .. (276.63,220) -- (165.38,220) .. controls (162.96,220) and (161,218.04) .. (161,215.63) -- cycle ;
\draw  [draw opacity=0][fill={rgb, 255:red, 152; green, 208; blue, 255 }  ,fill opacity=0.43 ] (166,244.38) .. controls (166,241.96) and (167.96,240) .. (170.38,240) -- (261.63,240) .. controls (264.04,240) and (266,241.96) .. (266,244.38) -- (266,260.63) .. controls (266,263.04) and (264.04,265) .. (261.63,265) -- (170.38,265) .. controls (167.96,265) and (166,263.04) .. (166,260.63) -- cycle ;
\draw  [draw opacity=0][fill={rgb, 255:red, 255; green, 246; blue, 170 }  ,fill opacity=0.7 ] (366,199.38) .. controls (366,196.96) and (367.96,195) .. (370.38,195) -- (481.63,195) .. controls (484.04,195) and (486,196.96) .. (486,199.38) -- (486,215.63) .. controls (486,218.04) and (484.04,220) .. (481.63,220) -- (370.38,220) .. controls (367.96,220) and (366,218.04) .. (366,215.63) -- cycle ;
\draw  [draw opacity=0][fill={rgb, 255:red, 255; green, 246; blue, 170 }  ,fill opacity=0.7 ] (350.38,244.38) .. controls (350.38,241.96) and (352.33,240) .. (354.75,240) -- (441.63,240) .. controls (444.04,240) and (446,241.96) .. (446,244.38) -- (446,260.63) .. controls (446,263.04) and (444.04,265) .. (441.63,265) -- (354.75,265) .. controls (352.33,265) and (350.38,263.04) .. (350.38,260.63) -- cycle ;
\draw   (203.6,46.8) .. controls (203.6,42.27) and (207.27,38.6) .. (211.8,38.6) .. controls (216.33,38.6) and (220,42.27) .. (220,46.8) .. controls (220,51.33) and (216.33,55) .. (211.8,55) .. controls (207.27,55) and (203.6,51.33) .. (203.6,46.8) -- cycle ;
\draw  [fill={rgb, 255:red, 255; green, 0; blue, 0 }  ,fill opacity=0.31 ] (170,86.8) .. controls (170,82.27) and (173.67,78.6) .. (178.2,78.6) .. controls (182.73,78.6) and (186.4,82.27) .. (186.4,86.8) .. controls (186.4,91.33) and (182.73,95) .. (178.2,95) .. controls (173.67,95) and (170,91.33) .. (170,86.8) -- cycle ;
\draw   (203.6,86.8) .. controls (203.6,82.27) and (207.27,78.6) .. (211.8,78.6) .. controls (216.33,78.6) and (220,82.27) .. (220,86.8) .. controls (220,91.33) and (216.33,95) .. (211.8,95) .. controls (207.27,95) and (203.6,91.33) .. (203.6,86.8) -- cycle ;
\draw   (240,86.8) .. controls (240,82.27) and (243.67,78.6) .. (248.2,78.6) .. controls (252.73,78.6) and (256.4,82.27) .. (256.4,86.8) .. controls (256.4,91.33) and (252.73,95) .. (248.2,95) .. controls (243.67,95) and (240,91.33) .. (240,86.8) -- cycle ;
\draw   (165,129.4) .. controls (165,124.87) and (168.67,121.2) .. (173.2,121.2) .. controls (177.73,121.2) and (181.4,124.87) .. (181.4,129.4) .. controls (181.4,133.93) and (177.73,137.6) .. (173.2,137.6) .. controls (168.67,137.6) and (165,133.93) .. (165,129.4) -- cycle ;
\draw   (135,129.4) .. controls (135,124.87) and (138.67,121.2) .. (143.2,121.2) .. controls (147.73,121.2) and (151.4,124.87) .. (151.4,129.4) .. controls (151.4,133.93) and (147.73,137.6) .. (143.2,137.6) .. controls (138.67,137.6) and (135,133.93) .. (135,129.4) -- cycle ;
\draw   (203.6,128.8) .. controls (203.6,124.27) and (207.27,120.6) .. (211.8,120.6) .. controls (216.33,120.6) and (220,124.27) .. (220,128.8) .. controls (220,133.33) and (216.33,137) .. (211.8,137) .. controls (207.27,137) and (203.6,133.33) .. (203.6,128.8) -- cycle ;
\draw   (245,128.2) .. controls (245,123.67) and (248.67,120) .. (253.2,120) .. controls (257.73,120) and (261.4,123.67) .. (261.4,128.2) .. controls (261.4,132.73) and (257.73,136.4) .. (253.2,136.4) .. controls (248.67,136.4) and (245,132.73) .. (245,128.2) -- cycle ;
\draw  [fill={rgb, 255:red, 255; green, 0; blue, 0 }  ,fill opacity=0.31 ] (256.8,168.2) .. controls (256.8,163.67) and (260.47,160) .. (265,160) .. controls (269.53,160) and (273.2,163.67) .. (273.2,168.2) .. controls (273.2,172.73) and (269.53,176.4) .. (265,176.4) .. controls (260.47,176.4) and (256.8,172.73) .. (256.8,168.2) -- cycle ;
\draw    (211.8,55) -- (211.8,78.6) ;
\draw    (211.8,95) -- (211.8,120.6) ;
\draw    (211.8,55) -- (178.2,78.6) ;
\draw    (175,94) -- (173.2,121.2) ;
\draw    (175,94) -- (143.2,121.2) ;
\draw    (211.8,55) -- (245,80) ;
\draw    (248.2,95) -- (253.2,120) ;
\draw    (253.2,136.4) -- (265,160) ;
\draw   (405,46.8) .. controls (405,42.27) and (408.67,38.6) .. (413.2,38.6) .. controls (417.73,38.6) and (421.4,42.27) .. (421.4,46.8) .. controls (421.4,51.33) and (417.73,55) .. (413.2,55) .. controls (408.67,55) and (405,51.33) .. (405,46.8) -- cycle ;
\draw   (365,86.8) .. controls (365,82.27) and (368.67,78.6) .. (373.2,78.6) .. controls (377.73,78.6) and (381.4,82.27) .. (381.4,86.8) .. controls (381.4,91.33) and (377.73,95) .. (373.2,95) .. controls (368.67,95) and (365,91.33) .. (365,86.8) -- cycle ;
\draw   (406.4,86.8) .. controls (406.4,82.27) and (410.07,78.6) .. (414.6,78.6) .. controls (419.13,78.6) and (422.8,82.27) .. (422.8,86.8) .. controls (422.8,91.33) and (419.13,95) .. (414.6,95) .. controls (410.07,95) and (406.4,91.33) .. (406.4,86.8) -- cycle ;
\draw  [fill={rgb, 255:red, 126; green, 211; blue, 33 }  ,fill opacity=0.31 ] (445,86.8) .. controls (445,82.27) and (448.67,78.6) .. (453.2,78.6) .. controls (457.73,78.6) and (461.4,82.27) .. (461.4,86.8) .. controls (461.4,91.33) and (457.73,95) .. (453.2,95) .. controls (448.67,95) and (445,91.33) .. (445,86.8) -- cycle ;
\draw   (431.4,126.8) .. controls (431.4,122.27) and (435.07,118.6) .. (439.6,118.6) .. controls (444.13,118.6) and (447.8,122.27) .. (447.8,126.8) .. controls (447.8,131.33) and (444.13,135) .. (439.6,135) .. controls (435.07,135) and (431.4,131.33) .. (431.4,126.8) -- cycle ;
\draw   (430,161.8) .. controls (430,157.27) and (433.67,153.6) .. (438.2,153.6) .. controls (442.73,153.6) and (446.4,157.27) .. (446.4,161.8) .. controls (446.4,166.33) and (442.73,170) .. (438.2,170) .. controls (433.67,170) and (430,166.33) .. (430,161.8) -- cycle ;
\draw   (466.4,126.8) .. controls (466.4,122.27) and (470.07,118.6) .. (474.6,118.6) .. controls (479.13,118.6) and (482.8,122.27) .. (482.8,126.8) .. controls (482.8,131.33) and (479.13,135) .. (474.6,135) .. controls (470.07,135) and (466.4,131.33) .. (466.4,126.8) -- cycle ;
\draw   (470,161.8) .. controls (470,157.27) and (473.67,153.6) .. (478.2,153.6) .. controls (482.73,153.6) and (486.4,157.27) .. (486.4,161.8) .. controls (486.4,166.33) and (482.73,170) .. (478.2,170) .. controls (473.67,170) and (470,166.33) .. (470,161.8) -- cycle ;
\draw    (413.2,55) -- (373.2,78.6) ;
\draw    (413.2,55) -- (414.6,78.6) ;
\draw    (413.2,55) -- (453.2,78.6) ;
\draw    (456.4,95) -- (474.6,118.6) ;
\draw    (453.2,95) -- (439.6,118.6) ;
\draw    (439.6,135) -- (438.2,153.6) ;
\draw    (474.6,135) -- (478.2,153.6) ;
\draw [color={rgb, 255:red, 0; green, 0; blue, 225 }  ,draw opacity=1 ] [dash pattern={on 0.84pt off 2.51pt}]  (211.8,46.8) .. controls (233.82,30.29) and (287.48,24.92) .. (331.25,27.31) .. controls (375.02,29.69) and (393.42,36.37) .. (413.2,46.8) ;
\draw [shift={(413.2,46.8)}, rotate = 27.8] [color={rgb, 255:red, 0; green, 0; blue, 225 }  ,draw opacity=1 ][fill={rgb, 255:red, 0; green, 0; blue, 225 }  ,fill opacity=1 ][line width=0.75]      (0, 0) circle [x radius= 1.34, y radius= 1.34]   ;
\draw [shift={(211.8,46.8)}, rotate = 323.13] [color={rgb, 255:red, 0; green, 0; blue, 225 }  ,draw opacity=1 ][fill={rgb, 255:red, 0; green, 0; blue, 225 }  ,fill opacity=1 ][line width=0.75]      (0, 0) circle [x radius= 1.34, y radius= 1.34]   ;
\draw [color={rgb, 255:red, 0; green, 0; blue, 225 }  ,draw opacity=1 ] [dash pattern={on 0.84pt off 2.51pt}]  (143.2,129.4) .. controls (183.2,99.4) and (313.2,85.6) .. (373.2,86.8) ;
\draw [shift={(373.2,86.8)}, rotate = 1.15] [color={rgb, 255:red, 0; green, 0; blue, 225 }  ,draw opacity=1 ][fill={rgb, 255:red, 0; green, 0; blue, 225 }  ,fill opacity=1 ][line width=0.75]      (0, 0) circle [x radius= 1.34, y radius= 1.34]   ;
\draw [shift={(143.2,129.4)}, rotate = 323.13] [color={rgb, 255:red, 0; green, 0; blue, 225 }  ,draw opacity=1 ][fill={rgb, 255:red, 0; green, 0; blue, 225 }  ,fill opacity=1 ][line width=0.75]      (0, 0) circle [x radius= 1.34, y radius= 1.34]   ;
\draw [color={rgb, 255:red, 0; green, 0; blue, 225 }  ,draw opacity=1 ] [dash pattern={on 0.84pt off 2.51pt}]  (173.2,129.4) .. controls (214.2,102.6) and (305.2,99.6) .. (414.6,86.8) ;
\draw [shift={(414.6,86.8)}, rotate = 353.33] [color={rgb, 255:red, 0; green, 0; blue, 225 }  ,draw opacity=1 ][fill={rgb, 255:red, 0; green, 0; blue, 225 }  ,fill opacity=1 ][line width=0.75]      (0, 0) circle [x radius= 1.34, y radius= 1.34]   ;
\draw [shift={(173.2,129.4)}, rotate = 326.83] [color={rgb, 255:red, 0; green, 0; blue, 225 }  ,draw opacity=1 ][fill={rgb, 255:red, 0; green, 0; blue, 225 }  ,fill opacity=1 ][line width=0.75]      (0, 0) circle [x radius= 1.34, y radius= 1.34]   ;
\draw [color={rgb, 255:red, 0; green, 0; blue, 225 }  ,draw opacity=1 ] [dash pattern={on 0.84pt off 2.51pt}]  (211.8,86.8) .. controls (297.2,119.6) and (376.2,121.6) .. (439.6,126.8) ;
\draw [shift={(439.6,126.8)}, rotate = 4.69] [color={rgb, 255:red, 0; green, 0; blue, 225 }  ,draw opacity=1 ][fill={rgb, 255:red, 0; green, 0; blue, 225 }  ,fill opacity=1 ][line width=0.75]      (0, 0) circle [x radius= 1.34, y radius= 1.34]   ;
\draw [shift={(211.8,86.8)}, rotate = 21.01] [color={rgb, 255:red, 0; green, 0; blue, 225 }  ,draw opacity=1 ][fill={rgb, 255:red, 0; green, 0; blue, 225 }  ,fill opacity=1 ][line width=0.75]      (0, 0) circle [x radius= 1.34, y radius= 1.34]   ;
\draw [color={rgb, 255:red, 0; green, 0; blue, 225 }  ,draw opacity=1 ] [dash pattern={on 0.84pt off 2.51pt}]  (211.8,128.8) .. controls (269.2,148.6) and (357.2,167.6) .. (438.2,161.8) ;
\draw [shift={(438.2,161.8)}, rotate = 355.9] [color={rgb, 255:red, 0; green, 0; blue, 225 }  ,draw opacity=1 ][fill={rgb, 255:red, 0; green, 0; blue, 225 }  ,fill opacity=1 ][line width=0.75]      (0, 0) circle [x radius= 1.34, y radius= 1.34]   ;
\draw [shift={(211.8,128.8)}, rotate = 19.03] [color={rgb, 255:red, 0; green, 0; blue, 225 }  ,draw opacity=1 ][fill={rgb, 255:red, 0; green, 0; blue, 225 }  ,fill opacity=1 ][line width=0.75]      (0, 0) circle [x radius= 1.34, y radius= 1.34]   ;
\draw [color={rgb, 255:red, 0; green, 0; blue, 225 }  ,draw opacity=1 ] [dash pattern={on 0.84pt off 2.51pt}]  (248.2,86.8) .. controls (305.2,104.6) and (403.2,104.6) .. (474.6,126.8) ;
\draw [shift={(474.6,126.8)}, rotate = 17.27] [color={rgb, 255:red, 0; green, 0; blue, 225 }  ,draw opacity=1 ][fill={rgb, 255:red, 0; green, 0; blue, 225 }  ,fill opacity=1 ][line width=0.75]      (0, 0) circle [x radius= 1.34, y radius= 1.34]   ;
\draw [shift={(248.2,86.8)}, rotate = 17.34] [color={rgb, 255:red, 0; green, 0; blue, 225 }  ,draw opacity=1 ][fill={rgb, 255:red, 0; green, 0; blue, 225 }  ,fill opacity=1 ][line width=0.75]      (0, 0) circle [x radius= 1.34, y radius= 1.34]   ;
\draw [color={rgb, 255:red, 0; green, 0; blue, 225 }  ,draw opacity=1 ] [dash pattern={on 0.84pt off 2.51pt}]  (253.2,128.2) .. controls (310.6,148) and (392.2,149.6) .. (478.2,161.8) ;
\draw [shift={(478.2,161.8)}, rotate = 8.07] [color={rgb, 255:red, 0; green, 0; blue, 225 }  ,draw opacity=1 ][fill={rgb, 255:red, 0; green, 0; blue, 225 }  ,fill opacity=1 ][line width=0.75]      (0, 0) circle [x radius= 1.34, y radius= 1.34]   ;
\draw [shift={(253.2,128.2)}, rotate = 19.03] [color={rgb, 255:red, 0; green, 0; blue, 225 }  ,draw opacity=1 ][fill={rgb, 255:red, 0; green, 0; blue, 225 }  ,fill opacity=1 ][line width=0.75]      (0, 0) circle [x radius= 1.34, y radius= 1.34]   ;

\draw (101,198.4) node [anchor=north west][inner sep=0.75pt]    {$S(\mathbf{F})$};
\draw (101,243.4) node [anchor=north west][inner sep=0.75pt]    {$S(\mathbf{G})$};
\draw (146,197.4) node [anchor=north west][inner sep=0.75pt]    {$($};
\draw (146,243.4) node [anchor=north west][inner sep=0.75pt]    {$($};
\draw (186,197.4) node [anchor=north west][inner sep=0.75pt]    {$($};
\draw (171,243.4) node [anchor=north west][inner sep=0.75pt]    {$($};
\draw (203,197.4) node [anchor=north west][inner sep=0.75pt]    {$)$};
\draw (188,243.4) node [anchor=north west][inner sep=0.75pt]    {$)$};
\draw (228,197.4) node [anchor=north west][inner sep=0.75pt]    {$($};
\draw (213,243.4) node [anchor=north west][inner sep=0.75pt]    {$($};
\draw (246,197.4) node [anchor=north west][inner sep=0.75pt]    {$)$};
\draw (231,243.4) node [anchor=north west][inner sep=0.75pt]    {$)$};
\draw (288,197.4) node [anchor=north west][inner sep=0.75pt]    {$($};
\draw (308,197.4) node [anchor=north west][inner sep=0.75pt]    {$($};
\draw (273,243.4) node [anchor=north west][inner sep=0.75pt]    {$($};
\draw (293,243.4) node [anchor=north west][inner sep=0.75pt]    {$($};
\draw (326,197.4) node [anchor=north west][inner sep=0.75pt]    {$)$};
\draw (313,243.4) node [anchor=north west][inner sep=0.75pt]    {$)$};
\draw (346,197.4) node [anchor=north west][inner sep=0.75pt]    {$)$};
\draw (331,243.4) node [anchor=north west][inner sep=0.75pt]    {$)$};
\draw (371,197.4) node [anchor=north west][inner sep=0.75pt]    {$($};
\draw (391,197.4) node [anchor=north west][inner sep=0.75pt]    {$($};
\draw (353,243.4) node [anchor=north west][inner sep=0.75pt]    {$($};
\draw (376,243.4) node [anchor=north west][inner sep=0.75pt]    {$($};
\draw (393,243.4) node [anchor=north west][inner sep=0.75pt]    {$)$};
\draw (451,197.4) node [anchor=north west][inner sep=0.75pt]    {$)$};
\draw (473,197.4) node [anchor=north west][inner sep=0.75pt]    {$)$};
\draw (413,243.4) node [anchor=north west][inner sep=0.75pt]    {$)$};
\draw (488,197.4) node [anchor=north west][inner sep=0.75pt]    {$)$};
\draw (453,243.4) node [anchor=north west][inner sep=0.75pt]    {$)$};
\draw (166,197.4) node [anchor=north west][inner sep=0.75pt]    {$\textcolor[rgb]{1,0,0}{(}$};
\draw (266,197.4) node [anchor=north west][inner sep=0.75pt]    {$\textcolor[rgb]{1,0,0}{)}$};
\draw (252,243.4) node [anchor=north west][inner sep=0.75pt]    {$\textcolor[rgb]{0.35,0.6,0.07}{(}$};
\draw (431,243.4) node [anchor=north west][inner sep=0.75pt]    {$\textcolor[rgb]{0.35,0.6,0.07}{)}$};
\draw (408,197.4) node [anchor=north west][inner sep=0.75pt]    {$\textcolor[rgb]{1,0,0}{(}$};
\draw (431,197.4) node [anchor=north west][inner sep=0.75pt]    {$\textcolor[rgb]{1,0,0}{)}$};
\draw (202,17.4) node [anchor=north west][inner sep=0.75pt]    {$\mathbf{F}$};
\draw (406,17.4) node [anchor=north west][inner sep=0.75pt]    {$\mathbf{G}$};
\draw (191,37.4) node [anchor=north west][inner sep=0.75pt]  [font=\scriptsize]  {$v_{1}$};
\draw (156,72.4) node [anchor=north west][inner sep=0.75pt]  [font=\scriptsize]  {$v_{2}$};
\draw (126,132.4) node [anchor=north west][inner sep=0.75pt]  [font=\scriptsize]  {$v_{3}$};
\draw (166,137.4) node [anchor=north west][inner sep=0.75pt]  [font=\scriptsize]  {$v_{4}$};
\draw (191,78.2) node [anchor=north west][inner sep=0.75pt]  [font=\scriptsize]  {$v_{5}$};
\draw (200,134.4) node [anchor=north west][inner sep=0.75pt]  [font=\scriptsize]  {$v_{6}$};
\draw (255,71.4) node [anchor=north west][inner sep=0.75pt]  [font=\scriptsize]  {$v_{7}$};
\draw (261,115.4) node [anchor=north west][inner sep=0.75pt]  [font=\scriptsize]  {$v_{8}$};
\draw (274,163.4) node [anchor=north west][inner sep=0.75pt]  [font=\scriptsize]  {$v_{9}$};
\draw (423,35.4) node [anchor=north west][inner sep=0.75pt]  [font=\scriptsize]  {$u_{1}$};
\draw (356,69.4) node [anchor=north west][inner sep=0.75pt]  [font=\scriptsize]  {$u_{2}$};
\draw (395,72.4) node [anchor=north west][inner sep=0.75pt]  [font=\scriptsize]  {$u_{3}$};
\draw (463,73.4) node [anchor=north west][inner sep=0.75pt]  [font=\scriptsize]  {$u_{4}$};
\draw (423,128.2) node [anchor=north west][inner sep=0.75pt]  [font=\scriptsize]  {$u_{5}$};
\draw (426,167.4) node [anchor=north west][inner sep=0.75pt]  [font=\scriptsize]  {$u_{6}$};
\draw (484,120.4) node [anchor=north west][inner sep=0.75pt]  [font=\scriptsize]  {$u_{7}$};
\draw (487,160.4) node [anchor=north west][inner sep=0.75pt]  [font=\scriptsize]  {$u_{8}$};

\end{tikzpicture}

%% file: s4_decomp.tex
\section{A Hierarchical Decomposition of Forests}
\label{sec:decomp}

In the following section we present the construction of the tree decomposition that satisfies all of the following properties:
\begin{enumerate}[(1)]
    \item\label{pr:depth_p} The hierarchical decomposition has $d = \Oh(\log_c n)$ levels.
    \item\label{pr:deg_p} For $t\in \fragmentoo{0}{d}$, each level-$t$ subforest is decomposed into $\Oh(c)$ level-$(t-1)$ subforests.
    \item\label{pr:size_p} Each level $t\in \fragmentco{0}{d}$ consists of $\Oh(n/c^t)$ subforests of size $\Oh(c^t)$ each.
    \item\label{pr:sub_p} For $t\in \fragmentco{0}{d}$, a uniformly random level-$t$ subforest can be sampled in $\Ohtilde(c^t)$ time w.h.p.
     \item\label{pr:dyn_p} A node update (insertion or deletion) to $\bF$ can be implemented by modifying $\Oh(d)$ subforests from the decomposition, in the amortized sense.
\end{enumerate}

We briefly describe how our decomposition compares with previous papers on \TED. Several of these, including~\cite{BGHS19,SS22,DGHKS23}, construct a single-level decomposition satisfying property~\ref{pr:size}.
A recent work~\cite{DGHKS25} shows how to achieve properties~\ref{pr:depth} and~\ref{pr:deg} for $c=2$.
The underlying decomposition can be generated lazily in $\Ohtilde(1)$ time per subforest provided that $\bF$ supports $\Ohtilde(1)$-time level ancestor and lowest common ancestor queries.
This assumption is valid in the dynamic setting~\cite{NS14}, but the constructed decomposition may change significantly upon an update.
Top trees~\cite{AHLT05} also satisfy properties~\ref{pr:depth} and \ref{pr:deg} for $c=2$, and they support dynamic updates in worst-case $\Oh(\log n)$ time.
Still, their update interface (inserting and deleting individual edges) is incompatible with ours (a node deletion redefines the parent of all its children).
We are not aware of previous hierarchical decompositions (static or dynamic) satisfying property~\ref{pr:size}, but a weaker version, with $\Oh(n \log_c n\;/\;c^t)$ subforests per level, can likely be achieved by appropriately assigning nodes to levels.
More importantly, we do not know of any result coming close to satisfying property~\ref{pr:sub}.

\subparagraph{Construction of a single level of our decomposition.} We first define how we construct a single level.

\begin{definition}\label{def:decomp}
    Given an $n$-node forest $\bF$ and a sequence of $b+1$ integers $\lambda = (\lambda_0, \ldots, \lambda_b) \in \fragment{0}{2n}^{b+1}$ such that $0 = \lambda_0 < \lambda_1 < \ldots < \lambda_b = 2n$, we define the set $P_{\lambda}(\bF)$ as follows:
    \begin{itemize}
        \item Partition $\fragmentco{0}{2n}$ into $b$ contiguous blocks $B_0, \ldots, B_{b-1}$, where $B_i = \fragmentco{\lambda_i}{\lambda_{i+1}}$, and initialize $P_{i,j}=\emptyset$ for every $0\leq i\leq j<b$.
        \item Consider a node $u\in \bF$. If $\Left(u)\in B_i$ and $\Right(u)\in B_j$ for distinct $i,j$, assign $u$ to $P_{i,j}$.
        \item Otherwise, $\Left(u),\Right(u)\in B_i$. Let $x'$ be the smallest position in $B_i$ such that $x'\geq \Right(u)$ and $\Other(x')\notin B_i$, if such a position exists. The node $u'$ underlying $x'$ is assigned by the preceding rule to $P_{i,j}$ or $P_{j,i}$ for some $j\neq i$; assign $u$ to the same set. If $x'$ does not exist, assign $u$ to $P_{i,i}$.
        \item For every node $u$ assigned to $P_{i,j}$, add both $\Left(u), \Right(u)$ to $P_{i,j}$. Finally, $P_{\lambda}(\bF)$ consists of all non-empty $P_{i,j}$. \qedhere
    \end{itemize}
\end{definition}

For later use, we give an equivalent description of this rule for individual positions. For $x\in B_i$, let $a_i(x)$ be the smallest $z\in B_i$ such that $z\geq x$ and $\Other(z)\notin B_i$, if it exists. The node underlying $x$ is assigned to $P_{i,i}$ if $a_i(x)$ does not exist, and otherwise to $P_{\min(i,j),\max(i,j)}$, where $\Other(a_i(x))\in B_j$. Indeed, if both occurrences of the node underlying $x$ lie in $B_i$, every position between them has its other occurrence in $B_i$, so \cref{def:decomp} uses $a_i(x)$ for both occurrences. If they lie in distinct blocks, then $a_i(x)=x$.

\begin{figure*}[htbp]
    \centering

    \begin{subfigure}[t]{0.48\textwidth}
  \centering
  \resizebox{\linewidth}{!}{\input{figures/decomp/block}}
  \caption{A 2-level decomposition of $\fragmentco{0}{2n}$ into blocks. Note that the lower level is a finer partition than the upper level, with $B'_0 = B_0\cup B_1$ and $B'_1 = B_2\cup B_3$.
   We marked all the nodes whose two occurrences do not appear in the same block. We assume the rest of the nodes are all contained within one block.}
  \label{fig:decomp:a}
\end{subfigure}
\hfill
\begin{subfigure}[t]{0.48\textwidth}
  \centering
    \raisebox{-0.2em}{\resizebox{\linewidth}{!}{\input{figures/decomp/partition}}}
  \caption{The induced 2-level decomposition of $\fragmentco{0}{2n}$ into bi-intervals and mono-intervals drawn in different colors.
   We have $P_{0,0}',P_{0,1}', P_{1,1}'$ in the upper level and $P_{0,0}, P_{0,1}, P_{1,1}, P_{1,3}, P_{2,2}, P_{2,3},P_{3,3}$ in the lower level. Note that $P_{0,0}' = P_{1,1}$, $P_{1,1}'=P_{3,3}$, and $P_{0,1}'$ decomposes into $P_{0,0},P_{0,1},P_{1,3},P_{2,2},P_{2,3}$.}
  \label{fig:decomp:b}
\end{subfigure}
    \caption{A 2-level decomposition of an $n$-node forest $\bF$ derived from a 2-level decomposition of $\fragmentco{0}{2n}$. }
    \label{fig:decomp}
\end{figure*}
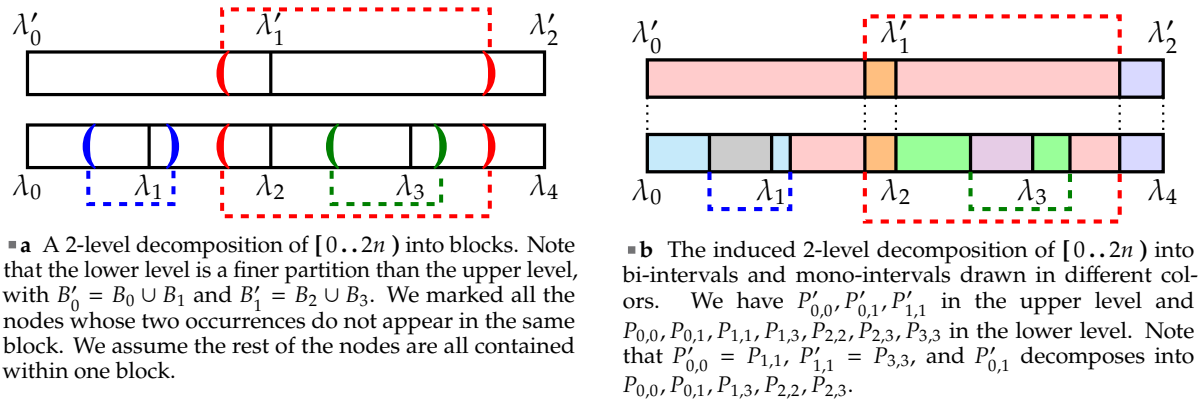

\subparagraph*{Basic Properties.} In the remainder of this section, we prove that $P_{\lambda}(\bF)$ achieves properties similar to a string decomposition into $b$ contiguous blocks. We start with some properties that are easy to verify.

\begin{proposition}\label{prp:decomp_intervals}
    Let $\bF, b$ and $\lambda$ be given as in \cref{def:decomp}.
    Then, for all $i \in \fragmentco{0}{b}$, $P_{i,i}\subseteq B_i$ is an interval, and for all $i,j \in \fragmentco{0}{b}$ with $i < j$, both $P_{i,j}\cap B_i$ and $P_{i,j}\cap B_j$ are intervals.
\end{proposition}

\begin{proof}
    If $x$ is the smallest position in $P_{i,i}$, then $a_i(x)$ does not exist. The same is true for every $y\in B_i$ with $y\geq x$, so $P_{i,i}$ is a suffix of $B_i$.
    Next, consider $x\leq y\leq z$ in $B_i$ such that $x,z\in P_{i,j}$. The positions $a_i(x),a_i(y),a_i(z)$ exist and appear in this order. Since the other occurrences of the first and last belong to $B_j$, \cref{fct:non_crossing} implies that the other occurrence of $a_i(y)$ also belongs to $B_j$. Thus, $y\in P_{i,j}$, and $P_{i,j}\cap B_i$ is an interval. The argument for $P_{i,j}\cap B_j$ is symmetric.
\end{proof}

We remark that \cref{prp:decomp_intervals} implies that we can see $P_\lambda(\bF)$ as a set of bi-intervals and mono-intervals of the form $\langle P_{i,j} \cap B_i, P_{i,j} \cap B_j \rangle$ and $\langle P_{i,i} \rangle$. In the remaining part of this paper, we regard $P_\lambda(\bF)$ as a set of bi-intervals and mono-intervals and use the same names $P_{i,j}$ for them.

\begin{fact}
    Let $\bF, b$ and $\lambda$ be given as in \cref{def:decomp}.
    Then, $P_\lambda(\bF)$ is a partition of $\langle \fragmentco{0}{2n} \rangle$.
\end{fact}

Further, we can show that for each $\mI \in P_\lambda(\bF)$, we have that $\bF_{\mI}$ is a balanced forest.

\begin{proposition}\label{lem:decomp_prop}
    Let $\bF, b$ and $\lambda$ be given as in \cref{def:decomp}.
    Then, every $\mI \in P_\lambda(\bF)$ defines a balanced subforest $\bF_{\mI}$ with at most $\max_{i \in \fragmentco{0}{b}} \{ \lambda_{i+1} - \lambda_i \}$ nodes.
\end{proposition}

\begin{proof}
    Balancedness follows directly from \cref{def:decomp}, which assigns each node, and hence both of its occurrences, to a single element of $P_\lambda(\bF)$. Moreover, every $P_{i,j}$ is contained in $B_i\cup B_j$. It therefore contains at most twice the maximum block length positions, or at most the maximum block length nodes.
\end{proof}

Next, we show the key property that makes $P_{\lambda}(\bF)$ behave similarly to a partition of a string: for any subset $S \subseteq P_{\lambda}(\bF)$, the number of blocks \emph{incident} to any $\mI \in S$, i.e. the number of indices $i \in \fragmentco{0}{b}$ such that $(\bigcup_{\mI \in S} U(\mI)) \cap B_i \neq \emptyset$, is roughly $|S|$. This allows us to think of the elements of $P_{\lambda}(\bF)$ and the blocks $B_i$ as being more or less interchangeable.

\begin{lemma}\label{lem:num_decomp}
    Let $\bF, b$ and $\lambda$ be given as in \cref{def:decomp}.
    For a subset $S \subseteq P_{\lambda}(\bF)$, let $m_S \coloneqq \abs{\{i \in \fragmentco{0}{b} \mid B_i \cap (\bigcup_{\mI \in S} U(\mI)) \neq \emptyset \}}$ be the number of blocks incident to any $\mI \in S$. Then, $|S| \leq 3m_S$.
    In particular, for $S = P_{\lambda}(\bF)$ we get $\abs{P_{\lambda}(\bF)} \leq 3b$.
\end{lemma}
\begin{proof}
    We assign every element of $P_\lambda(\bF)$ to one of its incident blocks. Assign $P_{i,i}$ to $B_i$. For $P_{i,j}$ with $i<j$, let $u$ be its highest spine node. If $u$ has no parent in $\bF$, assign $P_{i,j}$ to $B_i$. Otherwise, the parent $u'$ is not a spine node of $P_{i,j}$, so $\Left(u')\notin B_i$ or $\Right(u')\notin B_j$. In the former case, assign $P_{i,j}$ to $B_i$, and in the latter case to $B_j$, breaking ties in favor of $B_i$.

    At most three elements of $P_\lambda(\bF)$ are assigned to any block $B_i$. Indeed, suppose that $P_{i,j},P_{i,j'}\in P_\lambda(\bF)$ for $i<j<j'$. By \cref{fct:non_crossing}, a spine node of $P_{i,j'}$ is an ancestor of every spine node of $P_{i,j}$. Hence, the parent of the highest spine node of $P_{i,j}$ has its left occurrence in $B_i$ and its right occurrence outside $B_j$, so $P_{i,j}$ is assigned to $B_j$. Consequently, among the elements $P_{i,j}\in P_\lambda(\bF)$ with $j>i$, only the one with the largest possible $j$ can be assigned to $B_i$. Symmetrically, among the elements $P_{j,i}\in P_\lambda(\bF)$ with $j<i$, only the one with the smallest possible $j$ can be assigned to $B_i$. Together with $P_{i,i}$, these are at most three elements of $P_\lambda(\bF)$.

    Every $\mI\in S$ is assigned to a block incident to $\mI$. There are $m_S$ such blocks, and each receives at most three elements of $S$. Therefore, $|S|\leq 3m_S$. Taking $S=P_\lambda(\bF)$ yields $|P_\lambda(\bF)|\leq 3b$.
\end{proof}

\subparagraph*{Constructing Hierarchical Decompositions.}
Next, we show that $P_{\lambda}(\bF)$ also allows for an easy extension to a hierarchical decomposition of $\bF$.
To this end, we consider two sequences of $b+1$ integers $\lambda = (\lambda_0, \ldots, \lambda_{b})$
and $b'+1$ integers $\lambda' = (\lambda_0', \ldots, \lambda_{b'}')$
such that for every $i \in \fragmentco{0}{b'}$ there is $j \in \fragmentco{0}{b}$ such that $\lambda_i' = \lambda_j$. In other words, we have that $\lambda'$ induces a coarser partition of $\fragment{0}{2n}$ than $\lambda$, respecting the endpoints of the partition of $\lambda$.

\begin{lemma}\label{lem:mult_decomp}
    Let $\bF$ be an $n$-node forest, and let $b, \lambda$ and $b',\lambda'$ be given as in \cref{def:decomp}.
    Assume that $\{\lambda_i' : i\in\fragment{0}{b'}\}\subseteq\{\lambda_j : j\in\fragment{0}{b}\}$, and consider the blocks $\{B_i\}_i, \{B_i'\}_i$ used for the construction of $P_{\lambda}(\bF)$ and $P_{\lambda'}(\bF)$, respectively.
    Then, all of the following hold:
    \begin{enumerate}[(a)]
        \item For every $\mI' \in P_{\lambda'}(\bF)$, all $\mI \in P_{\lambda}(\bF)$ such that $U(\mI) \subseteq U(\mI')$ are a partition of $\mI'$; and
        \label{lem:mult_decomp:i}
        \item For every $\mI' \in P_{\lambda'}(\bF)$, the number of $\mI \in P_{\lambda}(\bF)$ such that $U(\mI) \subseteq U(\mI')$ is at most $6c$, where $c \coloneqq \max_{i \in \fragmentco{0}{b'}} \abs{\{j \in \fragmentco{0}{b} \mid B_j \subseteq B_{i}'\}}$.
        \label{lem:mult_decomp:ii}
    \end{enumerate}
\end{lemma}
\begin{proof}
    For every $i\in\fragmentco{0}{b}$, let $f(i)$ be the unique index such that $B_i\subseteq B_{f(i)}'$, and let $a_h'$ be defined as $a_h$ following \cref{def:decomp}, but with respect to the blocks $B_h'$. Consider $P_{i,j}\in P_\lambda(\bF)$ with $f(i)<f(j)$. For every $x\in P_{i,j}\cap B_i$, we have $a_{f(i)}'(x)=a_i(x)$ and $\Other(a_i(x))\in B_j\subseteq B_{f(j)}'$. The symmetric statement holds for $x\in P_{i,j}\cap B_j$. Hence, $U(P_{i,j})\subseteq U(P_{f(i),f(j)}')$.

    Now consider $P_{i,j}\in P_\lambda(\bF)$ with $f(i)=f(j)$. Every $z\in U(P_{i,j})$ satisfies $z,\Other(z)\in B_{f(i)}'$. Moreover, if $x,y\in U(P_{i,j})$ and $x<z<y$, then $z,\Other(z)\in B_{f(i)}'$; this is immediate when $i=j$ and follows from \cref{fct:non_crossing} when $i<j$. Consequently, $a_{f(i)}'(z)$ has the same value for every $z\in U(P_{i,j})$, or is undefined for every such $z$. The description following \cref{def:decomp} therefore yields an $\mI'\in P_{\lambda'}(\bF)$ such that $U(P_{i,j})\subseteq U(\mI')$. Since both decompositions partition $\fragmentco{0}{2n}$, this proves \ref{lem:mult_decomp:i}.

    For \ref{lem:mult_decomp:ii}, fix $\mI'\in P_{\lambda'}(\bF)$ and set $S\coloneqq\{\mI\in P_\lambda(\bF):U(\mI)\subseteq U(\mI')\}$. The set $U(\mI')$ intersects at most two blocks $B_h'$, each of which is the union of at most $c$ blocks $B_j$. Hence, $m_S\leq 2c$. By \cref{lem:num_decomp}, $|S|\leq 3m_S\leq 6c$.
\end{proof}

This allows us to define the following hierarchical tree decomposition of $\bF$.

\begin{definition}\label{def:partition_tree}
     Given an $n$-node forest $\bF$ and $c \in \mathbb{Z}_{\geq 2}$,
     we define the tree $T_c(\bF) = (V, E)$ as follows.
     Let $\lambda(c)$ be the sequence $ \lambda(c) = (\lambda_0(c), \ldots, \lambda_{\ceil{2n/c}}(c))$ such that $\lambda_i(c) = \min(i \cdot c, 2n)$. Then:

     \begin{itemize}
         \item For every $i \in \fragment{0}{\ceil{\log_c 2n}}$ and every $\mI \in P_{\lambda(c^i)}(\bF)$ the tree $T_c(\bF)$ contains a node $v$. We say that such a node $v$ \emph{corresponds} to the pair $(i, \mI)$.
         \item For every $i,j\in \fragment{0}{\ceil{\log_c 2n}}$
         and $\mI \in P_{\lambda(c^i)}(\bF)$, $\mJ \in P_{\lambda(c^j)}(\bF)$, where nodes $u,v$ correspond to $(i, \mI)$ and $(j, \mJ)$, respectively, there is an edge $(u,v)$ if and only if $U(\mJ) \subseteq U(\mI)$ and $j=i-1$.
         \qedhere
     \end{itemize}
\end{definition}

Thus, we can view $T_c(\bF) = (V, E)$ as the tree analogue of the hierarchical partitioning of a string into blocks whose lengths are powers of $c$. The main difficulty of working with a tree rather than a string is that we must work with bi-intervals instead of only intervals. Moreover, in the finer partitioning, the number of subsets per part increases from $c$ to at most $6c$.

\begin{lemma}\label{lmm:num_nodes_in_TcF}
    Let $\bF$ be an $n$-node forest and $c \geq 2$ be an integer. The partition tree $T_c(\bF)$ has $\ceil{\log_c(2n)} + 1$ levels and at most $6n = \Oh(n)$ nodes. The degree of $T_c(\bF)$ is at most $6c = \Oh(c)$.
\end{lemma}
\begin{proof}
    Let $d = \ceil{\log_c(2n)} + 1$ be the number of levels in $T_c(\bF)$, indexed by $i \in \fragmentco{0}{d}$. Level $i$ has $\ceil{2n/c^i}$ blocks of size $c^i$.
    From \cref{lem:num_decomp}, we know that the number of nodes in $T_c(\bF)$ on level $i$ is at most $3\ceil{2n/c^i}$, which we can coarsely bound with $12n/2^{i}$.

    We bound the number of nodes on levels 0, 1, and 2 with $n$ each, which is better than what $12n/2^i$ gives on those levels. Therefore, we can bound the total number of nodes in $T_c(\bF)$ with
    \[
    3n + \sum_{i = 3}^{d-1}\frac{12n}{2^i} = 3n + 12n\sum_{i=3}^{d-1}\frac{1}{2^i} \leq 6n. 
    \]
    The degree of $T_c(\bF)$ is proved by \cref{lem:mult_decomp}\ref{lem:mult_decomp:ii}.
\end{proof}

\subparagraph*{Sensitivity of $P_\lambda(\bF)$.}
We conclude this section by showing that $P_\lambda(\bF)$ has good ``sensitivity'', meaning that if we change $\lambda$ or the underlying tree only by a bit, $P_\lambda(\bF)$ does not change too much either.
This will be relevant for our dynamic algorithms.

First, we analyze the sensitivity under removing or adding one element to $\lambda$.

\begin{proposition}\label{def:partition_sensitivity}
    Let $\bF$ be an $n$-node forest, and let $b, \lambda$ and $b',\lambda'$ be given as in \cref{def:decomp}. Suppose $b'= b-1$, and suppose that for all
    $j \in \fragmentco{0}{b'}$ there is an $i \in \fragmentco{0}{b}$
    such that $\lambda_j' = \lambda_i$, i.e. $\lambda'=(\lambda_0,\ldots,\lambda_{k-1},\lambda_{k+1},\ldots,\lambda_b)$ for some $k\in\fragmentoo{0}{b}$.
    Then, the following two statements hold:
    \begin{enumerate}[(a)]
        \item The size of the symmetric difference $P_{\lambda}(\bF) \Delta P_{\lambda'}(\bF)$, i.e., $\abs{P_{\lambda'}(\bF) \setminus P_{\lambda}(\bF)} + \abs{P_{\lambda}(\bF) \setminus P_{\lambda'}(\bF)}$, is a constant that does not depend on $b$; and
        \label{it:partition_sensitivity:a}
        \item To know the elements in $P_{\lambda}(\bF) \Delta P_{\lambda'}(\bF)$, it suffices to read $\mS(\bF)$ in the block $\fragmentco{\lambda_{k-1}'}{\lambda_{k}'}$ and one other block which can be inferred after reading the former.
        In particular, the time-complexity to infer them is linear in the number of characters read from $\mS(\bF)$.\label{it:partition_sensitivity:b}
    \end{enumerate}
\end{proposition}

\begin{proof}
    We have $B_{k-1}'=B_{k-1}\cup B_k$. By the description following \cref{def:decomp}, every $\mI'\in P_{\lambda'}(\bF)$ satisfying $B_{k-1}\cap U(\mI')=\emptyset$ or $B_k\cap U(\mI')=\emptyset$ also belongs to $P_\lambda(\bF)$. (Indeed, if $U(\mI') \cap B_k = \emptyset$, then for every $x \in U(\mI') \cap B_{k-1}$ no $z \in \fragment{x}{a'_{k-1}(x)}$ has $\Other(z) \in B_k$, since all such $z$ are assigned to $\mI'$ together with $\Other(z)$; hence $a_{k-1}(x) = a'_{k-1}(x)$, or both are undefined, and all positions of $\mI'$ receive the same keys under $\lambda$ and $\lambda'$ up to renumbering the blocks. The case $U(\mI') \cap B_{k-1} = \emptyset$ is symmetric.)

    By \cref{prp:decomp_intervals}, there is at most one $\mJ\in P_{\lambda'}(\bF)$ such that both $B_{k-1}\cap U(\mJ)\neq\emptyset$ and $B_k\cap U(\mJ)\neq\emptyset$. If no such $\mJ$ exists, the two decompositions are equal. Otherwise, $|P_{\lambda'}(\bF)\setminus P_\lambda(\bF)|\leq 1$, and every $\mI\in P_\lambda(\bF)\setminus P_{\lambda'}(\bF)$ satisfies $U(\mI)\subseteq U(\mJ)$. Since $B_{k-1}'$ is the union of two blocks $B_{k-1}$ and $B_k$, \cref{lem:mult_decomp}\ref{lem:mult_decomp:ii} implies that there are at most $12$ such $\mI$. This proves \ref{it:partition_sensitivity:a}.

    For \ref{it:partition_sensitivity:b}, scan $B_{k-1}'=\fragmentco{\lambda_{k-1}'}{\lambda_k'}$ to determine $\mJ$. If $\mJ$ is a bi-interval, also scan the other block intersecting $U(\mJ)$. These one or two blocks determine $\mJ$ and every $\mI\in P_\lambda(\bF)$ with $U(\mI)\subseteq U(\mJ)$, and determine the entire symmetric difference. The work is linear in the number of positions scanned.
\end{proof}

Last but not least,  we analyze the sensitivity under removing or adding a node to $\bF$.

\begin{proposition}\label{def:partition_sensitivity2}
    Let $\bF$ be an $n$-node forest, and let $\bF'$ be a forest such that
    \[
        \mS(\bF') = \mS(\bF)\fragmentco{0}{\alpha} \circ v \circ \mS(\bF)\fragmentco{\alpha}{\beta} \circ v \circ \mS(\bF)\fragmentco{\beta}{2n}
    \]
    for $0\leq\alpha\leq\beta\leq 2n$ and a new node $v$. Define the order embedding of the old positions by $\iota(x)=x+\mathbf{1}_{x\geq\alpha}+\mathbf{1}_{x\geq\beta}$ for $x\in\fragmentco{0}{2n}$. Separately, define $g:\fragment{0}{2n}\to\fragment{0}{2n+2}$ on interval boundaries by
    \[
        g(t)=\begin{cases}
            0 & \text{if }t=0,\\
            2n+2 & \text{if }t=2n,\\
            t+\mathbf{1}_{\alpha<t}+\mathbf{1}_{\beta<t} & \text{otherwise}.
        \end{cases}
    \]
    Let $b,\lambda$ be given as in \cref{def:decomp}, and define $\lambda'=(g(\lambda_0),\ldots,g(\lambda_b))$. For an interval $I=\fragmentco{p}{q}$, set $g(I)=\fragmentco{g(p)}{g(q)}$, and apply $g$ componentwise to interval collections. If $P_{r,s}'$ denotes the set constructed from $\bF'$ and $\lambda'$, let
    $S=\{P_{r,s}\in P_\lambda(\bF):P_{r,s}'\in P_{\lambda'}(\bF')\text{ and }g(P_{r,s})=P_{r,s}'\}$. Then, all of the following hold:
    \begin{enumerate}
        \item The sets $N\coloneqq P_\lambda(\bF)\setminus S$ and $N'\coloneqq P_{\lambda'}(\bF')\setminus\{g(\mI):\mI\in S\}$ satisfy $|N|\leq 3$ and $|N'|\leq 3$.
        \item We can compute $N$ and $N'$ in time $\Oh(1)$ plus $\Oh(1)$ answers from the following queries on $P_{\lambda}(\bF)$ and $P_{\lambda'}(\bF')$:
        \begin{enumerate*}[(a)]
            \item Is the partition $P_{i,j}$ an element of $P_{\lambda}(\bF)$ (or $P_{\lambda'}(\bF')$)? If so, retrieve it. \label{query:partition_sensitivity2:a}
            \item Given a block $B_i$ in $P_{\lambda}(\bF)$ and an index $x \in B_{i}$, what is the smallest (resp. largest) index $z \geq x$ (resp. $z \leq x$) such that $z \in B_i$ and $\Other(z) \notin B_i$? \label{query:partition_sensitivity2:b}
        \end{enumerate*}
    \end{enumerate}
\end{proposition}
\begin{proof}
    The definition of $g$ gives $g(0)=0$ and $g(2n)=2n+2$, so the blocks $B_0',\ldots,B_{b-1}'$ cover all of $\fragmentco{0}{2n+2}$. Moreover, for every old position $x$ and every $h\in\fragmentco{0}{b}$, we have $x\in B_h$ if and only if $\iota(x)\in B_h'$, and $\Other'(\iota(x))=\iota(\Other(x))$, where $\Other'$ is the counterpart map of $\bF'$.

    Let $a_h$ be as in the description following \cref{def:decomp}. For $x\in B_h$, define $\kappa(x)=(h,h)$ if $a_h(x)$ does not exist, and define $\kappa(x)=(\min(h,t),\max(h,t))$ if $\Other(a_h(x))\in B_t$. Thus, the position $x$ belongs to $P_{r,s}$ precisely when $\kappa(x)=(r,s)$. Define $a_h'$ and $\kappa'$ analogously for $\bF'$ and the blocks $B_h'$. Let $\alpha\in B_i'$ and $\beta+1\in B_j'$. Since $\alpha\leq\beta$, we have $i\leq j$.

    \begin{description}
        \item[Case $i=j$.] The positions $\alpha$ and $\beta+1$ do not satisfy the condition in the definition of $a_i'$. Hence, $\kappa'(\iota(x))=\kappa(x)$ for every old position $x$. The validity of the insertion also implies that every occurrence in $\mS(\bF)\fragmentco{\alpha}{\beta}$ has its other occurrence in the same interval. Therefore, no position between the two insertion boundaries satisfies the condition in the definition of $a_i$.

        If $\beta<2n$, set $x=\beta$ and let $\kappa(x)=(c,d)$. Both new occurrences belong to $P_{c,d}'$, and the convention in the definition of $g$ adds them to $g(P_{c,d})$. Thus, every element of $P_\lambda(\bF)$ is in $S$, their images partition $\fragmentco{0}{2n+2}$, and $N=N'=\emptyset$. If $\beta=2n$, let $\kappa(2n-1)=(c,d)$. The map $g$ adds both new occurrences to $g(P_{c,d})$, whereas \cref{def:decomp} assigns $v$ to $P_{i,i}'$. Consequently,
        $N\subseteq\{P_{c,d},P_{i,i}\}\cap P_\lambda(\bF)$ and $N'\subseteq\{P_{c,d}',P_{i,i}'\}\cap P_{\lambda'}(\bF')$. In either subcase, $|N|,|N'|\leq2$.

        \item[Case $i<j$.] Set $p_i=\alpha$ and $p_j=\beta$. For each $h\in\{i,j\}$, set $x_h=p_h$ when $p_h<2n$ and $x_h=2n-1$ when $p_h=2n$, and write $\kappa(x_h)=(c_h,d_h)$. Notice that $x_h\in B_h$. Let $z_h$ be the largest $z\in B_h$ such that $z<p_h$ and $\Other(z)\notin B_h$, if it exists, and set $R_h=B_h\cap\fragmentco{z_h+1}{p_h}$ if it does and $R_h=B_h\cap\fragmentco{0}{p_h}$ otherwise.

        Every $x\in R_h$ satisfies $\kappa(x)=(c_h,d_h)$. Moreover, non-crossing and the validity of the insertion imply that $x\in R_h$ if and only if $\Other(x)\in R_h$. After the insertion, the first position at or after $\iota(x)$ whose other occurrence is outside $B_h'$ is $\alpha$ for $h=i$ and $\beta+1$ for $h=j$. Hence, $\kappa'(\iota(x))=(i,j)$. For every old position outside $R_i\cup R_j$, the first such position is the image under $\iota$ of the one used before the insertion, or neither exists. Its key therefore remains unchanged.

        It remains to account for the two positions inserted at the boundaries. By the definition of $g$, the position inserted at $p_h$ is added to the component of $g(P_{c_h,d_h})$ containing $\iota(x_h)$; when $p_h=2n$, it is added to the component ending at $2n+2$. On the other hand, both inserted occurrences belong to $P_{i,j}'$. We conclude that
        \[
            N\subseteq\{P_{c_i,d_i},P_{c_j,d_j},P_{i,j}\}\cap P_\lambda(\bF)
            \quad\text{and}\quad
            N'\subseteq\{P_{c_i,d_i}',P_{c_j,d_j}',P_{i,j}'\}\cap P_{\lambda'}(\bF').
        \]
        Each set on the right contains at most three elements, proving the first claim.
    \end{description}

    We finally prove the second claim. In the case $i=j$, one successor query from $\beta$ determines $(c,d)$ when $\beta<2n$; when $\beta=2n$, the same query is made from $2n-1$. In the case $i<j$, query for $z_h$ from $p_h-1$ whenever $B_h\cap\fragmentco{0}{p_h}$ is non-empty; otherwise $R_h$ is empty. Two successor queries from $x_i,x_j$ determine $(c_i,d_i),(c_j,d_j)$. Queries of type \ref{query:partition_sensitivity2:a} on the two decompositions then retrieve the at most three partitions on each side displayed above. Comparing their endpoints after applying $g$ determines exactly which of them belong to $N$ and $N'$. The number of queries and the additional work are constant.
\end{proof}

%% file: s5_approx.tex
\section{A Near-Linear-Time Approximation Algorithm for \TED}
\label{sec:near-linear}

In this section, we prove \cref{thm:approxted}.

\approxted*

This section begins by generalizing the \emph{Tree Distance} framework of \cite{AKO10,BCFN22} from strings to forests (\cref{subsec:td}). Then, as a warm-up, \cref{subsec:nkalgo} demonstrates that a direct application of this framework (mirroring the approach for strings) yields an algorithm whose runtime falls short of our near-linear goal. Finally, \cref{subsec:nalgo} introduces further structural insights to achieve the desired near-linear runtime.

\subparagraph*{New Notation:} Starting from this section, for a node $v$ in a partition tree defined in \cref{def:partition_tree}, we use $\mI_v$ to denote the mono-interval or bi-interval that $v$ is assigned to.

\subsection{Tree Distance (not to be Confused with Tree Edit Distance)}
\label{subsec:td}

Given two $n$-node forests $\bF, \bG$, let us consider the partition tree $T_{c}(\bF)$ and a node $v \in T_{c}(\bF)$. We define the notion of \emph{Tree Distance} ($\TD$) at $v$ with respect to some shifts. If $v$ represents a bi-interval, then we consider two shifts, one for the left component and one for the right. Otherwise, $v$ represents a single interval, and we only consider one shift.

\begin{definition}[Tree Distance $\TD$]\label{def:TD}
    Given two $n$-node forests $\bF, \bG$, consider the partition tree $T_{c}(\bF)$ and a node $v \in T_{c}(\bF)$. Denote the interval collection corresponding to $v$ as $\mI = \langle I_1, \ldots I_\ell \rangle$ for $\ell \in \{1,2\}$.
    We define $\TD_{v,s}(\bF, \bG)$ relative to a shift $s = \langle s_1, \ldots, s_{\ell} \rangle, s_i \in \fragment{-{2k}}{{2k}}$ as follows:
    \begin{itemize}
        \item If $v$ is a leaf, then
        \[
         \TD_{v,s}(\bF, \bG) \coloneqq \ted(\bF_{\mI}, \bG_{\mI, s}).
        \]
    \item If $v$ is an internal node with $B_v$ children $v_1, \ldots, v_{B_v}$, where $v_i$ has corresponding interval collection $\mI^{i} = \langle I^i_{1}, \ldots, I^{i}_{\ell_i} \rangle$
    with $\ell_i \in \{1,2\}$ intervals, then
    \begin{align*}
        \TD_{v,s}(\bF, \bG) &\coloneqq \sum_{i=1}^{B_v}\min_{r \in \fragment{-{2k}}{{2k}}^{\ell_i}} \left\{\TD_{v_i, r}(\bF, \bG)+\penaltyFunc(v,v_i,s, r) \right\},
    \end{align*}
    where $\penaltyFunc$ is the \emph{penalty term} defined as
    \[
        \penaltyFunc(v,v_i,s, r) \coloneqq 2\sum_{j=1}^{\ell_i} \abs{r_j - s_{\mathsf{pi}(i,j)}}.
    \]
    \end{itemize}
    We write $\TD(\bF, \bG) = \TD_{r,\langle 0\rangle}(\bF, \bG)$, where $r$ is the root of $T_{c}(\bF)$, a mono-interval. When the context is clear, we omit $\bF$ and $\bG$ and simply write $\TD_{v,s}$ instead of $\TD_{v,s}(\bF, \bG)$.
\end{definition}

Observe that in our construction of $T_{c}(\bF)$ in \cref{def:partition_tree}, the leaves are always bi-intervals whose left and right components form a single node. The value $\ted(\bF_{\mI}, \bG_{\mI, s})$ is easy to compute for leaves.

As we will see later on in the analysis of the algorithm, it suffices to only consider shifts in the range $\fragment{-{2k}}{{2k}}$ to solve the $(k,K)$-\GAPTED Problem with subpolynomial gap.

\subparagraph*{Upper and Lower Bounds for $\TD$.}
We can show that the Tree Distance is a good approximation of the Tree Edit Distance. First, let us show that $\TD(\bF, \bG)$ is lower-bounded by $\ted(\bF, \bG)$.
\begin{lemma}\label{lmm:TD_lb}
    Given two size-$n$ forests $\bF, \bG$, we have
    $\ted(\bF, \bG) \leq \TD(\bF, \bG)$.
\end{lemma}

\begin{proof}
     In fact, we will show the more general claim that $\ted(\bF_\mI, \bG_{\mI,s}) \leq \TD_{v,s}$ using induction,
    where $\mI = \langle I_1, \ldots, I_\ell \rangle$ is the interval collection corresponding to a node $v$ and the shift $s \in \fragment{-{2k}}{{2k}}^{\ell}$ is arbitrary. The lemma then follows by taking $\mI = \langle \fragmentco{0}{2n} \rangle$ and $s = \langle 0 \rangle$, i.e., the root with no shift.

    For the base case, if $v$ is a leaf, then we have $\ted(\bF_\mI, \bG_{\mI,s}) = \TD_{v,s}$ by definition.

    Now let us suppose $v$ is an internal node corresponding to an interval collection $\mI = \langle I_1, \ldots I_\ell \rangle$ for $\ell \in \{1,2\}$
    and let us consider a shift $s = \langle s_1, \ldots, s_{\ell} \rangle$ for $\mI$.
    Further, suppose $v$ has $B_v$ children $v_1, \ldots, v_{B_v}$ of $v$,
    where $v_i$ has corresponding interval collection $\mI^{i} = \langle I^i_{1}, \ldots, I^{i}_{\ell_i} \rangle$ with $\ell_i \in \{1,2\}$ intervals.
    Setting $s^i = \langle s_{\mathsf{pi}(i,1)}, \ldots, s_{\mathsf{pi}(i,\ell_i)} \rangle$,
    allows us to write
    \begin{align*}
        \ted(\bF_\mI, \bG_{\mI,s}) &\leq \sum_{i=1}^{B_v} \ted(\bF_{\mI^i},\bG_{\mI^i, s^i}) && \text{(\Cref{prop:ted_decomp})} \\
        &\leq \sum_{i=1}^{B_v} \min_{r \in \fragment{-{2k}}{{2k}}^{\ell_i}}
    \Big\{ \ted(\bF_{\mI^i}, \bG_{\mI^i, r}) + 2 \sum_{j}^{\ell_i} \abs{s_{\mathsf{pi}(i,j)} - r_j} \Big\}
    && \text{(\cref{prop:triangle_inequality})} \\
    &\leq \sum_{i=1}^{B_v} \min_{r \in \fragment{-{2k}}{{2k}}^{\ell_i}}
    \left\{ \TD_{v_i, r} + \penaltyFunc(v,v_i, s,r) \right\} = \TD_{v,s}. && \text{(inductive hypothesis)}
    \end{align*}
\end{proof}

Our next goal will be to prove that $\TD(\bF,\bG)$ can also be upper-bounded reasonably well by $\ted(\bF,\bG)$.

\begin{lemma}\label{lmm:TD_ub}
Given two size-$n$ forests $\bF, \bG$ such that $\ted(\bF, \bG) \leq k$,
we have
    $\TD(\bF, \bG) \leq 48\cdot cD\cdot \ted(\bF, \bG)$, where $D = \ceil{\log_c(2n)} + 1$ denotes the number of levels of the partition tree $T_{c}(\bF)$ (\cref{lmm:num_nodes_in_TcF}).
\end{lemma}
\begin{proof}
    Let $\ell_v$ be the number of intervals in the interval collection of $\mI_v$, and let $r$ be the root of $T_{c}(\bF)$.
    By expanding the recursive definition of $\TD$ in \cref{def:TD}, we get that
\begin{equation}
\TD(\bF, \bG) = \min_{\substack{s_{r} = \langle 0 \rangle \\ \text{all other $v$, $s_v \in \fragment{-{2k}}{{2k}}^{\ell_v}$}}}
\left(
\ \sum_{\substack{\text{internal} \\ \text{nodes $v$}}} \
\sum_{\substack{\text{child } w \\ \text{of } v}}
\penaltyFunc(v,w,s_v,s_w)
\right)
+ \sum_{\text{leaves } v} \ted(\bF_{\mI_v}, \bG_{\mI_v,s_v}),
\label{eq:TD_unravel}
\end{equation}
where the minimum is taken over all shifts $s_v \in \fragment{-{2k}}{{2k}}^{\ell_v}$ for all nodes (of appropriate length), with the exception that the shift at the root node is $\langle 0 \rangle$. It suffices to choose specific shifts for each node and argue that \cref{eq:TD_unravel} with our specified shifts plugged in is upper bounded by $48 \cdot cD\cdot \ted(\bF, \bG)$.

Fix an arbitrary optimal $\mA : \bF \onto \bG$ of cost at most $k$.
We choose our shifts as follows: for a node $v \in T_{c}(\bF)$ let  $\mI_v = \langle I_{v,1},\ldots, I_{v,\ell_v} \rangle$, where $I_{v,i} = \fragmentco{a_{v,i}}{b_{v,i}}$, and set $s_v = \langle \pi_{\mA}(a_{v,1})- a_{v,1}, \ldots,  \pi_{\mA}(a_{v,\ell_v})-a_{v,\ell_v} \rangle$ (for $\pi_{\mA}(\cdot)$ see \cref{def:ghost_mappings}). Note that, by \cref{clm:TD_ub_diff_ted}, we have $s_v \in \fragment{-{2k}}{{2k}}^{\ell_v}$.
Further, for a node $v \in T_{c}(\bF)$, we let $\mJ_v$ be the image of $\mI_v$ under $\pi_{\mA}$, i.e., $\mJ_v = \langle J_{v,1}, \ldots, J_{v,\ell_v} \rangle$, where $J_{v,i} = \fragmentco{\pi_{\mA}(a_{v,i})}{\pi_{\mA}(b_{v,i})}$.

\begin{claim}\label{clm:TD_ub_pen_ted}
    For an internal node $v \in T_{c}(\bF)$ and a child $w$ of $v$, we have $\penaltyFunc(v,w,s_v,s_w) \leq 8 \cdot \ted(\bF_{\mI_v}, \bG_{\mJ_v})$.
\end{claim}
\begin{claimproof}
    Consider $I_{w,i}$ for $i \in \fragment{1}{\ell_w}$.
    If $a_{w,i} \in I_{v,i}$, we can use \cref{clm:TD_ub_diff_ted} on $a_{w,i}$
    and $a_{v,j}$ to get $\abs{(a_{v,j} - \pi_{\mA}(a_{v,j})) - (a_{w,i} - \pi_{\mA}(a_{w,i}))} \leq 2 \cdot \ted(\bF_{\mI_v}, \bG_{\mJ_v})$.
    Applying this to all $\ell_w \leq 2$ intervals in $\mI_w$ yields $\penaltyFunc(v,w,s_v,s_w) \leq 8 \cdot \ted(\bF_{\mI_v}, \bG_{\mJ_v})$.
\end{claimproof}

Thus, \cref{clm:TD_ub_pen_ted} allows us to bound the first term in \cref{eq:TD_unravel}. We now bound the second term:
\begin{align}
    \ted(\bF_{\mI_v}, \bG_{\mI_v,s_v})
    &\leq \ted(\bF_{\mI_v}, \bG_{\mJ_v}) + \frac{1}{2}( \abs{\mI_v+s_{v}} - \abs{\mJ_v}) + 2 \textstyle\sum_{i=1}^{\ell_v} \abs{J_{v,i} \setminus  (I_{v,i}+s_v)}, \label{eq:TD_ub_pen_ted:1}  \\
    &\leq \ted(\bF_{\mI_v}, \bG_{\mJ_v}) + \frac{5}{2} \cdot \textstyle\sum_{i=1}^{\ell_v} ||J_{v,i}|  - |I_{v,i}+s_{v,i}||
    \label{eq:TD_ub_pen_ted:2} \\
    & \leq 6 \cdot \ted(\bF_{\mI_v}, \bG_{\mJ_v}) \label{eq:TD_ub_pen_ted:3}
\end{align}
where in \cref{eq:TD_ub_pen_ted:1}, we use \cref{prop:triangle_inequality};
in \cref{eq:TD_ub_pen_ted:2} we use that $\mI_v + s_v$ and $\mJ_v$ have the same starting points; and finally in \cref{eq:TD_ub_pen_ted:3} we use \cref{prop:size_lb}. Plugging in our specified shifts into \cref{eq:TD_unravel} we conclude:
\begin{align*}
\TD(\bF, \bG) &\leq
\left(
\sum_{\substack{\text{internal} \\ \text{nodes $v$}}}
\sum_{\substack{\text{child } w \\ \text{of } v}}
8 \cdot \ted(\bF_{\mI_v}, \bG_{\mJ_v})
\right)
+ \sum_{\text{leaves } v} 6 \cdot \ted(\bF_{\mI_v}, \bG_{\mJ_v}) && \text{(\cref{clm:TD_ub_pen_ted} and previous bound)}\\
&\leq 8\cdot 6c \sum_{\text{all $v$}} \ted(\bF_{\mI_v}, \bG_{\mJ_v}) && \text{($v$ has at most $6c$ children)}\\
&\leq 48\cdot cD\cdot \ted(\bF, \bG).  && \text{(\cref{clm:TD_ub_sum_ted})}
\end{align*}
\end{proof}

\subparagraph*{Remark:} Note that if we compute $\TD_{v,s}(\bF,\bG)$ for every node $v \in T_{c}(\bF)$ and shift $s \in \fragment{-{2k}}{{2k}}^{\ell_v}$, it would take $\Oh(nck^2)$ time: For every parent-child pair $(v,v_i)$ where $v$ has a shift
$s$ and $v_i$ has a shift $r$, computing $\min_{r\in \fragment{-{2k}}{{2k}}^{\ell_{v_i}}}\{\TD_{v_i,r}(\bF,\bG) + \penaltyFunc(v,v_i,s,r)\}$ for all $s$ takes $\Oh(k^2)$ time using \cref{lmm:range_min_query}.
In the next (sub)section, we will see that by using \emph{Precision Sampling} (\cref{lmm:precision_sampling}) in the process of computing $\TD$, we can shave off almost a $k$ factor in the running time, at a negligible cost to the approximation factor.

\subsection{Warm-up: An \texorpdfstring{$\Ohhat(nk)$}{Ô(nk)}-Time Algorithm}
\label{subsec:nkalgo}

In this (sub)section, we give an algorithm that does not yet achieve the desired $\Ohhat(n)$
time bound, but comes close to it and runs in $\Ohhat(nk)$ time.
Throughout this (sub)section, for a node $v \in T_{c}(\bF)$, let $\mI_v$ be its corresponding interval collection with $\ell_v$ intervals.

To achieve this, we crucially rely on \cref{lmm:precision_sampling}.

For a formulation and proof of this lemma, see \cite[Lemma~11 and Appendix~B]{BCFN22}.

\begin{lemma}[Precision Sampling Lemma \cite{AKO10}]\label{lmm:precision_sampling}
    Let $\varepsilon, \delta > 0$. There is a distribution $\mathcal{D}(\varepsilon, \delta)$ supported over the real interval $(0,1]$ with the following two properties:
    \begin{description}
        \item[Accuracy:] Let $A_1, \ldots, A_n \in \mathbb{R}_{\geq 0}$ and let $u_1, \ldots, u_n \sim \mathcal{D}(\varepsilon, \delta)$ be sampled independently. There is a recovery algorithm with the following guarantee: Given $(\alpha, \beta \cdot u_i)$-approximations $\widetilde{A}_i$ of $A_i$, the algorithm $((1+\varepsilon)\cdot \alpha, \beta)$-approximates $\sum_iA_i$ in time $\Oh(n \cdot  \varepsilon^{-2}\log(\delta^{-1}))$, with probability at least $1-\delta$ and for any parameters $\alpha \geq 1$ and $\beta \geq 0$.
        \item[Efficiency:] Sample $u \sim \mathcal{D}(\varepsilon, \delta)$. Then, for any $N \geq 1$, there is an event $E = E(u, N)$ happening with probability at least $1 - 1/N$, such that $\Ex{1/u}{E} \leq \Ohtilde(\varepsilon^{-2}\log(\delta^{-1})\log N)$. \lipicsEnd
    \end{description}
\end{lemma}

\begin{algorithm}[th]
    \KwInput{Forests $\bF, \bG$, a node $v$ in $T_{c}(\bF)$, and a sampling rate $r_v > 0$.}
    \KwOutput{$\Delta_{v,s}$ for all shifts $s \in \fragment{-{2k}}{{2k}}^{\ell_v}$.}

    \If{$\abs{\mI_v} \leq \frac{1}{r_v}$} { \label{alg:base:1}
        \textbf{return} $\Delta_{v,s} = 0$ for all $s \in \fragment{-{2k}}{{2k}}^{\ell_v}$; \label{alg:base:2}
    }
    \If {$v$ is a leaf node then} { \label{alg:base:3}
        \textbf{return} $\Delta_{v,s} = \ted(\bF_{\mI_v}, \bG_{\mI_v,s})$ for all $s \in \fragment{-{2k}}{{2k}}^{\ell_v}$; \label{alg:base:4}
    }
    \For {$i \in \fragment{1}{B_v}$} {  \label{alg:base:5}
    Let $v_i$ be the $i$-th child of $v$\; \label{alg:base:6}
    Sample $u_{v_i} \sim \mD(\varepsilon=(2\log(n))^{-1}, \delta = 0.01 \cdot (6n(4k+1)^2)^{-1})$, where $\mD$ comes from \cref{lmm:precision_sampling}\; \label{alg:base:7}
    Recursively compute $\Delta_{v_i, s}$ with rate $r_v/u_{v_i}$ for all $s \in \fragment{-{2k}}{{2k}}^{\ell_{v_i}}$\;\label{alg:base:8}
    Compute $\widetilde{\Delta}_{i,s} = \min_{s_i \in \fragment{-{2k}}{{2k}}^{\ell_{v_i}}} \left\{ \Delta_{v_i,s_i} + \penaltyFunc(v, v_i, s, s_i) \right\}$ for all $s \in \fragment{-{2k}}{{2k}}^{\ell_v}$ using \cref{lmm:range_min_query};} \label{alg:base:9}
    \For{$s \in \fragment{-{2k}}{{2k}}^{\ell_v}$} { \label{alg:base:10}
        Let $\Delta_{v,s}$ be the result of the Precision Sampling Lemma (\cref{lmm:precision_sampling}) applied to $\widetilde{\Delta}_{1,s}, \ldots, \widetilde{\Delta}_{B_v,s}$ with precisions $u_{v_1}, \ldots, u_{v_{B_v}}$; \label{alg:base:11}
    }
    \textbf{return} $\Delta_{v,s}$ for all $s \in \fragment{-{2k}}{{2k}}^{\ell_v}$. \label{alg:base:12}
    \caption{A baseline algorithm for approximating $\ted$ (or more precisely $\TD$).}
    \label{alg:nearlinear_baseline}
\end{algorithm}

Instead of calculating $\TD_{v,s}(\bF, \bG)$ exactly, our baseline algorithm approximates the Tree Distance at every node. In order to specify the extent of approximation, we assign two parameters to every node $v$:
\begin{itemize}
    \item The \emph{additive} accuracy $1/r_v$. If $v$ is the root, we set $r_v = 1000/K$. Otherwise if $v$ is a child of $w$ then we sample $u_v \sim \mD((2\log(n))^{-1}, 0.01 \cdot (6n(4k+1)^2)^{-1})$ and set $r_v = r_w/u_v$.
    \item The \emph{multiplicative} accuracy $\alpha_v$. We set $\alpha_v = 2 \cdot (1-(2\log n)^{-1})^{d_v}$, where $d_v$ is the depth of $v$. Note that $\alpha_v \leq 2$, and that $\alpha_v \geq 1$ holds for all $n$ larger than a suitable constant, since $d_v \leq \lceil \log_c 2n \rceil \leq \log n + 2$ and $(1 - (2\log n)^{-1})^{\log n + 2} \to e^{-1/2} > 1/2$; we assume this throughout.
\end{itemize}

To be consistent with previous works, we also define the \emph{Tree Distance Problem}.

\begin{definition}[Tree Distance Problem]
Given a node $v \in T_{c}(\bF)$ and shifts $s \in \fragment{-{2k}}{{2k}}^{\ell_v}$, compute a number $\Delta_{v,s}$ such that
\[
\frac{1}{\alpha_v}\TD_{v,s}(\bF, \bG) - \frac{1}{r_v} \leq \Delta_{v,s} \leq \alpha_v \TD_{v,s}(\bF, \bG) + \frac{1}{r_v}.
\]
i.e., $\Delta_{v,s}$ is an $(\alpha_v, 1/r_v)$-approximation of $\TD_{v,s}(\bF, \bG)$.
\end{definition}
\cref{alg:nearlinear_baseline} describes the steps to calculate $\Delta_{v,s}$ for all nodes $v$ and shifts $s \in \fragment{-{2k}}{{2k}}^{\ell_v}$. Note that the structure of our algorithm closely mirrors the definition of Tree Distance in \cref{def:TD}. \cref{alg:base:1} to \cref{alg:base:4} handles the base case. If $\mI_v$ is too short, then we set $\Delta_{v,s} = 0$ for all $s$ (\cref{alg:base:1} to \cref{alg:base:2}). If $v$ is a leaf node then we calculate the Tree Distance exactly (\cref{alg:base:3} to \cref{alg:base:4}). The remaining lines handle the recursive case where $v$ is an internal node with children $v_1, \ldots, v_{B_v}$. We sample a precision $u_{v_i}$ for every child $v_{i}$ (\cref{alg:base:7}), recursively compute $\Delta_{v_i,s}$ at each child for all shifts of the child (\cref{alg:base:8}), and choose the best shift for the child with respect to the parent shift (\cref{alg:base:9}). Finally, in \cref{alg:base:10} to \cref{alg:base:11} we obtain $\Delta_{v,s}$ by applying the Precision Sampling Lemma to combine the values at the children.

\begin{lemma}\label{lmm:one_call}
    For any node $v$ and shift $s \in \fragment{-{2k}}{{2k}}^{\ell_v}$, the value $\Delta_{v,s}$ computed in \cref{alg:nearlinear_baseline} $(\alpha_v, 1/r_v)$-approximates $\TD_{v,s}(\bF, \bG)$ with probability at least 0.9. Moreover, \cref{alg:nearlinear_baseline} runs in time $\Oh(nc(k+1)\cdot (\log(n))^{\Oh(\log_c(n))})$ with probability at least 0.98, provided that $K \geq k$.
\end{lemma}

\begin{proof}
We claim that $\Delta_{v,s}$ is an $(\alpha_v, 1/r_v)$-approximation of $\TD_{v,s}(\bF, \bG)$, with probability at least 0.9.
We prove this using structural induction over $v \in T_{c}(\bF)$.
\begin{itemize}
    \item In the base case where $\abs{\mI_v}$ is too short, i.e. \cref{alg:base:1,alg:base:2}, we simply set $\Delta_{v,s}$ to 0. This is valid because $\TD_{v,s}(\bF, \bG)$ is at most  $1/r_v$\footnote{To see this, set the shift of every child in accordance with the shift of its parent, so that across all recursive calls, all penalty terms are $0$.}, so $0$ is an $(\alpha_v, 1/r_v)$-approximation of $\TD_{v,s}(\bF, \bG)$.
    \item In the other base case where $v$ is a leaf node (\cref{alg:base:3,alg:base:4}), we calculate the Tree Distance exactly.
    \item Finally, in the recursive case (\cref{alg:base:5} to \cref{alg:base:12}), suppose we have successfully computed $\Delta_{v_i, s}$ for all children nodes $v_i$ of $v$ and shifts $s \in \fragment{-{2k}}{{2k}}^{\ell_{v_i}}$. Let us further unpack the definition of Tree Distance to understand what \cref{alg:nearlinear_baseline} is approximating at each step. Let us write
    \[
    \widetilde{\TD}_{i, s} \coloneqq \min_{s_i \in \fragment{-{2k}}{{2k}}^{\ell_{v_i}}} \left\{ \TD_{v_i, s_i}(\bF, \bG)+\penaltyFunc(v,v_i,s,s_i) \right\}.
    \]
    Then it is immediate that
    \[
    \TD_{v,s}(\bF, \bG) =  \sum_{i \in \fragment{1}{B_v}}{ \widetilde{\TD}_{i,s}}.
    \]
     The algorithm first computes an approximation $\widetilde{\Delta}_{i,s}$ of $\widetilde{\TD}_{i,s}$ (\cref{alg:base:9}), and then uses the approximations $\widetilde{\Delta}_{i,s}$ to approximate $\TD_{v,s}(\bF, \bG)$ (\cref{alg:base:11}).
     Since $\Delta_{v_i,s}$ is assumed to be an $(\alpha_{v_i}, 1/r_{v_i})$-approximation of $\TD_{v_i, s}$, some simple calculations show that $\widetilde{\Delta}_{i,s}$ computed in \cref{alg:base:9} is also an $(\alpha_{v_i}, 1/r_{v_i})$-approximation of $\widetilde{\TD}_{i,s}$ (thanks to $\alpha_{v_i} \geq 1$). Now we are ready to approximately evaluate $\sum_{i \in \fragment{1}{B_v}}{\widetilde{\TD}_{i,s}}$. Recall that $1/r_{v_i} = u_{v_i}/r_v$. By applying \cref{lmm:precision_sampling}, we get that $\Delta_{v,s}$ is an $((1+\varepsilon)\cdot \alpha_{v_i}, 1/r_v)$-approximation of $\TD_{v,s}(\bF, \bG)$. Since $(1+\varepsilon)\cdot \alpha_{v_i} \leq \alpha_v$, we have that $\Delta_{v,s}$ is an $(\alpha_v, 1/r_v)$-approximation of $\TD_{v,s}(\bF, \bG)$, as desired.
\end{itemize}

Now let us analyze the success probability of \cref{alg:nearlinear_baseline}. The only source of randomness is calling the Precision Sampling Lemma for all nodes and all at most $(4k+1)^2$ shifts per node. Since one call fails with probability at most $\delta = 0.01 \cdot (6n(4k+1)^2)^{-1}$, by a union bound over at most $6n$ nodes and $(4k+1)^2$ shifts each, all calls succeed with probability at least $0.99 \geq 0.9$.

Finally, let us analyze the running time of \cref{alg:nearlinear_baseline}. Let us call a node $v$ \emph{active} if $\abs{\mI_v} > \frac{1}{r_v}$, as otherwise we do not do any work. With probability at least 0.98, the number of active nodes in $T_{c}(\bF)$ is $\Oh(\frac{n}{K}(\log(n))^{\Oh(\log_c(n))})$ (Proof is delayed to \cref{lmm:num_active_nodes}). Since the time complexity of a single execution of \cref{alg:nearlinear_baseline} is $\Oh(k^2c\polylog(n))$ disregarding the recursive calls\footnote{Calculating $\widetilde{\Delta}_{i,s}$ in \cref{alg:base:9} can be cast as variants of a 2D range minimum query problem. We give an efficient $\Ohtilde(k^2)$ algorithm in \cref{lmm:range_min_query}.}, the total time complexity of our baseline algorithm is $\Oh(nck\cdot (\log(n))^{\Oh(\log_c(n))})$.
\end{proof}

\begin{lemma}\label{lmm:solve_GAPTED}
Let $k,K$ be two non-negative thresholds.
Then, given two $n$-node forests $\bF, \bG$,
\cref{alg:nearlinear_baseline}
solves the $(k,K)$-\GAPTED Problem satisfying $K \geq 194cDk$ with probability at least 0.9, where $c$ is the parameter in the partition tree $T_{c}(\bF)$, and $D = \Oh(\log_c(n))$ is the number of levels of $T_{c}(\bF)$. That is, it outputs \yes if $\ted(\bF, \bG) \leq k$, \no if $\ted(\bF, \bG) > K$ and any arbitrary \yes/\no otherwise.
\end{lemma}
\begin{proof}
    Let $r$ be the root of $T_{c}(\bF)$. We execute  \cref{alg:nearlinear_baseline} with parameters $k$ and $K+1$.
    We output \yes, whenever $\Delta_{r, \langle 0 \rangle} \leq 0.498(K+1)$, and \no otherwise.
     If $\ted(\bF, \bG) \leq k$, then by \cref{lmm:TD_ub} we have that $\TD(\bF, \bG) \leq 48cDk \leq 48(K+1)/194$. By \cref{lmm:one_call}, with probability at least 0.9, $\Delta_{r, \langle 0 \rangle}$ as computed by \cref{alg:nearlinear_baseline} is a $(2,0.001(K+1))$-approximation of $\TD(\bF, \bG)$. Condition on this event. Therefore, we have that $\Delta_{r, \langle 0 \rangle} \leq 2\cdot \frac{48(K+1)}{194} + 0.001(K+1) \leq 0.498(K+1)$. On the other hand, if $\ted(\bF,\bG) > K$, then $\TD(\bF,\bG) \geq K+1$ by \cref{lmm:TD_lb}. We then have that $\Delta_{r, \langle 0 \rangle}\geq (K+1)/2-0.001(K+1) = 0.499(K+1)$.
\end{proof}
\begin{corollary}\label{cor:static_approx_alg}
    For every $0 < \varepsilon < 1$, there is an algorithm that, given two $n$-node forests $\bF,
    \bG$, outputs an $(\log n)^{\Oh(1/\varepsilon)}$-approximation of $k \coloneqq \ted(\bF,\bG)$ in
    $\Oh((k+1)n^{1+\varepsilon})$ time, with high probability.
\end{corollary}
\begin{proof}
    First of all, we can test whether $\ted(\bF, \bG) = 0$ in $\Oh(n)$ time. Therefore, in the proof below, we assume $\ted(\bF, \bG) > 0$.

     Consider the partition tree $T_{c}(\bF)$ with $D = \Oh(\log_c n)$ levels. We are going to run \cref{alg:nearlinear_baseline} on multiple $(k_i,K_i)$-\GAPTED instances with $k_i = \frac{K_i}{194cD}$ and $K_i = 2^i$, for $i \in \fragmentco{0}{\ceil{\log n}}$. There are $m = \Oh(\log n)$ such threshold scales. By \cref{lmm:solve_GAPTED}, we know that a single run of \cref{alg:nearlinear_baseline} on the $(k_i, K_i)$-\GAPTED problem can be capped at
     \[
     T = C \cdot nc(k_i+1)\cdot (\log n)^{\Oh(\log_c n)}
     \]
     steps, where $C > 0$ is the constant hidden in the runtime of \cref{lmm:one_call}, while correctly solving the problem with probability at least $1 - 0.1 - 0.02 \geq 0.87$, as we know that \cref{alg:nearlinear_baseline} with probability at least $0.98$ stays within time $T$. Thus, at each threshold $(k_i, K_i)$, we run $\Oh(\log n)$ independent trials of \cref{alg:nearlinear_baseline} capped at $T$ steps and take the majority answer (a timeout counts as \no). We process the thresholds in increasing order of $i$ and stop at the first $K_i$ whose majority answer is \yes, which we return; if no threshold answers \yes, we return $2n$. Since $T$ grows geometrically in $i$ and the last processed threshold satisfies $k_i = \Oh(\ted(\bF,\bG) + 1)$, the total time is $\Oh(\log n)$ times the time spent on the last threshold.

     Using standard success amplification analysis and a union bound over all $m = \Oh(\log n)$ threshold scales, we get that the entire process succeeds with high probability.

    Since we are running \cref{alg:nearlinear_baseline} for $\Oh(\log n)$ threshold scales, with $\Oh(\log n)$ repetitions in every threshold scale, the running time remains $\Ohtilde((k+1)\cdot  nc (\log(n))^{\Oh(\log_c(n))})$. To see correctness, note that \cref{alg:nearlinear_baseline} indicating \no in the round before $K_i$ implies that $\ted(\bF, \bG) > K_{i-1}/194cD$, and \cref{alg:nearlinear_baseline} indicating \yes in round $K_i$ implies that $\ted(\bF, \bG) \leq K_i$. For $i = 0$, the initial test gives $\ted(\bF, \bG) \geq 1 = K_0$. If no round answers \yes, then $\ted(\bF, \bG) > K_{m-1}/194cD \geq n/388cD$, as $K_{m-1} = 2^{\ceil{\log n}-1} \geq n/2$, so the returned $2n$ satisfies $\ted(\bF, \bG) \leq 2n < 776cD \cdot \ted(\bF, \bG)$.
    Choosing the correct $c = (\log n)^{\Oh(1/\varepsilon)}$, we obtain that $K_i$ is an  $\Oh(cD) = (\log n)^{\Oh(1/\varepsilon)}$ approximation of $\ted(\bF, \bG)$, and the running time simplifies to $\Oh(kn^{1+\varepsilon})$.
\end{proof}

\subsection{A Near-Linear Time Algorithm}

The reason why the algorithm from the previous section fails to achieve the desired near-linear running time is because of the prohibitive number of $\Oh(k^2)$ shifts required for a single bi-interval $\mI_v = \langle I_1, I_2 \rangle$.
Fortunately, we can bypass this limitation by considering a tradeoff between the tree edit distance $\ted(\mathbf{F}_{\mI_v}, \mathbf{G}_{\mJ_v})$ (where $\mJ_v$ is the image of $\mI_v$ under a fixed optimal $\mA : \bF \onto \bG$) and $m_v$, the number of spine nodes within $\mI_v$.

\begin{description}
    \item[Case 1: $\ted(\mathbf{F}_{\mI_v}, \mathbf{G}_{\mJ_v}) \ll m_v$.]
     When the edit distance is significantly smaller than the number of spine nodes, many spine nodes will match exactly in an optimal alignment $\mathcal{A} : \mathbf{F}_{\mI_v} \onto \mathbf{G}_{\mJ_v}$. The hope is that it is easy to find such a spine node $v \in \mathbf{F}$ that matches exactly. Trying out the shift $s = \langle \Left(w) - \Left(v), \Right(w) - \Right(v) \rangle$ for reasonable choices of $w$ in $\mathbf{G}$ (which should be at most $\Ohtilde(k)$),
     hopefully yields a shift $s$ such that $\ted(\mathbf{F}_{\mI_v}, \mathbf{G}_{\mJ_v}) \approx \ted(\mathbf{F}_{\mI_v}, \mathbf{G}_{\mI_v,s})$.
    \item[Case 2: $\ted(\mathbf{F}_{\mI_v}, \mathbf{G}_{\mJ_v}) \gg m_v$.]
	     In this scenario, even if we were to ignore (i.e., delete) the spine nodes in an optimal alignment $\mA : \mathbf{F}_{\mI_v} \onto \mathbf{G}_{\mI_v, s}$ would only grow very little. But ignoring spine nodes makes everything easier: we can ``decouple'' $\mI_v$ into two mono-intervals, $\langle I_1 \rangle$ and $\langle I_2 \rangle$ and for each of these, we only need to consider $\mathcal{O}(k)$ shifts. The hope is that $\ted(\mathbf{F}_{\mI_v}, \mathbf{G}_{\mJ_v}) \approx \ted(\bF_{\langle I_1 \rangle}, \bG_{\langle I_1 \rangle, \langle s_1 \rangle}) + \ted(\bF_{\langle I_2 \rangle}, \bG_{\langle I_2 \rangle, \langle s_2 \rangle})$ for some $s_1,s_2 \in \fragment{-{2k}}{{2k}}$.
\end{description}
Thus, in both cases we hope to meet the running time target by limiting the shifts to $\Ohtilde(k)$ per bi-interval. As we will see later on in \cref{lmm:decoupling_ub}, when $\ted(\bF_{\mI_v}, \bG_{\mJ_v})$ is close to $m_v$, we can still adopt the above approach and obtain a good approximation.

\subparagraph*{The partition DAG $D_c(\bF)$.}
We need to change our Tree Distance framework a bit
to capture the fact that for a bi-interval, we also consider its two parts individually.
First, we define the \emph{partition DAG $D_c(\bF)$} based on $T_{c}(\bF)$ as follows:
\begin{itemize}
    \item We add to $D_c(\bF)$ all nodes of $T_{c}(\bF)$. Moreover, for each $v \in T_{c}(\bF)$ such that $\mI_v = \langle I_1, I_2 \rangle$ is a bi-interval, we add to $D_{c}(\bF)$ two more nodes, denoted by $\LeftD(v)$ and $\RightD(v)$, that represent $\langle I_1 \rangle$ and $\langle I_2 \rangle$. For any node $v \in D_{c}(\mathbf{F})$, we still let $\mI_v$ denote the corresponding interval collection and $\ell_v$ the number of intervals in it.
    \item The nodes that $v \in D_c(\bF)$ has an edge to, which we refer to as \emph{successors of $v$}, are defined as follows:
    \begin{itemize}
        \item  If $v \in T_{c}(\bF)$, then the successors are the children in $T_{c}(\bF)$.
        \item Otherwise, if $v$ is a new node and corresponds to $\langle I_1 \rangle$
        for some $\mI_u = \langle I_1, I_2 \rangle$ with $u \in T_{c}(\bF)$, we
        consider all successors $w$ of $u$ such that $U(\mI_w) \cap I_1 \neq \emptyset$. There can only be two cases: If $U(\mI_w) \subseteq I_1$, then we make $w$ a successor of $v$; Otherwise if $U(\mI_w) \setminus I_1 \neq \emptyset$ meaning that $\mI_w = \langle I_1', I_2' \rangle$ is a bi-interval with $I_1' \subseteq I_1$, we make $\langle I_1' \rangle$ a successor of $v$.
        We do the same if $v$ corresponds to $\langle I_2 \rangle$.

        It is not difficult to verify that also in this case, the successors of $v$ form a partition of $\mI_v$.
    \end{itemize}
\end{itemize}

\subparagraph*{Shift Selection and a New $\TD$ Definition.}
\label{subsec:nalgo}

We also introduce the notion of a \emph{shift selection} $S_v$.

\begin{definition}
    For a node $v \in D_{c}(\mathbf{F})$, a \emph{shift selection} $S_v$ is a subset of $\fragment{-{2k}}{{2k}}^{\ell_v}$. Whenever $\ell_v=1$, we further impose $S_v = \fragment{-2k}{2k}$, i.e., $S_v$ contains all shifts.
\end{definition}

This allows us to finally define the new notion of $\TD$.

\begin{definition}[Tree Distance (bis)]\label{def:alternative_TD}
    Suppose we are given two $n$-node forests $\bF,
    \bG$, the partition DAG $D_c(\bF)$, and a shift selection $S_v$ for every $v \in D_c(\bF)$ such that $\ell_v = 2$. Then, we define the following variants of Tree Distance for every node $v \in D_c(\bF)$ and $s = \langle s_1, \ldots s_{\ell_v} \rangle \in S_v$:
    \begin{itemize}
        \item If $v$ has no successors, then
            \[
                \TD_{v,s}(\bF, \bG) \coloneqq
                \begin{cases}
                    \ted(\bF_{\mI_v},\bG_{\mI_{v},s}) & \ell_v = 1\\
                    \min \left\{ \ted(\bF_{\mI_v},\bG_{\mI_{v},s}), \ \TD_{\LeftD(v),\langle s_1 \rangle}(\bF, \bG) + \TD_{\RightD(v),\langle s_2 \rangle}(\bF, \bG)\ \right\} & \ell_v = 2
                \end{cases}
            \]
    \item If $v$ has successors $v_1, \ldots, v_{B_v}$, then
    \[
        \TD_{v,s}(\bF, \bG) \coloneqq \textstyle \sum_{i=1}^{B_v} \TDrec_{v,s, v_i}(\bF, \bG),
    \]
    where
    \begin{align*}
        \TDrec_{v,s,u}(\bF, \bG) &\coloneqq
        \begin{cases}
            \TDsel_{v,s,u}(\bF, \bG) & \ell_u = 1\\
            \min \left\{ \TDsel_{v,s,u}(\bF, \bG) , \ \TDsel_{v,s, \LeftD(u)}(\bF, \bG) + \TDsel_{v,s,\RightD(u)}(\bF, \bG)\ \right\} & \ell_u = 2
        \end{cases}, \\
        \TDsel_{v,s,u}(\bF, \bG) &\coloneqq \min_{r \in S_{u}} \big\{ \ \TD_{u, r}(\bF, \bG)+ 2 \cdot ||s_v(u) - r||_1 \ \big\}.
    \end{align*}
    and $\penaltyFunc$ is defined in the same way as \cref{def:TD}.
    Here, for a node $u \in D_c(\bF)$ each of whose intervals $I_{u,j}$ is contained in an interval $I_{v,\mathsf{pi}(j)}$ of $\mI_v$ (this is the case for every successor $u$ of $v$, as well as for $\LeftD(u)$ and $\RightD(u)$ when $u$ is a bi-interval successor), we write $s_v(u) \coloneqq \langle s_{\mathsf{pi}(1)}, \ldots, s_{\mathsf{pi}(\ell_u)} \rangle$ for the shift of $u$ inherited from the shift $s$ of $v$, so that $2\|s_v(u) - r\|_1 = \penaltyFunc(v,u,s,r)$. Moreover, here $\min \emptyset \coloneqq +\infty$, so that $\TDsel_{v,s,u}(\bF,\bG) = +\infty$ whenever $S_u = \emptyset$. \qedhere
    \end{itemize}
\end{definition}

Note that in this last definition, the first case happens exactly for those nodes $v$ that were formerly leaves in $T_{c}(\bF)$ (and for which $\mI_v$ is necessarily a bi-interval), and for the nodes of the form $\langle I_1 \rangle$, $\langle I_2 \rangle$ where $\langle I_1, I_2 \rangle$ corresponds to a leaf in $T_{c}(\bF)$.

\subparagraph*{Upper and Lower Bounds for $\TD$ (bis).} We now prove new versions of \cref{lmm:TD_lb} and \cref{lmm:TD_ub}.

\begin{lemma}\label{lmm:decoupling_lb}
    Given two $n$-node forests $\bF,
    \bG$, the partition DAG $D_c(\bF)$, and a shift selection $S_v  \subseteq \fragment{-{2k}}{{2k}}^{\ell_v}$ for every $v \in D_c(\bF)$, we have $\ted(\bF,\bG) \leq \TD_{r,\langle 0 \rangle}(\bF, \bG)$ where $r$ represents the node of $D_c(\bF)$ with corresponding interval collection $\langle \fragmentco{0}{2n} \rangle$, i.e., the former root of $T_{c}(\bF)$.
\end{lemma}
\begin{proof}
    We prove the more general claim that $\ted(\bF_{\mI_v},\bG_{\mI_v,s}) \leq \TD_{v,s}(\bF, \bG)$ for every node $v \in D_c(\bF)$ and $s = \langle s_1, \ldots s_{\ell_v} \rangle \in S_v$ by structural induction over $D_c(\bF)$. Here, the structural induction for a node $v \in D_c(\bF)$ with $\ell_v = 2$ considers $\LeftD(v)$ and $\RightD(v)$ before $v$; this is clearly feasible.

For the base case, fix a node $v$ for which the first case of \cref{def:alternative_TD} applies.
If $\mI_v$ is a mono-interval, then by definition we have $\ted(\bF_{\mI_v},\bG_{\mI_v,s}) = \TD_{v,s}(\bF, \bG)$ (and note that in this case also $\ted(\bF_{\mI_v},\bG_{\mI_v,s}) = 1$, since all positions must be deleted).
If instead $\mI_v$ is a bi-interval, then
$\TD_{v,s}(\bF,\bG) = \min \left\{ \ted(\bF_{\mI_v},\bG_{\mI_{v},s}), \TD_{\LeftD(v),\langle s_1 \rangle}(\bF, \bG) + \TD_{\RightD(v),\langle s_2 \rangle}(\bF, \bG) \right\} = \min \{\ted(\bF_{\mI_v},\bG_{\mI_{v},s}), 2\} = \ted(\bF_{\mI_v},\bG_{\mI_{v},s})$ by the previous note.

Now assume that $v$ is an internal node. Let $s \in S_v$, and let $\{v_1, \dots, v_{B_v}\}$ be the successors of $v$ in $D_c(\bF)$.
Then, for every node
$u \in \{v_1, \ldots, v_{B_v}\} \cup \bigcup_{i \in \fragment{1}{B_v} \mid \ell_{v_i} = 2}\{\LeftD(v_i), \RightD(v_i)\}$,
we obtain
\begin{align}
        \ted(\bF_{\mI_u}, \bG_{\mI_u,s_{v}(u)}) &\leq \min_{r \in S_{u}}
    \left\{ \ \ted(\bF_{\mI_u}, \bG_{\mI_u,r}) + 2 \cdot || s_v(u) - r ||_1 \ \right \}\nonumber\\
    &\leq \min_{r \in S_u}\left\{ \ \TD_{u,r}(\bF, \bG) + 2 \cdot || s_v(u) - r ||_1 \ \right\}\nonumber\\
    &= \TDsel_{v,s,u}(\bF, \bG), \label{eq:decoupling_lb:1}
\end{align}
where we use \cref{prop:triangle_inequality} together with the inductive hypothesis. In the inequalities above we implicitly assumed $S_u \neq \emptyset$, but observe that $\ted(\bF_{\mI_u}, \bG_{\mI_u,s_{v}(u)}) \leq \TDsel_{v,s,u}(\bF, \bG)$ trivially holds when $S_u = \emptyset$, because then $\TDsel_{v,s,u}(\bF, \bG) = +\infty$.

Whenever $\ell_u = 1$, this immediately yields $\ted(\bF_{\mI_u}, \bG_{\mI_u,s_{v}(u)}) \leq \TDrec_{v,s,u}(\bF, \bG)$, as $\TDrec_{v,s,u}(\bF, \bG) = \TDsel_{v,s,u}(\bF, \bG)$ for $\ell_u = 1$.

On the other hand, when $\ell_u = 2$ note that $\mI_{\LeftD(u)}$ and $\mI_{\RightD(u)}$ form a partition of $\mI_u$. Hence, using \cref{prop:ted_decomp} and \cref{eq:decoupling_lb:1}, we get
\begin{align*}
    \ted(\bF_{\mI_u}, \bG_{\mI_u,s_v(u)}) &\leq \ted(\bF_{\mI_{\LeftD(u)}}, \bG_{\mI_{\LeftD(u)}, s_v(\LeftD(u))}) + \ted(\bF_{\mI_{\RightD(u)}}, \bG_{\mI_{\RightD(u)},s_v(\RightD(u))})\\
    &\leq \TDsel_{v,s, \LeftD(u)}(\bF, \bG) + \TDsel_{v,s,\RightD(u)}(\bF, \bG).
\end{align*}
Then using \cref{eq:decoupling_lb:1} again, we get that
\begin{align*}
    \ted(\bF_{\mI_u}, \bG_{\mI_u,s_v(u)}) \leq  \min \left\{ \TDsel_{v,s,u}(\bF, \bG) , \ \TDsel_{v,s, \LeftD(u)}(\bF, \bG) + \TDsel_{v,s,\RightD(u)}(\bF, \bG)\ \right\}
        = \TDrec_{v,s,u}(\bF, \bG).
\end{align*}

Finally, using \cref{prop:ted_decomp}, we conclude
\[
        \ted(\bF_{\mI_v}, \bG_{\mI_v,s})
        \leq \textstyle \sum_{i=1}^{B_v} \ted(\bF_{\mI_{v_i}},\bG_{\mI_{v_i}, s_v(v_i)}) \leq \sum_{i=1}^{B_v}\TDrec_{v,s,v_i}(\bF, \bG) = \TD_{v,s}(\bF, \bG). \qedhere
\]
\end{proof}

For the upper bound, we fix again an arbitrary optimal alignment $\mA : \bF \onto \bG$ of cost at most $k$, for each $v \in T_{c}(\bF)$, we let $\mJ_v$ be the image of $\mI_v$ under $\mA$, i.e., $\mJ_v \coloneqq \pi_{\mA}(\mI_v)$.
This time, for the upper bound to go through, we need the selected shifts to be good enough.

\begin{definition}\label{def:good_shifts}
    We say that the shift selection $S_v  \subseteq \fragment{-{2k}}{{2k}} \times \fragment{-{2k}}{{2k}}$ is \emph{good} (w.r.t. $\mA$) for $v \in D_{c}(\bF)$ such that $\ell_v=2$ if $S_v$ satisfies the following conditions:
    \begin{enumerate}[(i)]
        \item $\abs{S_v} = \Oh(k \log n)$;
        \label{it:good_shifts:i}
        \item If $\ted(\bF_{\mI_v}, \bG_{\mJ_v}) < m_v / 24c$ and $\ted(\bF, \bG) \leq k$, where $\mJ_v = \pi_{\mA}(\mI_v)$ is defined as in \cref{def:align_interval} and $m_v$ denotes the number of spine nodes in $\mI_v$, then there exists a spine node $x \in \bF_{\mI_v}$ such that $\mA$ matches $x$ and $(\mA(\Left(x)) - \Left(x), \mA(\Right(x)) - \Right(x)) \in S_v$. \qedhere
         \label{it:good_shifts:ii}
    \end{enumerate}
\end{definition}

Note that \cref{def:good_shifts}\ref{it:good_shifts:ii} is well-defined, because if $\ted(\bF,\bG) \leq k$, then $\mA$, by \cref{clm:TD_ub_diff_ted}, cannot match a parenthesis to another parenthesis more than a distance of $2k$ away.

\begin{lemma}\label{lmm:decoupling_ub}
    Consider $n$-node forests $\bF,
    \bG$ such that $\ted(\bF,\bG)\leq k$, an arbitrary optimal alignment $\mA : \bF \onto \bG$,
    the partition DAG $D_{c}(\bF)$, and a shift selection $S_v$ for every $v \in D_{c}(\bF)$ that is good whenever $\ell_v = 2$ w.r.t. $\mA$.
    Then, $\TD_{r,\langle 0 \rangle}(\bF,\bG) \leq 48c \cdot 2D \cdot  \ted(\bF,\bG)$, where $r$ represents the former root of $T_{c}(\bF)$ and $D$ is the number of levels of $T_{c}(\bF)$.
\end{lemma}
\begin{proof}
    We partition the nodes of $T_c(\bF)$ (and not the nodes of $D_c(\bF)$) in two sets:
    \begin{enumerate}[(a)]
        \item The set $A$ contains all $v \in T_{c}(\bF)$ such that $\ell_v=1$ holds or both $\ell_v=2, \ted(\bF_{\mI_v}, \bG_{\mJ_v}) < m_v/24c$ hold;
        \item The set $B$ contains all remaining nodes; that is all $v \in T_{c}(\bF)$ such that $\ell_v = 2$ and $\ted(\bF_{\mI_v}, \bG_{\mJ_v}) \geq m_v/24c$.
    \end{enumerate}
    To each node $v \in T_{c}(\bF)$ we assign a shift $s_v \in \fragment{-2k}{2k}^{\ell_v}$:
    \begin{itemize}
        \item If for $v$ both $v \in A, \ell_v = 1$ hold or both $v \in B, \ell_v = 2$ then let $s_v$ be defined as in the proof of \cref{lmm:TD_ub}.
        That is, we set $s_v = \langle \pi_{\mA}(a_{v,1})- a_{v,1}, \ldots,  \pi_{\mA}(a_{v,\ell_v})-a_{v,\ell_v} \rangle$ where $\mI_v = \langle \fragmentco{a_{v,1}}{b_{v,1}},\ldots, \fragmentco{a_{v,\ell_v}}{b_{v,\ell_v}} \rangle$.

        \item Otherwise, if $v \in A$ and $\ell_v = 2$, then $s_v$ is chosen to be the optimal shift that is guaranteed to be in $S_v$ by \cref{def:good_shifts}\ref{it:good_shifts:ii}.
    \end{itemize}

    Let $h_v$ denote the height of $v$ in $T_{c}(\bF)$, which is defined as the number of nodes on a longest path to a leaf node from $v$ (So leaf nodes have height 1).
    We prove using structural induction over $v \in T_{c}(\bF)$ the following two claims:
    \begin{itemize}
        \item For every node $v \in A$, we have
        \begin{align}
            &\TD_{v,s_v}(\bF, \bG) \leq 48c \cdot 2h_v\cdot \ted(\bF_{\mI_v}, \bG_{\mJ_v}). \label{eq:decoupling_ub1}
        \end{align}
        \item For every node $v \in T_{c}(\bF)$ such that $\ell_v=2$, we have
        \begin{align}
            &\TD_{\LeftD(v),\langle s_1 \rangle}(\bF, \bG) + \TD_{\RightD(v),\langle s_2 \rangle}(\bF, \bG) \leq 48c \cdot (2h_v-1)\cdot \ted(\bF_{\mI_v}, \bG_{\mJ_v}) + 2 m_v,\label{eq:decoupling_ub2}
        \end{align}
        where $s_v = \langle s_1, s_2 \rangle$.
        Note that if also $v \in B$, then this implies $\TD_{\LeftD(v),\langle s_1 \rangle}(\bF, \bG) + \TD_{\RightD(v),\langle s_2 \rangle}(\bF, \bG) \leq 48c \cdot 2h_v \cdot \ted(\bF_{\mI_v}, \bG_{\mJ_v})$ because $\ted(\bF_{\mI_v}, \bG_{\mJ_v}) \geq m_v/24c$.
    \end{itemize}
    Notice that \cref{lmm:decoupling_ub} follows from \cref{eq:decoupling_ub1} applied to $r$, because $\ell_r = 1$, and so $r \in A$.

    \subparagraph*{Base Case.}
    For the base case, let $v \in T_{c}(\bF)$ be a leaf node
    with $\mI_v = \langle \Left(x), \Right(x) \rangle$ for some node $x \in \bF$. Recall that all leaf nodes in $T_{c}(\bF)$ are bi-intervals and note that $m_v = 1, h_v = 1$. Let $s_v = \langle s_1, s_2 \rangle$.

    For \cref{eq:decoupling_ub1},  note that $v \in A$ if and only if $\ted(\bF_{\mI_v}, \bG_{\mJ_v}) = 0$ (as $m_v = 1$ and every positive cost is at least $1/2$), i.e., if and only if $\mA$ matches $x$. In this case, $s_v = \langle \mA(\Left(x)) - \Left(x), \mA(\Right(x)) - \Right(x) \rangle \in S_v$ and $\ted(\bF_{\mI_v}, \bG_{\mI_v, s_v}) = 0$, so \cref{eq:decoupling_ub1} holds by the definition of $\TD_{v,s_v}(\bF,\bG)$ in the base case.

    For \cref{eq:decoupling_ub2} note that, by definition
    $\TD_{\LeftD(v),\langle s_1 \rangle} + \TD_{\RightD(v),\langle s_2 \rangle}  = 2$, as we have four unbalanced parentheses. Since $48c \cdot (2h_v-1)\ted(\bF_{\mI_v},\bG_{\mJ_{v}}) + 2 \cdot m_v \geq 2$ because $m_v = 1$, \cref{eq:decoupling_ub2} also holds.

    \subparagraph*{Inductive Step.}
    For the inductive step, let $v \in T_{c}(\bF)$ be an internal node of $T_{c}(\bF)$, and let $v_1, \ldots, v_{B_v}$ be its children in $T_{c}(\bF)$. For the sake of brevity, in the following proof, we abbreviate $(s_v)_v(u)$ with $s_v(u)$.

    We first note that the same as in \cref{clm:TD_ub_pen_ted} still applies.

    \begin{claim}\label{claim:decoupling_ub:0}
    $2||s_v(u) - s_u||_1 \leq 8 \cdot \ted(\bF_{\mI_v}, \bG_{\mJ_v})$ for
    $u \in \{v_1, \ldots, v_{B_v}\} \cup \bigcup_{i \in \fragment{1}{B_v} \mid \ell_{v_i} = 2}\{\LeftD(v_i), \RightD(v_i)\}$.
    \end{claim}
    \begin{claimproof}
        Note that the $i$-th shift that we choose for the $i$-th interval $I_i$ in $\mI_v$ for $i \in \fragment{1}{\ell_v}$
        is of the form $(x, \pi'(x))$ for $x \in I_i$, with $\pi'$ as in \cref{clm:TD_ub_diff_ted}: $\pi' = \mA$ for shifts derived from a matched spine node, and $\pi' = \pi_{\mA}$ otherwise.
        Thus, any component of $s_v(u) - s_u$ will have the form $\abs{(x - \pi'(x)) - (x' - \pi'(x'))}$ for both $x,x'$ contained in $I_i$. Consequently, we use \cref{clm:TD_ub_diff_ted} again on all $\ell_u \leq 2$ components, to obtain the claim.
    \end{claimproof}

    Next, we prove the following.

    \begin{claim}\label{claim:decoupling_ub:1}
        Let $u \in \{v_1, \ldots, v_{B_v}\}$. Then, for all $u$,
        \begin{align}
            \TDrec_{v,s,u}(\bF, \bG) \leq 48c \cdot 2h_u\cdot \ted(\bF_{\mI_u}, \bG_{\mJ_u}) + 8 \cdot \ted(\bF_{\mI_v}, \bG_{\mJ_v}).\label{eq:intertarget_1_ub}
        \end{align}
        Moreover, for $u$ such that $\ell_u = 2$, we have
        \begin{align}
            \TDrec_{v,s,\LeftD(u)}(\bF, \bG) + \TDrec_{v,s,\RightD(u)}(\bF, \bG) \leq 48c \cdot (2h_u-1)\cdot \ted(\bF_{\mI_u}, \bG_{\mJ_u}) + 8 \cdot \ted(\bF_{\mI_v}, \bG_{\mJ_v}) + 2m_{u}.\label{eq:intertarget_2_ub}
        \end{align}

    \end{claim}
    \begin{claimproof}
        For $u \in A$, for which we always have $s_u \in S_u$, we can write
        \begin{align*}
            \TDsel_{v,s,u}(\bF, \bG)
            &\leq \min_{r \in S_{u}} \big\{ \ \TD_{u, r}(\bF, \bG)+ 2 \cdot ||s_v(u) - r||_1 \ \big\} && \text{(definition of $\TDsel$)}\\
            &\leq \TD_{u, s_u}(\bF, \bG) + 2 \cdot ||s_v(u) - s_u||_1 && \text{(set $r=s_u$)}\\
            &\leq 48c \cdot 2h_u\cdot \ted(\bF_{\mI_u}, \bG_{\mJ_u}) + 8\cdot \ted(\bF_{\mI_v}, \bG_{\mJ_v}) && \text{(induction hypothesis and \cref{claim:decoupling_ub:0})}
        \end{align*}
        On the other hand, for any $u \in T_c(\bF)$ such that $\ell_u = 2$, $s_u = \langle s_1, s_2 \rangle$, we have
        \begin{align}
            \MoveEqLeft \TDsel_{v,s,\LeftD(u)}(\bF, \bG) + \TDsel_{v,s,\RightD(u)}(\bF, \bG) \notag \\
            &\leq \min_{r \in S_{\LeftD(u)}} \big\{ \ \TD_{\LeftD(u), r}(\bF, \bG)+ 2||s_v(\LeftD(u)) - r||_1 \ \big\} + \min_{r' \in S_{\RightD(u)}} \big\{ \ \TD_{\RightD(u), r'}(\bF, \bG)+ 2||s_v(\RightD(u)) - r'||_1 \ \big\}  \label{eq:decoupling_ub:0:1} \\
            &\leq \TD_{\LeftD(u), s_1}(\bF, \bG) + \TD_{\RightD(u), s_2}(\bF, \bG) + 2 \cdot ||s_v(u) - s_u||_1 \label{eq:decoupling_ub:0:2} \\
            &\leq 48c \cdot (2h_u-1)\cdot \ted(\bF_{\mI_u}, \bG_{\mJ_u}) + 8\cdot \ted(\bF_{\mI_v}, \bG_{\mJ_v}) + 2m_u, \label{eq:decoupling_ub:0:3}
        \end{align}
        where in \cref{eq:decoupling_ub:0:1} we use the definition of $\TDsel$; in \cref{eq:decoupling_ub:0:2} we set $\langle r, r' \rangle$ to $s_u$ and use that by doing so $ ||s_v(u) - s_u||_1 = ||s_v(\LeftD(u)) - r||_1 + ||s_v(\RightD(u)) - r'||_1$; and in \cref{eq:decoupling_ub:0:3} we use the induction hypothesis and \cref{claim:decoupling_ub:0}.

        Now, \cref{eq:intertarget_2_ub} follows directly from \cref{eq:decoupling_ub:0:3}. It remains to prove \cref{eq:intertarget_1_ub}. Note that if $u$ is a mono-interval, then $\TDrec_{v,s,u}(\bF, \bG) = \TDsel_{v,s,u}(\bF, \bG)$, and \cref{eq:intertarget_1_ub} follows. If $u$ is a bi-interval in $A$, then we use $\TDrec_{v,s,u}(\bF, \bG) \leq \TDsel_{v,s,u}(\bF, \bG)$, and \cref{eq:intertarget_1_ub} follows for the same reason. Finally, for a bi-interval $u \in B$, we use $\TDrec_{v,s,u}(\bF, \bG) \leq \TDrec_{v,s,\LeftD(u)}(\bF, \bG) + \TDrec_{v,s,\RightD(u)}(\bF, \bG) \leq 48c \cdot 2h_u\cdot \ted(\bF_{\mI_u}, \bG_{\mJ_u}) + 8 \cdot \ted(\bF_{\mI_v}, \bG_{\mJ_v})$, because $\ted(\bF_{\mI_u}, \bG_{\mJ_u}) \geq m_u/24c$.

    \end{claimproof}

    We conclude the proof of the induction step, by showing \cref{eq:decoupling_ub1} and \cref{eq:decoupling_ub2}.

    \begin{claim}\label{claim:decoupling_ub:2}
        If $v \in A$, then \cref{eq:decoupling_ub1} holds.
    \end{claim}
    \begin{claimproof}
        We can verify directly that:
        \begin{align*}
            \TD_{v,s_{v}}(\bF,\bG)
            &= \sum_{i=1}^{B_v} \TDrec_{v,s_v,v_i}(\bF, \bG) && \text{(definition of $\TD$)}\\
            &\leq 48c \cdot \max_{i\in \fragment{1}{B_v}}2h_{v_i} {\cdot} \sum_{v_i}\ted(\bF_{\mI_{v_i}},
             \bG_{\mJ_{v_i}}) + 8B_v\cdot \ted(\bF_{\mI_{v}},
             \bG_{\mJ_{v}}) && \text{(\cref{claim:decoupling_ub:1})}\\
            &= (48c\cdot \max_{i\in \fragment{1}{B_v}} 2h_{v_i} + 8B_v)\cdot \ted(\bF_{\mI_{v}},
             \bG_{\mJ_{v}}) && \text{(\cref{clm:TD_ub_sum_ted})}\\
            &\leq 48c \cdot 2h_v \cdot \ted(\bF_{\mI_{v}},
             \bG_{\mJ_{v}}), && (6c \geq B_v, h_v = \max_i h_{v_i} +1)
         \end{align*}
          which yields the claim.
    \end{claimproof}

    \begin{claim}
        If $v \in T_c(\bF)$ and $\ell_v = 2$, then \cref{eq:decoupling_ub2} holds.
    \end{claim}
    \begin{claimproof}
    Let $\mI_v = \langle I_1, I_2\rangle$ and $s_v = \langle s_1, s_2 \rangle$.

    We begin by partitioning $\fragment{1}{B_v}$ into the three sets $Q \cup L \cup R$, as follows. Set $L$ and $R$ to contain all $i \in \fragment{1}{B_v}$ such that $U(\mI_{v_i}) \subseteq I_1$ and $U(\mI_{v_i}) \subseteq I_2$, respectively. Moreover, set $Q \subseteq \fragment{1}{B_v}$ to contain all $i$ such that $\ell_{v_i}=2$, $U(\mI_{v_i}) \cap I_1 \neq \emptyset$ and $U(\mI_{v_i}) \cap I_2 \neq \emptyset$. Note that for such $i \in Q$, we have $\sum_{i \in Q} m_{v_i} = m_v$.

    Observe that the successors of $\LeftD(v)$ are all $v_i$ for $i \in L$ together with all $\LeftD(v_i)$ for $i \in Q$.
    By the definition of $\TD$, we have:
    \begin{equation}
        \label{eq:decoupling_ub:2:0}
        \begin{aligned}
            \TD_{\LeftD(v),\langle s_1 \rangle}(\bF, \bG)
            &= \sum_{i \in L} \TDrec_{\LeftD(v),\langle s_1 \rangle ,v_i}(\bF, \bG) + \sum_{i \in Q} \TDrec_{\LeftD(v),\langle s_1 \rangle ,\LeftD(v_i)}(\bF, \bG) \\
            &= \sum_{i \in L} \TDrec_{v,s_v,v_i}(\bF, \bG) + \sum_{i \in Q} \TDrec_{v,s_v,\LeftD(v_i)}(\bF, \bG),
        \end{aligned}
    \end{equation}
    where in the last equation we use that in the definitions of $\TDrec$ and $\TDsel$ we only ever consider $s_1$ among the two components of $s_v$.
    Similarly, the successors of $\RightD(v)$ are all $v_i$ for $i \in R$ together with all $\RightD(v_i)$ for $i \in Q$. An analogous equation to \cref{eq:decoupling_ub:2:0} holds for $\TD_{\RightD(v),\langle s_2 \rangle}(\bF, \bG)$.

    Putting everything together, we get:
     \begin{align*}
        \MoveEqLeft \TD_{\LeftD(v),\langle s_1 \rangle}(\bF, \bG) + \TD_{\RightD(v),\langle s_2 \rangle}(\bF, \bG) \\
        &= \sum_{i \in \fragment{1}{B_v} \setminus Q} \TDrec_{v,s_v,v_i}(\bF, \bG) + \sum_{i \in Q} \left( \TDrec_{v,s_v,\LeftD(v_i)}(\bF, \bG)+\TDrec_{v,s_v,\RightD(v_i)}(\bF, \bG) \right) && \text{(\cref{eq:decoupling_ub:2:0})} \\
        &\leq \sum_{i \in \fragment{1}{B_v}} \left(48c \cdot 2h_{v_i}\cdot \ted(\bF_{\mI_{v_{i}}}, \bG_{\mJ_{v_{i}}}) + 8\ted(\bF_{\mI_{v}},
         \bG_{\mJ_{v}}) \right) + 2 \sum_{i \in Q} m_{v_i} && \text{(\cref{claim:decoupling_ub:1})} \\
        &\leq (48c \cdot \max_{i \in \fragment{1}{B_v}}\{2h_{v_i}\} + 8B_v) \cdot  \ted(\bF_{\mI_{v}}, \bG_{\mJ_{v}})+ 2 m_v && \text{(\cref{clm:TD_ub_sum_ted})}\\
        &\leq 48c \cdot (2h_v-1)\cdot \ted(\bF_{\mI_{v}}, \bG_{\mJ_{v}}) + 2m_v,
     \end{align*}
     where we used similar computations as in \cref{claim:decoupling_ub:1}.
\end{claimproof}

This concludes the proof of \cref{lmm:decoupling_ub}.
\end{proof}

We will adapt \cref{alg:nearlinear_baseline} to obtain the near-linear-time algorithm.  First, we use a randomized routine (\cref{alg:good_shift}) to select a good shift selection for every $v$ representing a bi-interval. To do this, we first sample $\lceil \gamma \cdot \log n \rceil$ spine nodes from $\bF_{\mI_v}$ uniformly at random \cref{ln:good:2}. For every sampled spine node $x$, we look for all possible nodes it can align with, and add the induced shift to $S_v$. So in \cref{ln:good:4}, we look for all opening parentheses in $\bG$ that are within a distance of at most $2k$ from the opening parenthesis of $x$, and if the two corresponding closing parentheses are also within a distance of no more than $2k$ (\cref{ln:good:5}), then we include this shift (\cref{ln:good:6}).
\begin{algorithm}[th]
    \KwInput{A node $v$ in $D_{c}(\bF)$ such that $\ell_v = 2$ and a constant $\gamma$.}
    \KwOutput{A good shift selection $S_v$.}
    Let $S_v$ be an empty set\label{ln:good:1}\;
    Sample $\ceil{\gamma \cdot \log n}$ spine nodes from $\bF_{\mI_v}$ uniformly at random\label{ln:good:2}\;
    \For{every sampled spine node $x$\label{ln:good:3}} {
        \For{every $j \in \fragment{\Left(x)-2k}{\Left(x)+2k} \cap \fragmentco{0}{2n}$ with $j = \Left(w)$ for some $w \in \bG$\label{ln:good:4}}{
        \If{$\abs{\Other(j) - \Right(x)} \leq {2k}$\label{ln:good:5}}{
        $S_v = S_v \cup \{\langle j-\Left(x), \Other(j)-\Right(x)\rangle\}$\label{ln:good:6}\;}
        }
    }
   \textbf{return } $S_v$.\label{ln:good:7}
    \caption{A randomized procedure to compute a good shift selection for bi-intervals.}
    \label{alg:good_shift}
\end{algorithm}

\begin{lemma}\label{lmm:good_shift_selection}
    Suppose $\ted(\bF, \bG) \leq k$.
    Fix an optimal alignment $\mA : \bF \onto \bG$. Then, for any $v \in D_c(\bF)$ with $\ell_v = 2$, \cref{alg:good_shift} returns a shift selection that is good for $v$ w.r.t. $\mA$ with probability at least $1 - n^{-4\gamma}$.
\end{lemma}
\begin{proof}
    First, notice that $S_v$ as output by \cref{alg:good_shift} always satisfies the size requirement: At most $\Oh(\log n)$ spine nodes are sampled, each contributing at most $4k+1$ shifts. Thus, $\abs{S_v} = \Oh((k+1) \log n)$.

    Assume $\ted(\bF_{\mI_v}, \bG_{\mJ_v}) < m_v / 24c$.
    Since a deleted and a substituted node cost at least $1/2$, fewer than $2\ted(\bF_{\mI_v}, \bG_{\mJ_v}) < m_v/12c \leq m_v/24$ spine nodes are deleted or substituted (using $c \geq 2$); hence more than $\frac{23}{24} m_v$ spine nodes are matched.
    The probability that we sample one of them is at least $1 - (\frac{1}{24})^{\ceil{\gamma \log n}} \geq 1-n^{-4\gamma}$. Note for any such spine node $v$ that is matched, we have $\abs{\mA(\Left(v))-\Left(v)} \leq {2k}$ and $\abs{\mA(\Right(v))-\Right(v)}\leq {2k}$ by  \cref{clm:TD_ub_diff_ted} with $\pi' = \mA$.
\end{proof}
Next, we need to set the additive and multiplicative accuracies for each node in $D_{c}(\bF)$, if $v \in D_{c}(\bF)$ also exists in $T_{c}(\bF)$, then the multiplicative accuracy $\alpha_v$ is still defined as in \cref{alg:nearlinear_baseline}. However, for the additive accuracy $1/r_v$, if $v$ is the root of $T_{c}(\bF)$, then we set $r_v = 1000/K$; Otherwise if $v$ is a child of $w$ in $T_{c}(\bF)$, we set $r_v = 2r_w/u_v$, where $u_v$ is sampled in the same way as in \cref{alg:nearlinear_baseline}.
For the new nodes $\LeftD(v), \RightD(v)$ in $D_{c}(\bF)$ that represent the left and right parts of a bi-interval, we set $\alpha_{\LeftD(v)} = \alpha_{\RightD(v)} = \alpha_v$ and $r_{\LeftD(v)} = r_{\RightD(v)} = r_v$.

Finally, we can give the near-linear-time algorithm in \cref{alg:nearlinear}.
First, we compute a shift selection $S_v$ for $v$ if $v$ is a bi-interval (\cref{ln:nearlinear:1}). From \cref{ln:nearlinear:2} to \cref{ln:nearlinear:8}, we handle the base case, where if $\mI_v$ is too short (\cref{ln:nearlinear:2} to \cref{ln:nearlinear:3}), we directly set $\Delta_{v,s} = 0$, and if $v$ is a leaf node (\cref{ln:nearlinear:4} to \cref{ln:nearlinear:8}), we calculate the Tree Distance exactly. In \cref{ln:nearlinear:recurse} to \cref{ln:nearlinear:18}, we handle the recursive case where $v$ is an internal node in $T_c(\bF)$. First, on \cref{ln:nearlinear:14} to \cref{ln:nearlinear:15}, we recursively compute $\Delta_{v_i,s}$ for all children $v_i$, $s \in S_{v_i}$ (as well as $\Delta_{\LeftD(v_i),s}, s \in S_{\LeftD(v_i)}$ and $\Delta_{\RightD(v_i),s}, s \in S_{\RightD(v_i)}$ if $\ell_{v_i} = 2$). Then, on \cref{ln:nearlinear:combine}, we call \cref{alg:combine} to obtain $\Delta_{v,s}$ for all $s \in S_v$, as well as $\Delta_{\LeftD(v),s}, s \in S_{\LeftD(v)}$ and $\Delta_{\RightD(v),s}, s \in S_{\RightD(v)}$ if $\ell_{v} = 2$. Note that \cref{alg:combine} exactly follows the definition of Tree Distance (\cref{def:alternative_TD}). It is worth remarking that the recursion in the main algorithm (\cref{alg:nearlinear}) is done on $T_c(\bF)$, but the output also includes Tree Distance values for $\LeftD(v) \in D_c(\bF)$ and $\RightD(v) \in D_c(\bF)$, for every $v \in T_c(\bF), \ell_v = 2$, which are combined in \cref{alg:combine} in the evaluation of the Tree Distance value for the parent of $v$ in $T_c(\bF)$.

\begin{algorithm}[!t]
    \KwInput{A node $v$ in $D_{c}(\bF)$, a sampling rate $r_v > 0$, the precisions $u_{v_i}$ of its successors, a shift selection $S_v$, and for each successor $v_i$ of $v$ in $D_{c}(\bF)$ the shift selection $S_{v_i}$ and value $\Delta_{v_i,r}$ for all shifts $r \in S_{v_i}$. If $\ell_{v_i} = 2$, then also $\Delta_{\LeftD(v_i),r}$ for all $r \in S_{\LeftD(v_i)} = \fragment{-2k}{2k}$ and $\Delta_{\RightD(v_i),r}$ for all $r \in S_{\RightD(v_i)} = \fragment{-2k}{2k}$.}
    \KwOutput{The value $\Delta_{v,s}$ for all shifts $s \in S_{v}$.}

    \For {successor $v_i$ of $v$ in $D_{c}(\bF)$}{
        \If {$\ell_{v_i} = 1$}{
            Set $\Delta_{v,s,v_i}^{\mathsf{rec}} = \min_{r \in S_{v_i}} \left\{ \Delta_{v_i,r} + 2||s_{v}(v_i) - r||_1 \right\}$ for $s \in S_v$ via \cref{lmm:range_min_query}\;
            }
        \Else{
            Compute $\Delta_{v,s,v_i}^{\mathsf{sel}} = \min_{r \in S_{v_i}} \left\{ \Delta_{v_i,r} + 2||s_{v}(v_i) - r||_1 \right\}$ for $s \in S_v$ via \cref{lmm:range_min_query} (if $S_{v_i} = \emptyset$, set $\Delta_{v,s,v_i}^{\mathsf{sel}} = +\infty$)\;
            Compute $\Delta_{v,s,\LeftD(v_i)}^{\mathsf{sel}} = \min_{r \in S_{\LeftD(v_i)}} \left\{ \Delta_{\LeftD(v_i),r} + 2||s_{v}(\LeftD(v_i)) - r||_1 \right\}$ for $s \in S_v$ via \cref{lmm:range_min_query}\;
            Compute $\Delta_{v,s,\RightD(v_i)}^{\mathsf{sel}} = \min_{r \in S_{\RightD(v_i)}} \left\{ \Delta_{\RightD(v_i),r} + 2||s_{v}(\RightD(v_i)) - r||_1 \right\}$ for $s \in S_v$ via \cref{lmm:range_min_query}\;
            Set $\Delta_{v,s,v_i}^{\mathsf{rec}} = \min \{\Delta_{v,s,v_i}^{\mathsf{sel}}, \Delta_{v,s,\LeftD(v_i)}^{\mathsf{sel}} + \Delta_{v,s,\RightD(v_i)}^{\mathsf{sel}}\}$ for $s \in S_v$\;
        }
    }

    For each $s \in S_v$, apply the Precision Sampling Lemma (\cref{lmm:precision_sampling}) on $\{\Delta_{v,s,v_i}^{\mathsf{rec}} \}_{i}$ to get $\Delta_{v,s}$\;

    \textbf{return } $\{\Delta_{v,s}\}_{s \in S_v}$\;

    \caption{Algorithm implementing \cref{def:alternative_TD}.}
    \label{alg:combine}
\end{algorithm}

\begin{algorithm}[!t]
    \KwInput{A node $v$ in $T_{c}(\bF)$ and corresponding rate $r_v > 0$.}
    \KwOutput{A set $S_v \subseteq \fragment{-2k}{2k}^{\ell_v}$ and $\Delta_{v,s}$ for all shifts $s \in S_v$. If $\ell_v = 2$, then also $\Delta_{\LeftD(v),s}$ for all $s \in S_{\LeftD(v)} = \fragment{-2k}{2k}$ and $\Delta_{\RightD(v),s}$ for all $s \in S_{\RightD(v)} = \fragment{-2k}{2k}$.}

    \If{$\ell_v = 2$}{
        Compute $S_v$ using \cref{alg:good_shift}\; \label{ln:nearlinear:1}
        Set $S_{\LeftD(v)} = S_{\RightD(v)} = \fragment{-2k}{2k}$\;
    } \Else {
        Set $S_v = \fragment{-2k}{2k}$\;
    }

    \If{$\abs{\mI_v} \leq \frac{1}{r_v}$} {  \label{ln:nearlinear:2}
        \textbf{return} $\Delta_{v,s} = 0$ for all $s \in S_v$, and if $\ell_v = 2$, also $\Delta_{\LeftD(v),s} = 0$ for $s \in S_{\LeftD(v)}$ and $\Delta_{\RightD(v),s} = 0$ for $s \in S_{\RightD(v)}$; \label{ln:nearlinear:3}
    }
    \If {$v$ is a leaf node then} { \label{ln:nearlinear:4}
        \If {$\ell_v = 1$ then}{ \label{ln:nearlinear:5}
        \textbf{return} $\Delta_{v,s} = \ted(\bF_{\mI_v}, \bG_{\mI_v,s})$ for all $s \in S_v$\; \label{ln:nearlinear:6}
        }
        \Else { \label{ln:nearlinear:7}
        \textbf{return} $\Delta_{v,s} = \min \{\ted(\bF_{\mI_v}, \bG_{\mI_v,s}), \ted(\bF_{\mI_{\LeftD(v)}}, \bG_{\mI_{\LeftD(v)},\langle s_1 \rangle}) + \ted(\bF_{\mI_{\RightD(v)}}, \bG_{\mI_{\RightD(v)},\langle s_2 \rangle})\}$ for $s \in S_v$, $\Delta_{\LeftD(v),s} = 1$ for $s \in S_{\LeftD(v)}$, and $\Delta_{\RightD(v),s} = 1$ for $s \in S_{\RightD(v)}$ \;\label{ln:nearlinear:8}
        }
    }

    \For {each child $v_i$ of $v$ in $T_{c}(\bF)$}{\label{ln:nearlinear:recurse}
        Sample $u_{v_i} \sim \mD(\varepsilon=(2\log(n))^{-1}, \delta = 0.01 \cdot (18n(4k+1)\lceil \gamma \log n \rceil)^{-1})$\; \label{ln:nearlinear:14}
        Recursively call \cref{alg:nearlinear} on $v_i$ with rate $2r_v/u_{v_i}$ for all $s_{v_i} \in S_{v_i}$\;\label{ln:nearlinear:15}
    }

    Call \cref{alg:combine} on $v$. If $\ell_v = 2$ also call \cref{alg:combine} on $\LeftD(v)$ and $\RightD(v)$\label{ln:nearlinear:combine}\;

    \textbf{return } $\Delta_{v,s}$ for $s \in S_v$, and if $\ell_v = 2$ also $\Delta_{\LeftD(v),s}$ for $s \in S_{\LeftD(v)}$ and $\Delta_{\RightD(v),s}$ for $s \in S_{\RightD(v)}$\; \label{ln:nearlinear:18}

    \caption{A near-linear-time algorithm for approximating $\ted$ (or more precisely $\TD$).}
    \label{alg:nearlinear}
\end{algorithm}

\begin{lemma}
    For any node $v \in D_{c}(\bF)$ and shift $s \in S_v$, $\Delta_{v,s}$ computed in \cref{alg:nearlinear} $(\alpha_v, 1/r_v)$-approximates $\TD_{v,s}(\bF, \bG)$ with probability at least 0.9. Moreover, \cref{alg:nearlinear} runs in time $\Oh(nc\cdot (\log(n))^{\Oh(\log_c(n))})$ with probability at least 0.98, provided that $K \geq k$.
\end{lemma}
\begin{proof}
    First, we prove that, for every $v \in D_c(\bF)$ and every $s \in S_v$, the value $\Delta_{v,s}$ is an $(\alpha_v,1/r_v)$-approximation of $\TD_{v,s}(\bF,\bG)$.
    We proceed by structural induction on $D_c(\bF)$. We choose an induction order in which every successor $v_i$ of an internal node $v$ precedes $v$. Moreover, whenever $\ell_{v_i}=2$, we require $\LeftD(v_i)$ and $\RightD(v_i)$ to precede $v$ as well. It is clear that such an ordering exists.

    Some of the values $\Delta_{v,s}$ associated with different nodes $v$ may be computed during the same invocation of \cref{alg:nearlinear}; in particular, this may occur for $v$, $\LeftD(v)$, and $\RightD(v)$. Nevertheless, we may treat the computations associated with these nodes separately for the purposes of the induction.

    For the sake of brevity, sometimes we write $\TD_{v,s}$ instead of $\TD_{v,s}(\bF, \bG)$.
    \begin{description}
        \item[Base case.] Suppose $v \in D_c(\bF)$ is solved directly by \cref{alg:nearlinear}. This can happen in two places: On the one hand, if $|\mI_v| \leq 1/r_v$, then we directly return $\Delta_{v,s} = 0$ for all $s \in S_v$ on \cref{ln:nearlinear:3}. This is valid because $\TD_{v,s} \leq |\mI_v| \leq 1/r_v$ for all $v \in D_c(\bF)$\footnote{To see this, use the split branch $\TDsel_{v,s,\LeftD(u)} + \TDsel_{v,s,\RightD(u)}$ for every bi-interval successor $u$ and the inherited shift for every mono-interval successor (these selections are complete), so that all penalties vanish; the leaf terms then sum to at most $|\mI_v|$.}, and $0$ is trivially an $(\alpha_v, 1/r_v)$-approximation of $\TD_{v,s}$. On the other hand, if $\Delta_{v,s}$ is solved in the if-block on \cref{ln:nearlinear:4}, then $\Delta_{v,s} = \TD_{v,s}$ is computed exactly.

        \item[Recursive case.] Suppose \cref{alg:nearlinear} reaches the successors of $v$.
        Let $v \in D_c(\bF)$ have successors $\{v_i\}_i$.
        For the inductive hypothesis, we assume that every successor $v_i$ satisfies
        \begin{align}
           \frac{1}{\alpha_{v_i}}\TD_{v_i, r}(\bF, \bG) - 1/r_{v_i} \leq \Delta_{v_i, r} \leq \alpha_{v_i}\TD_{v_i, r}(\bF, \bG) + 1/r_{v_i}.\label{eq:correctness_IH}
        \end{align}
        for all $r \in S_{v_i}$. Whenever $\ell_{v_i}=2$, we also assume the symmetric for $\LeftD(v_i)$ and $\RightD(v_i)$.
        Since $\Delta_{v,s}$ is the result of combining $\{\Delta^{\mathsf{rec}}_{v,s,v_i}\}_i$ (\cref{alg:combine}) using Precision Sampling (\cref{lmm:precision_sampling}), our goal is therefore to show that for all $v_i$,
        \begin{align}
            \frac{1}{\alpha_{v_i}}\TDrec_{v,s,v_i}(\bF, \bG)- u_{v_i}/r_{v} \leq \Delta^{\mathsf{rec}}_{v,s,v_i} \leq \alpha_{v_i}\TDrec_{v,s,v_i}(\bF, \bG) + u_{v_i}/r_{v}.\label{eq:correctness_rec}
        \end{align}
        Then, after applying Precision Sampling (\cref{lmm:precision_sampling}), we have the guarantee that $\Delta_{v,s}$ is a $((1+\varepsilon)\alpha_{v_i}, 1/r_v)$-approximation of $\TD_{v,s}(\bF, \bG)$, and since $(1+\varepsilon)\alpha_{v_i} \leq \alpha_v$, $\Delta_{v,s}$ is automatically an $(\alpha_{v}, 1/r_v)$-approximation of $\TD_{v,s}(\bF, \bG)$ as desired. Note that the precisions $\{u_{v_i}\}_i$ associated with the successors of $v$ are mutually independent. Indeed, for any $w,w' \in T_c(\bF)$, at most one of the nodes $\LeftD(w')$, $\RightD(w')$, and $w'$ can be a successor of $w$. Since these are the only nodes whose precisions may be dependent, the claim follows.

        Thus, our goal now is to prove \cref{eq:correctness_rec}.
        From \cref{eq:correctness_IH}, it immediately follows that
        \begin{align}
            \frac{1}{\alpha_{v_i}}\TDsel_{v,s,v_i}(\bF, \bG)- 1/r_{v_i} \leq \Delta^{\mathsf{sel}}_{v,s,v_i} \leq \alpha_{v_i}\TDsel_{v,s,v_i}(\bF, \bG) + 1/r_{v_i}.\label{eq:correctness_sel}
        \end{align}
        due to $\alpha_{v} \geq 1$ for all $v \in D_c(\bF)$. Note that if $\ell_{v_i} = 1$, then we are done because $\Deltarec_{v,s,v_i} = \Deltasel_{v,s,v_i}$ (in \cref{alg:combine}) and $\TDrec_{v,s,v_i} = \TDsel_{v,s,v_i}$ in the definition (\cref{def:alternative_TD}), and $1/r_{v_i} = u_{v_i}/2r_v < u_{v_i}/r_v$. So the more interesting case is when $\ell_{v_i} = 2$. We split into four cases:
        \begin{itemize}
            \item If $\Deltarec_{v,s,v_i} = \Deltasel_{v,s,v_i}$ and $\TDrec_{v,s,v_i} = \TDsel_{v,s,v_i}$, then this case is exactly the same as $\ell_{v_i} = 1$.
            \item If $\Deltarec_{v,s,v_i} = \Deltasel_{v,s,\LeftD(v_i)} +\Deltasel_{v,s,\RightD(v_i)}$ and $\TDrec_{v,s,v_i} = \TDsel_{v,s,\ld(v_i)} + \TDsel_{v,s,\rd(v_i)}$, then
            \begin{align}
                \Deltarec_{v,s,v_i} &\leq \alpha_{v_i}(\TDsel_{v,s,\ld(v_i)} + \TDsel_{v,s,\rd(v_i)})+2/r_{v_i} && \text{(inductive hypothesis)}\nonumber\\
                &= \alpha_{v_i}(\TDsel_{v,s,\ld(v_i)} + \TDsel_{v,s,\rd(v_i)})+u_{v_i}/r_{v} && \text{(we set $r_{v_i} = 2r_v/u_{v_i}$)}\nonumber\\
                &= \alpha_{v_i}\TDrec_{v,s,v_i} + u_{v_i}/r_v. && \text{(by assumption)}\label{eq:correctness_mixed1}
            \end{align}
            The lower bound $\Deltarec_{v,s,v_i} \geq 1/\alpha_{v_i}\TDrec_{v,s,v_i} - u_{v_i}/r_v$ is obtained in the exact same way.
            \item If $\Deltarec_{v,s,v_i} = \Deltasel_{v,s,v_i}$ and $\TDrec_{v,s,v_i} = \TDsel_{v,s,\ld(v_i)} + \TDsel_{v,s,\rd(v_i)}$, then
            \begin{align*}
                \Deltarec_{v,s,v_i} = \Deltasel_{v,s,v_i} \leq \Deltasel_{v,s,\LeftD(v_i)} +\Deltasel_{v,s,\RightD(v_i)} \leq \alpha_{v_i}\TDrec_{v,s,v_i} + u_{v_i}/r_v
            \end{align*}
            as in \cref{eq:correctness_mixed1}. The lower bound follows directly from \cref{eq:correctness_sel}.
            \item If $\Deltarec_{v,s,v_i} = \Deltasel_{v,s,\LeftD(v_i)} +\Deltasel_{v,s,\RightD(v_i)}$ and $\TDrec_{v,s,v_i} = \TDsel_{v,s,v_i}$, this case is similar to the previous case.
        \end{itemize}
        This concludes the correctness analysis.
    \end{description}

    Now, let us analyze the success probability of \cref{alg:nearlinear}. As guaranteed by \Cref{lmm:good_shift_selection}, a single call of \cref{alg:good_shift} returns a good shift selection for the node it is called on with probability at least $1 - n^{-4\gamma}$. Setting $\gamma = 2$ and union bounding over at most $6n$ nodes in $T_c(\bF)$, we get that all bi-interval nodes have a good shift selection with probability at least $1-n^{-6}$, assuming $n > 6$.
    For every node $v \in T_{c}(\bF)$, if $v$ represents a mono-interval, the Precision Sampling Lemma is called on it $4k+1$ times; if $v$ represents a bi-interval, at most $3(4k+1)\lceil \gamma \log n \rceil$ times. Hence the total number of calls is at most $18n(4k+1)\lceil \gamma \log n \rceil$, and with $\delta = 0.01 \cdot (18n(4k+1)\lceil \gamma \log n \rceil)^{-1}$ all calls succeed with probability at least $0.99$.

    To see the running time, we conclude that the number of active nodes in \cref{alg:nearlinear} is still $\Oh(n/K\cdot (\log n)^{\Oh(\log_{c}(n))})$ with probability at least 0.98, using a similar proof as \cref{lmm:num_active_nodes}.\footnote{More precisely, we have set $r_v = 2r_w/u_v$ where $v$ is a child of $w$ in $T_c(\bF)$, instead of setting $r_v = r_w/u_v$ as in \cref{alg:nearlinear_baseline}. Tracing through the calculations of \cref{lmm:num_active_nodes}, we get that $\Ex{r_v}{E} \leq (2\log n)^{\Oh(d)}/K$. The total number of active nodes across $T_{c}(\bF)$ is bounded by $\frac{100n}{K}(2\log n)^{\Oh(\log_c n)}$ with probability at least 0.98.} The running time of a single execution (ignoring recursive calls) is $\Oh(kc\polylog(n))$ now, thanks to our sparsified shift selection. Enumerating the spine nodes of $\bF_{\mI_v}$ for the sampling in \cref{alg:good_shift} costs $\Oh(|\mI_v|)$ per node, and constructing $T_c(\bF)$ costs $\Oh(nD)$; both sum to $\Oh(nD)$ and are absorbed in the bound. Therefore, the total time complexity of \cref{alg:nearlinear} is $\Oh(nc\cdot (\log(n))^{\Oh(\log_c(n))})$
\end{proof}

\approxted
\begin{proof}
Fix thresholds $k$ and $K \geq 388cDk$. Run \cref{alg:nearlinear} with parameters $k$ and $K+1$, and output \yes if and only if $\Delta_{r,\langle 0 \rangle} \leq 0.498(K+1)$. If $\ted(\bF,\bG) \leq k$, then, whenever all shift selections are good (probability at least $1 - n^{-6}$ by \cref{lmm:good_shift_selection}), \cref{lmm:decoupling_ub} gives $\TD_{r,\langle 0 \rangle}(\bF,\bG) \leq 96cDk \leq 96(K+1)/388$, so $\Delta_{r,\langle 0 \rangle} \leq 2 \cdot 96(K+1)/388 + 0.001(K+1) \leq 0.498(K+1)$; if $\ted(\bF,\bG) > K$, then $\Delta_{r,\langle 0 \rangle} \geq 0.499(K+1)$ by \cref{lmm:decoupling_lb}. A single trial thus succeeds whenever the good-shift event, all Precision Sampling calls (probability at least $0.9$), and the active-node bound (probability at least $0.98$) hold simultaneously, i.e., with probability at least $1 - 0.1 - 0.02 - n^{-6} > 1/2$. We cap every trial at its time bound $\Oh(nc(\log n)^{\Oh(\log_c n)})$ and run the amplified threshold search of \cref{cor:static_approx_alg} with $k_i = K_i/(388cD)$. Choosing $c = (\log n)^{\Theta(1/\delta)}$ yields a $(\log n)^{\Oh(1/\delta)}$-approximation in $\Oh(n^{1+\delta})$ time with high probability.
\end{proof}

\subparagraph*{Remark:}
It is worth mentioning that the algorithm in \cref{thm:approxted}
can compute values more general than $\ted(\bF,\bG)$.
It can approximate $\ted(\bF_{\mI},\bG_{\mI})$ for any balanced mono-interval or disjoint bi-interval $\mI$ such that $\bF_{\mI}$ is a balanced sub-forest. The time needed to do this does not depend on $n$ anymore, but on $\abs{\mI}$ and only requires that we know the content of $\bF_{\mI}$ and $\bG_{\mI}$.

If $\mI$ is a mono-interval, the claim is straightforward: we can simply trim all elements of $\bF$ that lie outside $\mI$. Note that it is not necessary for $\bG_{\mI}$ to be balanced. In all proofs concerning the properties of tree distance, we never require $\bG$ to be balanced, so we can similarly trim $\bG_{\mI}$ at the bi-order level.
On the other hand, if $\mI$ is a bi-interval, it suffices to pad $\bF_{\mI}$ and $\bG_{\mI}$ outside with the padding symbol $\$$ of \cref{sec:preliminaries}, which is never aligned. In this setting, we can treat the two parts of $\mI$ as “unreachable” from each other: any shift from the left part that would enter the right part instead encounters $\$$, rather than the actual characters from the right part. Then, it is straightforward to show that the $(k,K)$-\GAPTED problem for $\ted(\bF_{\mI}, \bG_{\mI})$ with $K \geq 388cDk$ can still be solved in $\Oh(|\mI|c\cdot \log(|\mI|)^{\Oh(\log_c|\mI|)})$ time with high constant probability.

\subsection{Deferred Proofs}

\begin{lemma}\label{lmm:range_min_query}
    Given an integer grid $\mathbb{Z} \times \mathbb{Z}$, let  $P$ be a set of points in the grid, where each $p\in P$ has an associated value $v_p$. Let $Q$ be a set of query points in the grid. There is an algorithm that calculates $A_q = \min_{p} \{v_p + 2||p-q||_1\}$ for all $q \in Q$ in total time $\Ohtilde(\abs{P} + \abs{Q})$.
\end{lemma}
\begin{proof}
    It is well known (see \cite{ASR00,C88}) that a set of weighted points $P'$ on a grid can be preprocessed in $\Oh(|P'| \log |P'|)$ time to support the following queries efficiently: given a rectangle on the grid, report the point with minimum weight contained within it, if any.

    In the preprocessing stage, we prepare four such range minimum query data structures on all $p=(p_1,p_2) \in P$, where the goal is to minimize
    \begin{itemize}
        \item $v^{++}_{p} = v_p + 2p_1 + 2p_2$
        \item $v^{+-}_{p} = v_p + 2p_1 - 2p_2$
        \item $v^{-+}_{p} = v_p - 2p_1 + 2p_2$
        \item $v^{--}_{p} = v_p - 2p_1 - 2p_2$
    \end{itemize}
    respectively. This stage takes $\Ohtilde(\abs{P})$ time.

    In the query stage, for a certain $q \in Q$, we split $P$ into groups based on their relative positions to $q$:
    \begin{itemize}
        \item $P_{++} = \{p \in P \mid p_1 \geq q_1, p_2 \geq q_2\}$
        \item $P_{+-} = \{p \in P \mid p_1 \geq q_1, p_2 \leq q_2\}$
        \item $P_{-+} = \{p \in P \mid p_1 \leq q_1, p_2 \geq q_2\}$
        \item $P_{--} = \{p \in P \mid p_1 \leq q_1, p_2 \leq q_2\}$
    \end{itemize}
	    Then, we query for $p_{++} \in P_{++}$ that minimizes $v^{++}_{p}$,  $p_{+-} \in P_{+-}$ that minimizes $v^{+-}_{p}$, $p_{-+} \in P_{-+}$ that minimizes $v^{-+}_{p}$, and $p_{--} \in P_{--}$ that minimizes $v^{--}_{p}$. We set
        \[
        A_q = \min\{v^{++}_p+(-2q_1-2q_2), v^{+-}_p+(-2q_1+2q_2), v^{--}_p+(2q_1+2q_2), v^{-+}_p+(2q_1-2q_2)\}.
        \]
         Each query takes $\Ohtilde(1)$ time. Thus, the overall query time complexity is $\Ohtilde(\abs{Q})$.
\end{proof}

\begin{lemma}\label{lmm:num_active_nodes}
The  number of active nodes in \cref{alg:nearlinear_baseline} is $\Oh(n/K\cdot (\log n)^{\Oh(\log_{c}(n))})$ with probability at least 0.98.
\end{lemma}
\begin{proof}
    Let us choose the parameter $N$ from the Precision Sampling Lemma (\cref{lmm:precision_sampling}) to be $100\cdot (6n)$, where $6n$ is an upper bound on the number of nodes in $T_c(\bF)$. Thus, for any node $v$ and any $u_v$ sampled from $D(\varepsilon = (2\log n)^{-1}, \delta = 0.01 \cdot (6k^2n)^{-1})$, there exists an event $E_v$ happening with probability at least $1-1/600n$, such that
    $\Ex{1/u_v}{E_v} \leq \Ohtilde(\varepsilon^{-2}\log(\delta^{-1})\log N) \leq \polylog(n)$. Let $E$ be the event $\bigcap_v E_v$. By a union bound argument over at most $6n$ nodes in $T_c(\bF)$, $E$ happens with probability at least $0.99$. From now on, we condition on the event $E$.

    For any node $v$, consider the path from the root to $v$: $v_0, v_1, \ldots, v_d = v$. By the definition of $r_v$, $r_v = \frac{1000}{K}\cdot (u_{v_1}\cdot \ldots u_{v_d})^{-1}$. Since all the $u_{v_1}, \ldots, u_{v_d}$ are sampled independently from each other, we have that
    \begin{align*}
      \Ex{r_v}{E} = \frac{1000}{K} \cdot \prod_{i=1}^{d}\Ex{1/u_{v_i}}{E_{v_i}} \leq \frac{(\log n)^{\Oh(d)}}{K}
    \end{align*}
    Using the definition that a node is active if  $\abs{\bF_{\mI_v}} > \frac{1}{r_v}$, we have that
    \begin{align*}
      \Prob[v \text{ is active} | E] = \Prob[r_v > \frac{1}{\abs{\bF_{\mI_{v}}}} | E] \leq \Ex{r_v}{E}\cdot \abs{\bF_{\mI_{v}}} \leq \frac{\abs{\bF_{\mI_{v}}}(\log n)^{\Oh(d)}}{K}
    \end{align*}
    by Markov's inequality.
    Thus, the expected number of active nodes in one level is
    \[
    \sum_{v \text{ in one level}} \frac{\abs{\bF_{\mI_{v}}}(\log n)^{\Oh(d)}}{K} = \frac{n}{K}(\log n)^{\Oh(d)}.
    \]
    Taking $d$ to be the deepest level $\log_{c}(n)$, by another Markov's inequality, we get that with probability at least 0.99, the total number of active nodes across the $T_c(\bF)$ is bounded by $\frac{100n}{K}(\log n)^{\Oh(\log_cn)}$. The total success probability is at least 0.98.
\end{proof}

%% file: s6_sublinear_revised.tex
\section{A Simple Sublinear Algorithm for \TED}\label{sec:sublinear}

In this section, we prove \cref{thm:sublinearted}
and \cref{thm:linearted}. We restate here our sublinear model.

\defproblemwlist
{Sublinear $(k,K)$-Gap Tree Edit Distance ($(k,K)$-\GAPTED)}
{Oracle access to the bi-order traversals $\mS(\bF), \mS(\bG)$ of two $n$-node forests $\bF,\bG$ via these queries:
\begin{itemize}
    \item Given $x \in \fragmentco{0}{2n}$, return the label and the first and second occurrences of the node at $\mS(\bF)\position{x}$.
    \item Given $y \in \fragmentco{0}{2n}$, return the label and the first and second occurrences of the node at $\mS(\bG)\position{y}$.
\end{itemize}
}
{\yes if $\ted(\bF, \bG) \leq k$, \no if $\ted(\bF, \bG) > K$ and any arbitrary answer \yes/\no otherwise.}
\begin{algorithm}[H]
    \KwInput{Oracle access to two size-$n$ forests $\bF, \bG$, and three thresholds $k, K, \tau \in \mathbb{Z}_{+}$.}
    \KwOutput{\yes if $\ted(\bF, \bG) \leq k$; \no with probability at least $1-e^{-1}$ if $\ted(\bF, \bG) > K$.}
    \If{$K \geq 2n$}{\textbf{return} \yes\;}
    Let $\rho \coloneqq \frac{168\tau}{K}$\; 
    \For{$\ell \in \fragment{\ceil{\log (\tau/2)}}{\floor{\log (2 \rho n)}}$}{ \label{ln:sublinear:for1}
        \For{$r \in \fragmentco{0}{ \ceil{2\rho \cdot \ceil{\frac{2n}{2^\ell}}}}$}{ \label{ln:sublinear:for2}
            Sample u.a.r. a block $B$ out of the blocks $\{B_i\}_i$ 
             used to construct $P_{\lambda(2^{\ell})}(\bF)$ in \cref{def:decomp}\;  \label{ln:sublinear:3}
             Construct the subset $S \subseteq P_{\lambda(2^{\ell})}(\bF)$ containing all $\mI \in P_{\lambda(2^{\ell})}(\bF)$ such that $U(\mI) \cap B \neq \emptyset$\;\label{ln:sublinear:4}
             \For{$\mI \in S$}{ \label{alg:sublinear:return}
                Call the $(21k,\tau)$-\GAPTED oracle on $\bF_\mI, \bG_\mI$;\\ 
                 \If {oracle returns \no} {
                    \textbf{return} \no\;
                }
            }
        }
    }
    \textbf{return} \yes\;
    
    \caption{Reduction of $(k,K)$-\GAPTED to $(21k,\tau)$-\GAPTED with $21k \leq \tau \leq K/168$.} 
    \label{alg:sublinear}
\end{algorithm}

As in \cite{GKKS22}, we prove a reduction from a $(k,K)$-\GAPTED instance to multiple smaller ones. 

Our reduction is given in \cref{alg:sublinear}. If $K\geq 2n$, it returns \yes immediately. Otherwise, we first define the parameter $\rho \coloneqq \frac{168\tau}{K}$.
Then, we examine all \emph{levels} $\ell$ in $T_2(\bF)$ for $\ell \in \fragment{\ceil{\log (\tau/2)}}{\floor{\log (2 \rho n)}}$ (\cref{ln:sublinear:for1}),
where level $\ell$ contains all nodes of $T_2(\bF)$ originating from $P_{\lambda(2^{\ell})}(\bF)$.
We sample $\ceil{2\rho \cdot \ceil{2n/2^{\ell}}}$ blocks $B_i$ used for the construction of $P_{\lambda(2^{\ell})}(\bF)$ as defined in \cref{def:decomp} (\cref{ln:sublinear:for2} and \cref{ln:sublinear:3}).
Subsequently, in \cref{ln:sublinear:4}, for each sampled block $B_i$, we find all $\mI \in P_{\lambda(2^{\ell})}(\bF)$ such that $U(\mI) \cap B_i \neq \emptyset$.
To accomplish this, we first execute queries of our sublinear model in $\mS(\bF)$ within $B_i$. This allows us to identify all $j$ such that $P_{j,i}$ and $P_{i,j}$ are in $P_{\lambda(2^{\ell})}(\bF)$ (see \cref{def:decomp}), and we can then query $\mS(\bF)$ in all such corresponding blocks $B_j$. This provides sufficient information to find the required $\mI$.
Finally, for each such $\mI$ satisfying $U(\mI) \cap B_i \neq \emptyset$,
we call an algorithm to solve $(21k,\tau)$-\GAPTED on $\bF_{\mI}$ and $\bG_{\mI}$.
If any of these calls return \no, we interrupt and return \no.
Otherwise, if all calls complete without interruption, we return \yes.
\begin{lemma}\label{lmm:sublinear_reduction}
    Let $k, K, \tau \in \mathbb{Z}_{+}$ be such that $21k \leq \tau \leq K/168$.
    \Cref{alg:sublinear} solves the instance of $(k,K)$-\GAPTED
    by using an expected $\Oh(n/K)$ calls to an oracle solving $(21k,\tau)$-\GAPTED.
    The expected total size of the forests involved in the calls is $\Oh\left(\frac{\tau n \log n}{K}\right)$.
    For $K<2n$, the reduction takes expected $\Oh\left(\frac{\tau n \log n}{K}\right)$ time, excluding execution of the gap oracle, and makes the same expected number of queries to the input forests.
    For $K\geq 2n$, it returns \yes in $\Oh(1)$ time without querying either forest.
    Assuming that every gap-oracle answer obeys its promise, the reduction errs with probability at most $e^{-1}$.
\end{lemma}

\begin{proof}
Henceforth, for a node $v \in T_2(\bF)$, we let $\mI_v$ be its associated mono- or bi-interval.

\subparagraph*{Correctness.}
Since $\ted(\bF,\bG)\leq 2n$, returning \yes is valid whenever $K\geq 2n$.
Henceforth, assume $K<2n$.
Let us first assume $\ted(\bF, \bG) \leq k$. Then, the correctness follows directly from this claim.
\begin{claim}
Given two $n$-node forests $\bF, \bG$, and any node $v \in T_2(\bF)$, we have $\ted(\bF_{\mI_v}, \bG_{\mI_v}) \leq 21\cdot \ted(\bF, \bG)$.
\end{claim}
\begin{claimproof}
    Let $\ell_v$ be the number of intervals in $\mI_v$.
    Further, for $\mI_v = \langle \fragmentco{a_{v,1}}{b_{v,1}}, \ldots, \fragmentco{a_{v, \ell_v}}{b_{v, \ell_v}}\rangle$, 
    we define another interval collection $\mJ_{v} = \langle \fragmentco{\pi_{\mA}(a_{v,1})}{\pi_{\mA}(b_{v,1})}, \ldots, \fragmentco{\pi_{\mA}(a_{v,\ell_v})}{\pi_{\mA}(b_{v,\ell_v})}\rangle$,
    just as in \cref{def:align_interval}.
    Then, we have 
    \begin{align}
        \ted(\bF_{\mI_v},\bG_{\mI_v}) &\leq \ted(\bF_{\mI_v}, \bG_{\mJ_v}) + \frac{5}{2}\sum_{i=1}^{\ell_v}(\abs{a_{v,i}-\pi_{\mA}(a_{v,i})} + \abs{b_{v,i}-\pi_{\mA}(b_{v,i})})\label{eq:sublinear_1}\\
    &\leq \ted(\bF_{\mI_v}, \bG_{\mJ_v}) + \frac{5}{2}\cdot 4 \cdot 2\ted(\bF,\bG)\label{eq:sublinear_2}\\
    &\leq \ted(\bF_{\mI_v}, \bG_{\mJ_v}) + 20 \cdot \ted(\bF, \bG)\nonumber\\
    &\leq 21 \cdot \ted(\bF, \bG), \label{eq:sublinear_4}
    \end{align}
    where \cref{eq:sublinear_1} is due to \cref{prop:triangle_inequality}; \cref{eq:sublinear_2} is due to $\mI_v$ having at most two intervals and \cref{clm:TD_ub_diff_ted}; and \cref{eq:sublinear_4} is implied by \cref{clm:TD_ub_sum_ted}.
\end{claimproof}

Therefore, all calls to the $(21k, \tau)$-\GAPTED oracle on \Cref{alg:sublinear:return} will return \textsc{yes}, and \Cref{alg:sublinear} will return \textsc{yes} as well.

Next, let us assume $\ted(\bF, \bG) > K$.
To show that we return \no with probability at least $1-e^{-1}$,
we first prove \cref{claim:sublinear}.

\begin{claim}\label{claim:sublinear}
    Mark each node $v \in T_{2}(\bF)$ that satisfies $\ted(\bF_{\mI_{v}}, \bG_{\mI_{v}}) > \tau$.
    Now, given a node $v \in T_{2}(\bF)$, we say that it is at \emph{level $\ell$} if $v$ comes from $P_{\lambda(2^{\ell})}(\bF)$. Let $S_{\ell} \coloneqq \{v \mid \text{$v$ is marked and at level $\ell$}\}$.
	
    If $\ted(\bF, \bG) > \tau$, then
    \[
        \ted(\bF, \bG) \leq c\tau \cdot \sum_{\ell=0}^{\lceil \log(2n)\rceil}\abs{S_\ell}
    \]
    where $c=2 \cdot 6 = 12$ is the maximum number of children of any node in $T_{2}(\bF)$. 
\end{claim}
    
\begin{claimproof}
    Let us prove the more general claim that for any $0\leq\ell \leq \lceil \log(2n)\rceil$,
    \begin{align}
        \sum_{v \in S_\ell}\ted(\bF_{\mI_v}, \bG_{\mI_v}) \leq c\tau \cdot \sum_{q = 0}^{\ell}\abs{S_q}.\label{eq:sub_helper}
    \end{align}

    Note that when $\ted(\bF,\bG) > \tau$, we have that $S_{\lceil \log(2n)\rceil}$ consists only of the root node, and the claim follows by setting $\ell = \lceil \log(2n)\rceil$ in \cref{eq:sub_helper}.
    Let us prove \cref{eq:sub_helper} using induction. 
    \subparagraph*{Base Case.} If $\ell < \lceil \log(\tau/2) \rceil$, then we have $S_\ell = \emptyset$ because $\abs{\mI_v} < \tau$ for any $v$ at level $\ell$ (as $\abs{\mI_v} \leq 2\cdot 2^{\ell} < \tau$ by \cref{lem:decomp_prop}), so no node at level $\ell$ is marked, and \cref{eq:sub_helper} holds vacuously.
    \subparagraph*{Inductive Step.} Assume $\ell \geq \lceil \log(\tau/2)\rceil$. Let us define $\hat{S}_{\ell-1} \coloneqq \{\text{$u \in T_2(\bF)$} \mid \text{$u$ is a child of some $v \in S_\ell$}\}$. Note that $\abs{\hat{S}_{\ell-1}} \leq c\abs{S_\ell}$.
    \begin{align}
        \sum_{v \in S_\ell}\ted(\bF_{\mI_v},\bG_{\mI_v}) &\leq \sum_{u \in \hat{S}_{\ell-1}}\ted(\bF_{\mI_u}, \bG_{\mI_u})\label{eq:sublinear_helper_1}\\
        &\leq \sum_{u \in \hat{S}_{\ell-1}\setminus S_{\ell-1}}\ted(\bF_{\mI_u}, \bG_{\mI_u}) + \sum_{u \in \hat{S}_{\ell-1}\cap S_{\ell-1}}\ted(\bF_{\mI_u}, \bG_{\mI_u})\nonumber\\
        &\leq \sum_{u \in \hat{S}_{\ell-1}\setminus S_{\ell-1}}\ted(\bF_{\mI_u}, \bG_{\mI_u}) + \sum_{u \in S_{\ell-1}}\ted(\bF_{\mI_u}, \bG_{\mI_u})\nonumber\\
        &\leq \tau \abs{\hat{S}_{\ell-1} \setminus S_{\ell-1}} + c\tau \sum_{q=0}^{\ell-1}\abs{S_q}\label{eq:sublinear_helper_2}\\
        &\leq c\tau \abs{S_\ell} +  c\tau \sum_{q=0}^{\ell-1}\abs{S_q}\nonumber\\
        &\leq c\tau \sum_{q=0}^{\ell}\abs{S_q}\nonumber
    \end{align}
	    where \cref{eq:sublinear_helper_1} and \cref{eq:sublinear_helper_2} are due to \cref{prop:ted_decomp} and the inductive hypothesis.
\end{claimproof}

Combining \cref{claim:sublinear} and $\ted(\bF, \bG) > K$, we obtain 
\begin{align}
    \sum_{\ell=0}^{\ceil{\log(2n)}} \abs{S_{\ell}} \geq K/(c\tau) = \frac{14}{\rho}. \label{eq:sublinear:2}
\end{align}
However, what we need for our proof is a bound that only involves
nodes from the sampled levels, i.e. $\ell \in \fragment{\ceil{\log (\frac{\tau}{2})}}{\floor{\log (2 \rho n)}}$.
For that purpose, notice that for any node $v$ at level $\ceil{\log (\frac{\tau}{2})}-1$ and lower, we have $\abs{\mI_{v}} \leq \tau$. Thus $\ted(\bF_{\mI_{v}}, \bG_{\mI_{v}}) \leq \tau$, and no node at such level can be marked, i.e. $\sum_{\ell=0}^{\ceil{\log (\frac{\tau}{2})}-1} \abs{S_{\ell}} = 0$.
On the other hand, the number of marked nodes at level $\ell \in \fragmentoc{\floor{\log (2 \rho n)}}{\ceil{\log (2n)}}$ can be coarsely bounded by the overall number of nodes in such levels, since there are few of them.
Let $L=\floor{\log(2\rho n)}$ and $H=\ceil{\log(2n)}$.
For $\ell\leq H$, we have $2n/2^\ell\geq 1/2$, so $\ceil{2n/2^\ell}\leq 2\cdot 2n/2^\ell$.
Moreover, $2^L>\rho n$. Using \cref{lem:num_decomp}, we obtain
\begin{align}
    \sum_{\ell=\floor{\log (2 \rho n)}+1}^{\ceil{\log (2n)}} \abs{S_{\ell}} &\leq  \sum_{\ell=\floor{\log (2 \rho n)}+1}^{\ceil{\log (2n)}} \abs{P_{\lambda(2^{\ell})}(\bF)}
    \leq 
    \sum_{\ell=\floor{\log (2 \rho n)}+1}^{\left\lceil\log (2n)\right\rceil} 3\cdot \left\lceil \frac{2n}{2^\ell}\right\rceil\nonumber\\
    &\leq 6\sum_{\ell=L+1}^{H}\frac{2n}{2^\ell}
    \leq \frac{12n}{2^L}
    \leq \frac{12}{\rho}
    \label{eq:sublinear:3}
\end{align}
Combining \cref{eq:sublinear:2} and \cref{eq:sublinear:3}, we get
\begin{equation}
    \sum_{\ell=\ceil{\log (\frac{\tau}{2})}}^{\floor{\log (2 \rho n)}} \abs{S_{\ell}} 
    = \left(  \sum_{\ell=0}^{\ceil{\log (2n)}} \abs{S_{\ell}} \right) 
    - \left(  \sum_{\ell=0}^{\ceil{\log (\tau/2)}-1} \abs{S_{\ell}} \right)
    - \left(  \sum_{\ell=\floor{\log (2 \rho n )}+1}^{\ceil{\log (2n)}} \abs{S_{\ell}} \right)
    \geq \frac{14}{\rho} - 0 - \frac{12}{\rho} > \frac{3}{2\rho}.\label{eqn:sublinear:sum_marked_nodes}
\end{equation}

Note that for each level $\ell$ and each sampled block $B_i$, we read all $\mI \in P_{\lambda(2^{\ell})}(\bF)$ such that $U(\mI) \cap B_i \neq \emptyset$.
Thus, the probability that we encounter at least one $\mI \in S_{\ell}$ is at least $m_{S_{\ell}}/\ceil{2n/2^\ell}$. Recall that $m_{S_\ell}$ is the number of blocks at level $\ell$ intersecting any member of $S_\ell$.
By \cref{lem:num_decomp}, we have $m_{S_{\ell}} \geq |S_\ell|/3$. Thus, summing over all levels $\ell \in \fragment{\ceil{\log (\frac{\tau}{2})}}{\floor{\log (2 \rho n)}}$, we have 
\begin{align}
    \sum_{\ell=\ceil{\log (\frac{\tau}{2})}}^{\floor{\log (2 \rho n)}} m_{S_\ell} \geq \sum_{\ell=\ceil{\log (\frac{\tau}{2})}}^{\floor{\log (2 \rho n)}} {\abs{S_\ell}/3} \geq \frac{1}{2\rho}.
\end{align}
by plugging in \cref{eqn:sublinear:sum_marked_nodes}.

Now we are ready to bound the error probability. At a single iteration in a single level, the probability that the algorithm fails to sample one of the $m_{S_\ell}$ blocks of interest is 
$1-m_{S_\ell}/\ceil{2n/2^\ell} \leq \exp(-m_{S_\ell}/\ceil{2n/2^\ell})$

Set $m_\ell=\ceil{2n/2^\ell}$ and $r_\ell=\ceil{2\rho m_\ell}$.
Thus, by independence of the block samples, the probability that this event happens across all iterations and all levels is at most
\[
\prod_{\ell=\ceil{\log (\frac{\tau}{2})}}^{\floor{\log (2 \rho n)}}
\exp\left(-\frac{r_\ell m_{S_\ell}}{m_\ell}\right)
\leq \exp\left(-2\rho\sum_{\ell=\ceil{\log (\frac{\tau}{2})}}^{\floor{\log (2 \rho n)}}m_{S_\ell}\right)
\leq e^{-1}.
\] 
Therefore, we conclude that with probability at least $1-e^{-1}$, the algorithm correctly returns \no. This concludes the correctness of \cref{alg:sublinear}.

\subparagraph*{Complexity.}

Next, let us show that \cref{alg:sublinear} uses an expected $\Oh(n/K)$ gap-oracle calls.
Recall that $m_\ell=\ceil{2n/2^\ell}$ is the number of blocks and $r_\ell=\ceil{2\rho m_\ell}$ is the number of samples at level $\ell$.
For a block $B_j$, let $d_\ell(j)$ be the number of interval collections in $P_{\lambda(2^{\ell})}(\bF)$ incident to $B_j$.
Every interval collection is incident to at most two blocks. Hence, by \cref{lem:num_decomp},
\[
    \sum_{j=0}^{m_\ell-1}d_\ell(j)
    \leq 2\abs{P_{\lambda(2^{\ell})}(\bF)}\leq 6m_\ell.
\]
Thus, a uniformly sampled block exposes at most six interval collections in expectation.
If $Q$ is the total number of gap-oracle calls, we have that
\begin{align*}
    \mathbb{E}[Q]
    &\leq 6\sum_{\ell=\ceil{\log (\tau/2)}}^{\floor{\log (2\rho n)}}r_\ell
    =\Oh\left(\frac{\rho n}{\tau}\right)
    =\Oh\left(\frac{n}{K}\right).
\end{align*}

Now let us bound the expected total length of $\mI$ involved in the oracle calls. In one sampling iteration in some level $\ell$, the expected length of $\mI$ incident to the sampled block is at most $\sum_{v \in P_{\lambda(2^{\ell})}(\bF)}\frac{2}{\ceil{2n/2^\ell}}\abs{\mI_{v}} = {\frac{2\cdot2n}{\ceil{2n/2^\ell}}}$. This is because any $\mI$ spans at most 2 blocks, so the probability of reading $\mI$ is at most $2/\ceil{2n/2^\ell}$. Summing over all iterations across all levels (using $r_\ell \leq 2\rho m_\ell + 1$ and $4n/m_\ell \leq 2^{\ell+1} \leq 4\rho n$ for $\ell \leq \floor{\log(2\rho n)}$), we get that 
\begin{align*}
    \Oh\left(\sum_{\ell=\ceil{\log (\frac{\tau}{2})}}^{\floor{\log (2 \rho n)}}{\frac{2\cdot2n}{\ceil{2n/2^\ell}}\cdot r_\ell}\right) = \Oh\left(\sum_{\ell=\ceil{\log (\frac{\tau}{2})}}^{\floor{\log (2 \rho n)}}{\rho n}\right) = \Oh\left(\rho n \log(n)\right) = \Oh\left(\frac{\tau n \log(n)}{K}\right).
\end{align*}

For a sampled block $B_j$, constructing its incident collections requires reading at most $1+d_\ell(j)$ blocks (which is at most $7$ in expectation over the sampled block, by $\sum_j d_\ell(j) \leq 6m_\ell$), each of length at most $2^\ell$.
Consequently, the expected total construction time is
\[
    \Oh\left(\sum_{\ell=\ceil{\log(\tau/2)}}^{\floor{\log(2\rho n)}}r_\ell2^\ell\right)
    =\Oh(\rho n\log n)
    =\Oh\left(\frac{\tau n\log n}{K}\right).
\]
This also covers the loop overhead in the nontrivial case $K<2n$.
\end{proof}

Applying this last reduction with the correct parameters yields \cref{thm:sublinearted}
and \cref{thm:linearted}. This concludes our proof.

\sublinearted
\begin{proof}
Handle $n\leq 2$ directly.
If $\tau\geq n/\ceil{\log(2n)}^2$, we return $2n+1$ as the answer directly.
This gives the required approximation because
$1+\ted(\bF,\bG)\leq 2n+1\leq 3\tau\ceil{\log(2n)}^2$.
Henceforth, assume $n\geq 3$ and $\tau<n/\ceil{\log(2n)}^2$.

\subparagraph{The algorithm.} 
Choose $g=2^{\Oh(\sqrt{\log n\log\log n})}$ for a sufficiently large
hidden constant.
For each \emph{scale} $i=\ceil{\log(2n)}-1,\ldots,0$, set
\[
    k_i=2^{i+1},\qquad
    K_i=\ceil{168k_i g\tau},\qquad
    \tau_i=\ceil{k_i\sqrt{g}}.
\]

These satisfy $21k_i\leq\tau_i\leq K_i/168$, as required by
\cref{lmm:sublinear_reduction}.

At every threshold scale $(k_i, K_i)$, we run the reduction to $(21k_i, \tau_i)$-\GAPTED (\cref{lmm:sublinear_reduction}) independently $\Theta(\log n)$ times, where in each \emph{run} we:
\begin{itemize}
    \item we return \yes immediately when $K_i \geq 2n$;
    \item we return \yes once the runtime exceeds $M \coloneqq \left\lceil\frac{Cn}{\tau\ceil{\log(2n)}^2} \right\rceil$, for some sufficiently large constant $C$. 
\end{itemize}
Let the answer of each scale be the majority of the $\Theta(\log n)$ answers from the runs. The algorithm stops at the first \no scale $i$ and outputs $1 + K_{i+1}$ as the final approximation; if none occurs, the algorithm outputs $1+ K_0$.

Now we analyze each piece in the algorithm and set their parameters.

\subparagraph{One reduction run.}
In each run, we implement the $(21k_i, \tau_i)$-\GAPTED oracle call using \cref{thm:approxted}, with success probability amplified to $1 -1/(100M)$. {The use of \cref{thm:approxted} on $\ted(\bF_{\mI}, \bG_{\mI})$ for balanced $\bF_{\mI}$ is justified by the remark after \cref{thm:approxted} in \cref{sec:near-linear}.}
On a nonempty input of size $s = |\mI|$, one local oracle call takes $\Oh(s\cdot2^{\Oh(\sqrt{\log n\log\log n})})$ time.\footnote{Analogous to \cref{lmm:solve_GAPTED} but adapted for the improved algorithm, this gap version yields the same guarantees but requires $K/k \geq 388cD$, the gap constant in the proof of \cref{thm:approxted} and runs in $n (\log n)^{\Oh(\log_c n)}$ time. Note that this adaptation supports computations of the form $\ted(\bF_{\mI_v},\bG_{\mI_v})$, as detailed in the remark following \cref{thm:approxted}. Choose $c = (\log n)^{\Oh(1/\varepsilon)}$, $\varepsilon=\Theta(\sqrt{\log\log n/\log n})$, and $g$ so that $\tau_i/(21k_i)\geq\sqrt{g}/21$ exceeds $388cD$. Amplification costs $\Oh(\log M)$ repetitions, which are absorbed
into $2^{\Oh(\sqrt{\log n \log\log n})}$.}{ Note that the $2^{\Oh(\sqrt{\log n \log\log n})}$ time bound is deterministic after capping each of the $\Oh(\log M)$ repetitions used for amplification at the time bound of \cref{thm:approxted}, which holds with probability at least $0.98$ in each repetition; a timeout counts as a wrong answer of that repetition.}

For $K_i<2n$, \cref{lmm:sublinear_reduction} bounds both the expected
reduction time and total subforest size by
$\Oh(\tau_i n\log n/K_i)$. Let $T_i$ be the uncapped running time of a single run at scale $i$, including the reduction and all its local oracle calls. We have 
\[
    \mathbb{E}[T_i]
    = \Oh\left(\frac{\tau_i n\log n}{K_i}\cdot 2^{\Oh(\sqrt{\log n\log\log n})}\right)
    =\Oh\left(1+\frac{n\log n}{\tau\sqrt{g}}2^{\Oh(\sqrt{\log n\log\log n})}\right)
    =\Oh\left(\frac{n}{\tau\ceil{\log(2n)}^2}\right).
\]
The last bound uses sufficiently large $g$ and
$n/(\tau\ceil{\log(2n)}^2)>1$.
Choose $C$ large enough that $\mathbb{E}[T_i]\leq M/100$.%
\footnote{
To account explicitly for amplification's dependence on $C$,
the bound can be written as
$C_0(1+\log C)n/(\tau\ceil{\log(2n)}^2)$
for an absolute $C_0$ independent of $C$.
Choose $C\geq 100C_0(1+\log C)$, which is possible because
$C$ grows faster than $\log C$.
}
Markov's inequality then bounds timeout probability by $1/100$.
Since each local oracle call costs at least one step, a capped run
completes at most $M$ such calls; their total error probability
is at most $1/100$.
Thus, a run errs on a promised input with probability at most
$e^{-1}+2/100<1/2$.

\subparagraph{Combining runs.}
Via amplification and a union bound over all scales, we get that each scale answer obeys the $(k_i, K_i)$-\GAPTED promise w.h.p. Assume that this is the case.
The first scale returns \yes since
$k_{\ceil{\log(2n)}-1}\geq 2n$.
At the first \no scale $i$, the current and preceding answers
imply $k_i<\ted(\bF, \bG)\leq K_{i+1}$, so
\[
    1+\ted(\bF, \bG)\leq 1+K_{i+1}
    \leq 336k_i g\tau+2
    =\Oh\bigl(g\tau(1+\ted(\bF, \bG))\bigr).
\]
If no \no occurs, then $\ted(\bF, \bG)\leq K_0$ and $1+K_0=\Oh(g\tau)$,
giving the same approximation guarantee.

There are $\Oh(\log n)$ scales and $\Oh(\log n)$ reduction runs per scale,
each taking at most $M$ steps.
The total time is therefore $\Oh(\log^2 n\cdot M)=\Oh(n/\tau)$.
\end{proof}

%% file: s7_dynamic_approximate.tex
\section{Dynamic Exact and Approximate Algorithms}\label{sec:dyn}

In this section, we prove \cref{thm:dynapproxted} and \cref{thm:dynexactted}.

\dynapproxted*

\dynexactted*

\subparagraph*{Conventions for the Dynamic Setting.}
All dynamic algorithms in this section start from \emph{empty} forests $\bF$ and $\bG$. Hence, every data structure we maintain is initialized empty, and no separate initialization procedure is needed; an arbitrary initial instance can be loaded through one insertion per node.

Throughout, $n$ and $n'$ denote the \emph{current} numbers of nodes of $\bF$ and $\bG$, respectively (so $\mS(\bF)$ has length $2n$ and the root of $T(\bF)$ represents $\langle \fragmentco{0}{2n} \rangle$). As is standard for dynamic algorithms, we assume that an upper bound $N \geq 2$ on $n + n'$ at all times is known in advance, and that the total number of updates is $\poly(N)$. For the randomized algorithms, we assume an oblivious adversary, i.e., the sequence of updates does not depend on the random choices of the algorithm; the stated success probabilities then hold for the whole sequence of updates. Thresholds $k$ are positive integers; for $k = 0$, we run the algorithms with threshold $1$, whose output determines whether $\ted(\bF, \bG) = 0$. While $\bF$ is empty, $T(\bF)$ consists of a single root representing $\langle \fragmentco{0}{0} \rangle$; the statements below assume $n \geq 1$, and $\log n$ stands for $\max(1, \log n)$.

``With high probability'' means with probability at least $1 - N^{-\gamma}$ for an arbitrarily large constant $\gamma$.

\subparagraph*{Structure of the Section.}
This section is structured as follows:
\begin{description}
    \item[\cref{subsec:restructure_tree}:] We first outline how to modify bi-intervals, mono-intervals, and the tree decomposition $T_c(\bF)$ to adapt them to the dynamic setting, yielding our main result, \cref{thm:final_decomp_tree}.
    \item[\cref{subsec:rec_approx}:] Next, we formally define the windows and the local tree edit distances considered in our dynamic scheme, and we explain how to sparsify them to approximate many distances using only a few.
    \item[\cref{subsec:fvsf}:] We then prove how to maintain these distances when only distances within $\bF$ are considered (\cref{thm:fvsf}).
    \item[\cref{subsec:sub_F_all_G}:] This, together with the machinery for finding exact matches of $\bF$ in $\bG$, allows us to maintain the distances between $\bF$ and $\bG$ (\cref{thm:fvsg}). We also give proofs of \cref{thm:dynexactted} and \cref{thm:dynapproxted}.
    \item[\cref{sec:omitted}:] Finally, we provide the proofs for several theorems that were deferred, as they are standalone results whose immediate proof would distract from the main narrative.
\end{description}

\subsection{Restructuring Bi-Intervals and Tree Decomposition}
\label{subsec:restructure_tree}

The goal of this (sub)section is to construct a dynamic decomposition tree $T(\cdot)$, which will act as the point of access to the input forests $\bF$ and $\bG$ in our dynamic algorithm. We formally state its properties in the theorem below.

\begin{restatable}{theorem}{finaldecomptree}\label{thm:final_decomp_tree}
    Let $\bF$ be a forest that undergoes node insertions and deletions. Then, there exists a dynamic decomposition tree $T(\bF)$ with the following properties:
    \begin{enumerate}[(i)]
        \item \label{it:final_tree_1}  $T(\bF)$ has depth $d = \Oh(\log n)$. Every internal node has at most two children. All leaves are at the same depth.
        \item \label{it:final_tree_2} Every node represents a mono-interval or a bi-interval{, and every $\bF_{\mI_v}$ is a balanced subforest}. The root of $T(\bF)$ represents the entire interval collection $\langle \fragmentco{0}{2n} \rangle$; the leaves represent singleton nodes in $\bF$. Nodes on the same level form a partition of $\langle \fragmentco{0}{2n} \rangle$. The interval collections represented by the children of a non-leaf node $v$ form a partition of the interval collection represented by $v$.
        \item \label{it:final_tree_3}For each internal node $v \in T(\bF)$ and child $u$ of $v$ such that $\mI_v =\langle I_{v,1}, \ldots, I_{v,\ell_v} \rangle$ and $\mI_u =\langle I_{u,1}, \ldots, I_{u,\ell_u} \rangle$ we have that for each $I_{u,j} = \fragmentco{a'}{b'}$ there is an $I_{v,i} = \fragmentco{a}{b}$ such that $a = a'$ or $b = b'$ (see also \cref{lem:make_binary} and \cref{fig:rel}). In the former case, we say that \emph{$I_{u,j}$ is a prefix of $I_{v,i}$}; In the latter case, we say that \emph{$I_{u,j}$ is a suffix of $I_{v,i}$}.
        \item \label{it:final_tree_4} Upon an update in $\bF$, the \emph{affected nodes} (nodes that are newly created in memory after the update, or nodes that have a new node in their subtree{, or, for a deletion, nodes containing an occurrence of the deleted node}) are bounded by $\Oh(\log N)$ amortized per update.
    \item \label{it:final_tree_5}The amortized update time of $T(\bF)$ under node insertions and deletions in $\bF$ is $\Oh(\log^2 N)$. \qedhere
    \end{enumerate}
\end{restatable}

To arrive at \cref{thm:final_decomp_tree}, we first show how to modify bi-intervals such that they consistently reference the same content of $\mS(\bF)$ despite changes to $\bF$. We then demonstrate how to make the decomposition tree, $T_4(\bF)$, dynamic, which we call $\bar{T}(\bF)$. Finally, we show how to modify $\bar{T}(\bF)$ to ensure that its degree is at most two. This yields the final dynamic tree decomposition $T(\bF)$ that our algorithms rely on.

\subparagraph*{Maintaining Bi-Intervals using Pointers.}

In the dynamic setting, we must manage a changing sequence of forests, denoted $\bF^0, \bF^1, \bF^2, \ldots$, where $\bF^{t}$ is the forest after $t$ updates. This, in turn, changes the derived sequence of bi-order traversal strings: $\mS(\bF^0), \mS(\bF^1), \mS(\bF^2), \ldots$. Up until now, an ordered interval collection $\mI$, e.g. $\mI = \langle \fragmentco{x}{x'} , \fragmentco{y}{y'}\rangle$, is defined by positional indices (because the intervals $\fragmentco{x}{x'} $ and $\fragmentco{y}{y'}$ are defined themselves as a subset of $\fragmentco{0}{2n}$).
 Any insertion or deletion to $\bF$ occurring before $x, x', y$, or $ y'$ causes them to refer to different content in the updated sequence $\mS(\bF^t)$.

To address this, we will employ a concept similar to pointers.
The following paragraph will introduce a data structure used under the hood. Our goal is to provide a brief introduction and then immediately abstract it so the reader can continue without needing to go into its technical details.

On a technical level, we use \emph{identified strings}, already used in \cite{KK22} and \cite{KMS23}\footnote{There, they are called \emph{labelled strings}, but to avoid confusion with node labels, we change the name.}.  An \emph{identified string} over a set of \emph{identifiers} $\Sigma$ is a string over a set of $\Sigma \times  \mathbb{Z}_{\geq 0}$. For $c \coloneqq (a,\iota) \in \Sigma \times  \mathbb{Z}_{\geq 0}$, we say that $\val(c) \coloneqq a$ of $c$  is the \emph{value} of $c$ and $\id(c) \coloneq \iota$ is the  \emph{identifier} of $c$. For an identified string $S \in (\Sigma \times  \mathbb{Z}_{\geq 0})^{*}$, we define the $\ids(S) \coloneqq  \{\id(S\position{i}) \mid i \in \fragmentco{0}{\abs{S}}\}$.
We say that $S$ is \emph{uniquely identified} if $\abs{\ids(S)} =  \abs{S}$; equivalently, for each identifier $\iota \in \ids(S)$,  there exists a unique position $i \in \fragmentco{0}{\abs{S}}$ such that $\iota = \id(S\position{i})$.

\begin{lemma}[{\cite[Lemma 7.3]{KMS23}}] \label{lem:uniq_strs}
    There is a data structure maintaining a uniquely identified
    string $S \in (\Sigma \times  \mathbb{Z}_{\geq 0})^{*}$ using the following interface:

    \begin{description}
        \item[$\mathsf{insert}(S, i, a, \iota)$:] Given $i \in \fragment{0}{\abs{S}}$ and $\iota \in \mathbb{Z}_{\geq 0} \setminus \ids(S)$, set $S \coloneq S\fragmentco{0}{i} \circ (a, \iota) \circ \fragmentco{i}{\abs{S}}$.
        \item[$\mathsf{delete}(S, i)$:] Given $i \in \fragmentco{0}{\abs{S}}$, set $S \coloneq S\fragmentco{0}{i} \circ S\fragmentoo{i}{\abs{S}}$.
        \item[$\id(S, i)$:] Given $i \in\fragmentco{0}{\abs{S}}$, return $\id(S\position{i})$.
        \item[$\val(S, i)$:]  Given $i \in\fragmentco{0}{\abs{S}}$, return $\val(S\position{i})$.
        \item[$\mathsf{pos}(S, \iota)$:]  Given $\iota \in \ids(S)$, return $i \in \fragmentco{0}{\abs{S}}$ such that $\iota = \id(S\position{i})$.
    \end{description}
    Assuming that labels belong to $\fragmentco{0}{L}$, each operation can be implemented in $\Oh(\log L)$ time.
    \lipicsEnd
\end{lemma}

This allows us to maintain a forest $\bF$ by associating to each node $v$ two unique identifiers (drawn from a counter, so that all identifiers are bounded by $\poly(N)$ and each operation of \cref{lem:uniq_strs} takes $\Oh(\log N)$ time). We store $\mS(\bF)$ as a uniquely identified string, and we use the two unique identifiers associated with $v$ instead of the first and second occurrences of $v$ in $\mS(\bF)$. Consequently, all indexing and structural elements are defined and operated upon using these identifiers rather than the raw positional indices of the string. This includes:
\begin{enumerate}[(1)]
    \item All indices within the overall bounds $\fragmentco{0}{2n}$, fragments $\fragmentco{x}{x'}$ over $\fragmentco{0}{2n}$, mono-intervals, and bi-intervals. Throughout, a nonempty interval $\fragmentco{a}{b}$ is stored as the pair $(\mathsf{id}(a), \mathsf{id}(b-1))$ of identifiers of its first and last \emph{included} positions (four identifiers for a bi-interval; the root $\langle \fragmentco{0}{0} \rangle$ of the empty forest stores none), and $b = \mathsf{pos}(\mathsf{id}(b-1)) + 1$ is recomputed on demand; in particular, neither an excluded endpoint nor $2n$ is ever an identifier. A stored interval always consists of the positions from its first to its last stored occurrence, so an update changes its content only by inserting an occurrence strictly between them, by deleting one of them, or by replacing a stored identifier, which we do only when maintaining the decomposition tree.
    \item Breakpoints $\lambda$ in $P_{\lambda}(\bF)$ from \cref{def:decomp}.
\end{enumerate}

The benefit of using uniquely identified strings is that they allow us to track the parts of a forest whose contents \emph{do not} change upon an update, but are only shifted. To see this with an example, consider a mono-interval $\langle \fragmentco{a}{b} \rangle$ of interest that is represented by node $v \in \bF$. Suppose an update occurs and a new node is inserted to the beginning of $\mS(\bF)$. Then, the positions of the contents at $\bF_{\mI_v}$ are now located in $\langle \fragmentco{a+2}{b+2} \rangle$, assuming we are to use raw positional indices. However, by letting $v$ reference the identifiers of the two end points of the subforest of interest, the notation $\bF_{\mI_v}$ references the same content cleanly throughout the updates, as long as the updates do not modify the subforest of interest.

For the sake of clear discussion, we often use an identifier $x$ and its corresponding position in the string $\mathsf{pos}(x)$ interchangeably. This is justified because the identifier is sufficient to uniquely determine the position in the identified string $\mS(\bF)$. This ``hand-waving'' is also acceptable from a computational point of view, as converting between an identifier and its position takes only logarithmic time.

\subparagraph*{Maintaining a Decomposition Tree under Updates on $\bF$.}
Next, we describe how to make the decomposition tree $T_{c}(\bF)$ (from \cref{def:partition_tree}) dynamic.
Similar to \cite{KMS23}, to achieve this, we employ weight-balanced B-trees first introduced in \cite{ArgeV03}:
Given an ordered set $S$ and an integer parameter $b$, such a tree maintains the set in its leaves and supports updates in $\Oh(\log_b n)$ amortized time while satisfying the following properties:

\begin{enumerate}[(1)]
    \item All leaves of $T$ have the same depth $d = \log_b|T| + \Theta(1)$.
    \item The root has between $2$ and $4b$ children, while all other internal nodes have between $b/4$ and $4b$ children.
    \item \label{it:btree_weight} For any non-root node $v$ at depth $d_v$, the number of leaves in its subtree is between $b^{d-d_v}/2$ and $2b^{d-d_v}$.
    \item \label{it:btree_property}If a non-root node $v$ at depth $d_v$ is split, or two non-root nodes are merged, at least $\Omega(b^{d-d_v})$ updates must occur within its subtree before another rebalancing of $v$ is required. If the root in a tree containing $M$ elements is split, then at least $\Omega(M)$ updates must occur within its subtree before another rebalancing of the root is required.
\end{enumerate}

Now, we will briefly explain how a weight-balanced B-tree maintains the ordered set $S$ under element insertions and deletions. When a new element is inserted to or deleted from leaf $u$, the nodes on the leaf-to-root path from $u$ may no longer satisfy Property~\ref{it:btree_weight}, and merging and splitting of nodes are done starting from $u$ and working towards the root to rebalance the B-tree.
From now on, we will refer to the process of splitting and merging nodes as \emph{rebalancing}.

Note that we can think of every node in the weight-balanced B-tree as an interval in the ordered set it maintains, so every level of the weight-balanced B-tree naturally forms a contiguous partition, which we can use as the sequence of breakpoints $\lambda$ on top of which we construct our hierarchical decomposition tree $T(\bF)$ in Section~\ref{sec:decomp}. We will therefore hinge on the dynamic nature of the weight-balanced B-trees to make our decomposition tree dynamic.

\begin{lemma}\label{lem:partition_query_index}
    Let $\mS(\bF)$ be maintained by a constant-degree B-tree
    of height $\Oh(\log N)$ with constant-size leaf blocks.
    Suppose each nonempty interval collection $P_{r,s}$ as defined in \cref{def:decomp}
at each level is already stored using the first and last
occurrence identifiers of each of its intervals.

    We can support the following queries in $\Oh(\log N)$ time:
    \begin{enumerate}[(a)]
        \item Given a level and the identifiers of blocks
        $B_r,B_s$ with $r\le s$, retrieve $P_{r,s}$ or report
        that it is empty.
        \item Given a block $B$ and $x\in B$, find the first
        $z\ge x$ or the last $z\le x$ in $B$ satisfying
        $\Other(z)\notin B$, or report that none exists.
        The block may be specified by its first and last
        occurrence identifiers.
    \end{enumerate}
    Upon an update to $\bF$, i.e., an insertion or a deletion of a node, the data structure used to answer the queries can be updated in $\Oh(\log N)$ amortized time.
\end{lemma}

\begin{proof}
    We first assign to every node of the B-tree a unique identifier, which it keeps for as long as it exists. Every nonempty interval collection is stored in a balanced dictionary, keyed by the ordered pair of identifiers of the B-tree nodes whose blocks are $B_r$ and $B_s$; for a mono-interval, we simply repeat the identifier of its block. As there are $\Oh(N\log N)$ interval collections in total, query~(a), as well as every insertion, deletion, or modification of an entry, takes $\Oh(\log N)$ time.

    For query~(b), we additionally keep the identifiers of $\mS(\bF)$ in an order-maintenance list \cite{Bender02}, which compares two identifiers in constant time and supports insertions and deletions in amortized constant time. With each identifier $x$, we store its twin $\Other(x)$ and a pointer to the leaf block containing $x$. Moreover, for every B-tree node $u$, let $R_u$ denote the interval of $\mS(\bF)$ covered by the subtree of $u$; at $u$, we store the first and the last identifier of $R_u$ together with
    \[
        m_u=\min_{x\in R_u}\Other(x)
        \qquad\text{and}\qquad
        M_u=\max_{x\in R_u}\Other(x),
    \]
    where the minimum and the maximum refer to the order of occurrence in $\mS(\bF)$.

    Now, let a block $B$ and $x\in B$ be given, and consider the suffix of $B$ that starts at $x$ (the case of the prefix ending at $x$ is symmetric). The root-to-leaf paths towards the two endpoints of this suffix decompose it into $\Oh(\log N)$ intervals $R_u$ of B-tree nodes $u$ and at most two partial leaf blocks. Since $B$ is contiguous, the interval $R_u$ contains a position $z$ with $\Other(z)\notin B$ if and only if $m_u\notin B$ or $M_u\notin B$, which we test by two comparisons against the first and the last identifier of $B$. We therefore scan the pieces of the decomposition from left to right, checking a partial leaf block position by position. At the first piece $R_u$ that contains such a position, we descend from $u$, always into the leftmost child whose interval contains one, until we reach a leaf block, where we find the sought $z$ directly. If no piece contains such a position, we report that none exists. Both the scan and the descent visit $\Oh(\log N)$ nodes and spend constant time at each of them, so query~(b) takes $\Oh(\log N)$ time.

    It remains to describe how the data structure is maintained under updates. First, consider the insertion or deletion of a node of $\bF$, before the B-tree performs any rebalancing. We insert the two identifiers of the node into the B-tree and the order-maintenance list (resp. delete them from the B-tree and, afterwards, from the list), set their twin and leaf-block pointers, and recompute the first and last identifiers as well as $m_u$ and $M_u$ along the two leaf-to-root paths. As leaf blocks have constant size and nodes have constant degree, this takes constant time per node, hence $\Oh(\log N)$ time along both paths, and the list operations add amortized constant time. Second, a split or a merge in the B-tree changes constantly many B-tree nodes. We update the leaf-block pointers of the constantly many identifiers concerned and recompute the stored identifiers at the changed nodes and at their ancestors, which again takes $\Oh(\log N)$ time. Third, whenever an interval collection is created, deleted, or modified, we update its entry in the dictionary; if its pair of block identifiers changes, we delete the old entry and insert the new one. Each such change takes $\Oh(\log N)$ time; these changes are not charged to the maintenance above but are accounted for separately, in the proof of \cref{lem:dynamic_decomp}.
\end{proof}

\begin{lemma}\label{lem:dynamic_decomp}
    Let $\bF$ be a forest that undergoes insertions and deletions.
    Then, we can maintain sequences
    $\lambda^0,\ldots,\lambda^{d+1}$ and sets
    $P_{\lambda^0}(\bF),\ldots,P_{\lambda^{d+1}}(\bF)$
    (defined in \cref{def:decomp}), where $d=\Oh(\log n)$
    is the current height of the weight-balanced B-tree
    with $b=4$, with leaf blocks at height $0$ containing
    one or two occurrences, such that all of the following hold:
    \begin{enumerate}[(a)]
        \item For $i\in\fragment{0}{d}$, all elements of
        $\lambda^{i+1}$ also belong to $\lambda^i$;
        \label{lem:dynamic_decomp:1}
        \item $\lambda^{d+1}=(0,2n)$ and
        $\lambda^0=(0,1,\ldots,2n-1,2n)$; and
        \label{lem:dynamic_decomp:2}
        \item For $i\in\fragment{1}{d}$, any adjacent elements
        of $\lambda^i$ satisfy
        \[
            \lambda^i_{j+1}-\lambda^i_j
            \in\fragment{4^{i-1}/2}{2\cdot4^{i-1}}.
        \]
        \label{lem:dynamic_decomp:3}
    \end{enumerate}

    Access to these partitions is provided through a tree
    $\bar T(\bF)$, whose vertices are the pairs $(i,\mI)$
    with $0\leq i\leq d+1$ and
    $\mI\in P_{\lambda^i}(\bF)$, and which has an edge from
    $(i+1,\mI')$ to $(i,\mI)$ whenever
    $U(\mI)\subseteq U(\mI')$.
    All updates take amortized time
    $\Oh(\log^2 N)=\Ohtilde(1)$.
\end{lemma}

\begin{proof}
    We first show how a weight-balanced B-tree naturally
    induces our decomposition tree from Section~\ref{sec:decomp}.
    Maintain a weight-balanced B-tree with $b=4$ over the identifiers of $\mS(\bF)$. Each B-tree node
    covers the interval of $\mS(\bF)$ consisting of the identifiers stored
    in the leaf blocks of its subtree.

    For a fixed height $i\in\fragment{0}{d}$, let the nodes
    ordered from left to right be $u_1,\ldots,u_\ell$.
    If the subtree of $u_j$ covers
    $\fragmentco{x_{j-1}}{x_j}$, define
    $\lambda^{i+1}=(x_0,\ldots,x_\ell)$, where $x_0=0$
    and $x_\ell=2n$.
    The sequences $\lambda^1,\ldots,\lambda^{d+1}$ induce
    the corresponding partitions of \cref{def:decomp}.
    However, the blocks induced by $\lambda^1$ might not represent individual
    identifiers in $\mS(\bF)$, because a leaf block can contain two
    identifiers. We therefore manually add the extra sequence
    $\lambda^0=(0,1,\ldots,2n-1,2n)$.

    Properties~\ref{lem:dynamic_decomp:1},
    \ref{lem:dynamic_decomp:2}, and
    \ref{lem:dynamic_decomp:3} now follow from the B-tree
    structure. Further note that each B-tree block has at most $16$ child blocks, and each
    leaf block has at most two singleton blocks. Thus
    \cref{lem:mult_decomp} gives constant degree for
    $\bar T(\bF)$, including its bottom and root levels.
    A root split or contraction adds or removes the top level;
    the level indices of surviving lower vertices do not change.

    The B-tree implicitly maintains the sequences
$\lambda^1,\ldots,\lambda^{d+1}$. $P_{\lambda^0}$ corresponds directly to the actual nodes in $\bF$, which we manage directly.

    When a node is inserted or deleted in $\bF$\footnote{From a technical point of view, inserting (resp. deleting) a node in $\bF$ corresponds to two insertions (resp. deletions) of parentheses in $\mS(\bF)$.}, before performing any rebalancing in the weight-balanced B-tree, we use \cref{def:partition_sensitivity2} to update $P_{\lambda^i}(\bF)$ for all $i \in \fragment{1}{d+1}$ and relink the nodes in $\bar{T}(\bF)$, which takes amortized $\Oh(\log^2 N)$ time, including the helper data structures from \cref{lem:partition_query_index}. Here, an occurrence inserted at a breakpoint $\lambda^i_j < 2n$ with $i \geq 1$ is inserted as the first occurrence in the first leaf block of the B-tree subtree whose block starts at $\lambda^i_j$, and one inserted at $2n$ joins the last block, matching the boundary map $g$ of \cref{def:partition_sensitivity2} (at level $0$, each inserted occurrence forms a new singleton block). For deletion, first transfer a deleted occurrence that
ends a nonfinal block to the following block, moving
the boundary one position to the left.
Discard empty blocks and then reverse the insertion
convention above. There are constantly many such
boundary adjustments per level, each handled by
constantly many queries of
\cref{lem:partition_query_index}. The parts in $S$ keep their vertices, and their stored identifier pairs are updated to represent $g(\mI)$ for $\mI \in S$: a stored identifier changes only if the part starts at an inserted occurrence, ends at $2n$ while an occurrence is inserted there, or loses its first or last position by a deletion, and it is then replaced by the identifier of the inserted, resp.\ neighboring, occurrence; constantly many interval collections per level are concerned, so this costs $\Oh(\log n)$ identifier operations. Note that now the partitions $\{P_{\lambda^i}(\bF)\}_{i=0}^{d+1}$ correctly reflect the current nodes in $\bF$.

    If the weight-balanced B-tree performs rebalancing, then we need to update the partitions $\{P_{\lambda^i}(\bF)\}_{i=1}^{d+1}$ accordingly. If a B-tree node at height $i-1$, where
$1\leq i\leq d+1$, is split into two, then if the original sequence of break points for level $i$ is $\lambda^i = (\lambda_0^i, \lambda_1^i, \ldots, \lambda_k^i, \ldots, \lambda_j^i)$ for some $j$, then the new sequence of breakpoints grows by one: $(\lambda^{i})' = (\lambda_0^i, \lambda_1^i, \ldots, \lambda_k^i, \lambda_{\bar{k}}^{i}, \lambda_{k+1}^i, \ldots, \lambda_j^i)$. Therefore, we can use \cref{def:partition_sensitivity}\ref{it:partition_sensitivity:b} to update $P_{\lambda^i}(\bF)$, which takes time $\Oh(|\lambda_{k+1}^i - \lambda_{k}^i|\log N)$. If a new root is created in the weight-balanced B-tree, then we set $(d+1,\langle \fragmentco{0}{2n}\rangle)$ to be the new root of our decomposition tree. Finally, we need to update the parent and child pointers for every newly created $\mI \in P_{(\lambda^i)'}(\bF) \setminus P_{\lambda^i}(\bF)$, and according to \cref{def:partition_sensitivity}\ref{it:partition_sensitivity:a}, it takes $\Oh(\log N)$ time to relink the nodes in $\bar{T}(\bF)$.
    The case where two nodes are merged by the weight-balanced B-tree is analogous.

    We now analyze the amortized running time.
Before rebalancing, constantly many helper queries,
dictionary changes, and relinking operations are needed
per level. Including maintenance of the helper data
structures, this costs $\Oh(\log^2 N)$ amortized time.

A rebalancing operation at height $h$ scans
$\Oh(4^h)$ positions and requires at most $\Oh(4^h)$
dictionary changes. Including identifier conversions,
block lookups, relinking, and helper maintenance,
its cost is $\Oh(4^h\log N)$.
By Property~\ref{it:btree_property}, this is charged to
$\Omega(4^h)$ updates within the affected subtree,
giving amortized cost $\Oh(\log N)$ per height.
Summing over $\Oh(\log N)$ possible heights, and
including the work before rebalancing and the underlying
B-tree updates, gives $\Oh(\log^2 N)$ amortized time.
\end{proof}

\subparagraph*{The Decomposition Tree $T(\bF)$.}
To make $\bar{T}(\bF)$ have degree at most two, we take each node $v \in \bar{T}(\bF)$ and replace it with a gadget $T_v(\bF)$ that is a little binary tree with leaves the former children of $v$. The following proposition captures the construction of the gadget.

\begin{figure*}[!t]
     \centering
     \begin{subfigure}[b]{0.35\textwidth}
         \centering
         \scalebox{0.8}{\input{figures/dynamic/mono-to-mono-mono}}
         \caption{}
         \label{fig:rel:a}
     \end{subfigure}
     \hfill
     \begin{subfigure}[b]{0.55\textwidth}
         \centering
          \scalebox{0.8}{\input{figures/dynamic/bi-to-bi-bi}}
         \caption{}
         \label{fig:rel:b}
     \end{subfigure}
     \\[15pt]
     \begin{subfigure}[b]{0.35\textwidth}
         \centering
         \scalebox{0.8}{\input{figures/dynamic/mono-to-bi}}
         \caption{}
         \label{fig:rel:c}
     \end{subfigure}
     \hfill
     \begin{subfigure}[b]{0.55\textwidth}
         \centering
           \scalebox{0.8}{\input{figures/dynamic/bi-to-bi-mono}}
         \caption{}
         \label{fig:rel:d}
     \end{subfigure}
     \\[15pt]

    \caption{
        The figure shows the possible parent-child relations from Property \ref{it:make_binary:4} of \cref{lem:make_binary}. The intervals of the first child $u_1$ and second child $u_2$ (if there is any) are displayed in blue and red. The intervals of the parent $v$ can be inferred from the union of the intervals of the children.
     }
     \label{fig:rel}
\end{figure*}
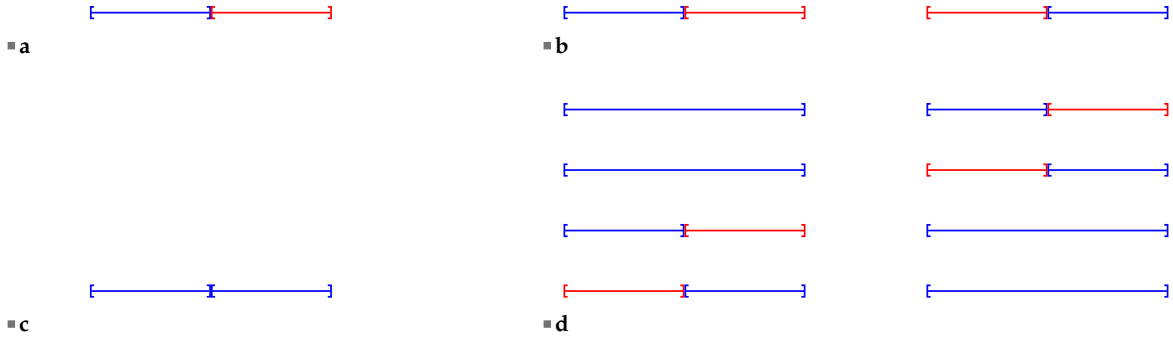

\begin{lemma}\label{lem:make_binary}
Let $v \in \bar{T}(\bF)$ be an internal node with children $v_1, \ldots, v_c$.
Then, given access to the interval collections $\mI_v$ and $\{\mI_{v_i}\}_{i=1}^{c}$, we can construct a tree $T_v(\bF)$ in $\Oh(c)$ time that satisfies all of the following:
\begin{enumerate}[(i)]
    \item $T_v(\bF)$ is a binary tree with $\Oh(c)$ nodes;
    \label{it:make_binary:1}
    \item $v$ is the root of $T_v(\bF)$, $v_1, \ldots, v_c$ are its leaves, and every other internal node $u$ is also associated with a mono-interval or bi-interval $\mI_u$;
    \label{it:make_binary:2}
    \item For each node $u$ in $T_v(\bF)$, if $\mI_u$ is a bi-interval, it must contain at least one spine node; and
    \label{it:make_binary:3}
    \item For each internal node $u \in T_v(\bF)$, the interval collections of its children, $\mI_{u_1}, \ldots, \mI_{u_z}$, form a partition of $\mI_{u}$ according to one of the following five types (see  \cref{fig:rel}):
    \begin{description}
         \item[\emph{Mono-to-Mono/Mono (\cref{fig:rel:a}):}] $u$ has a mono-interval $\mI_u = \langle I_1 \rangle$ and two children $u_1, u_2$ with mono-intervals $\mI_{u_1}  = \langle I_1^1 \rangle,\mI_{u_2}  = \langle I_1^2 \rangle$;
        \item[\emph{Mono-to-Bi (\cref{fig:rel:c}):}] $u$ has a mono-interval $\mI_u = \langle I_1 \rangle$ and a single child $u_1$ with a bi-interval $\mI_{u_1} = \langle I_1^1, I_2^1 \rangle$;
        \item[\emph{Bi-to-Bi/Mono (\cref{fig:rel:d}):}] $u$ has a bi-interval $\mI_u = \langle I_1, I_2 \rangle$ and two children, $u_1$ and $u_2$, such that $\mI_{u_1} = \langle I_1^1, I_2^1 \rangle$  and $\mI_{u_2} = \langle I_1^2 \rangle$, satisfying $I_1^1 \subseteq I_1$ and $I_2^1 \subseteq I_2$;
        \item[\emph{Bi-to-Bi/Bi (\cref{fig:rel:b}):}] $u$ has a bi-interval $\mI_u = \langle I_1, I_2 \rangle$ and two children, $u_1$ and $u_2$, where both $\mI_{u_1} = \langle I_1^1, I_2^1 \rangle$ and $\mI_{u_2} = \langle I_1^2, I_2^2 \rangle$ are bi-intervals, satisfying $I_1^1, I_1^{2}  \subseteq I_1$ and $I_2^1,I_2^2 \subseteq I_2$.
        \item[\emph{Identity:}] $u$ has a single child $u_1$ with $\mI_{u_1} = \mI_u$.
    \end{description}
    \label{it:make_binary:4}
\end{enumerate}
\end{lemma}

\begin{proof}
We construct the desired tree $T_v(\bF)$ recursively. At each step, we define a node $u$ and recursively construct the subtree rooted at $u$. The recursive invariant is defined by a node $u$, its associated mono- or bi-interval $\mI_u$, and a subset $V_u \subseteq \{v_i\}_{i=1}^{c}$ of the original children as the leaves of $u$, such that the collection of intervals $\{\mI_{w}\}_{w \in V_{u}}$ forms a partition of $\mI_u$. Moreover, we will maintain the invariant that $\mI_{u}$ is a bi-interval only if $\mI_u$ contains a spine node.
Recall that, by \cref{lem:decomp_prop}, each subforest $\bF_{\mI_w}$ with $w \in V_u$ is balanced, and note that each bi-interval $\mI_w$ contains a spine node, namely any node assigned to it by the first rule of \cref{def:decomp}. We will repeatedly use the following consequence of \cref{fct:non_crossing}: if $\mI_w = \langle J_1, J_2 \rangle$ is a bi-interval with $w \in V_u$ and $G$ denotes the set of positions strictly between $J_1$ and $J_2$, then every other child $w'' \in V_u$ satisfies $U(\mI_{w''}) \subseteq G$ or $U(\mI_{w''}) \cap (J_1 \cup G \cup J_2) = \emptyset$. Indeed, otherwise $\mI_{w''}$ would be a bi-interval with one interval inside $G$ and the other one outside of $J_1 \cup G \cup J_2$, and a spine node of $\mI_{w''}$ would cross a spine node of $\mI_w$.

We start the recursion at the root $v$, where $V_v = \{v_i\}_{i=1}^{c}$ and $\mI_v$ is given (note that all invariants hold for $v$).
For a generic internal node $u$, we introduce its children $u_1, \ldots, u_z$ where $z \in \{1, 2\}$ must conform to one of the five types specified in Property \ref{it:make_binary:4}.
We associate each child $u_i$ with an interval collection $\mI_{u_i}$ and a non-empty subset $V_{u_i}$ such that the family of sets $\{V_{u_i}\}_{i=1}^z$ forms a partition of $V_u$ (we will explain how after this paragraph). We then recursively construct the subtrees of $u_1, \ldots, u_z$. The base case for the recursion is when $\abs{V_{u}}=1$. In this case, let $v_i$ be the unique element of $V_u$. If $u \neq v$ and $\mI_u = \mI_{v_i}$, we identify $u$ with $v_i$, making it a leaf of $T_v(\bF)$. Otherwise, $u$ keeps its own vertex and receives the single child $v_i$: through an Identity edge if $\mI_u = \mI_{v_i}$, and through a Mono-to-Bi edge if $\mI_u = \langle I_1 \rangle$ and $\mI_{v_i} = \langle J_1, J_2 \rangle$ with $J_1 \cup J_2 = I_1$; by the partition and spine-node invariants, no other case arises.

Suppose we are at an internal node $u$ with interval $\mI_u$ and set $V_u$ such that $\abs{V_u} \geq 2$. We construct the children $u_1, \ldots, u_z$ based on $\mI_u$ and $V_u$:
\begin{itemize}
    \item If $\mI_u = \langle I_1 \rangle$ is a mono-interval, then we take the first $w \in V_{u}$ appearing on $I_1$. If $w$ is such that $\mI_w = \langle J_1 \rangle$ is a mono-interval, then we set $z = 2$ and we choose $u$ and $u_1,u_2$ to follow a Mono-to-Mono/Mono relation by setting $V_{u_1} = \{w\}$, $V_{u_2} = V_{u} \setminus \{w\}$, $I_1^1 = J_1$,  and $I_1^2 = I_1 \setminus J_1$.
    Otherwise, if $w$ is such that $\mI_w = \langle J_1, J_2 \rangle$ is a bi-interval, then we set $z = 1$ and we choose $u$ and $u_1$ to follow a Mono-to-Bi relation by setting $V_{u_1} = V_u$ and $I_1^1 = J_1$, $I_2^1 = I_1 \setminus J_1$ (a spine node of $\mI_w$ is also a spine node of $\mI_{u_1}$).

    \item If $\mI_u = \langle I_1, I_2 \rangle$ is a bi-interval, we consider any $w \in V_{u}$
    such that $\mI_{w} = \langle J_1, J_2 \rangle$ satisfies $J_1 \subseteq I_1$ and $J_2 \subseteq I_2$. Note that such $w$ must exist, because of the spine node invariant: if $s$ is a spine node of $\mI_u$, then the child containing $\Left(s)$ also contains $\Right(s)$, as its subforest is balanced. We perform a case distinction depending on whether there is another distinct $w'  \in V_{u}$ such that $\mI_{w'} = \langle J_1', J_2' \rangle$ satisfies $J_1' \subseteq I_1$ and $J_2' \subseteq I_2$:
    \begin{itemize}
        \item If such $w'$ exists, assume without loss of generality that $J_1'$ appears after $J_1$ on $I_1$. We set $z = 2$ and we choose $u$ and $u_1,u_2$ to follow a Bi-to-Bi/Bi relation. If $\mI_{u} = \langle \fragmentco{i_1}{i_1'}, \fragmentco{i_2}{i_2'} \rangle$
        and $\mI_{w} = \langle \fragmentco{j_1}{j_1'}, \fragmentco{j_2}{j_2'} \rangle$, then we set $\mI_{u_1} = \langle \fragmentco{i_1}{j_1'}, \fragmentco{j_2}{i_2'} \rangle$ and $\mI_{u_2} = \langle \fragmentco{j_1'}{i_1'}, \fragmentco{i_2}{j_2} \rangle$.
        Moreover, we set $V_{u_2} = \{w'' \in V_u \mid U(\mI_{w''}) \subseteq \fragmentco{j_1'}{j_2}\}$ and $V_{u_1} = V_u \setminus V_{u_2}$. Here, $\fragmentco{j_1'}{j_2}$ is the set of positions strictly between $J_1$ and $J_2$, so by the observation above (applied to $w$), every $w'' \in V_{u_1} \setminus \{w\}$ satisfies $U(\mI_{w''}) \subseteq \fragmentco{i_1}{j_1} \cup \fragmentco{j_2'}{i_2'}$. Hence, the children in $V_{u_1}$ and $V_{u_2}$ partition $\mI_{u_1}$ and $\mI_{u_2}$, respectively. Both sets are non-empty: $w \in V_{u_1}$, and $w' \in V_{u_2}$ because $J_1' \subseteq \fragmentco{j_1'}{i_1'}$ (in particular, $J_2'$ lies before $J_2$, so both intervals of $\mI_{u_2}$ are non-empty). Finally, spine nodes of $\mI_w$ and $\mI_{w'}$ are spine nodes of $\mI_{u_1}$ and $\mI_{u_2}$, respectively.
        \item If not, by $\abs{V_u} \geq 2$, we set $z = 2$ and we choose $u$ and $u_1,u_2$ to follow a Bi-to-Bi/Mono relation. Note that there must be another $\hat{w} \in V_u$ such that one of the intervals $\hat{J}$ of $\mI_{\hat{w}}$ appears as first or last on some $I_j$ for $j \in \{1,2\}$ (otherwise $J_1 = I_1$ and $J_2 = I_2$, contradicting $\abs{V_u} \geq 2$). Since $\hat{w}$ is not of the same form as $w$, all intervals of $\mI_{\hat{w}}$ lie in $I_j$. At this point, we can set $\mI_{u_2}$ to be the smallest interval $X$ in $I_j$ to include all intervals of $\hat{w}$, which by the previous discussion is a prefix or suffix of $I_j$. We set $\mI_{u_2} = \langle X \rangle$ and $\mI_{u_1}$ to include as intervals $I_{j} \setminus X$ and $I_{3-j}$ (ordered according to the value of $j$). Moreover, we set $V_{u_2} = \{w'' \in V_u \mid U(\mI_{w''}) \subseteq X\}$ and $V_{u_1} = V_u \setminus V_{u_2}$. If $\mI_{\hat{w}}$ is a mono-interval, then $X = U(\mI_{\hat{w}})$; otherwise, $X$ consists of the two intervals of $\mI_{\hat{w}}$ and the positions strictly between them, so by the observation above (applied to $\hat{w}$), every other child lies entirely inside or entirely outside of $X$. Hence, the children in $V_{u_1}$ and $V_{u_2}$ partition $\mI_{u_1}$ and $\mI_{u_2}$, respectively. Both sets are non-empty: $\hat{w} \in V_{u_2}$, and $w \in V_{u_1}$ since $J_{3-j} \subseteq I_{3-j}$ lies outside of $X$. In particular, $J_j \subseteq I_j \setminus X$, so $I_j \setminus X$ is a non-empty interval and a spine node of $\mI_w$ is a spine node of $\mI_{u_1}$.
    \end{itemize}
\end{itemize}

In every case, the sets $V_{u_1}$ and $V_{u_2}$ (resp.\ $V_{u_1}$) are non-empty and partition $V_u$, the children in $V_{u_i}$ partition $\mI_{u_i}$, and every newly created bi-interval contains a spine node, so the invariants are maintained. The recursion terminates because $|V_{u_i}| < |V_u|$ whenever $z = 2$, and a Mono-to-Bi step ($z = 1$) is always followed by a step with $z = 2$, as $u_1$ is then a bi-interval node with $|V_{u_1}| \geq 2$.
Clearly, properties \ref{it:make_binary:2} and \ref{it:make_binary:3} and \ref{it:make_binary:4} are satisfied. For \ref{it:make_binary:1} notice that the number of degree one nodes in $T_v(\bF)$ is bounded by the number of bi-intervals among $\{\mI_{v_i}\}_{i=1}^{c}$, plus one for the root if $c = 1$: each Mono-to-Bi step is triggered by a bi-interval child $w$, which from then on belongs to bi-interval nodes only. This, together with the fact that the tree has $c$ leaves, yields that the size of $T_{v}(\bF)$ is at most $\Oh(c)$.
\end{proof}

Now we are ready to put together the pieces for \cref{thm:final_decomp_tree}.

\finaldecomptree*

\begin{proof}
    First, consider the dynamic tree $\bar{T}(\bF)$ from \cref{lem:dynamic_decomp} and the mini binary tree gadget $T_v(\bF)$ from \cref{lem:make_binary}.
    $T(\bF)$ is obtained by replacing every node $v \in \bar{T}(\bF)$ and the edges to its children with $T_v(\bF)$. Note that simply by doing the replacement, we fail to satisfy that: all leaves are at the same depth; and nodes on the same level form a partition of $\langle \fragmentco{0}{2n} \rangle$ because the mini binary tree gadgets do not satisfy them. To remedy this issue, we will choose a large enough constant $D$. Then, take every leaf $w$ of every gadget $T_v(\bF)$ whose depth $i$ is strictly smaller than the $D$, and iteratively add a new node representing the same interval collection as $w$ as the child of $w$ on depth $i+1$ (an Identity edge in the sense of \cref{lem:make_binary}\ref{it:make_binary:4}), until the last node we added reaches depth $D$. In this way, every gadget $T_v(\bF)$ has the same depth for all $v \in \bar{T}(\bF)$, and nodes on the same level of every gadget $T_v(\bF)$ form a partition of $\mI_v$. The choice of $D$ is valid because according to Property~\ref{lem:dynamic_decomp:3} of the weight-balanced B-tree and \cref{lem:mult_decomp} Item~\ref{lem:mult_decomp:ii}, the number of children of every internal node in $\bar{T}(\bF)$ is some constant throughout the updates. Therefore, the size and depth of every gadget $T_v(\bF)$ are also constants, which can be bounded by a large enough number $D$. Thus, with our modification of the gadgets, $T(\bF)$ is a decomposition tree of $\bF$, and it is immediate to see that Properties~\ref{it:final_tree_1}, \ref{it:final_tree_2}, and \ref{it:final_tree_3} are satisfied.

    To prove Properties~\ref{it:final_tree_4} and \ref{it:final_tree_5}, we first describe how $T(\bF)$ is dynamically maintained: Upon an update, first update $\bar{T}(\bF)$ as in \cref{lem:dynamic_decomp}. Then for each new node $v \in \bar{T}(\bF)$, we recompute the gadgets $T_v(\bF)$.

    Recall that a node is new if it is either newly instantiated after the update to $\bF$, or it contains a newly instantiated node in its subtree{, or, for a deletion, it contains an occurrence of the deleted node (a deletion may instantiate nothing, while the lowest surviving ancestor of the deleted leaf loses a child, so its gadget must be rebuilt)}. A new node can be instantiated in the following two ways:
    \begin{itemize}
        \item When a node is inserted or deleted in $\bF$, before performing any rebalancing in the weight-balanced B-tree, \cref{def:partition_sensitivity2} is called to update the partitions $\{P_{\lambda^i}(\bF)\}$.
        \item When the weight-balanced B-tree rebalances, \cref{def:partition_sensitivity} is called to update the partitions $\{P_{\lambda^i}(\bF)\}$.
    \end{itemize}
    Before rebalancing, an insertion or a deletion incurs a constant number of affected nodes at every height in the worst case, and hence $\Oh(\log n)$ many affected nodes in $\bar{T}(\bF)$, by \cref{def:partition_sensitivity2}. During rebalancing, a B-tree split or merge changes
constantly many vertices, by
\cref{def:partition_sensitivity}.
Including their ancestors, this affects $\Oh(\log n)$
vertices in total.
To bound this contribution over the update sequence,
we use $\log n\leq\log N$. Since we can charge a rebalancing at height $h$ to $\Omega(4^h)$ updates, the total amortized number of affected nodes per update is
    \[
    \Oh(\log N) + \sum_{h\geq0}\Oh(\frac{\log N}{4^h}) = \Oh(\log N).
    \]

    For each new node, we need to rebuild the mini binary tree gadgets, each requiring a constant number of identifier operations, and thus $\Oh(\log N)$ time. Therefore, the number of new nodes in $T(\bF)$ is just a constant factor more than the number of new nodes in $\bar{T}(\bF)$, which is still $\Oh(\log N)$ amortized per update, and the amortized update time of $T(\bF)$ is just a constant factor more than the update time of $\bar{T}(\bF)$, which is still $\Oh(\log ^2 N)$.
\end{proof}

\subsection{Allowed Windows and Oracles}
\label{subsec:rec_approx}
For each node $v \in T(\bF)$, we maintain the values of $\ted(\bF_{\mI_v}, \bG_{\mJ})$, or approximations thereof, for a collection of carefully chosen intervals $\mJ$. In this (sub)section, we first introduce the notations required to define the specific intervals $\mJ$ for which we maintain these values. Then, we formalize, through the concept of \emph{oracles}, how we can recompute or otherwise utilize these values.

Regarding $T(\bF)$, we introduce the following notation for its associated quantities, which will be used throughout the rest of the proof.

\begin{definition}
    Let $v \in T(\bF)$ be a node, and let $h_v$ denote its height. We denote by $\mI_v$ the sequence of intervals associated with $v$, and by $\ell_v$ the number of intervals it contains, where $I_{v,i}$ is the $i$-th interval of $\mI_v$ for $i \in \fragment{1}{\ell_v}$. Furthermore, we define the endpoints $a_{v,i}$ and $b_{v,i}$ such that $I_{v,i} = \fragmentco{a_{v,i}}{b_{v,i}}$.
\end{definition}

 We also fix $\varepsilon$ such that $0 < \varepsilon \leq 1/100$.

\subparagraph*{Allowed Intervals.}
We begin by defining the set of allowed lengths for our chosen intervals $\mJ$.

\begin{definition}\label{def:dyn_allowed_length}
Let
\[
    \Delta \coloneqq
    \bigl\{-\lfloor(1+\varepsilon/2)^i\rfloor \bigr\}_{i  \in \mathbb{Z}_{\geq 0}}
    \cup \fragment{-\lceil 2/\varepsilon \rceil}{\lceil 2/\varepsilon \rceil} \cup \bigl\{ \lfloor(1+\varepsilon/2)^i\rfloor \bigr\}_{i \in \mathbb{Z}_{\geq 0}}.
\]
Moreover, given an interval length $q$ (so simply a non-negative number), we define the following two sets:
\[
    \sL^{k}(q) \coloneq \fragment{\max(q - 4k, 0)}{q + 4k} \quad \text{ and } \quad
    \tilde{\sL}^{k}(q) \coloneq \{ \ q + s \in \sL^{k}(q) \mid s \in \Delta \}.
\]

Lastly, for lengths $q,q'$ with $q' \in \sL^{k}(q)$ and an upper bound $h \geq q$, define the function $\mathsf{Round}^k(q, q',h)$ as
\[
    \mathsf{Round}^k(q, q',h) \coloneqq
    \begin{cases}
        \max\{\tilde{q} \in \tilde{\sL}^{k}(q) \mid \tilde{q} \leq \min(q',h) \} & q' \geq q, \\
        \min\{\tilde{q} \in \tilde{\sL}^{k}(q) \mid \tilde{q} \geq q'\} & q' < q.
    \end{cases}
\]
\end{definition}

Note that $\mathsf{Round}^k(q,q',h)$ is well defined, as $q \in \tilde{\sL}^{k}(q)$ and $q \leq h$. Moreover, $\min(q, q') \leq \mathsf{Round}^k(q,q',h) \leq \max(q, \min(q', h))$. In all applications, $q$ and $q'$ are the lengths of an interval $I$ and of an interval $J \in W^{k}(I)$ (introduced below), or of corresponding parts thereof, so that indeed $q' \in \sL^{k}(q)$. When $k$ is clear from the context, we drop the superscript and write $\sL(q)$, $\tilde{\sL}(q)$, and $\mathsf{Round}(q,q',h)$.

\begin{proposition}\label{rmk:rounding_small_gap}
     Key properties of $\mathsf{Round}^k(q,q',h)$ that we will use are the following:
     \begin{enumerate}[(i)]
        \item For any $q' \in \sL^{k}(q)$ such that $h \geq q' \geq 0$, we have $|q'-\tilde{q}| \leq \varepsilon \cdot |q'-q|$, where $\tilde{q} \coloneqq \mathsf{Round}^k(q,q',h)$.
        \label{it:rmk:rounding_small_gap:i}
         \label{it:rmk:rounding_small_gap:ii}
        \item Let $I = \fragmentco{a}{b}$, $J = \fragmentco{a'}{b'}$ be such that $|a-a'|,|b-b'| \leq 2k$. Then, the rounded interval $\tilde{J} = \fragmentco{a'}{a'+\tilde{q}}$, where $\tilde{q} = \mathsf{Round}^k(|I|, |J|,h)$ for any $h \geq |I|$, stays within bounds; formally, $|b - a'-\tilde{q}| \leq 2k$.
        \label{it:rmk:rounding_small_gap:iii}
        \item Similarly, for $I, J, \tilde{q}$ as defined in \ref{it:rmk:rounding_small_gap:iii}, if we consider the rounded interval $\tilde{J} = \fragmentco{b'-\tilde{q}}{b'}$ aligned to a suffix of $J$, we have $|a - b'+\tilde{q}| \leq 2k$.
        \label{it:rmk:rounding_small_gap:iv}
     \end{enumerate}
\end{proposition}
\begin{proof}
    We start with the proof of \ref{it:rmk:rounding_small_gap:i}. We only prove the case $q \leq q' \leq h$, where, because we round down, it suffices to show $q'-\tilde{q} \leq \varepsilon \cdot (q'-q)$; the case $0 \leq q' < q$ follows by a symmetric argument.
    Let $d \coloneqq q' - q \geq 0$. We distinguish two cases:
    \begin{itemize}
        \item If $0 \leq d \leq \lceil 2/\varepsilon \rceil$,
        then $d \in \Delta$. So $\tilde{q} = q'$ and $q' - \tilde{q} = 0 \leq \varepsilon \cdot d$.
        \item If $d > \lceil 2/\varepsilon \rceil \geq 2/\varepsilon$,
        let $p$ be the largest non-negative integer such that
        $\lfloor (1+\varepsilon/2)^{p} \rfloor \leq d$.
        Then $\tilde{q} \geq q + \lfloor (1+\varepsilon/2)^{p} \rfloor$, because $q + \lfloor (1+\varepsilon/2)^{p} \rfloor \in \tilde{\sL}^{k}(q)$ by $\lfloor (1+\varepsilon/2)^{p} \rfloor \leq d \leq 4k$.
        By the maximality of $p$, we have
        $\lfloor (1+\varepsilon/2)^{p+1} \rfloor > d$,
        which gives $(1+\varepsilon/2)^{p+1} > d$.
        Thus, $\lfloor (1+\varepsilon/2)^{p} \rfloor
        \geq (1+\varepsilon/2)^{p} - 1
        > d/(1+\varepsilon/2) - 1$.
        We conclude that
        \[
            q' - \tilde{q}
            \leq d - \lfloor (1+\varepsilon/2)^{p} \rfloor
            < d - d/(1+\varepsilon/2) + 1
            = (\varepsilon/2) \cdot d/(1+\varepsilon/2) + 1
            \leq (\varepsilon/2) \cdot d + 1.
        \]
        Since $d > 2/\varepsilon$, we have $1 < (\varepsilon/2) \cdot d$, and therefore
        $q' - \tilde{q} < (\varepsilon/2) \cdot d + (\varepsilon/2) \cdot d
        = \varepsilon \cdot d$.
    \end{itemize}

    We proceed to prove \ref{it:rmk:rounding_small_gap:iii}.
    If $|J| \geq |I|$, then for $\tilde{q} \coloneqq \mathsf{Round}^k(|I|, |J|, h)$ we round down, so $|I| \leq \tilde{q} \leq |J|$. It follows that
    $-2k \leq a'-a = a'+|I| - b \leq a'+\tilde{q}-b \leq a'+|J|-b = b'-b \leq 2k$,
    and thus $|a'+\tilde{q}-b| \leq 2k$.
    If instead $|J| < |I|$, we round up, so $|J| \leq \tilde{q} \leq |I|$. Similarly,
    $-2k \leq b' - b = a'+|J|-b \leq a' + \tilde{q} - b \leq a' + |I| - b = a' - a \leq 2k$.

    The proof of \ref{it:rmk:rounding_small_gap:iv} is analogous to that of \ref{it:rmk:rounding_small_gap:iii}.
\end{proof}

This allows us, given an interval $I$, to define the set of intervals $W^{k}(I)$ whose endpoints fall within $2k$ of the endpoints of $I$, and $\widetilde{W}^{k}(I)$, a sparsified subset of $W^{k}(I)$ where only lengths in $\tilde{\sL}^{k}(|I|)$ upper bounded by $|I|/\varepsilon$ are retained.

\begin{definition}
For an interval $I = \fragmentco{a}{b}$, we define the two sets:
\begin{align*}
    W^{k}(I)
    \coloneqq \{\ \fragmentco{a'}{b'} \mid |a'-a| \leq 2k, |b'-b| \leq 2k \ \}
    \ \text{ and } \
     \widetilde{W}^{k}(I)
    \coloneqq \{\ \tilde{I} \in W^{k}(I) \mid
    |\tilde{I}| \in \tilde{\sL}(|I|), |\tilde{I}| \leq |I|/\varepsilon \ \}.
\end{align*}
Moreover, for an interval collection $\mI = \langle I_1, \ldots, I_{\ell} \rangle$, we define:
\[
    W^{k}(\mI) \coloneqq \{\ \mJ = \langle J_1, \ldots, J_{\ell} \rangle \mid J_j \in W^{k}(I_{j}) \ \} \quad \text{ and } \quad
     \widetilde{W}^{k}(\mI) \coloneqq \{\ \mJ = \langle J_1, \ldots, J_{\ell} \rangle \mid J_j \in \widetilde{W}^{k}(I_{j}) \ \}. \qedhere
\]
\end{definition}

\begin{remark}\label{rmk:wki}
We make a few remarks regarding the definitions just introduced:
\begin{enumerate}[(i)]
    \item All $J \in W^{k}(I)$ satisfy $|J| \in \sL(|I|)$.
    \label{it:rmk:wki:i}
    \item For each interval $|I|$, we have $|W^{k}(I)| \leq \Oh(k^2)$. Thus, for each $v \in T(\bF)$, we have $|W^{k}(\mI_v)| \leq \Oh(k^4)$ because $\ell_v \leq 2$.
    \label{it:rmk:wki:ii}
    \item For each interval $|I|$, we have $|\widetilde{W}^{k}(I)| \leq \Oh(k\varepsilon^{-1} \log N)$ because a similar bound holds for $|\tilde{\sL}(|I|)|$. So, for each $v \in T(\bF)$, we have $|\widetilde{W}^{k}(\mI_v)| \leq \Oh(k^{\ell_v}\varepsilon^{-\ell_v} \log^{\ell_v} N)$.
    \label{it:rmk:wki:iii}\qedhere
\end{enumerate}
\end{remark}

Finally, we introduce notations that are more specific to the decomposition tree $T(\bF)$.

\begin{definition}\label{def:dyn_allowed_windows}
    Let $v \in T(\bF)$ be a non-root node that is the $i$-th child of its parent $w$.
    \begin{itemize}
        \item For an interval $J = \fragmentco{x}{y}$, we define
        \[
             W_{J}^{k}(I_{v,j})
             \coloneqq
             \begin{cases}
                 \{ \ \fragmentco{x}{b} \in  W^{k}(I_{v,j}) \mid b \leq y \ \} & \text{if }a_{v,j}=a_{w,\mathsf{pi}(i,j)}\\
                 \{ \ \fragmentco{a}{y} \in  W^{k}(I_{v,j}) \mid x \leq a \ \} & \text{else if } b_{v,j}=b_{w,\mathsf{pi}(i,j)}
             \end{cases},
        \]
        \item For $\mJ = \langle J_1, \ldots, J_{\ell_w} \rangle$, we define $W^{k}_{\mJ}(\mI_v) \coloneqq \{\  \langle J_1', \ldots, J_{\ell_v}' \rangle \in W^{k}(\mI_v) \mid J_j' \in W^{k}_{J_{\mathsf{pi}(i,j)}}(I_{v,j}) \ \}$.
    \item If $\mI_v$ is a bi-interval (i.e., $\ell_v = 2$), then for a spine node $z$ in $\bF_{\mI_v}$, we define:
    \begin{align*}
        \tilde{\sL}_{v,z}
        \coloneqq
        \big\{ \
        (q_1,q_1',q_2, q_2') \mid & \
        q_1 \in \tilde{\sL}(\Left(z)-a_{v,1}), q_1' \in \tilde{\sL}(b_{v,1} - \Left(z)), q_2 \in \tilde{\sL}(\Right(z)-a_{v,2}), q_2' \in \tilde{\sL}(b_{v,2} - \Right(z)), \\
        & \ q_1,q_1' \leq |I_{v,1}|/\varepsilon, q_2,q_2'  \leq |I_{v,2}|/\varepsilon, q_1', q_2' \geq 1
        \ \big\}
    \end{align*}
    allowing us to set for $q=(q_1,q_1',q_2,q_2') \in \tilde{\sL}_{v,z}$:
    \begin{align*}
         \widetilde{W}_{z,q}^{k}(\mI_v) &\coloneqq \{ \ \langle J_1, J_2 \rangle \in W^{k}(\mI_v) \mid
         z' \in \bG,
         \ J_1 = \fragmentco{\Left(z')-q_1}{\Left(z')+q_1'}, J_2 = \fragmentco{\Right(z')-q_2}{\Right(z')+q_2'}
         \ \}.
    \end{align*}
    Finally, $\widetilde{W}_{z}^{k}(\mI_v) \coloneqq \bigcup_{q \in \tilde{\sL}_{v,z}} \widetilde{W}_{z,q}^{k}(\mI_v)$. \qedhere
    \end{itemize}
\end{definition}

We observe that for each non-root node $v \in T(\bF)$ that is the $i$-th child of its parent $w$, the way we constructed $T(\bF)$ guarantees that for all $j \in \fragment{1}{\ell_v}$, either $a_{v,j}=a_{w,\mathsf{pi}(i,j)}$ or $b_{v,j}=b_{w,\mathsf{pi}(i,j)}$. Thus, $I_{v,j}$ is ``anchored'' to either a prefix or a suffix of $I_{w,\mathsf{pi}(i,j)}$. The subset $W^{k}_J(I_{v,j})$ contains all intervals that are contained within $J$ that preserve this same type of anchoring with respect to $J$. (The extension to interval collections is then $W^{k}_{\mJ}(\mI_v)$.)

The set $\widetilde{W}_z^{k}(\mI_v)$ is even a more sparsified version of $\widetilde{W}^k(\mI_v)$, but is defined strictly for bi-intervals $\mI_v$. The primary distinction lies in how the allowed lengths for each $|J_i|$ are determined. Rather than sparsifying the total length of $I_{v,i}$ directly, $I_{v,i}$ is split at the position of the spine node $z$. We then independently compute the sparsified sets of lengths $\tilde{\sL}(\cdot)$ for the resulting left and right sub-intervals. The allowed lengths for $|J_i|$ are formed by taking the pairwise sums of the lengths from these two sets. Finally, the exact positions of $J_1$ and $J_2$ are shifted such that the split points coincide with $\Left(z')$ and $\Right(z')$ for a node $z' \in \bG$. The condition $q_1', q_2' \geq 1$ places both occurrences of $z'$ inside $U(\mJ)$.

\subparagraph*{Approximation Oracles.}
In our dynamic algorithms, we would like to maintain, for each $v \in T(\bF)$ and $\mJ \in W^{k}(\mI_v)$, a value $t$ sandwiched between $\ted(\bF_{\mI_v}, \bG_{\mJ})$ and $\ted^{(k)}(\bF_{\mI_v}, \bG_{\mJ})$. In the approximate algorithm, however, this will not always be possible. Instead, we only need to maintain approximations to these values for a subset of $\mJ$, which still allows us to recover slightly weaker approximations for all $\mJ$. We formalize this concept through an \emph{approximation oracle}.

\begin{definition}\label{def:approx_oracle}
    For a node $v \in T(\bF)$, we say that an oracle $O_v$ is an \emph{$\alpha$-approximation oracle for node $v$} if it answers queries of the following form: given any $\mJ \in W^{k}(\mI_v)$, it returns a value $t$ satisfying
    \[
        \ted(\bF_{\mI_v}, \bG_{\mJ}) \leq t \leq \alpha \cdot \ted^{(k)}(\bF_{\mI_v}, \bG_{\mJ}).
    \]
    Moreover, an $\alpha$-approximation oracle for $v$ is called \emph{extended} if it also answers queries of the form: Given $i \in \fragment{1}{\ell_v}$ and $\mJ \in W^{k}(\langle I_{v,i} \rangle)$, it returns a value $t$ satisfying $\ted(\bF_{\langle I_{v,i} \rangle}, \bG_{\mJ}) \leq t \leq \alpha \cdot \ted^{(k)}(\bF_{\langle I_{v,i} \rangle}, \bG_{\mJ})$.
\end{definition}

So, simply put, an oracle is nothing other than an algorithm having access to some data stored locally at some node $v \in T(\bF)$ and that utilizes this data to answer the queries described above.

In the next lemma, we show that when aiming for an approximation, an oracle can indeed provide slightly weaker approximations for all $\mJ \in W^{k}(\mI_v)$ by only storing the values for a subset of $W^{k}(\mI_v)$. To provide a more concise proof and exposition, we must also introduce a distance measure between bi-intervals and the notation of \emph{good spine node samples}.

\begin{definition}
    For two interval collections $\mI = \langle I_1, \ldots, I_\ell \rangle$ and $\mJ = \langle J_1, \ldots, J_\ell \rangle$, we define their distance $\dist(\mI,\mJ)$ as $\sum_{i=1}^{\ell} |a_i - a_i'|$, where $I_i = \fragmentco{a_i}{b_i}$ and $J_i = \fragmentco{a_i'}{b_i'}$.
\end{definition}

\begin{definition}\label{def:good_spine_sample}
    For a node $v \in T(\bF)$ with $\ell_v = 2$, we say that a subset $Z_v$ of spine nodes in $\bF_{\mI_v}$ is a \emph{good spine node sample} if:
    \begin{enumerate}[(i)]
        \item $|Z_v| = \Oh(\varepsilon^{-1} \log N)$;
        \item \label{it:good_spine_sample:ii} For every $\mJ \in W^{k}(\mI_v)$, if there exists an optimal $k$-bounded alignment $\mA : \bF_{\mI_v} \onto \bG_{\mJ}$ that aligns more than an $\varepsilon$-fraction of spine nodes in $\bF_{\mI_v}$, then at least one such alignment $\mA$ aligns a node $z \in Z_v$. \qedhere
    \end{enumerate}
\end{definition}

In the remainder of this subsection and the next, we assume that each node $v$ has such a good spine node sample $Z_v$.

We remark that for any $\gamma$, a simple way to construct $Z_v$ with probability at least $1 - \Oh(k^4 N^{-\gamma})$ is to choose a uniformly random set of spine nodes from $\bF_{\mI_v}$ of size $|Z_v| = \lceil \gamma \varepsilon^{-1} \log N \rceil$ (or all spine nodes of $\bF_{\mI_v}$ if there are fewer). This is because if such an alignment $\mA$ exists for a given $\mJ \in W^{k}(\mI_v)$, then for this fixed alignment, the probability that none of the spine nodes aligned by $\mA$ belong to $Z_v$ is at most $(1 - \varepsilon)^{|Z_v|} \leq e^{-\varepsilon \cdot |Z_v|} \leq N^{-\gamma}$. Therefore, with probability at least $1 - N^{-\gamma}$, at least one spine node aligned by $\mA$ is present in $Z_v$. A union bound over all $\mJ \in W^{k}(\mI_v)$ yields the final bound. We also note that by choosing $\gamma$ to be a sufficiently large constant, we can afford to union bound over polynomially many more subforests of $\bG$.

\begin{lemma}\label{lem:round}
    For any constant $\gamma > 0$ and any node $v \in T(\bF)$, it is possible to construct an extended $\alpha \cdot (1+21\varepsilon)$-approximation oracle for $v$ with probability at least $1 - N^{-\gamma}$, by storing the following values:
    \begin{enumerate}[(a)]
        \item For each $i \in \fragment{1}{\ell_v}$, $\mJ \in \widetilde{W}^{k}(\langle I_{v,i} \rangle)$ a value $t_{v,i}(\mJ)$ s.t. $\ted(\bF_{\langle I_{v,i} \rangle}, \bG_{\mJ}) \leq t_{v,i}(\mJ) \leq \alpha \cdot \ted^{(k)}(\bF_{\langle I_{v,i} \rangle}, \bG_{\mJ})$;
        \item Whenever $\ell_v=2$, for a good spine node sample $Z_v$, and for each $z \in Z_v$ and $\mJ \in \widetilde{W}^{k}_z(\mI_v)$, a value $t_{v}(\mJ)$ s.t. $\ted(\bF_{\mI_v}, \bG_{\mJ}) \leq t_{v}(\mJ) \leq \alpha \cdot \ted^{(k)}(\bF_{\mI_v}, \bG_{\mJ})$.
    \end{enumerate}
    Assuming \cref{ln:approx_oracle_bi:nn} of \cref{alg:approx_oracle_bi} can run in time $\Oh(\sT)$, the oracle answers queries in time $\Oh(\sT \cdot \varepsilon^{-5} \log^5 N)$.
\end{lemma}

\begin{algorithm}[t!]
    \KwInput{A node $v \in T(\bF)$, an index $i \in \fragment{1}{\ell_v}$, and an interval collection $\mJ = \langle J \rangle \in W^{k}(I_{v,i})$.}
    \KwOutput{A value $t$.}
    Let $\mJ' = \langle J' \rangle$, where $J' \in \widetilde{W}^k(I_{v,i})$ has the same starting point as $J$ but is of length $\mathsf{Round}^k(|I_{v,i}|, |J|, |I_{v,i}|/\varepsilon)$\; \label{ln:approx_oracle_mono:Jp}
    \textbf{return} $t \coloneqq t_{v,i}(\mJ') + \max(3/2 \cdot (|J'|-|J|), 1/2 \cdot (|J|-|J'|))$\;
    \caption{An oracle $O_v$ for the mono-interval case.}
    \label{alg:approx_oracle_mono}
\end{algorithm}

\begin{proof}
    \Cref{alg:approx_oracle_mono} and \cref{alg:approx_oracle_bi} are the two implementations for the oracle for the cases where $v$ is a mono-interval and bi-interval, i.e. $\ell_v = 1$ and $\ell_v = 2$, respectively. Note that \cref{alg:approx_oracle_bi} internally uses \cref{alg:approx_oracle_mono} itself.

    \paragraph*{Correctness of \cref{alg:approx_oracle_mono}, i.e., the Mono-interval Case.}
    We start by showing the correctness of \cref{alg:approx_oracle_mono}.
    More specifically, we show that on input $v, i, \mJ = \langle J \rangle$, the value $t$ returned by \cref{alg:approx_oracle_mono} satisfies
    \begin{align}
        \ted(\bF_{\mI}, \bG_{\mJ}) \leq t \leq (1 + 8\varepsilon) \cdot \alpha \cdot \ted^{(k)}(\bF_{\mI}, \bG_{\mJ}), \label{eq:round:1}
    \end{align}
    where we abbreviate $\mI = \langle I_{v,i} \rangle$.

    To this end, we first show that the selected $\mJ'$ in \cref{ln:approx_oracle_mono:Jp} is valid. That is, for $\mJ' = \langle J' \rangle$, where $J' = \fragmentco{a'}{b'}$, we require $J' \in \widetilde{W}^k(I_{v,i})$. This means that we need to show that $|a'-a_{v,i}|, |b'-b_{v,i}| \leq 2k$, $b'-a' \in \tilde{\sL}(|I_{v,i}|)$ and $|J'| \leq |I_{v,i}|/\varepsilon$. To establish the first it is sufficient to use \cref{rmk:rounding_small_gap}\ref{it:rmk:rounding_small_gap:iii},
    for the second we use $b'-a' = \mathsf{Round}^k(|I_{v,i}|, |J|, |I_{v,i}|/\varepsilon) \in \tilde{\sL}(|I_{v,i}|)$, and for the third $\mathsf{Round}^k(q,q',h) \leq \max(q, \min(q',h)) \leq h$

    Regarding the length $|J'| = b'-a'$, note that either $|J| > |I_{v,i}|/\varepsilon$, or, by \cref{rmk:rounding_small_gap}\ref{it:rmk:rounding_small_gap:i}, we have $||J| - |J'|| \leq \varepsilon \cdot ||J| - |I_{v,i}||$.
    Moreover, if $|J| > |I_{v,i}|/\varepsilon$, then $J' \subseteq J$; otherwise, $J$ and $J'$ are nested according to their lengths: $J' \subseteq J$ if $|J| \geq |J'|$, and $J \subseteq J'$ otherwise.

    \subparagraph*{Lower Bound of \cref{eq:round:1}.} For the lower bound, applying \cref{prop:triangle_inequality} directly yields
    \begin{align*}
        \ted(\bF_{\mI}, \bG_{\mJ}) &\leq \ted(\bF_{\mI}, \bG_{\mJ'}) + 2 \cdot |J' \setminus J| + 1/2 \cdot (|J| - |J'|) \\
        &\leq t_{v,i}(\mJ') + \max(3/2 \cdot (|J'|-|J|), 1/2 \cdot (|J|-|J'|)) = t,
    \end{align*}
    where we use the fact that either $J \subseteq J'$ or $J' \subseteq J$, which implies $|J' \setminus J| \leq |J'| - |J|$ or $|J' \setminus J| = 0$, respectively.

    \subparagraph*{Upper Bound of \cref{eq:round:1}.}
    First, we consider the case $|J| \leq |I_{v,i}|/\varepsilon$.
    By \cref{prop:size_lb} we get
    \begin{align}\label{eq:round:mono:1}
        ||J|-|J'||
        \leq \varepsilon \cdot ||I_{v,i}| - |J||
        \leq \varepsilon 2 \cdot \ted^{(k)} (\bF_{\mI}, \bG_{\mJ}).
    \end{align}
    Moreover, using a similar argument to the lower bound:
    \begin{align}
        \ted^{(k)}(\bF_{\mI}, \bG_{\mJ'}) \leq \ted^{(k)}(\bF_{\mI}, \bG_{\mJ}) + \max(3/2 \cdot (|J|-|J'|), 1/2 \cdot (|J'|-|J|)). \label{eq:round:mono:2}
    \end{align}
    This means that we can prove the upper bound for $t$ as follows:
    \begin{align}
        \MoveEqLeft t = t_{v,i}(\mJ')  + \max(3/2 \cdot (|J'|-|J|), 1/2 \cdot (|J|-|J'|)) \nonumber \\
        &\leq \alpha \cdot \ted^{(k)}(\bF_{\mI}, \bG_{\mJ'}) + \max(3/2 \cdot (|J'|-|J|), 1/2 \cdot (|J|-|J'|)) \label{eq:round:mono:3}\\
        &\leq \alpha \cdot \ted^{(k)}(\bF_{\mI}, \bG_{\mJ}) +  2 \alpha \cdot ||J|-|J'|| \label{eq:round:mono:4} \\
        &\leq \alpha \cdot \ted^{(k)}(\bF_{\mI}, \bG_{\mJ}) + 4\alpha \varepsilon \cdot \ted^{(k)} (\bF_{\mI}, \bG_{\mJ})
        \leq \alpha (1 + 4\varepsilon) \cdot \ted^{(k)}(\bF_{\mI}, \bG_{\mJ}).\label{eq:round:mono:5}
    \end{align}
    where \cref{eq:round:mono:3} is due to the upper bound on $t_{v,i}(\mJ')$; \cref{eq:round:mono:4} is due to \cref{eq:round:mono:2}, $\alpha \geq 1$, and a case distinction whether $|J|\geq |J'|$ or not; \cref{eq:round:mono:5} is due to \cref{eq:round:mono:1}.

    Now, suppose $|J| > |I_{v,i}|/\varepsilon$. From $\varepsilon < 1/2$ follows
    \begin{align}\label{eq:round:mono:6}
        |I_{v,i}| \leq 2|I_{v,i}| - 2 \varepsilon |I_{v,i}| \leq 2\varepsilon \cdot ||I_{v,i}| - |J|| \leq 4\varepsilon \cdot \ted^{(k)} (\bF_{\mI}, \bG_{\mJ}).
    \end{align}
     This means that we can prove the upper bound $t$ as follows:
    \begin{align}
        t_{v,i}(\mJ')  + 1/2 \cdot (|J|-|J'|)
        &\leq \alpha \cdot \ted^{(k)}(\bF_{\mI}, \bG_{\mJ}) + \alpha/2 \cdot (|J'|-|J|) + 2\alpha \cdot |I_{v,i}| +  1/2 \cdot (|J|-|J'|) \label{eq:round:mono:6b} \\
        &\leq \alpha \cdot \ted^{(k)}(\bF_{\mI}, \bG_{\mJ}) + \alpha 8\varepsilon \cdot \ted^{(k)} (\bF_{\mI}, \bG_{\mJ})
        \leq \alpha (1 + 8\varepsilon) \cdot \ted^{(k)}(\bF_{\mI}, \bG_{\mJ}) \label{eq:round:mono:7},
    \end{align}
    where \cref{eq:round:mono:6b} is due to \cref{prop:triangle_inequality}, $|J| \geq |J'|$; \cref{eq:round:mono:7} is due to \cref{eq:round:mono:6}, $\alpha \geq 1$.

    \begin{algorithm}[t!]
        \KwInput{A node $v \in T(\bF)$ such that $\ell_v=2$ and interval collection $\mJ = \langle J_1, J_2 \rangle \in W^{k}(\mI_v)$.}
        \KwOutput{A value $t$.}
        Let $t_1,t_2$ be the result of \cref{alg:approx_oracle_mono} on inputs ($v$, $1$ , $\langle J_1 \rangle$) and ($v$, $2$ , $\langle J_2 \rangle$), respectively\;  \label{alg:approx_oracle_bi:decouple}
        $t \leftarrow t_1 + t_2$\;
        \For{$z \in Z_v$}{ \label{alg:approx_oracle_bi:for_z}
            \For{$q = (q_1,q_1',q_2,q_2') \in \tilde{\sL}_{v,z}$}{ \label{alg:approx_oracle_bi:for_q}
                Set $\mJ' = \langle J_1', J_2' \rangle$, where $J_i' \coloneqq \fragmentco{a_i'}{a_i'+q_i+q_i'}$ for $J_i = \fragmentco{a_i'}{b_i'}$\; \label{ln:approx_oracle_bi:jp}
                Find $ \mJ'' = \langle J_1'', J_2'' \rangle \in \widetilde{W}^k_{z,q}(\mI_v)$ such that $\mathsf{dist}(\mJ', \mJ'') \leq (1+\varepsilon) \cdot \mathsf{dist}(\mJ', \mJ^*)$, where $\mJ^* \coloneqq \arg \min_{\mJ \in \widetilde{W}^k_{z,q}(\mI_v)} \mathsf{dist}(\mJ', \mJ)$\; \label{ln:approx_oracle_bi:nn}
                \If{such $\mJ''$ exists}{
                    $t \leftarrow \min \left\{ \ t \ , \ t_{v}(\mJ'') + 2 \cdot \mathsf{dist}(\mJ',\mJ'') + 5/2 \cdot \sum_{i=1}^{2} ||J_i| - |J_i'|| \ \right\}$\; \label{ln:approx_oracle_bi:return}
                }
            }
        }
        \textbf{return} $t$\;
        \caption{An oracle $O_v$ for the bi-interval case.}
        \label{alg:approx_oracle_bi}
    \end{algorithm}

    \paragraph*{Correctness of \cref{alg:approx_oracle_bi}, i.e., the Bi-interval Case.}
    We show that on input $v$ such that $\ell_v=2$ and $\mJ \in W^{k}(\mI_v)$, the value $t$ returned by \cref{alg:approx_oracle_bi} satisfies
    \begin{align}
        \ted(\bF_{\mI_v}, \bG_{\mJ}) \leq t \leq (1 + 21\varepsilon) \cdot \alpha \cdot \ted^{(k)}(\bF_{\mI_v}, \bG_{\mJ}). \label{eq:round:2}
    \end{align}

    As a sanity check of \cref{alg:approx_oracle_bi}, notice that in \cref{alg:approx_oracle_bi:decouple} we have that $\mJ \in W^{k}(\mI_v)$ implies $J_1 \in W^k(I_{v,1})$,  $J_2 \in W^k(I_{v,2})$,
    so the calls to \cref{alg:approx_oracle_mono} are valid.

    \subparagraph*{Lower Bound of \cref{eq:round:2}.}
     For the lower bound, by \cref{prop:ted_decomp} and \cref{eq:round:1}, we observe that $\ted(\bF_{\mI_v}, \bG_{\mJ}) \leq \ted(\bF_{\langle I_{v,1} \rangle}, \bG_{\langle J_1 \rangle}) + \ted(\bF_{\langle I_{v,2} \rangle}, \bG_{\langle J_2 \rangle}) \leq t_1+t_2$. Moreover, for any bi-intervals $\mJ'$ and $\mJ''$ such that \cref{ln:approx_oracle_bi:return} is executed, we can apply \cref{prop:triangle_inequality} twice to obtain
    \begin{align*}
        \ted(\bF_{\mI_v}, \bG_{\mJ}) &\leq \ted(\bF_{\mI_v}, \bG_{\mJ'}) + 5/2 \cdot \textstyle \sum_{i=1}^{2} ||J_i|-|J_i'|| \\
        &\leq \ted(\bF_{\mI_v}, \bG_{\mJ''}) + 5/2 \cdot \textstyle \sum_{i=1}^{2} ||J_i|-|J_i'|| + 2 \cdot \dist(\mJ',\mJ''),
    \end{align*}
    where, in the second application, we use the fact that $\mJ'$ and $\mJ''$ have corresponding intervals of the same length. (This allows us to use the bound relevant to the shift version in \cref{prop:triangle_inequality}.)
    This covers all possible initializations of $t$.

    \subparagraph*{Upper Bound of \cref{eq:round:2}.}
    Let $m_v$ be the number of spine nodes in $\bF_{\mI_v}$.
    If there exists an optimal $k$-bounded alignment $\mA : \bF_{\mI_v} \onto \bG_{\mJ}$ that aligns more than $\varepsilon m_v$ spine nodes in $\bF_{\mI_v}$, then we fix one such alignment $\mA$ that aligns a node $z \in Z_v$. Otherwise, we fix an arbitrary optimal $k$-bounded alignment.

    \begin{claim}\label{clm:round:1}
        If $\mA$ aligns no more than $\varepsilon m_v$ spine nodes, then $t \leq \alpha(1+13\varepsilon) \cdot \ted^{(k)}(\bF_{\mI_v}, \bG_{\mJ})$.
    \end{claim}
    \begin{claimproof}
        Since $\mA$ aligns no more than $\varepsilon m_v$ spine nodes, it must delete at least $m_v - \floor{\varepsilon m_v} = \ceil{(1-\varepsilon) m_v}$ spine nodes, and has therefore cost at least $\cost(\mA) \geq \ceil{(1-\varepsilon) m_v} \geq (1-\varepsilon)m_v$, which, when rearranged, yields $m_v \leq \cost(\mA)/(1-\varepsilon) \leq 2 \cost(\mA)$, because $\varepsilon \leq 1/2$.
        Therefore, the alignment $\mA'$, obtained from $\mA$ by removing all the aligned spine nodes, has a cost of $\cost(\mA') \leq \cost(\mA) + 2 \varepsilon m_v \leq (1 + 4\varepsilon) \cdot \cost(\mA)$.
        Since $\mA'$ does not align any spine nodes, we can decompose it into two alignments $\mA' = \mA_1' \cup \mA_2'$, where $\mA_1' : \bF_{\langle I_{v,1} \rangle} \onto \bG_{\langle J_1 \rangle}$ and $\mA_2' : \bF_{\langle I_{v,2} \rangle} \onto \bG_{\langle J_2 \rangle}$, such that $\cost(\mA_1') + \cost(\mA_2') = \cost(\mA')$. By \cref{eq:round:1}, we find that the returned value $t$ satisfies:
        \begin{align*}
            t
            &\leq t_1 + t_2
            \leq \alpha(1+8\varepsilon) \cdot (\ted^{(k)}(\bF_{\langle I_{v,1} \rangle}, \bG_{\langle J_1 \rangle}) + \ted^{(k)}(\bF_{\langle I_{v,2} \rangle}, \bG_{\langle J_2 \rangle}))
            \leq \alpha(1+8\varepsilon) \cdot (\cost(\mA_1') + \cost(\mA_2'))\\
            &= \alpha(1+8\varepsilon) \cdot \cost(\mA')
            \leq \alpha(1+8\varepsilon)(1+4\varepsilon) \cdot \cost(\mA)
            \leq \alpha(1+13\varepsilon) \cdot \ted^{(k)}(\bF_{\mI_v}, \bG_{\mJ}),
        \end{align*}
        where in the last step we use $\varepsilon \leq 1/100$.
    \end{claimproof}

    \begin{claim}\label{clm:round:2}
        If for some $i \in \fragment{1}{\ell_v}$, we have $|J_i| \geq |I_{v,i}|/\varepsilon$, then $t \leq \alpha(1+17\varepsilon) \cdot \ted^{(k)}(\bF_{\mI_v}, \bG_{\mJ})$.
    \end{claim}
    \begin{claimproof}
        Let $R$ be the set of all $i \in \fragment{1}{\ell_v}$ such that $|J_i| \geq |I_{v,i}|/\varepsilon$.
        We consider the alignment $\mA'$, obtained from $\mA$ by removing $(x,y)$ and $(\Other(x), \Other(y))$ for all $(x,y) \in \mA$ such that $x \in I_{v,i}$ for some $i \in R$.
        Just as in \cref{eq:round:mono:6}, we have $\sum_{i \in R} |I_{v,i}| \leq 2\varepsilon \sum_{i \in R} (|J_i| - |I_{v,i}|) \leq 4\varepsilon \cdot \ted^{(k)}(\bF_{\mI_v}, \bG_{\mJ})$ by \cref{prop:size_lb} (the per-$i$ bound alone gives only $16\varepsilon$ below when $|R| = 2$).
        This allows us to bound the cost as $\cost(\mA') \leq \cost(\mA) + 2 \cdot \sum_{i \in R} |I_{v,i}| \leq \cost(\mA) + 8\varepsilon \cdot \ted^{(k)} (\bF_{\mI_v}, \bG_{\mJ})$.
        Moreover, just as in \cref{clm:round:1}, we can decompose $\mA' = \mA_1' \cup \mA_2'$ with $\mA_1' : \bF_{\langle I_{v,1} \rangle} \onto \bG_{\langle J_1 \rangle}$ and $\mA_2' : \bF_{\langle I_{v,2} \rangle} \onto \bG_{\langle J_2 \rangle}$ such that $\cost(\mA_1') + \cost(\mA_2') = \cost(\mA')$ (as a side note, $\mA_i'=\emptyset$ for all $i \in R$).
        Using the same chain of inequalities as in \cref{clm:round:1}, we obtain $t \leq \alpha(1+8\varepsilon)(1+8\varepsilon) \cdot \cost(\mA) \leq \alpha(1+17\varepsilon) \cdot \cost(\mA) \leq \alpha(1+17\varepsilon) \cdot \ted^{(k)}(\bF_{\mI_v}, \bG_{\mJ})$.
    \end{claimproof}

    \begin{claim}
        If $\mA$ aligns more than $\varepsilon m_v$ spine nodes and for all $i \in \fragment{1}{\ell_v}$ we have $|J_i| \leq |I_{v,i}|/\varepsilon$, then $t \leq \alpha(1+21\varepsilon) \cdot \ted^{(k)}(\bF_{\mI_v}, \bG_{\mJ})$.
    \end{claim}
    \begin{claimproof}
        Let $J_i = \fragmentco{a_i'}{b_i'}$ for $i \in \fragment{1}{\ell_v}$.

        Let $z \in Z_v$ be a spine node in $\bF_{\mI_v}$ and $z '\in \bG$ such that $\mA$ aligns $z$ to $z'$. Note that because $\mA$ is $k$-bounded, it follows that $|\Left(z)-\Left(z')|, |\Right(z)-\Right(z')| \leq 2k$. Henceforth, we consider the iteration of the loop in \cref{alg:approx_oracle_bi:for_z} corresponding to this $z$.

        Since $\fragmentco{a_1'}{\Left(z')} \subseteq J_1$ and $|J_1| \leq |I_{v,1}|/\varepsilon$, we have $|\Left(z') - a_1'| \leq |I_{v,1}|/\varepsilon$; symmetrically, $|b_1' - \Left(z')| \leq |I_{v,1}|/\varepsilon$ and $|\Right(z') - a_2'|, |b_2' - \Right(z')| \leq |I_{v,2}|/\varepsilon$, so that in each of the four calls to $\mathsf{Round}^k$ below, the second argument is at most the third. Moreover, the first two arguments differ by at most $4k$, as $|\Left(z) - \Left(z')|, |\Right(z) - \Right(z')| \leq 2k$ and $|a_{v,i} - a_i'|, |b_{v,i} - b_i'| \leq 2k$, so the second argument lies in $\sL^{k}$ of the first, as required in \cref{def:dyn_allowed_length}. Thus, by \cref{rmk:rounding_small_gap}\ref{it:rmk:rounding_small_gap:i}, the rounded length $q_1 \coloneqq \mathsf{Round}^{k}(\Left(z) - a_{v,1}, \Left(z') - a_{1}', |I_{v,1}|/\varepsilon)$ satisfies $|\Left(z') - a_{1}' - q_1| \leq \varepsilon |\Left(z') - a_{1}' - (\Left(z) - a_{v,1})|$. For the sake of conciseness, we define the abbreviations $\eta_1 \coloneqq |\Left(z') - a_{1}' - q_1|$ and $\mu_1 \coloneqq |\Left(z') - a_{1}' - (\Left(z) - a_{v,1})|$.

        Similarly, for the rounded lengths $q_1',q_2,q_2'$ and shorthands $\eta_1',\eta_2,\eta_2'$ and $\mu_1',\mu_2,\mu_2'$ defined as
        \begin{align*}
            &q_1' = \mathsf{Round}^{k}(b_{v,1}-\Left(z), b_{1}' - \Left(z'), |I_{v,1}|/\varepsilon), \\
            & q_2 = \mathsf{Round}^{k}(\Right(z)-a_{v,2}, \Right(z') - a_{2}', |I_{v,2}|/\varepsilon), \\
            & q_2' = \mathsf{Round}^{k}(b_{v,2} - \Right(z), b_{2}' - \Right(z'), |I_{v,2}|/\varepsilon), \\
            & \eta_1' \coloneqq |b_1' - \Left(z') - q_1'|,
            && \mu_1' \coloneqq |b_1' - \Left(z') - b_{v,1} +\Left(z)|, \\
            & \eta_2 \coloneqq |\Right(z') - a_{2}' - q_2|,
            && \mu_2 \coloneqq |\Right(z') - a_{2}' - \Right(z)+a_{v,2}|, \\
            & \eta_2' \coloneqq |b_2' - \Right(z') - q_2'|,
            && \mu_2' \coloneqq |b_2' - \Right(z') - b_{v,2} + \Right(z)|,
        \end{align*}
        we have $\eta_1 \leq \varepsilon \mu_1$, $\eta_1' \leq \varepsilon \mu_1'$, $\eta_2 \leq \varepsilon \mu_2$ and $\eta_2' \leq \varepsilon \mu_2'$, respectively. Moreover, $q_1', q_2' \geq 1$, as $\mathsf{Round}^{k}(q,q',h) \geq \min(q,q')$ and $b_1' - \Left(z'), b_2' - \Right(z') \geq 1$ because $\Left(z') \in J_1$ and $\Right(z') \in J_2$; hence $q \in \tilde{\sL}_{v,z}$.
        Abbreviate also $\mu \coloneqq \sum_{i=1}^{2} \mu_i + \mu_{i}'$ and $\eta \coloneqq \sum_{i=1}^{2} \eta_i + \eta_{i}'$.
        Henceforth, we consider the iteration of the loop at \cref{alg:approx_oracle_bi:for_z} for such $q_1,q_1',q_2,q_2'$.

        We continue by analyzing how the aforementioned abbreviations play a role in bounding the penalty term returned alongside $t_{v}(\mJ'')$ in \cref{ln:approx_oracle_bi:return} of \cref{alg:approx_oracle_bi}. First, we observe that $||J_1|-|J_1'|| = |b_{1}'-a_{1}'-q_1-q_1'| \leq |(b_{1}'-\Left(z')-q_1')-(a_{1}' -\Left(z')+q_1)| \leq \eta_1 + \eta_1'$, so by applying a similar bound for $||J_2|-|J_2'|| \leq \eta_2 + \eta_2'$, we obtain
        \begin{align}
            \textstyle \sum_{i=1}^{2} ||J_i|-|J_i'|| \leq \eta.
            \label{eq:round:bi:1}
        \end{align}

        To bound $\dist(\mJ',\mJ'')$, we consider $\mL = \langle L_1, L_2 \rangle$, where $L_1 \coloneqq \fragmentco{\Left(z') - q_1}{\Left(z') + q_1'}$ and $L_2 \coloneqq \fragmentco{\Right(z') - q_2}{\Right(z') + q_2'}$.

        We first bound $\dist(\mJ', \mL)$. To this end, we observe that $J_i' = \fragmentco{a_i'}{a_i' + q_i+q_i'}$ for $i \in \fragment{1}{\ell_v}$. Thus, $\dist(\mJ', \mL) = |\Left(z')-q_1-a_1'| + |\Right(z')-q_2-a_2'|= \eta_1 + \eta_2 \leq \eta$.

        Our next goal is to prove that
        \begin{align}
            \dist(\mJ', \mJ'') \leq (1+\varepsilon) \cdot \dist(\mJ', \mJ^*) \leq (1+\varepsilon) \cdot \dist(\mJ', \mL) \leq (1+\varepsilon) \cdot \eta.
            \label{eq:round:bi:11}
        \end{align}
        To do so, it suffices to show that $\mL \in \widetilde{W}^{k}_{z,q}(\mI_v)$; that is, $L_1 = \fragmentco{\Left(\hat{z}) - q_1}{\Left(\hat{z}) + q_1'}$ and $L_2 = \fragmentco{\Right(\hat{z}) - q_2}{\Right(\hat{z}) + q_2'}$ for some $\hat{z} \in \bG$, and $\mL \in W^{k}(\mI_v)$. While the first condition follows from choosing $\hat{z}=z'$, for the second we recall that $|\Left(z)-\Left(z')|, |\Right(z)-\Right(z')|, |a_{v,1}-a_1'|, |b_{v,1}-b_1'|, |a_{v,2}-a_2'|, |b_{v,2}-b_2'| \leq 2k$ all hold. This allows us to apply \cref{rmk:rounding_small_gap}\ref{it:rmk:rounding_small_gap:iii} and \cref{rmk:rounding_small_gap}\ref{it:rmk:rounding_small_gap:iv} to demonstrate that $|\Left(z') + q_1'-b_{v,1}|, |\Right(z') + q_2'-b_{v,2}| \leq 2k$ and $|\Left(z') - q_1-a_{v,1}|, |\Right(z') - q_2 - a_{v,2}| \leq 2k$, respectively.

        This finally brings us to
        \begin{align*}
            \dist(\mL,\mJ'') \leq \dist(\mL, \mJ') + \dist(\mJ', \mJ'') \leq (2+\varepsilon) \dist(\mJ', \mL) \leq (2+\varepsilon)\eta,
        \end{align*}
        which we can combine with \cref{prop:triangle_inequality} to get
        \begin{align}
             \ted^{(k)}(\bF_{\mI_v},\bG_{\mJ''}) \leq \ted^{(k)}(\bF_{\mI_v},\bG_{\mL}) + 2 \cdot \mathsf{dist}(\mL,\mJ'') \leq \ted^{(k)}(\bF_{\mI_v},\bG_{\mL}) + (4+2\varepsilon) \eta,
             \label{eq:round:bi:3}
        \end{align}
        In particular, the returned value $t$ satisfies
        \begin{align*}
            \MoveEqLeft t_{v}(\mJ'') + 2\mathsf{dist}(\mJ',\mJ'') + 5/2 \cdot \textstyle \sum_{i=1}^{2} ||J_i|-|J_i'|| \\
            &\leq \alpha \ted^{(k)}(\bF_{\mI_v},\bG_{\mJ''}) + 2(1+\varepsilon) \eta + 5/2 \cdot \eta \\
            & \leq \alpha \ted^{(k)}(\bF_{\mI_v},\bG_{\mL}) + \alpha(4+2\varepsilon)\eta + 2(1+\varepsilon) \eta + 5/2 \cdot \eta, \\
            & \leq \alpha \ted^{(k)}(\bF_{\mI_v},\bG_{\mJ}) + \alpha \cdot 3/2 \cdot \eta + \alpha(4+2\varepsilon)\eta + 2(1+\varepsilon) \eta + 5/2 \cdot \eta \\
            &\leq \alpha \ted^{(k)}(\bF_{\mI_v},\bG_{\mJ}) + \alpha \cdot 21/2 \cdot\eta
        \end{align*}
        where we first use the guarantees on $t_{v}(\mJ'')$,
        \cref{eq:round:bi:1}, and \cref{eq:round:bi:11}; then \cref{eq:round:bi:3}; then \cref{prop:triangle_inequality} for $\mL$ and $\mJ$, whose corresponding endpoints differ by $\eta_i$ and $\eta_i'$, so that $|L_i \triangle J_i| \leq \eta_i + \eta_i'$ and $\ted^{(k)}(\bF_{\mI_v},\bG_{\mL}) \leq \ted^{(k)}(\bF_{\mI_v},\bG_{\mJ}) + \frac{1}{2}(|\mL| - |\mJ|) + 2\sum_{i=1}^{2} |J_i \setminus L_i| \leq \ted^{(k)}(\bF_{\mI_v},\bG_{\mJ}) + \frac{3}{2}\eta$; and finally $\alpha \geq 1$ and we loosely bound $\varepsilon \leq 1/8$, so that $\alpha(10 + 4\varepsilon) \leq 21\alpha/2$.

        It remains to lower bound $\ted^{(k)}(\bF_{\mI_v},\bG_{\mJ})$ by $\eta$.
        Since $\mA$ is non-crossing on $I_{v,1} \times J_1$ and contains $(\Left(z), \Left(z'))$, a pair $(x,y) \in \mA \cap (I_{v,1} \times J_1)$ satisfies $x < \Left(z)$ if and only if $y < \Left(z')$; hence at most $\min(\Left(z) - a_{v,1}, \Left(z') - a_1')$ pairs lie in $\fragmentco{a_{v,1}}{\Left(z)} \times \fragmentco{a_1'}{\Left(z')}$, so at least $\mu_1$ positions of these two pieces are deleted, each costing $1/2$. The same holds for the pieces to the right of $\Left(z)$ and $\Left(z')$, and for the two pieces of $I_{v,2}$ and $J_2$ around $\Right(z)$ and $\Right(z')$; the four pieces are pairwise disjoint on each side, so:
        \begin{align}
            \cost(\mA) \geq 1/2 \cdot \textstyle \sum_{i=1}^{2} \mu_i + \mu_{i}' = \mu/2 \geq \varepsilon^{-1} \eta / 2.
        \end{align}
        We conclude $t \leq \alpha \ted^{(k)}(\bF_{\mI_v},\bG_{\mJ}) + 21/2 \cdot \alpha \eta \leq \alpha (1 + 21\varepsilon) \ted^{(k)}(\bF_{\mI_v},\bG_{\mJ})$.
    \end{claimproof}

    This concludes the proof of \cref{lem:round}.
\end{proof}

In \cref{sec:omitted} we give a dynamic algorithm for the query operation required by \cref{alg:approx_oracle_bi}.

\begin{restatable*}{lemma}{findNN}\label{lem:findNN}
Let $\bF$ and $\bG$ be forests that undergo updates. There exists a data structure that supports the following queries: Given $k$, $v \in T(\bF)$ such that $\ell_v = 2$, $z \in Z_v$ and $q=(q_1,q_1',q_2,q_2') \in \tilde{\sL}_{v,z}$, and a query bi-interval $\mJ' = \langle J_1', J_2' \rangle$ with $|J_i'|=q_i+q_i'$, output (or report that it does not exist) a bi-interval
\[
    \mJ'' = \langle J_1'', J_2'' \rangle \in \widetilde{W}^k_{z,q}(\mI_v) \text{ such that } \mathsf{dist}(\mJ', \mJ'') \leq (1+\varepsilon) \cdot \mathsf{dist}(\mJ', \mJ^*),
\]
where $\mJ^* \coloneqq \arg \min_{\mJ \in \widetilde{W}^k_{z,q}(\mI_v)} \mathsf{dist}(\mJ', \mJ)$. Updates and queries take time $\Oh(\log^2 N)$ and $\Oh(\varepsilon^{-1} \log^3 N + \varepsilon^{-2} \log^2 N)$.
\end{restatable*}

In the next lemma, we show how to obtain such a value $t$ by calling the approximation oracles of the children of $v$, rather than querying the oracle of node $v$ directly. We can think of this as recomputing a value that the oracle for $v$ uses to answer queries itself, which can be derived from the oracles of its children.
We present two methods of recomputation: a slow one that preserves the approximation, and a faster one that incurs an additional error.

\begin{lemma}\label{lem:combine}
    Suppose $v \in T(\bF)$ is a node with $B$ children, $v_1, \ldots, v_B$, and assume that for each child $v_i$ we have access to an $\alpha$-approximation oracle answering queries in time $\sT$. For $j \in \fragment{1}{\ell_v}$, let $C_j$ denote the set of all pairs $(i,\hj)$ with $i \in \fragment{1}{B}$ and $\hj \in \fragment{1}{\ell_{v_i}}$ such that $\mathsf{pi}(i,\hj)=j$. Furthermore, if $|C_j|=2$, let $C_j'$ contain $(i,\hj) \in C_j$; such that $I_{v_i,\hj}$ is a prefix of $I_{v,j}$, otherwise let $C_j' = \emptyset$. Then:
    \begin{enumerate}[(i)]
        \item \cref{alg:dyn_combine} computes in time $\Oh(\sT k^2)$ a value $t$ such that $\ted(\bF_{\mI_v}, \bG_{\mJ}) \leq t \leq \alpha \cdot \ted^{(k)}(\bF_{\mI_v}, \bG_{\mJ})$;
        \label{it:combine:i}
        \item Moreover, if we substitute $\sL$ with $\tilde{\sL}$ at \cref{line:combine}, then \cref{alg:dyn_combine} computes in time $\Oh(\sT \varepsilon^{-2} \log^2 N)$ a value $t$ such that $\ted(\bF_{\mI_v}, \bG_{\mJ}) \leq t \leq \alpha(1+20\varepsilon) \cdot \ted^{(k)}(\bF_{\mI_v}, \bG_{\mJ})$.
        \label{it:combine:ii}
    \end{enumerate}
    Moreover, if $\ell_v=2$ and all the oracles of the children are extended:
    \begin{enumerate}[(i), resume]
        \item \cref{alg:combine:mono} computes in time $\Oh(\sT \varepsilon^{-1} \log N)$ a value $t$ that satisfies $\ted(\bF_{\langle I_{v,j} \rangle}, \bG_{\mJ}) \leq t \leq \alpha(1+20\varepsilon) \cdot \ted^{(k)}(\bF_{\langle I_{v,j} \rangle},\bG_{\mJ})$.
        \label{it:combine:iii}
    \end{enumerate}
\end{lemma}

\begin{algorithm}[t!]
    \KwInput{An interval collection $\mJ \in W^{k}(\mI_v)$.}
    \KwOutput{A value $t$ that satisfies $\ted(\bF_{\mI_v}, \bG_{\mJ}) \leq t \leq \alpha \cdot \ted^{(k)}(\bF_{\mI_v}, \bG_{\mJ})$.}
    $t \leftarrow \infty$\;
    \For(\nllabel{line:combine}){$\{\mJ_{v_i}\}_i$ partition of $\mJ$ s.t. $\mJ_{v_i} = \langle J_{v_i,1}, \ldots, J_{v_i,\ell_{v_i}} \rangle \in W_{\mJ}^{k}(\mI_{v_i})$ for $i \in \fragment{1}{B}$ and $|J_{v_i,\hj}| \in \sL(|I_{v_i,\hj}|) \cup \{|J_j|\}$ for $(i,\hj) \in \bigcup_j C_j'$}
    {
        For all $i \in \fragment{1}{B}$ let $t_i$ be obtained by querying the oracle of $v_i$ on $\mI_{v_i}$ and $\mJ_{v_i}$\;
        $t \leftarrow \min(t, \sum_{i=1}^{B} t_i)$\;
    }
    \textbf{return} $t$\;
    \caption{Computing $t$ via $\alpha$-approximation oracles at the children of $v$.}
     \label{alg:dyn_combine}
\end{algorithm}

\begin{algorithm}[t!]
    \KwInput{An index $j \in \fragment{1}{\ell_v}$ and an interval collection $\mJ  = \langle J \rangle \in W^{k}(\langle I_{v,j} \rangle)$.}
    \KwOutput{A value $t$ that satisfies $\ted(\bF_{\langle I_{v,j} \rangle}, \bG_{\mJ}) \leq t \leq \alpha \cdot \ted^{(k)}(\bF_{\langle I_{v,j} \rangle}, \bG_{\mJ})$.}
    $t \leftarrow \infty$\;
    \For{$\{J_{v_i,\hj}\}_{(i,\hj) \in C_j}$ partition of $J$ s.t. $J_{v_i,\hj} \in W_{J}^{k}(I_{v_i,\hj})$ for $(i,\hj) \in C_j$ and $|J_{v_i,\hj} \in \tilde{\sL}(|I_{v_i,\hj}|) \cup \{|J_j|\}$ for $(i,\hj) \in C_j'$}
    { \label{line:combine:iter}
        For each $(i,\hj) \in C_j$, let $t_{(i,\hj)}$ be the value obtained by querying the extended oracle of $v_i$ on $\bF_{\langle I_{v_i, \hj} \rangle}$ and $\bG_{\langle J_{v_i,\hj} \rangle}$ if $\ell_{v_i} = 2$, and by querying the normal oracle otherwise\;
        $t \leftarrow \min(t, \sum_{(i,\hj) \in C_j} t_{(i,\hj)})$\;
    }
    \textbf{return} $t$\;
    \caption{Computing $t$ via extended $\alpha$-approximation oracles at the children of $v$.}
    \label{alg:combine:mono}
\end{algorithm}

\begin{proof}
    We first provide a better understanding of \cref{alg:dyn_combine}, which will also clarify its runtime.

    The central point to be understood here is that each  $\{\mJ_{v_i}\}_i$ that forms a partition of $\mJ = \langle \fragmentco{a_{1}'}{b_{1}'}, \ldots, \fragmentco{a_{\ell_v}'}{b_{\ell_v}'} \rangle$ with $\mJ_{v_i} = \langle J_{v_i, 1}, \ldots, J_{v_i, \ell_{v_i}} \rangle \in W^{k}_{\mJ}(\mI_{v_i})$
    can be identified with $|\{C_j \mid |C_j|=2 , j \in \fragment{1}{\ell_v}\}|$ many positions. Therefore, in \cref{line:combine}, we also quantify over at most two positions.

    To see this, consider any $j \in \fragment{1}{\ell_v}$. If $|C_j| = 1$ (and therefore $|C_j'|=0$), then for the unique $(i,\hj) \in C_j$ we have $J_{v_i,\hj} = \fragmentco{a_{j}'}{b_{j}'}$.
    Otherwise, if $|C_j| = 2$ (and therefore $|C_j'|=1$), there are exactly two distinct pairs, $(i,\hj) \in C_j'$ and $(i',\hj') \in C_j \setminus C_j'$, such that $I_{v_i, \hj}$ and $I_{v_{i'},\hj'}$ are a prefix and a suffix of $I_{v,j}$,
     respectively. In this case the valid choices are exactly
    $J_{v_i, \hj} = \fragmentco{a_{j}'}{x}$ and $J_{v_{i'}, \hj'} = \fragmentco{x}{b_{j}'}$ with
    \[
        x \in \fragment{a_{j}'}{b_{j}'} \cap \fragment{b_{v_i,\hj}-2k}{b_{v_i,\hj}+2k},
    \]
    where $b_{v_i,\hj} = a_{v_{i'},\hj'}$ since $I_{v_i,\hj}$ and $I_{v_{i'},\hj'}$ partition $I_{v,j}$.
    (Notice at this point that the condition $|J_{v_i,\hj}| \in \sL(|I_{v_i,\hj}|)$ is superfluous, as it is already encoded in $\mJ_{v_i} \in W^{k}_{\mJ}(\mI_{v_i})$.)
    Thus, for any $j \in \fragment{1}{\ell_v}$, we either do not quantify at all, or we quantify over at most $4k+1$ positions (which already explains the runtime for \ref{it:combine:i}).

    For \ref{it:combine:ii}, we impose the additional restriction that $|J_{v_i,\hj}| \in \tilde{\sL}(|I_{v_i,\hj}|) \cup \{|J_j|\}$, which is no longer a superfluous condition. This limits the possible lengths, and thus the possible positions of $x$, to $\Oh(\varepsilon^{-1} \log N)$.

    Finally, \cref{alg:combine:mono} is a restriction of \ref{it:combine:ii} to only the $j$-th part of the bi-interval, meaning we are quantifying over at most one $x$ across $\Oh(\varepsilon^{-1} \log N)$ positions.
    Notice that when retrieving values from the oracles of the children for $(i,\hj) \in C_j$, the interval $I_{i,\hj}$ might not be a full mono-interval, but rather just one side of a bi-interval (e.g., if $\ell_{v_{i}} = 2$). In such cases, we need to call the extended oracles.\footnote{As a side note, observe that for all $(i,\hj) \in C_j$, the indices $i$ are distinct; a quick look at \cref{fig:rel} confirms this is the case for $\ell_v=2$.}

    We now move on to prove the claimed lower and upper bounds on $t$.

    \subparagraph*{Lower Bound.}
    To establish the lower bounds for \ref{it:combine:i} and \ref{it:combine:ii}, we examine the iteration in \cref{line:combine:iter} where $t$ is assigned its final returned value.
    Since each $t_i$ satisfies $t_i \geq \ted(\bF_{\mI_{v_i}}, \bG_{\mJ_{v_i}})$, the final returned value satisfies
    \[
        \textstyle t = \sum_{i=1}^{B} t_i \geq \sum_{i=1}^{B} \ted(\bF_{\mI_{v_i}}, \bG_{\mJ_{v_i}}) \geq \ted(\bF_{\mI}, \bG_{\mJ}),
    \]
    where the final inequality follows from \cref{prop:ted_decomp} and the fact that the collection $\{\mJ_{v_i}\}_i$ forms a partition of $\mJ$.

    The lower bound for \ref{it:combine:iii} follows essentially the same argument. The oracles guarantee that $t_{(i,\hj)} \geq \ted(\bF_{\langle I_{v_i, \hj} \rangle}, \bG_{\langle J_{v_i,\hj} \rangle})$. Because the collection $\{J_{v_i,\hj}\}_{(i,\hj) \in C_j}$ partitions $J$, we can again apply \cref{prop:ted_decomp} to establish the lower bound for $t$.

    \subparagraph*{Upper Bound.}
    To establish the upper bounds for \ref{it:combine:i} and \ref{it:combine:ii}, we consider an optimal $k$-bounded alignment $\mA : \bF_{\mI} \onto \bG_{\mJ}$ and define $\mJ_{v_i}^* \coloneqq \pi_{\mA}(\mI_{v_i})$ as in \cref{def:align_interval}. Furthermore, let $\mJ_{v_i}^* = \langle J_{v_i,1}^*, \ldots, J_{v_i,\ell_{v_i}}^*\rangle$.
    Note that the collection $\{\mJ_{v_i}^* \}_{i=1}^{B}$ forms a partition of $\mJ$.
    By combining the fact that $\mJ \in W^{k}(\mI_v)$ with \cref{clm:ghost_kbounded}, we obtain $\mJ_{v_i}^* \in W^{k}(\mI_{v_i})$ for each $i \in \fragment{1}{B}$.

    For \ref{it:combine:i}, since we iterate over all such partitions, the specific partition $\{\mJ_{v_i}^* \}_{i=1}^{B}$ will be evaluated in some iteration of \cref{line:combine:iter}.
    Following that iteration, by \cref{clm:TD_ub_sum_ted}, we have:
    \[
        t \leq \textstyle\sum\nolimits_{i=1}^{B} t_i \leq \sum\nolimits_{i=1}^{B} \alpha \cdot \ted^{(k)}(\bF_{\mI_{v_i}}, \bG_{\mJ_{v_i}^*}) = \alpha \cdot \ted^{(k)}(\bF_{\mI_{v}}, \bG_{\mJ}).
    \]

    For \ref{it:combine:ii}, we do not directly consider the partition $\mJ^*$, but rather a partition $\mJ'$ that is close enough.
    We consider the partition $\{\mJ'_{v_i}\}_{i}$ of $\mJ$ such that $|J_{v_i,\hj}'| = \min(r_j, |J_j|)$ for each $j \in \fragment{1}{\ell_v}$ and $(i,\hj) \in C_j'$, where $r_j \coloneqq \mathsf{Round}^{k}(|I_{v_i,\hj}|, |J_{v_i,\hj}^*|, +\infty)$ (note that $|J_{v_i,\hj}^*| \in \sL^{k}(|I_{v_i,\hj}|)$, as required in \cref{def:dyn_allowed_length}, because $\mJ_{v_i}^* \in W^{k}(\mI_{v_i})$).
    This effectively means that for each $j \in \fragment{1}{\ell_v}$ such that $|C_j| = 2$, we shift the quantified position $x$ by $|J_{v_i,\hj}'| - |J_{v_i,\hj}^*|$ compared to where it was with $\{\mJ_{v_i}^*\}_{i}$.

    Next, we must prove that $\{\mJ'_{v_i}\}_{i}$ is considered in some iteration of \cref{line:combine:iter}.
    Clearly, for each $j \in \fragment{1}{\ell_v}$ such that $C_{j}' \neq \emptyset$, we have $|J_{v_i,\hj}'| \in \tilde{\sL}(|I_{v_i,\hj}|) \cup \{|J_j|\}$, so it remains to prove that $\mJ_{v_i}' \in W_{\mJ}^{k}(\mI_{v_i})$ for $i \in \fragment{1}{B}$.

    Fix a $j$ such that $C_j' \neq \emptyset$, and let $(i,\hj) \in C_j'$. Let $J_j^* = \fragmentco{a_{j}'}{x^*}$ and $J_j' = \fragmentco{a_{j}'}{x'}$.
    We need to prove that the quantified position $x'$ stays within $2k$ of $b_{v_i,\hj}$. More specifically, we need to show that $x' = x^* + |J_{v_i,\hj}'| - |J_{v_i,\hj}^*|$ satisfies $|b_{v_i,\hj}-x'| \leq 2k$, given that $|b_{v_i,\hj}-x^*| \leq 2k$ (which holds because $\mJ_{v_i}^* \in W^{k}(\mI_{v_i})$ for all $i \in \fragment{1}{B}$).
    While doing this, we will also show that $||J_{v_i,\hj}'| - |J_{v_i,\hj}^*|| \leq \varepsilon ||J_{v_i,\hj}^*| - |I_{v_i,\hj}||$.

    To this end, we proceed with a case distinction:
    \begin{itemize}
        \item If $|J_{v_i,\hj}'| = r_j$, then from \cref{rmk:rounding_small_gap}\ref{it:rmk:rounding_small_gap:i}, it follows that $|J_{v_i,\hj}'|$ satisfies $||J_{v_i,\hj}'| - |J_{v_i,\hj}^*|| \leq \varepsilon ||J_{v_i,\hj}^*| - |I_{v_i,\hj}||$.
        Moreover, from \cref{rmk:rounding_small_gap}\ref{it:rmk:rounding_small_gap:iii}, we also have that $|x'-b_{v_i,\hj}| \leq 2k$.

        \item Otherwise, $|J_{v_i,\hj}'| = |J_j| < r_j$ and $x'=b_j'$. Since $|J_{v_i,\hj}^*| \leq |J_j|$, it follows that we must have ``rounded down'' in $\mathsf{Round}(\cdot)$. Thus, $|J_{v_i,\hj}'| \geq |J_{v_i,\hj}^*|$, and therefore we have:
        \[
            ||J_{v_i,\hj}'| - |J_{v_i,\hj}^*|| = |J_{v_i,\hj}'| - |J_{v_i,\hj}^*| \leq r_j - |J_{v_i,\hj}^*| \leq \varepsilon ||J_{v_i,\hj}^*| - |I_{v_i,\hj}||.
        \]
        This ensures that $|x'-b_{v_i,\hj}| \leq 2k$, because $-2k \leq x^* - b_{v_i,\hj} \leq x' - b_{v_i,\hj} = b_j' -  b_{v_i,\hj} \leq 2k$.
    \end{itemize}

    Proceeding, we observe that by \cref{prop:size_lb} and \cref{clm:TD_ub_sum_ted}, we have
    \begin{align}
        | |J_{v_i,\hj}^*| - |I_{v_i,\hj}| | / 2 \leq \ted^{(k)}(\bF_{\mI_{v_i}},\bG_{\mJ_{v_i}^*}) \leq \ted^{(k)}(\bF_{\mI_v},\bG_{\mJ}).
    \end{align}
    Moreover, note that for $j \in \fragment{1}{\ell_v}$ such that $|C_j| = 2$, and for $(i',\hj') \in C_j \setminus C_j'$, we have:
    \[
        | |J_{v_{i'},{\hj'}}'| - |J_{v_{i'},{\hj'}}^*|| = |(b_j'-x') - (b_j' - x^*)| = |x'-x^*| = | |J_{v_i,\hj}'| - |J_{v_i,\hj}^*||.
    \]
    For all other $(i,\hj) \notin \bigcup_j C'_j$, we simply have $|J_{v_i,\hj}^*| = |J_{v_i,\hj}'|$.

    Putting everything together and using \cref{prop:triangle_inequality}, we conclude
    \begin{align*}
        t \leq \textstyle\sum_{i=1}^{B} t_i
        &\leq \textstyle \sum_{i=1}^{B} \alpha\ted^{(k)}(\bF_{\mI_{v_i}}, \bG_{\mJ_{v_i}'}) \\
        &\leq \textstyle \sum_{i=1}^{B} \alpha\left(\ted^{(k)}(\bF_{\mI_{v_i}}, \bG_{\mJ_{v_i}^*}) + 5/2 \cdot \sum_{\hj \in \fragment{1}{\ell_{v_i}}} ||J_{v_i,\hj}'| - |J_{v_i,\hj}^*|| \right) \\
        &\leq \sum_{i=1}^{B} \alpha(\ted^{(k)}(\bF_{\mI_{v_i}}, \bG_{\mJ_{v_i}^*}) + 2 \cdot 5/2 \cdot \sum_{j \in \fragment{1}{\ell_v}, (i,\hj) \in C_j'} \varepsilon||J_{v_i,\hj}^*|-|I_{v_i,\hj}||) \\
        &\leq \alpha (\ted^{(k)}(\bF_{\mI_{v}}, \bG_{\mJ}) + 20\varepsilon \cdot \ted(\bF_{\mI_{v}},\bG_{\mJ_{v}^*}))
        \leq (1+20\varepsilon)\alpha \cdot \ted^{(k)}(\bF_{\mI_{v}}, \bG_{\mJ}).
    \end{align*}

    Regarding \cref{alg:combine:mono} and \ref{it:combine:iii}, we can use an almost identical proof. However, this time in \cref{clm:TD_ub_sum_ted}, instead of using the fact that each $\bF_{\mI_{v_i}}$ is balanced, we use the stronger condition that there is no node $z \in \bF_{\mI_v}$ with $\Left(z)$ and $\Right(z)$ contained in different intervals $I_{v_i,\hj}$ for $(i,\hj) \in C_j$.
    \qedhere
\end{proof}

 \subsection{Maintaining Distances within $\bF$}
\label{subsec:fvsf}
In this subsection, we demonstrate how to maintain distances (or approximations) between $\bF_{\mI_v}$ and $\bF_{\mJ}$ for $\mJ \in W^{k}(\mI_v)$ (or $\mJ \in \widetilde{W}^{k}(\mI_v)$ in the approximate setting) under updates to $\bF$.

To this end, we introduce the notation and bounds necessary to characterize the information that must be recomputed following updates.

\begin{definition}
    For two intervals $X = \fragmentco{a}{b}$ and $X' = \fragmentco{a'}{b'}$, we define their \emph{convex hull} as $\mathsf{co}(X,X') \coloneqq \fragmentco{\min(a,a')}{\max(b,b')}$.
\end{definition}

\begin{restatable*}{lemma}{relevant}\label{lem:relevant_bound}
Given a position $x$, consider the following sets of nodes
\[
    A^{\bullet}(x) = \{v \in T(\bF) \mid x \in U(\mI_v)\}
    \quad \text{ and } \quad
    A^{\circ}_{k}(x) = \{v \in T(\bF) \mid \fragment{x-2k}{x+2k} \cap U(\mI_v) \neq \emptyset\}.
\]
Then, the sets $A^{\bullet}(x),A^{\circ}_{k}(x)$ are either empty,
or the nodes in $A^{\bullet}(x)$ form a single root-to-leaf path in $T(\bF)$ and the nodes in $A^{\circ}_{k}(x)$ form a connected subtree in $T(\bF)$ rooted at the root of $T(\bF)$.
Moreover, the following bounds hold:
\begin{enumerate}[(i)]
    \item  $|A^{\bullet}(x)| = \Oh(\log n)$ and $|A^{\circ}_k(x)| = \Oh(k \log n)$;
    \label{it:relevant_bound:i}
    \item The following bounds hold for any $x$:
    \[
        \sum_{\substack{v \in  A^{\circ}_{k}(x) \\ i \in \fragment{1}{\ell_v}}}
        |\{ J \in \widetilde{W}^{k}(I_{v,i}) \mid x \in J \text{ or } (x \in \mathsf{co}(J, I_{v,i}) \text{ and }|\mathsf{co}(J, I_{v,i})| > 2k)\}| = \Oh(k\varepsilon^{-2} \log^2 N).
    \]
    \label{it:relevant_bound:ii}

    \item The following bounds hold for any $x$:
    \[
        \sum_{\substack{v \in  A^{\circ}_{k}(x) \mid \ell_v = 2 \\ z \in Z_v}}  |\{ \mJ=\langle J_1, J_2 \rangle \in \widetilde{W}^{k}_z(\mI_v), x \in U(\mJ) \text{ or } \exists_i \ x \in \mathsf{co}(J_i, I_{v,i}) \text{ and }  |\mathsf{co}(J_i, I_{v,i})| > 2k\}| = \Oh(k\varepsilon^{-6} \log^6 N).
    \]
    \label{it:relevant_bound:iii}
    \item For $v \in T(\bF)$ with $\ell_v = 2$, $z \in Z_v$, $q = (q_1, q_1', q_2, q_2') \in \tilde{\sL}_{v,z}$, and $z' \in \bG$, let $\mJ_{q,z'} \coloneqq \langle \fragmentco{\Left(z') - q_1}{\Left(z') + q_1'}, \fragmentco{\Right(z') - q_2}{\Right(z') + q_2'} \rangle$, so that $\widetilde{W}^{k}_{z,q}(\mI_v) = \{\mJ_{q,z'} \mid z' \in \bG\} \cap W^{k}(\mI_v)$. For any $x$,
    \[
        \small
        \sum_{\substack{v \in T(\bF) \mid \ell_v = 2 \\ z \in Z_v,\ q \in \tilde{\sL}_{v,z}}} |\{ z' \in \bG \mid \mJ_{q,z'} \in W^{k}(\mI_v),\ x \in U(\mJ_{q,z'}) \}| = \Oh(k\varepsilon^{-6} \log^6 N),
    \]
    where only nodes $v \in A^{\circ}_k(x)$ contribute. Moreover, for any set $A$ of nodes with $\ell_v = 2$, the triples $(z, q, z')$ counted here for $v \in A$ can be listed in $\Oh((|A| \varepsilon^{-5} \log^{5} N + k \varepsilon^{-6} \log^{6} N) \log N)$ time, using \cref{lem:uniq_strs}. \qedhere
    \label{it:relevant_bound:iv}
\end{enumerate}
\end{restatable*}

We postpone the proof to \cref{sec:omitted}. This brings us to \cref{thm:fvsf}.

\begin{restatable}{theorem}{fvsf}\label{thm:fvsf}
Let $\bF$ be a forest that undergoes updates.
Then:
\begin{enumerate}
    \item  There is a deterministic dynamic algorithm that maintains a value $t_{v}(\mJ)$ for each $v \in T(\bF)$ and $\mJ \in W^{k}(\mI_v)$, which satisfies
    \begin{align}
        \ted(\bF_{\mI_v}, \bF_{\mJ}) \leq t_{v}(\mJ) \leq \ted^{(k)}(\bF_{\mI_v}, \bF_{\mJ}). \label{inv:fvsf:1}
    \end{align}
    The (amortized) update time of this algorithm is $\Oh(k^7 \log^2 N)$.
    \label{it:fvsf:1}
    \item There exists a randomized dynamic algorithm that maintains a value $\tilde{t}_{v,i}(\mJ)$ for each $v \in T(\bF)$, $i \in \fragment{1}{\ell_v}$, and $\mJ \in \widetilde{W}^{k}(\langle I_{v,i} \rangle)$, which satisfies
    \begin{align}
        \ted(\bF_{\langle I_{v,i} \rangle}, \bF_{\mJ}) \leq \tilde{t}_{v,i}(\mJ) \leq (1 + 46\varepsilon)^{h_v} \cdot \ted^{(k)}(\bF_{\langle I_{v,i} \rangle}, \bF_{\mJ}), \label{inv:fvsf:2}
    \end{align}
    where $h_v$ is the height of node $v$ in $T(\bF)$.
    Moreover, for each $v \in T(\bF)$ with $\ell_v = 2$, the algorithm maintains a good spine node sample $Z_v$, and for each $\mJ \in \widetilde{W}_z^{k}(\mI_v)$, a value $\tilde{t}_{v}(\mJ)$ that satisfies
    \begin{align}
         \ted(\bF_{\mI_v}, \bF_{\mJ}) \leq \tilde{t}_{v}(\mJ) \leq (1 + 46\varepsilon)^{h_v} \cdot \ted^{(k)}(\bF_{\mI_v}, \bF_{\mJ}). \label{inv:fvsf:3}
    \end{align}
    The algorithm succeeds with high probability and has an (amortized) update time of $\Oh(k \varepsilon^{-15} \log^{17} N)$. \qedhere
    \label{it:fvsf:2}
\end{enumerate}
Furthermore, any maintained value can be accessed in time $\Oh(\log N)$.
\end{restatable}

\begin{proof}
    In both algorithms, we always maintain the tree $T(\bF)$ and the data structure from \cref{lem:findNN} (with $\bF$ in the role of $\bG$).

    \subparagraph*{The invariant.}
    We regard the statements of \cref{thm:fvsf}\eqref{it:fvsf:1} and \cref{thm:fvsf}\eqref{it:fvsf:2} as the invariants maintained at all times. Furthermore, we define the ``invariant for $v$'' as the collection of all statements that apply to a single node $v$.

    Note that it is easy to maintain the invariant regarding the good spine node sample $Z_v$. Whenever a node $v$ with $\ell_v = 2$ is created, or whenever all values at a node $v$ with $\ell_v = 2$ are recomputed from scratch, i.e., for $v \in A^{\bullet} \cup P_{\mathrm{new}}$ below, we choose a uniformly random set $Z_v$ of $\lceil \gamma \varepsilon^{-1} \log N \rceil$ spine nodes from $\bF_{\mI_v}$ (or all of them if there are fewer) for a sufficiently large constant $\gamma$. A uniformly random spine node of $\bF_{\mI_v}$, i.e., a uniformly random point of $S \cap (I_{v,1} \times I_{v,2})$ for the point set $S$ of \cref{lem:find_NN_pre}, is drawn in $\Oh(\log^3 N)$ time by range counting and rank selection in a dynamic two-dimensional range tree with subtree counts over $S$ (which uses $\Oh(\log^2 N)$ comparisons per operation, each taking $\Oh(\log N)$ time by \cref{lem:uniq_strs}); this cost is dominated by the recomputation of the values at $v$.
    By the conventions at the beginning of this section, only $\poly(N)$ updates are processed in total and the sequence of updates does not depend on the random choices; a union bound over all nodes $v$, all thresholds $k' \leq 8k$, all $\mJ \in W^{k'}(\mI_v)$, all $\poly(N)$ target forests arising over time (the states of $\bF$ here and of $\bG$ in \cref{thm:fvsg}), and all points in time thus shows that, with high probability, every $Z_v$ satisfies condition \ref{it:good_spine_sample:ii} of \cref{def:good_spine_sample} for each of these thresholds and targets whenever it is used. Therefore, for the rest of this proof, we simply assume that $Z_v$ is always a good spine node sample.

    \begin{claim}\label{clm:fvsf:recompute}
        Consider $v \in T(\bF)$ and suppose that the invariant holds for all children of $v$.
        Then, we compute any of the values $t_{v}(\mJ)$ or values $\tilde{t}_{v,i}(\mJ)$ or $\tilde{t}_{v}(\mJ)$ in time $\Oh(k^2 \log N)$ and time $\Oh(\varepsilon^{-9}\log^{11} N)$.
    \end{claim}
    \begin{claimproof}
        If $v$ is a leaf, we compute $\ted(\bF_{\mI_v}, \bF_{\mJ})$ exactly to determine the values for $t_{v}(\mJ)$ and $\tilde{t}_{v}(\mJ)$. Since $\bF_{\mI_v}$ represents a single node in $\bF$ in this case, it suffices to read the entire content of $\mS(\bF)$ within the bi-interval $\mJ$ in $\Oh(|\mJ|)$ time. We then check whether $\bF_{\mJ}$ contains a node with the same label as $\bF_{\mI_v}$ to see if an exact match is possible; if not, we determine whether a valid substitution exists.
        On the other hand, we set $\tilde{t}_{v,i}(\mJ)$ directly to $\ted(\bF_{\langle I_{v,i} \rangle}, \bF_{\mJ}) = 1/2 + |\mJ|/2$, because $\langle I_{v,i} \rangle$ corresponds to a single parenthesis and nothing can be matched.
        For the runtime in this case notice that $|\mJ| \leq \Oh(k)$ for the exact algorithm and $|\mJ| \leq \Oh(\varepsilon^{-1})$ for the approximate one.

        Conversely, if $v$ is not a leaf, we use the fact that the invariant holds for all children of $v$ in $\bF$. We treat the children as approximation oracles (as defined in \cref{def:approx_oracle}). Specifically, we use standard oracles with $\alpha = 1$ in the exact case, and extended oracles with $\alpha = (1+46\varepsilon)^{h_v-1}$ in the approximate case. In the exact case, we already store all the values the oracles need to return; in the approximate case, we implement the oracles using \cref{lem:round} (note that we are storing exactly what is required by \cref{lem:round}). To implement \cref{ln:approx_oracle_bi:nn} of \cref{alg:approx_oracle_bi} in \cref{lem:round}, we use the data structure we maintain in \cref{lem:findNN}.
        This setup allows us to apply \cref{lem:combine} to compute any of the desired values $t_v(\mJ)$, $\tilde{t}_{v}(\mJ)$, and $\tilde{t}_{v,i}(\mJ)$.
        Note that in the approximate case, the approximation factor increases to $(1+46\varepsilon)^{h_v-1} (1+20\varepsilon)(1+21\varepsilon) \leq (1+46\varepsilon)^{h_v}$, since $\varepsilon \leq 1/100$.

        This means that the runtime needed to compute $t_v(\mJ)$ is given by \cref{lem:combine}\ref{it:combine:i}, plus an additional $\Oh(\log N)$ access time to retrieve the values stored for the children. For $\tilde{t}_{v}(\mJ)$ and $\tilde{t}_{v,i}(\mJ)$, we combine the runtimes of \cref{lem:combine}\ref{it:combine:ii},\ref{it:combine:iii}, \cref{lem:round}, and \cref{lem:findNN} to obtain a time complexity of $\Oh(\log N \cdot (\varepsilon^{-1}\log^{3} N + \varepsilon^{-2} \log^{2} N) \cdot \varepsilon^{-5}\log^{5} N \cdot \varepsilon^{-2}\log^{2} N) = \Oh(\varepsilon^{-9}\log^{11} N)$, where we again account for the time required to access the children's stored values.
    \end{claimproof}

    \subparagraph*{Storing values.}
    We store the values $t_{v}(\mJ)$, $\tilde{t}_{v,i}(\mJ)$, and $\tilde{t}_{v}(\mJ)$ in data structures that depend on the identifiers. More precisely, recall that through \cref{lem:uniq_strs}, we associate two unique identifiers, $\iota_z$ and $\iota_z'$, with each node $z \in \bF$ to represent the first and second occurrences of $z$ in $\mS(\bF)$. Consequently, at any point in time while $z$ exists, we maintain the invariants $\mathsf{pos}(\iota_z) = \Left(z)$ and $\mathsf{pos}(\iota_z') = \Right(z)$.
    Since a window $\mJ \in W^{k}(\mI_v)$ may start outside of $\fragmentco{0}{2n}$ (in the padding, within distance $2k$ of $\mI_v$) or at position $2n$, we additionally keep $8k+1$ padding characters $\$$ with permanent identifiers in front of and after $\mS(\bF)$, which suffices for every window set used in this section (the largest radius, $2 \cdot 4k$, occurs in \cref{thm:has_match}). Padding characters are inserted at initialization and never deleted, so a padding identifier keeps its offset from the respective end of the string, and windows starting in the padding are treated below like all other windows. Window ends are recorded through lengths rather than through identifiers.

    Let $\Lambda$ be the set of all identifiers (including those of the padding characters). In the exact case, we store the values in a balanced binary tree whose key set is $S \subseteq (\Lambda \times \Lambda \times V(T(\bF)) \times \mathbb{Z}_{\geq 0} \times \mathbb{Z}_{\geq 0}) \cup (\Lambda \times V(T(\bF)) \times \mathbb{Z}_{\geq 0})$\footnote{Whenever we mention balanced binary trees on tuples, we sort them in lexicographical order, and whenever the key set comprises different incomparable sets, we maintain separate trees. Furthermore, whenever a set, such as the node set $V(T(\bF))$, lacks a natural ordering, we assume a global mapping of its elements to integers.}. In the approximate case, the key set changes to $S \subseteq (\Lambda \times \Lambda \times V(T(\bF)) \times \mathbb{Z}_{\geq 0} \times \mathbb{Z}_{\geq 0}) \cup (\Lambda \times V(T(\bF)) \times \mathbb{Z}_{\geq 0} \times \mathbb{Z}_{\geq 0})$.

    This allows us to store $t_{v}(\mJ)$, $\tilde{t}_{v,i}(\mJ)$, and $\tilde{t}_{v}(\mJ)$ as follows:
\begin{itemize}
    \item In the exact setting, for $v \in T(\bF)$ and $\mJ = \langle \fragmentco{a_1'}{b_1'}, \ldots, \fragmentco{a_{\ell_v}'}{b_{\ell_v}'} \rangle \in {W}^{k}(\mI_v)$, we assign a unique identifier $\mathsf{id}(v, \mJ) \coloneqq (\mathsf{id}(a_1'), \ldots, \mathsf{id}(a_{\ell_v}'), v, b_1'-a_1', \ldots, b_{\ell_v}'-a_{\ell_v}')$.

    Now, for $\mJ \in W^{k}(\mI_v)$, we use $\mathsf{id}(v, \mJ)$ to index an element in the stored key set $S$, and we store the value $t_{v}(\mJ)$ under the key $(\mathsf{id}(a_1'), \ldots, \mathsf{id}(a_{\ell_v}'), v, b_1'-a_1', \ldots, b_{\ell_v}'-a_{\ell_v}')$.
    It is easy to see that, accounting for identifier/index conversion and tree traversals, the retrieval time of $t_{v}(\mJ)$ is $\Oh(\log N)$.

    \item In the approximate setting, for $v \in T(\bF)$, $i \in \fragment{1}{\ell_v}$, and $\mJ = \langle \fragmentco{a'}{b'} \rangle \in \widetilde{W}^k(\langle I_{v,i} \rangle)$, we also assign a unique identifier: $\mathsf{id}(v, i, \mJ) \coloneqq (\mathsf{id}(a'), v, i, b'-a')$. This is again used to locate where $\tilde{t}_{v,i}(\mJ)$ is stored in the tree.

    \item In the approximate setting, for $v \in T(\bF)$, $z \in Z_v$, and $\mJ = \langle \fragmentco{a_1'}{b_1'}, \fragmentco{a_2'}{b_2'} \rangle \in \widetilde{W}^k_z(\mI_v)$, we also assign a unique identifier: $\mathsf{id}(v, \mJ) \coloneqq (\mathsf{id}(a_1'), \mathsf{id}(a_2'), v, b_1'-a_1', b_2'-a_2')$. We use this identifier to index $\tilde{t}_{v}(\mJ)$ in the tree as well.
\end{itemize}
     We let $\Gamma_{v}$ denote the corresponding set of identifiers. In the exact setting, we define $\Gamma_{v} \coloneqq \{\mathsf{id}(v,\mJ) \mid \mJ \in {W}^{k}(\mI_v)\}$, while in the approximate setting, we define $\Gamma_{v} \coloneqq \{\mathsf{id}(v,\mJ) \mid \mJ \in \widetilde{W}^{k}_z(\mI_v), z \in Z_v\} \cup \{\mathsf{id}(v,i,\mJ) \mid \mJ \in \widetilde{W}^{k}(\langle I_{v,i}\rangle), i \in \fragment{1}{\ell_v}\}$.

    \subparagraph*{Handling updates.}
    Next, we describe the procedure for handling updates to $\bF$. Let $y$ and $y'$ denote the indices where the update occurs. Specifically, if the update is a substitution or deletion of a node $w \in \bF$, we set $y = \Left(w)$ and $y' = \Right(w)$. Otherwise, if a new node $w$ is inserted, we choose $y$ and $y'$ such that the new bi-order traversal is given by $\mS(\bF)\fragmentco{0}{y} \circ w \circ \mS(\bF)\fragmentco{y}{y'} \circ w \circ \mS(\bF)\fragmentco{y'}{2n}$.

    Let also $\bar{\bF}$ represent the tree $\bF$ after the update. Furthermore, we apply a ``bar'' to all previously established notation to indicate its dependence on the updated tree $\bar{\bF}$ (e.g., $\bar{\mI}_v$ and $\bar{\mJ}$). In particular, $\bar y \coloneqq y$ and $\bar y' \coloneqq y'+1$, $y'-1$, or $y'$ for an insertion, deletion, or substitution, respectively, so that the update is reverted by an update of $\bar{\bF}$ at $\bar y, \bar y'$ (for an insertion, these are the positions of the inserted occurrences in $\mS(\bar{\bF})$; for a deletion, the positions at which the deleted occurrences are reinserted).

    Let $P_{\mathrm{new}}$ denote the set of affected nodes in $T(\bar{\bF})$. We observe that $P_{\mathrm{new}}$ is either empty or forms a connected subtree of $T(\bar{\bF})$ that contains the root.

    Similarly, let $P_{\mathrm{old}}$ denote the set of remaining nodes in $T(\bar{\bF})$, that is, nodes that were already present in $T(\bF)$ and do not contain any new nodes in their subtrees (meaning their height also remains unchanged).

    We also define two more sets $A^{\circ}$ and $A^{\bullet}$:
    \[
        A^{\bullet} \coloneq (A^{\bullet}(y) \cup A^{\bullet}(y')) \cap P_{\mathrm{old}}
        \quad \text{ and } \quad A^{\circ} \coloneqq (A^{\circ}(y) \cup A^{\circ}(y')) \cap P_{\mathrm{old}},
    \]
    where $A^{\bullet}(\cdot)$ and $A^{\circ}(\cdot)$ are defined as in \cref{lem:relevant_bound}. To maintain the desired values up-to-date for all $v \in P$, we use the following three claims.

    \begin{claim}\label{clm:fvsf:0}
        Let $v \in P_{\mathrm{old}} \setminus A^{\circ}$ and $i \in \fragment{1}{\ell_v}$. Then, for $J \in W^{k}(I_{v,i})$, we have $\mathsf{co}(I_{v,i}, J) \cap \{y, y'\} = \emptyset$.
    \end{claim}
    \begin{claimproof}
        Since $v \notin A^{\circ}$, the conditions $I_{v,i} \cap \fragment{y-2k}{y+2k} = \emptyset$ and $I_{v,i} \cap \fragment{y'-2k}{y'+2k} = \emptyset$ imply that $\fragmentco{a_{v,i}-2k}{b_{v,i}+2k} \cap \{y,y'\} = \emptyset$.
        Since $J \in W^{k}(I_{v,i})$ must satisfy $J \subseteq \fragmentco{a_{v,i}-2k}{b_{v,i}+2k}$, we immediately get $\mathsf{co}(I_{v,i},J) \cap \{y, y'\} = 0$.
    \end{claimproof}

    \begin{claim}\label{clm:fvsf:1}
        Let $v \in P_{\mathrm{old}} \setminus A^{\bullet}$. Let $\mJ = \langle J_1, \ldots, J_{\ell_v} \rangle \in W^{k}(\mI_v)$, $\bar{\mJ} = \langle \bar{J}_1, \ldots, \bar{J}_{\ell_v} \rangle \in W^{k}(\bar{\mI}_v)$ with $\mathsf{id}(v,\mJ)=\mathsf{id}(v,\bar{\mJ})$ and $U(\mJ) \cap \{y,y'\} = \emptyset$. If $v \in A^{\circ} \setminus A^{\bullet}$ and the update is a deletion, we further assume that for every $i \in \fragment{1}{\ell_v}$ either $\{y,y'\} \cap \mathsf{co}(I_{v,i},J_i) = \emptyset$ or $|\mathsf{co}(I_{v,i},J_i)| \leq 2k$.
        Then, we have
        \[
            \ted(\bF_{\mI_v}, \bF_{\mJ}) = \ted(\bar{\bF}_{\bar{\mI}_v}, \bar{\bF}_{\bar{\mJ}}) \quad \text{and} \quad \ted^{(k)}(\bF_{\mI_v}, \bF_{\mJ}) \leq \ted^{(k)}(\bar{\bF}_{\bar{\mI}_v}, \bar{\bF}_{\bar{\mJ}}).
        \]
        Similarly, for every $i \in \fragment{1}{\ell_v}$ and $\mJ = \langle J \rangle \in W^{k}(\langle I_{v,i} \rangle)$, $\bar{\mJ} = \langle \bar{J} \rangle \in W^{k}(\langle \bar{I}_{v,i} \rangle)$ such that $\mathsf{id}(\mJ,i,v) = \mathsf{id}(\bar{\mJ},i,v)$ and $J \cap \{y,y'\} = \emptyset$ (and also $\{y,y'\} \cap \mathsf{co}(I_{v,i},J) = \emptyset$ or $|\mathsf{co}(I_{v,i},J)| \leq 2k$ if $v \in A^{\circ} \setminus A^{\bullet}$ and the update is a deletion), the same relations hold.
    \end{claim}
    \begin{claimproof}
        We only prove the first part of the statement; the second follows from an almost analogous proof.

        For the proof, it suffices to show that any alignment (respectively, $k$-bounded alignment) $\bar{\mA} : \bar{\bF}_{\bar{\mI}_v} \onto \bar{\bF}_{\bar{\mJ}}$ has a corresponding alignment (respectively, $k$-bounded alignment) $\mA : \bF_{\mI_v} \onto \bF_{\mJ}$ of the same cost. For the non-$k$-bounded case, we must also show the converse.

        A natural bijection between these two alignments is established by preserving identifiers.
        This bijection works because the set of identifiers underlying $\bF_{\mI_v}$ is the same as $\bar{\bF}_{\bar{\mI}_v}$ because $v \notin A^{\bullet}$, and the one underlying $\bF_{\mJ}$ is the same as $\bar{\bF}_{\bar{\mJ}}$ because $U(\mJ) \cap \{y,y'\} = \emptyset$.
        It is clear that this mapping preserves conditions \ref{def:TED_on_biintervals:i}, \ref{def:TED_on_biintervals:ii}, and \ref{def:TED_on_biintervals:iii} of \cref{def:TED_on_biintervals}. It remains to argue that condition \ref{def:TED_on_biintervals:iv} of \cref{def:TED_on_biintervals} is preserved when mapping $k$-bounded alignments from $\bar{\mA}$ to $\mA$.
        More specifically, we need to prove that:
        \begin{align}
            |\bar{x} - \bar{x}'| \leq 2k \quad \text{implies} \quad |x - x'| \leq 2k, \label{eq:fvsf:1}
        \end{align}
        where $(\bar{x},\bar{x}') \in \bar{\mA}$ and $(x,x') \in \mA$, with $\bar{x},\bar{x}'$ and $x,x'$ corresponding to positions pointing to the same identifiers before and after the update.
        Note that for $v \notin A^{\circ}$ by \cref{clm:fvsf:0}, we have $\mathsf{co}(I_{v,i}, J_i) \cap \{y, y'\} = \emptyset$ for all $i \in \fragment{1}{\ell_v}$. This means that no update positions fall in-between $I_{v,i}$ and $J$ and the relative distance $|\bar{x} - \bar{x}'| = |x - x'|$ must be preserved.
        On the other hand for $v \in A^{\circ} \setminus A^{\bullet}$,
        for updates that are not deletions, the distance $ |\bar{x} - \bar{x}'|$ can only increase w.r.t. $|x - x'|$ because either new identifiers are inserted between $x,x'$ or it stays as it is. On the other side for deletions, if $\{y,y'\} \cap \mathsf{co}(I_{v,i},J_i) = \emptyset$, then no deleted position lies between $x$ and $x'$, so $|\bar{x} - \bar{x}'| = |x - x'|$; otherwise $x, x' \in \mathsf{co}(I_{v,i},J_i)$ and $|\mathsf{co}(I_{v,i},J_i)| \leq 2k$ give $|x - x'| < 2k$, so \cref{eq:fvsf:1} holds.
    \end{claimproof}

    \begin{claim}\label{clm:fvsf:2}
        We have $\Gamma_v = \bar{\Gamma}_v$ for $v \in P_{\mathrm{old}} \setminus A^{\circ}$. Furthermore, in the approximate setting, $\sum_{v \in A^{\circ} \setminus A^{\bullet}} |\Gamma_v \triangle \bar{\Gamma}_v| = \Oh(k \varepsilon^{-6} \log^{6} N)$, and all these sets can be computed in $\Oh(k \varepsilon^{-6} \log^{7} N)$ total time, using $\Oh(1)$ words of auxiliary information per identifier that we rebuild in $\Oh(k \varepsilon^{-5} \log^{6} N)$ time whenever all values at a node are recomputed from scratch.
    \end{claim}
    \begin{claimproof}
        Notice that by the symmetry of the updates (insertions revert deletions), it suffices to prove that $\Gamma_v \subseteq \bar{\Gamma}_v$ for $v \in P_{\mathrm{old}} \setminus A^{\circ}$, and, for the latter part of the statement, to bound and compute $\Gamma_v \setminus \bar{\Gamma}_v$ (for $\bar{\Gamma}_v \setminus \Gamma_v$, we apply the same argument to the reverted update of $\bar{\bF}$ at $\bar y, \bar y'$).

        First, consider the exact setting. Let $\mathsf{id} \in \Gamma_v$ and suppose $\mathsf{id} = (\iota_1, \ldots, \iota_{\ell_v}, v, q_1,\ldots, q_{\ell_v})$. Let $x_i$ and $\bar{x}_i$ be the positions of $\iota_i$ before and after the update, respectively. Thus, $\mathsf{id}$ corresponds to $\mJ = \langle \fragmentco{x_1}{x_1+q_1}, \ldots, \fragmentco{x_{\ell_v}}{x_{\ell_v}+q_{\ell_v}} \rangle \in W^{k}(\mI_v)$, which satisfies $|x_{i}-a_{v,i}| \leq 2k$ and $|x_{i}+q_i-b_{v,i}| \leq 2k$ for all $i \in \fragment{1}{\ell_v}$. It follows from \cref{clm:fvsf:0} that $\mathsf{co}(I_{v,i}, \fragmentco{x_{i}}{x_{i}+q_{i}}) \cap \{y,y'\} = \emptyset$ for all $i \in \fragment{1}{\ell_v}$, meaning that the update occurs neither within $I_{v,i}$ or $\fragmentco{x_{i}}{x_{i}+q_{i}}$, nor in between them. Consequently, $|\bar{x}_i-\bar{a}_{v,i}| = |x_{i}-a_{v,i}| \leq 2k$ and $|\bar{x}_i+q_i-\bar{b}_{v,i}| = |x_{i}+q_i-b_{v,i}| \leq 2k$. This immediately implies that $\mathsf{id} \in \bar{\Gamma}_v$.  More generally, notice that the condition $\mathsf{co}(J_i, I_{v,i}) \cap \{y,y'\} = \emptyset$, along with $|\bar{x}_i-\bar{a}_{v,i}| = |x_{i}-a_{v,i}|$ and $|\bar{x}_i+q_i-\bar{b}_{v,i}| = |x_{i}+q_i-b_{v,i}|$ for all $i \in \fragment{1}{\ell_v}$ and $J_i \in W^{k}(I_{v,i})$, also yields $\Gamma_v \subseteq \bar{\Gamma}_v$ in the approximate setting, because relative distances are preserved and updates fall outside the intervals.

        For the latter part of the statement, consider the approximate setting and $v \in A^{\circ} \setminus A^{\bullet}$. The update leaves the identifiers of $\mI_v$, and hence $Z_v$ and $\tilde{\sL}_{v,z}$, unchanged, which is all that the following argument uses; it therefore applies equally to the reverted update of $\bar{\bF}$ at $\bar y, \bar y'$. A mono-interval identifier $\mathsf{id} = (\iota, v, i, m) \in \Gamma_v \setminus \bar{\Gamma}_v$ with $x = \mathsf{pos}(\iota)$ satisfies $x \in \{y, y'\}$, $|x - a_{v,i}| \geq 2k-1$, or $|x + m - b_{v,i}| \geq 2k-1$, since $\iota$ can only be deleted at $y$ or $y'$ and each update position changes each of these distances by at most one; this leaves $\Oh(1)$ starts $x$ per length $m$, hence $\Oh(\varepsilon^{-1} \log N)$ such identifiers.

        Now consider a triple $(z, q, z')$ with $z \in Z_v$, $q \in \tilde{\sL}_{v,z}$, and $z' \in \bF$, and let $\mJ_{q,z'} = \langle J_1, J_2 \rangle$ with $J_i = \fragmentco{a_i'}{b_i'}$ be as in \cref{lem:relevant_bound}\ref{it:relevant_bound:iv}. If $\mJ_{q,z'} \in W^{k}(\mI_v)$, we say that the triple \emph{generates} $\mathsf{id}(v, \mJ_{q,z'}) \in \Gamma_v$, so that $\Gamma_v$ consists of the mono-interval identifiers and the generated identifiers, and likewise for $\bar{\Gamma}_v$ with respect to $\bar{\bF}$. A triple generating $\mathsf{id} \in \Gamma_v$ generates the same identifier after the update unless
        \begin{enumerate}[(a)]
            \item \label{it:clm:fvsf:2:a} $U(\mJ_{q,z'}) \cap \{y, y'\} \neq \emptyset$, or
            \item \label{it:clm:fvsf:2:b} $|a_i' - a_{v,i}| \geq 2k-1$ or $|b_i' - b_{v,i}| \geq 2k-1$ for some $i \in \{1,2\}$.
        \end{enumerate}
        Indeed, otherwise $\mathsf{id}(a_1')$, $\mathsf{id}(a_2')$, and $z'$ survive, no update position lies in $\fragmentoc{a_1'}{\Left(z')} \subseteq J_1$ or $\fragmentoc{a_2'}{\Right(z')} \subseteq J_2$ (recall $q_1', q_2' \geq 1$), so $q_1, q_2$ and the lengths $q_i + q_i'$ are preserved, and each update position changes each distance $|a_i' - a_{v,i}|$, $|b_i' - b_{v,i}|$ by at most one, so they stay at most $2k$. We call a triple \emph{listed} if it satisfies \ref{it:clm:fvsf:2:a} or \ref{it:clm:fvsf:2:b} before or after the update (for $\bar{\mJ}_{q,z'}$, $\bar{\mI}_v$, and $\bar y, \bar y'$); thus every bi-interval identifier of $\Gamma_v \triangle \bar{\Gamma}_v$ is generated by a listed triple. For fixed $(z, q)$, condition \ref{it:clm:fvsf:2:b} fixes $\Left(z')$ or $\Right(z')$ up to $\Oh(1)$ choices, so $\Oh(|Z_v| \cdot \max_{z \in Z_v} |\tilde{\sL}_{v,z}|) = \Oh(\varepsilon^{-5} \log^{5} N)$ triples per node satisfy \ref{it:clm:fvsf:2:b}, i.e., $\Oh(k \varepsilon^{-5} \log^{6} N)$ over the $\Oh(k \log n)$ nodes of $A^{\circ}$ (\cref{lem:relevant_bound}\ref{it:relevant_bound:i}), and $\Oh(k \varepsilon^{-6} \log^{6} N)$ triples satisfy \ref{it:clm:fvsf:2:a} by \cref{lem:relevant_bound}\ref{it:relevant_bound:iv}, applied to $T(\bF)$ with $x \in \{y, y'\}$ and to $T(\bar{\bF})$ with $x \in \{\bar y, \bar y'\}$. With the mono-interval identifiers, this proves $\sum_{v \in A^{\circ} \setminus A^{\bullet}} |\Gamma_v \triangle \bar{\Gamma}_v| = \Oh(k \varepsilon^{-6} \log^{6} N)$.

        For the computation, we store with every generated identifier in $\Gamma_v$ the number of triples generating it. These counters are rebuilt whenever all values at $v$ are recomputed from scratch, by enumerating the $\sum_{z \in Z_v} |\tilde{\sL}_{v,z}| \cdot (4k+1) = \Oh(k \varepsilon^{-5} \log^{5} N)$ candidate triples in $\Oh(\log N)$ time each via \cref{lem:uniq_strs}, i.e., in $\Oh(k \varepsilon^{-5} \log^{6} N)$ time. Upon an update, we list the triples satisfying \ref{it:clm:fvsf:2:a} by \cref{lem:relevant_bound}\ref{it:relevant_bound:iv} with $A = A^{\circ} \setminus A^{\bullet}$, applied to $T(\bF)$ with $x \in \{y, y'\}$ and to $T(\bar{\bF})$ with $x \in \{\bar y, \bar y'\}$, and the triples satisfying \ref{it:clm:fvsf:2:b} by trying the $\Oh(1)$ positions for $\Left(z')$ or $\Right(z')$ for each $(z, q)$, in $\bF$ and in $\bar{\bF}$, in $\Oh(\log N)$ time each. Each listed triple is processed once (a dictionary on the triples discards duplicates): we decrement the counter of the identifier it generates before the update, if any, and increment the counter of the identifier it generates after the update, if any, creating the counter if needed. Since unlisted triples generate the same identifier before and after the update, the counters are now correct for $\bar{\bF}$, and the bi-interval identifiers of $\Gamma_v \setminus \bar{\Gamma}_v$ (resp.\ $\bar{\Gamma}_v \setminus \Gamma_v$) are exactly those whose counter dropped to $0$ (resp.\ rose from $0$). The mono-interval identifiers of $\Gamma_v \triangle \bar{\Gamma}_v$ are found by testing, for each of the $\Oh(\varepsilon^{-1} \log N)$ lengths $m$, the $\Oh(1)$ starts above (likewise in $\bar{\bF}$ with $\bar y, \bar y'$). By \cref{lem:relevant_bound}\ref{it:relevant_bound:iv} and $|A^{\circ}| = \Oh(k \log n)$, the total time is $\Oh(k \varepsilon^{-6} \log^{7} N)$.
    \end{claimproof}

    By \cref{clm:fvsf:2}, for all nodes $v \in P_{\mathrm{old}} \setminus A^{\circ}$, the set $\Gamma_v$ of identifiers for the values we maintain remains unchanged. Furthermore, by \cref{clm:fvsf:1}, for any such $v$, if we leave the values stored at $\mathsf{id}(v, \mJ)$ as they are, they remain valid for $\mathsf{id}(v, \bar{\mJ})$ (the lower bound stays the same, and the upper bound only becomes looser). Consequently, we do not need to update any stored information for the nodes in $P_{\mathrm{old}} \setminus A^{\circ}$. (Note that the height remaining unchanged is crucial for establishing this in the approximate case.)

    It remains to update the values for the nodes $v \in A^{\circ} \cup P_{\mathrm{new}}$. To identify the set $A^{\circ}$, it suffices to execute a DFS on $T(\bF)$. We then update the maintained values by traversing $A^{\circ} \cup P_{\mathrm{new}}$ bottom-up, applying the necessary updates as we visit each node in increasing order of height.

    Now, consider an arbitrary node $v \in A^{\circ} \cup P_{\mathrm{new}}$ visited during this process. In the exact algorithm, we remove all values indexed by $\Gamma_v$ (so effectively for all $\mJ \in W^{k}(\mI_v)$), and then for each $\mJ \in W^{k}(\mI_v)$, we recompute the values that will be stored in $\bar{\Gamma}_v$ via \cref{clm:fvsf:recompute}.
    In the approximate setting, we only perform this full recomputation when $v \in A^{\bullet} \cup P_{\mathrm{new}}$. Therefore, it remains to specify our procedure for the approximate case when $v \in A^{\circ} \setminus A^{\bullet}$. Let us now assume this is the case.

    For any such $v$, we first compute $\Gamma_v \triangle \bar{\Gamma}_v$ as in \cref{clm:fvsf:2}. We remove all values indexed by $\Gamma_v \setminus \bar{\Gamma}_v$. Next, we use \cref{clm:fvsf:recompute} to compute the values (with respect to $\bar{\bF}$) for all identifiers in $\bar{\Gamma}_v \setminus \Gamma_v$. We also recompute $\tilde{t}_{v,i}(\bar{\mJ})$ and $\tilde{t}_{v}(\bar{\mJ})$ for all $\mathsf{id}(v,\bar{\mJ}) \in \bar{\Gamma}_v \cap \Gamma_v$ and $\mathsf{id}(v,i,\bar{\mJ}) \in \bar{\Gamma}_v \cap \Gamma_v$ such that, before the update, the corresponding $\mJ$ or $(i,\mJ)$ did not satisfy the premise of \cref{clm:fvsf:1}.

    \subparagraph*{Runtime for Exact Algorithm.}
    Let us first analyze the exact case. In this setting, the total number of recomputed entries is asymptotically upper-bounded by $\sum_{v \in A^{\circ} \cup P_{\mathrm{new}}} |W^{k}(\mI_v)|$, because for nodes $v \in A^{\circ} \cup P_{\mathrm{new}}$, we recompute everything from scratch. \Cref{clm:fvsf:recompute} states that each recomputation requires $\Oh(k^2 \log N)$ time. By \cref{lem:relevant_bound}, we have $|A^{\circ}| = \Oh(k \log n)$, and for each $v$, we know that $|W^{k}(\mI_v)| = \Oh(k^4)$. Thus, the total running time contribution from computations for nodes $v \in A^{\circ}$ is $\Oh(|A^{\circ}| \cdot k^4 \cdot k^2 \log N) = \Oh(k \log n \cdot k^4 \cdot k^2 \log N) = \Oh(k^7 \log^2 N)$.

    On the other hand, for the newly added nodes, we know that on average $|P_{\mathrm{new}}|$ has size $\Oh(\log n)$. Therefore, we invoke \cref{clm:fvsf:recompute} not more than $\Oh(k^4 \log n)$ times for these nodes, resulting in a total time contribution of $\Oh(k^6 \log^2 N)$. Consequently, the overall amortized update time in the exact setting is dominated by the $A^{\circ}$ computations, yielding $\Oh(k^7 \log^2 N)$, as claimed.

    \subparagraph*{Runtime for Approximate Algorithm.}
    In the approximate case, there are additional contributions to the running time. For each node $v \in A^{\bullet} \cup P_{\mathrm{new}}$, we recompute $\sum_{i=1}^{\ell_v} |\widetilde{W}^{k}(I_{v,i})| + \sum_{z \in Z_v} |\widetilde{W}^{k}_z(\mI_v)|$ entries. Note that this latter quantity is bounded by $\Oh(k\varepsilon^{-1}\log^{1} N + k \varepsilon^{-5} \log^5 N) = \Oh(k \varepsilon^{-5} \log^5 N)$.

    Since $|A^{\bullet}| = \Oh(\log n)$ by \cref{lem:relevant_bound}\ref{it:relevant_bound:i} and $|P_{\mathrm{new}}| = \Oh(\log n)$ on average, and since by \cref{clm:fvsf:recompute} the recomputation time per entry is $\Oh(\varepsilon^{-9} \log^{11} N)$, the total contribution from these nodes is $\Oh(k \varepsilon^{-14} \log^{17} N)$.

    For the remaining nodes $v \in A^{\circ} \setminus A^{\bullet}$, note that the number of recomputed entries (exactly those excluded by \cref{clm:fvsf:1}: some $x \in \{y,y'\}$ lies in $U(\mJ)$, or the update is a deletion and for some $i$ we have $x \in \mathsf{co}(J_i, I_{v,i})$ and $|\mathsf{co}(J_i, I_{v,i})| > 2k$) is bounded by \cref{lem:relevant_bound}\ref{it:relevant_bound:ii} and \cref{lem:relevant_bound}\ref{it:relevant_bound:iii}\footnote{Notice that these bounds apply to a larger superset of the entries we want to actually bound.}, plus the $\Oh(k \varepsilon^{-6} \log^{6} N)$ identifiers entering or leaving $\Gamma_v$ over all $v \in A^{\circ} \setminus A^{\bullet}$, which are found in $\Oh(k \varepsilon^{-6} \log^{7} N)$ time by \cref{clm:fvsf:2}; rebuilding the auxiliary information of \cref{clm:fvsf:2} at the amortized $\Oh(\log n)$ nodes of $A^{\bullet} \cup P_{\mathrm{new}}$ costs $\Oh(k \varepsilon^{-5} \log^{7} N)$ more. The recomputed entries are listed in $\Oh(\log N)$ time each: the windows containing $y$ or $y'$ by \cref{lem:relevant_bound}\ref{it:relevant_bound:iv} and, for mono-intervals, by the at most $\min(m, 4k+1)$ starts in $\fragmentoc{x - m}{x}$ per length $m$; for a deletion, the windows with $|\mathsf{co}(J_i, I_{v,i})| > 2k$ by trying the $\Oh(|I_{v,i}|)$ starts (ends) at distance more than $2k - |I_{v,i}|$ from $a_{v,i}$ ($b_{v,i}$) if $|I_{v,i}| \leq \varepsilon k$ and all $4k+1$ starts otherwise, which are the quantities counted in the proof of \cref{lem:relevant_bound}. Since by \cref{clm:fvsf:recompute} the recomputation time is $\Oh(\varepsilon^{-9} \log^{11} N)$, we obtain an overall runtime of $\Oh(k \varepsilon^{-15} \log^{17} N)$.
\end{proof}

\subsection{Handling any Updates in $\bF$ or $\bG$}
\label{subsec:sub_F_all_G}

We now present the final subsection of our dynamic algorithm. The key theorem of this section, \cref{thm:fvsg}, allows us to maintain quantities similar to those in \cref{thm:fvsf}, but this time comparing $\bF$ and $\bG$ rather than $\bF$ against itself.

Here, \cref{thm:fvsf} plays a crucial role: rather than recomputing all terms for $\bF$ and $\bG$ from scratch, we will copy some of the existing values maintained by \cref{thm:fvsf}. To achieve this, we require a data structure to efficiently determine when this reuse is possible. That is, when a subforest of $\bG$ is equivalent to a subforest of $\bF$.

\begin{restatable*}{lemma}{hasmatch}\label{thm:has_match}
Let $\bF$ and $\bG$ undergo updates.
\begin{enumerate}
    \item \label{it:has_match:1} There is a deterministic dynamic algorithm that supports the following $\Oh(\log^{3} N)$-time queries: given $v \in T(\bF)$ and an interval collection $\mJ$ consisting of $\ell_v$ intervals, determine whether there exists an $\mL \in W^{4k}(\mI_v)$ such that $\bF_{\mL} \equiv \bG_{\mJ}$ and, if so, return such an $\mL$.
    The amortized update time of this algorithm is $\Oh(k^5 \log^{4} N)$.

    \item \label{it:has_match:2} There is a deterministic dynamic algorithm that supports the following $\Oh(\log^{3} N)$-time queries: given $v \in T(\bF)$ and $i \in \fragment{1}{\ell_v}$ (resp. $v \in T(\bF)$ and $z \in Z_v$) alongside a mono-interval $\mJ$ (resp. a bi-interval $\mJ$), determine whether there exists an $\mL \in \widetilde{W}^{4k}(\langle I_{v,i} \rangle)$ (resp. $\mL \in \widetilde{W}^{4k}_z(\mI_v)$) such that $\bF_{\mL} \equiv \bG_{\mJ}$ and, if so, return such an $\mL$.
    Here, $Z_v$ is a set of spine nodes of $\bF_{\mI_v}$ of size $\Oh(\varepsilon^{-1} \log N)$, which is redrawn exactly when $v$ is created or the content of $\bF_{\mI_v}$ changes.
    The amortized update time of this algorithm is $\Oh(k \varepsilon^{-6} \log^{9} N)$. \qedhere
\end{enumerate}
\end{restatable*}

The proof is deferred to \cref{sec:omitted}, as it employs a standard combination of techniques introduced here and in \cite{DGHKS25}. This brings us to \cref{thm:fvsg}.

\begin{restatable}{theorem}{fvsg}\label{thm:fvsg}
Let $\bF$ and $\bG$ undergo updates.
Then:
\begin{enumerate}
    \item  There is a deterministic dynamic algorithm that either reports that $\ted(\bF, \bG) > k$ or maintains a value $t_{v}(\mJ)$ for each $v \in T(\bF)$ and $\mJ \in W^{k}(\mI_v)$, which satisfies
    \begin{align}
        \ted(\bF_{\mI_v}, \bG_{\mJ}) \leq t_{v}(\mJ) \leq \ted^{(k)}(\bF_{\mI_v}, \bG_{\mJ}).
    \end{align}
    The (amortized) update time of this algorithm is $\Oh(k^7 \log^{4} N)$.
    \label{it:fvsg:1}
    \item There exists a randomized dynamic algorithm that either reports that $\ted(\bF, \bG) > k$ or maintains a value $\tilde{t}_{v,i}(\mJ)$ for each $v \in T(\bF)$, $i \in \fragment{1}{\ell_v}$, and $\mJ \in \widetilde{W}^{k}(\langle I_{v,i} \rangle)$, which satisfies
    \begin{align}
        \ted(\bF_{\langle I_{v,i} \rangle}, \bG_{\mJ}) \leq \tilde{t}_{v,i}(\mJ) \leq (1 + 46\varepsilon)^{h_v} \cdot \ted^{(k)}(\bF_{\langle I_{v,i} \rangle}, \bG_{\mJ})
    \end{align}
    where $h_v$ is the height of node $v$ in $T(\bF)$.
    Moreover, for each $v \in T(\bF)$ with $\ell_v = 2$, the algorithm maintains a good spine node sample $Z_v$, and for each $\mJ \in \widetilde{W}_z^{k}(\mI_v)$, a value $\tilde{t}_{v}(\mJ)$ that satisfies
    \begin{align}
         \ted(\bF_{\mI_v}, \bG_{\mJ}) \leq \tilde{t}_{v}(\mJ) \leq (1 + 46\varepsilon)^{h_v} \cdot \ted^{(k)}(\bF_{\mI_v}, \bG_{\mJ}).
    \end{align}
    The algorithm succeeds with high probability and has an (amortized) update time of $\Oh(k \varepsilon^{-15} \log^{19} N)$. \qedhere
    \label{it:fvsg:2}
\end{enumerate}
Furthermore, any maintained value can be accessed in (amortized) time $\Oh(\log^{3} N)$.
\end{restatable}
\begin{proof}
The algorithm proceeds in \emph{epochs}, which are blocks of $k$ updates. Throughout the process, we maintain the following data structures:
\begin{itemize}
    \item The trees $T(\bF)$ and for the approximate case also the data structure from \cref{lem:findNN}. These data structures are always kept up-to-date;
    \item The algorithm from \cref{thm:fvsf}, operating with a higher threshold of $8k$ rather than $k$ (for the exact case, we use \cref{thm:fvsf}\eqref{it:fvsf:1}; for the approximate case, we use \cref{thm:fvsf}\eqref{it:fvsf:2}). This data structure reflects the exact state of $\bF$ at the beginning of the current epoch. At the end of the epoch, all $k$ updates performed on $\bF$ are applied to this data structure in bulk, ensuring it is fully updated and ready for the start of the next epoch;
    \item The algorithm from \cref{thm:has_match}, where $\bF$ is maintained at its state from the beginning of the epoch and updated in bulk at the end (as described above), while $\bG$ is continuously kept up-to-date with incoming updates.
\end{itemize}

\subparagraph*{The invariant.}
Let $\ul{\bF}$ denote the state of $\bF$ at the beginning of the epoch. Generally, we apply an underline to all notations referring to the state of the system at the start of the epoch (e.g., $\ul{\mI}_v$, $\ul{h}_v$).

For a given step $c \in \fragmentco{0}{k}$, consider any node $v$ in $\ul{\bF}$ that \emph{survives up to update $c$} in $\bF$ (i.e., its state immediately prior to update $c+1$ of the epoch). This means the node is neither rebalanced nor removed, its height remains constant, and the content referenced by $\mI_v$ remains unchanged. For any such surviving node, let $\ul{\mI}_v$ denote its corresponding bi-interval in $\ul{\bF}$, noting that $\bF_{\mI_v} \equiv \ul{\bF}_{\ul{\mI}_v}$ and $h_v = \ul{h}_v$.
In the approximate case, for any node $v$ that survives until update $c$, we consider its good spine sample to be the same as that of $v$ in \cref{thm:fvsf} for $\ul{\bF}$. If $v \in T(\bF)$ was not present in $T(\ul{\bF})$ or the content of $\bF_{\mI_v}$ changed since the beginning of the epoch, we initialize $Z_v$ from scratch by sampling\footnote{Condition \ref{it:good_spine_sample:ii} of \cref{def:good_spine_sample} depends on the target forest and on the threshold: on $\bG$ through the subforests $\bG_{\mJ}$, and on $k$ through the window set $W^{k}(\mI_v)$ and the optimal $k$-bounded alignments. Within an epoch, $Z_v$ is used with target $\ul{\bF}$ and threshold $8k$ (in \cref{thm:fvsf}), and with target $\bG$ after $c$ updates and threshold $2k-c$ for every $c \in \fragmentco{0}{k}$. The union bound in the proof of \cref{thm:fvsf} ranges over all these $\poly(N)$ (target, threshold) pairs, so, for a sufficiently large constant $\gamma$, a single sample $Z_v$ satisfies condition \ref{it:good_spine_sample:ii} for all of them simultaneously with high probability.}.

At the beginning of each epoch, we initialize a counter $\Phi_0 = \gamma \cdot k^6 \log n$ (resp. $\gamma \cdot k^2\varepsilon^{-6} \log^{6} N$ in the approximate setting) for a sufficiently large constant $\gamma$. We refer to this counter as the \emph{potential} (more generally, we denote it by $\Phi_c$ prior to the $c+1$-th update). We leave $T(\bF)$ unchanged and do not initialize any additional data structures. During the epoch, we decrease the potential upon certain updates (which will be detailed later). If the potential ever drops below $0$, we halt the process. For the remainder of the epoch, we simply report that $\ted(\bF,\bG) > k$ and wait for the next epoch to begin.

Otherwise, before the $c+1$-th update in the epoch, where $c \in \fragmentco{0}{k}$, we maintain the following invariants.
\begin{description}
    \item[Invariant for exact algorithm:]
    For each $v \in T(\bF)$ and $\mJ \in W^{2k-c}(\mI_v)$, at least one of these holds:
    \begin{enumerate}[(a)]
    \item We store a value $t_{v}(\mJ)$ such that $\ted(\bF_{\mI_v}, \bG_{\mJ}) \leq t_{v}(\mJ) \leq \ted^{(2k-c)}(\bF_{\mI_v}, \bG_{\mJ})$.
    \label{it:inv:a}
    \item Node $v$ survives until update $c$, and there is $\mL \in W^{4k}(\ul{\mI}_v)$ such that $\bG_{\mJ} \equiv \ul{\bF}_{\mL}$.
    \label{it:inv:b}
    \item Neither of the above holds.
    \label{it:inv:c}
    \end{enumerate}
    Additionally, the invariant guarantees that if $\ted(\ul{\bF}, \ul{\bG}) \leq 2k$, then the number of pairs $(v,\mJ)$ for which Invariant~\ref{it:inv:c} holds is at most $\Phi_c$.

    Proving that this final statement holds for $c = 0$ ensures correctness whenever the potential is exhausted. If this occurs, then $\ted(\ul{\bF}, \ul{\bG}) > 2k$, which implies $\ted(\bF, \bG) > 2k-c \geq k$.

    \item[Invariant for approximate algorithm:]
    For each $v \in T(\bF)$, $i \in \fragment{1}{\ell_v}$, and $\mJ \in \widetilde{W}^{2k-c}(\langle I_{v,i}\rangle)$ (resp. for each $v \in T(\bF)$ and $\mJ \in \widetilde{W}^{2k-c}_z(\mI_v)$ where $z \in Z_v$), at least one of these holds:
    \begin{enumerate}[(a)]
    \item We store a value $\tilde{t}_{v,i}(\mJ)$ (resp. $\tilde{t}_{v}(\mJ)$) that satisfies $\ted(\bF_{\langle I_{v,i} \rangle}, \bG_{\mJ}) \leq \tilde{t}_{v,i}(\mJ) \leq  (1 + 46\varepsilon)^{h_v} \cdot \ted^{(2k-c)}(\bF_{\langle I_{v,i} \rangle}, \bG_{\mJ})$ (resp. $\ted(\bF_{\mI_v}, \bG_{\mJ}) \leq \tilde{t}_{v}(\mJ) \leq (1 + 46\varepsilon)^{h_v} \cdot \ted^{(2k-c)}(\bF_{\mI_v}, \bG_{\mJ})$);
    \item Node $v$ survives up to update $c$ and there is $\mL \in \widetilde{W}^{4k}(\langle \ul{I}_{v,i} \rangle)$ (resp. $\mL \in \widetilde{W}^{4k}_z(\ul{\mI}_v)$) with $\bG_{\mJ} \equiv \ul{\bF}_{\mL}$.
    \item Neither of the above holds.
    \end{enumerate}
    Similarly to before, part of the invariant is that if $\ted(\ul{\bF}, \ul{\bG}) \leq 2k$, then the number of triples $(v,i,\mJ)$ and pairs $(v,\mJ)$ for which Invariant~\ref{it:inv:c} holds is at most $\Phi_c$.
\end{description}

At the start of an epoch, we discard all values stored during the previous epoch (each was inserted at cost $\Omega(\log N)$, to which its removal is charged): they are only guaranteed up to $\ted^{(k+1)}$, not up to $\ted^{(2k)}$, so initially only Invariants~\ref{it:inv:b} and \ref{it:inv:c} occur, as assumed in \cref{clm:inv:init}. During an epoch, no value is removed: a value whose window leaves $W^{2k-c}(\mI_v)$, $\widetilde{W}^{2k-c}(\langle I_{v,i} \rangle)$, or $\widetilde{W}^{2k-c}_z(\mI_v)$ only because the threshold decreased stays stored, is never accessed (every access is for a window of these families at the current threshold), and is discarded at the end of the epoch.

We observe that if we are in Case~\ref{it:inv:b} of the invariants, then we have indirect access to a value satisfying Invariant~\ref{it:inv:a} in both the exact and approximate case in time $\Oh(\log^{3} N)$. We achieve this by:
\begin{itemize}
    \item Query \cref{thm:has_match} to get $\mL \in W^{4k}(\ul{\mI}_v)$ (resp. $\mL \in \widetilde{W}^{4k}(\langle \ul{I}_{v,i} \rangle)$ and $\mL \in \widetilde{W}^{4k}_z(\ul{\mI}_v)$)
    with $\bG_{\mJ} \equiv \ul{\bF}_{\mL}$;
    \item Use \cref{thm:fvsf} to get a value $t$ that satisfies $\ted(\ul{\bF}_{\ul{\mI}_v}, \ul{\bF}_{\mL}) \leq t \leq \ted^{(8k)}(\ul{\bF}_{\ul{\mI}_v}, \ul{\bF}_{\mL})$
    (resp. a value $\tilde{t}$ that satisfies $\ted(\ul{\bF}_{\langle \ul{I}_{v,i} \rangle}, \ul{\bF}_{\mL}) \leq \tilde{t} \leq (1 + 46\varepsilon)^{\ul{h}_v} \cdot \ted^{(8k)}(\ul{\bF}_{\langle \ul{I}_{v,i} \rangle}, \ul{\bF}_{\mL})$ and $\ted(\ul{\bF}_{\ul{\mI}_v}, \ul{\bF}_{\mL}) \leq \tilde{t} \leq (1 + 46\varepsilon)^{\ul{h}_v} \cdot \ted^{(8k)}(\ul{\bF}_{\ul{\mI}_v}, \ul{\bF}_{\mL})$);
    \item Use this value as $t_{v}(\mJ)$ (resp. $\tilde{t}_{v,i}(\mJ)$ and $\tilde{t}_{v}(\mJ)$).
\end{itemize}
The following claim demonstrates that $t$ indeed satisfies $\ted(\bF_{\mI_v}, \bG_{\mJ}) \leq t \leq \ted^{(2k-c)}(\bF_{\mI_v}, \bG_{\mJ})$, and thus can be stored to ensure that Invariant~\ref{it:inv:a} holds for $v,\mJ$. It also establishes the corresponding bound for $\tilde{t}$ in the approximate case.

\begin{claim}\label{clm:fvsg:survive}
    Consider a node $v$ that survives up to update $c$.
    Then, for $\mJ \in W^{2k-c}(\mI_v)$ and $\mL \in W^{4k}(\ul{\mI}_v)$ with $\bG_{\mJ} \equiv \ul{\bF}_{\mL}$, we have:
    \[
        \ted(\bF_{\mI_v}, \bG_{\mJ}) = \ted(\ul{\bF}_{\ul{\mI}_v}, \ul{\bF}_{\mL}) \quad \text{ and } \quad
        \ted^{(8k)}(\ul{\bF}_{\ul{\mI}_v}, \ul{\bF}_{\mL}) \leq \ted^{(2k-c)}(\bF_{\mI_v}, \bG_{\mJ}).
    \]
    Moreover, for every $i \in \fragment{1}{\ell_v}$, $\mJ \in W^{2k-c}(\langle I_{v,i} \rangle)$ and $\mL \in W^{4k}(\langle \ul{I}_{v,i} \rangle)$ with $\bG_{\mJ} \equiv \ul{\bF}_{\mL}$, we have:
    \[
        \ted(\bF_{\langle I_{v,i} \rangle}, \bG_{\mJ}) = \ted(\ul{\bF}_{\langle \ul{I}_{v,i} \rangle}, \ul{\bF}_{\mL}) \quad \text{ and } \quad
        \ted^{(8k)}(\ul{\bF}_{\langle \ul{I}_{v,i} \rangle}, \ul{\bF}_{\mL}) \leq \ted^{(2k-c)}(\bF_{\langle I_{v,i} \rangle}, \bG_{\mJ}).
    \]
\end{claim}

\begin{claimproof}
    We present the proof for the first part only; the reader can verify that this argument implicitly establishes the second.
    Abbreviate $\ell_v$ with $\ell$.
    Let $\mJ = \langle \fragmentco{a_1'}{b_1'}, \ldots, \fragmentco{a_{\ell}'}{b_{\ell}'} \rangle$ and $\mL = \langle \fragmentco{\ul{a}_1'}{\ul{b}_1'}, \ldots, \fragmentco{\ul{a}_{\ell}'}{\ul{b}_{\ell}'} \rangle$.
    We observe that any alignment $\mA : \bF_{\mI_v} \onto \bG_{\mJ}$ can be transformed into an alignment $\underline{\mA} : \ul{\bF}_{\ul{\mI}_v} \onto \ul{\bF}_{\mL}$ by adding the element $(x+\ul{a}_{v,i} - a_{v,i}, y+\ul{a}_{i}'-a_{i}')$ to $\underline{\mA}$ for each $(x,y) \in \mA$ such that $x \in I_{v,i}$.
    Because $\bG_{\mJ} \equiv \ul{\bF}_{\mL}$ and $\bF_{\mI_v} \equiv \ul{\bF}_{\ul{\mI}_v}$, these alignments have the same cost.
    A similar transformation also works in the other direction (by subtracting $\ul{a}_{v,i} - a_{v,i}$and $\ul{a}_{i}'-a_{i}'$ from $x$ and $y$, respectively). Consequently, we derive that $\ted(\bF_{\mI_v}, \bG_{\mJ}) = \ted(\ul{\bF}_{\ul{\mI}_v}, \ul{\bF}_{\mL})$.

    For the $k$-boundedness, notice that the first transformation also works when $\mA$ is $2k-c$-bounded and $\underline{\mA}$ is $8k$-bounded. Indeed, from $|a_{v,i}- a_i'| \leq 4k-2c$, $|\ul{a}_{v,i} - \ul{a}_{i}'| \leq 8k$ and $|y-x| \leq 4k-2c$ follows
    \[
        |(x+\ul{a}_{v,i} - a_{v,i}) - (y+\ul{a}_{i}'-a_{i}')| \leq |x-y| + |\ul{a}_{v,i} - \ul{a}_{i}'| + |a_{v,i}- a_i'| \leq 16k,
    \]
    meaning $\underline{\mA}$ is appropriately bounded as well.
\end{claimproof}

As a final guarantee provided by the invariant, for any $v \in T(\bF)$ and $\mJ \in W^{2k-c}(\mI_v)$ (resp. $v \in T(\bF)$, $i \in \fragment{1}{\ell_v}$, $\mJ \in \widetilde{W}^{2k-c}\langle I_{v,i} \rangle)$ and $v \in T(\bF)$, $z \in Z_v$, $\mJ \in \widetilde{W}^{2k-c}_z(\mI_v)$), we can always re-establish the validity of the invariant by computing a value as in Invariant~\ref{it:inv:a}, provided that $v$ is a leaf, or that the invariant holds for each child $u$.

\begin{claim}\label{clm:fvsg:recompute}
    Consider $v \in T(\bF)$ and $\mJ \in W^{2k-c}(\mI_v)$ (resp. $v \in T(\bF)$, $i \in \fragment{1}{\ell_v}$, $\mJ \in \widetilde{W}^{2k-c}(\langle I_{v,i} \rangle)$ and $v \in T(\bF)$, $z \in Z_v$, $\mJ \in \widetilde{W}^{2k-c}_z(\mI_v)$). Suppose that either $v$ is a leaf, or for each child $u$ of $v$, all pairs $(u,\mJ')$ (resp. $(u,i',\mJ)$ and $(u,z',\mJ')$) satisfy Invariant~\ref{it:inv:a}, \ref{it:inv:b}, or \ref{it:inv:c}.

    Then, in time $\Oh((1 + \phi) \cdot k^2 \log^{3} N)$ (resp. $\Oh((1 + \phi) \cdot \varepsilon^{-9} \log^{13} N)$), we can compute a value $t_{v}(\mJ)$ (resp. $\tilde{t}_{v,i}(\mJ)$ and $\tilde{t}_{v}(\mJ)$) that satisfies Invariant~\ref{it:inv:a}, decreasing the potential $\Phi_c$ by $\phi$ in the process.
\end{claim}
\begin{claimproof}
    Ideally, we would like to proceed exactly as in \cref{clm:fvsf:recompute}, viewing each child as an oracle so that we can apply \cref{lem:combine} and \cref{lem:round}. However, unlike in \cref{clm:fvsf:recompute}, where values are stored explicitly, each child oracle stores the necessary values for recomputation differently.

    Here, we store the values explicitly only when Invariant~\ref{it:inv:a} holds. If Invariant~\ref{it:inv:b} holds (which we can verify via a query to \cref{thm:has_match}), we use the previously described process to obtain a proxy value. Finally, it is possible that neither checking the explicit storage nor querying \cref{thm:has_match} succeeds, meaning that Invariant~\ref{it:inv:c} holds. In this case, we decrease the potential $\Phi_c$ by one and apply this claim recursively to obtain the desired value.

    In the exact case, note that we incur exactly $1+\phi$ times the runtime of \cref{lem:combine}, plus the time required to retrieve a value, which can now be as costly as $\Oh(\log^{3} N)$, $\Oh((1+\phi) k^2 \log^{3} N)$. In the approximate case, summing up all the contributions, we obtain $\Oh(\log^{3} N \cdot (\varepsilon^{-1}\log^{3} N + \varepsilon^{-2} \log^{2} N) \cdot \varepsilon^{-5}\log^{5} N \cdot \varepsilon^{-2}\log^{2} N) = \Oh((1+\phi) \varepsilon^{-9}\log^{13} N)$, as each of the $1+\phi$ recomputations costs $\Oh(\varepsilon^{-9}\log^{13} N)$.
\end{claimproof}

\subparagraph*{Storing values.} We describe before the $c+1$-th update how we store values when Invariant~\ref{it:inv:a} holds.

Similarly to \cref{thm:fvsf}, we store the values $t_{v}(\mJ)$, $\tilde{t}_{v,i}(\mJ)$, and $\tilde{t}_{v}(\mJ)$ in data structures that depend on the identifiers. This time, the identifiers we consider are with respect to $\bG$.

Let $\Lambda$ be the set of all such identifiers (including those of the $8k+1$ padding characters that we keep, as for $\mS(\bF)$, in front of and after $\mS(\bG)$). In the exact case, we store the values in a balanced binary tree whose key set is $S \subseteq (\Lambda \times \Lambda \times V(T(\bF)) \times \mathbb{Z}_{\geq 0} \times \mathbb{Z}_{\geq 0}) \cup (\Lambda \times V(T(\bF)) \times \mathbb{Z}_{\geq 0})$. In the approximate case, the key set changes to $S \subseteq (\Lambda \times \Lambda \times V(T(\bF)) \times \mathbb{Z}_{\geq 0} \times \mathbb{Z}_{\geq 0}) \cup (\Lambda \times V(T(\bF)) \times \mathbb{Z}_{\geq 0} \times \mathbb{Z}_{\geq 0})$.
This allows us to store $t_{v}(\mJ)$, $\tilde{t}_{v,i}(\mJ)$, and $\tilde{t}_{v}(\mJ)$ as follows:
\begin{itemize}
    \item In the exact setting, for $v \in T(\bF)$ and $\mJ = \langle \fragmentco{a_1'}{b_1'}, \ldots, \fragmentco{a_{\ell_v}'}{b_{\ell_v}'} \rangle \in W^{k}(\mI_v)$, we assign a unique identifier $\mathsf{id}(v, \mJ) \coloneqq (\mathsf{id}(a_1'), \ldots, \mathsf{id}(a_{\ell_v}'), v, b_1'-a_1', \ldots, b_{\ell_v}'-a_{\ell_v}')$.

    Now, for $\mJ \in W^{2k-c}(\mI_v)$, we use $\mathsf{id}(v, \mJ)$ to index an element in the stored key set $S$, and we store the value $t_{v}(\mJ)$ under the key $(\mathsf{id}(a_1'), \ldots, \mathsf{id}(a_{\ell_v}'), v, b_1'-a_1', \ldots, b_{\ell_v}'-a_{\ell_v}')$ if and only if Invariant~\ref{it:inv:a} holds for $v$ and $\mJ$.
    We remark that this ``if and only if'' condition strictly applies to $\mJ \in W^{2k-c}(\mI_v)$. There might be some $\mJ \in W^{2k}(\mI_v) \setminus W^{2k-c}(\mI_v)$ for which we still store a value in memory, but this value will be unused; a similar situation occurs in the approximate case.

    \item In the approximate setting, for $v \in T(\bF)$, $i \in \fragment{1}{\ell_v}$, and $\mJ = \langle \fragmentco{a'}{b'} \rangle \in \widetilde{W}^{2k-c}(\langle I_{v,i} \rangle)$, we also assign a unique identifier: $\mathsf{id}(v, i, \mJ) \coloneqq (\mathsf{id}(a'), v, i, b'-a')$. This is again used to locate where $\tilde{t}_{v,i}(\mJ)$ is stored in the tree, and the value is stored there if and only if Invariant~\ref{it:inv:a} holds for $v$, $i$, and $\mJ$.

    \item In the approximate setting, for $v \in T(\bF)$, $z \in Z_v$, and $\mJ = \langle \fragmentco{a_1'}{b_1'}, \fragmentco{a_2'}{b_2'} \rangle \in \widetilde{W}^{2k-c}_z(\mI_v)$, we also assign a unique identifier: $\mathsf{id}(v, \mJ) \coloneqq (\mathsf{id}(a_1'), \mathsf{id}(a_2'), v, b_1'-a_1', b_2'-a_2')$. We use this identifier to index $\tilde{t}_{v}(\mJ)$ in the tree as well.
\end{itemize}
We let $\Gamma_{v}^{2k-c}$ denote the corresponding set of identifiers. In the exact setting, we define $\Gamma_{v}^{2k-c} \coloneqq \{\mathsf{id}(v,\mJ) \mid \mJ \in W^{2k-c}(\mI_v)\}$, while in the approximate setting, we define $\Gamma_{v}^{2k-c} \coloneqq \{\mathsf{id}(v,\mJ) \mid \mJ \in \widetilde{W}^{2k-c}_z(\mI_v), z \in Z_v\} \cup \{\mathsf{id}(v,i,\mJ) \mid \mJ \in \widetilde{W}^{2k-c}(\langle I_{v,i}\rangle), i \in \fragment{1}{\ell_v}\}$. Note that this set includes all $\mJ$, not just those for which Invariant~\ref{it:inv:a} holds. More generally, $\Gamma_{v}^{2k-c}$ is another representation of the set for which we want to maintain any of the invariants.

\subparagraph*{Handling updates.}
Fix $c \in \fragmentco{0}{k}$, and let $\bF,\bG$ denote the trees before the $c+1$-th update of the epoch.

Next, we describe the procedure for handling updates to $\bF$ and $\bG$. Let $y$ and $y'$ denote the indices where the update occurs. Specifically, if the update is a substitution or deletion in $\bF$ of a node $w \in \bF$, we set $y = \Left(w)$ and $y' = \Right(w)$. Otherwise, if a new node $w$ is inserted in $\bF$, we choose $y$ and $y'$ such that the new bi-order traversal is given by $\mS(\bF)\fragmentco{0}{y} \circ w \circ \mS(\bF)\fragmentco{y}{y'} \circ w \circ \mS(\bF)\fragmentco{y'}{2n}$.
Define $y,y'$ similarly if the update is in $\bG$.

Just as in \cref{thm:fvsf}, let $\bar{\bF},\bar{\bG}$ represent the trees $\bF,\bG$ after the update. Again, we apply a ``bar'' to all previously established notation to indicate its dependence on the updated tree $\bar{\bF}$ (e.g., $\bar{\mI}_v$ and $\bar{\mJ}_{v,q}$).

Similar to \cref{thm:fvsf}, let $P_{\mathrm{new}}$ denote the set of affected nodes in $T(\bar{\bF})$, and let $P_{\mathrm{old}}$ denote the remaining set of nodes in $T(\bar{\bF})$.
We define two additional sets, $A^{\circ}$ and $A^{\bullet}$, which satisfy $A^{\bullet} \subseteq A^{\circ}$, as follows:
\[
    A^{\bullet} \coloneqq (A^{\bullet}(y) \cup A^{\bullet}(y')) \cap P_{\mathrm{old}}
    \quad \text{ and } \quad
    A^{\circ} \coloneqq (A^{\circ}_{2k-c}(y) \cup A^{\circ}_{2k-c}(y')) \cap P_{\mathrm{old}},
\]
where $A^{\bullet}$ and $A^{\circ}_{2k-c}$ are defined as in \cref{lem:relevant_bound}.

Next, we present the three key lemmas that help us re-establish the invariant after the update.

\begin{claim}\label{clm:fvsg:0}
        Let $v \in P_{\mathrm{old}} \setminus A^{\circ}$ and $i \in \fragment{1}{\ell_v}$. Then, for $J \in W^{2k-c}(I_{v,i})$, we have $\mathsf{co}(I_{v,i}, J) \cap \{y, y'\} = \emptyset$.
    \end{claim}
    \begin{claimproof}
        The proof is the same as \cref{clm:fvsf:0}.
\end{claimproof}

\begin{claim}\label{clm:fvsg:1}
    Let $v \in P_{\mathrm{old}} \setminus A^{\bullet}$. Let $\mJ = \langle J_1, \ldots, J_{\ell_v} \rangle \in W^{2k-c}(\mI_v)$ and $\bar{\mJ} = \langle \bar{J}_1, \ldots, \bar{J}_{\ell_v} \rangle \in W^{2k-c-1}(\bar{\mI}_v)$ with $\mathsf{id}(v, \mJ) = \mathsf{id}(v, \bar{\mJ})$ and $U(\mJ) \cap \{y,y'\} = \emptyset$.
    Then, we have
    \[
        \ted(\bF_{\mI_v}, \bG_{\mJ}) = \ted(\bar{\bF}_{\bar{\mI}_v}, \bar{\bG}_{\bar{\mJ}}), \quad \ted^{(2k-c)}(\bF_{\mI_v}, \bG_{\mJ}) \leq \ted^{(2k-c-1)}(\bar{\bF}_{\bar{\mI}_v}, \bar{\bG}_{\bar{\mJ}}) \quad \text{and} \quad \bG_{\mJ} \equiv \bar{\bG}_{\bar{\mJ}}.
    \]
    Similarly, for every $i \in \fragment{1}{\ell_v}$ and $\mJ = \langle J \rangle \in W^{2k-c}(\langle I_{v,i} \rangle)$ and  $\bar{\mJ} = \langle \bar{J} \rangle \in W^{2k-c-1}(\langle \bar{I}_{v,i} \rangle)$ such that $\mathsf{id}(\mJ,i,v) = \mathsf{id}(\bar{\mJ},i,v)$ and $J \cap \{y,y'\} = \emptyset$, the same relations hold.
\end{claim}
\begin{claimproof}
    Notice that this claim is similar to \cref{clm:fvsf:1}, with the difference that $\mJ$ and $\bar{\mJ}$ are considered with respect to $2k-c$ and $2k-c-1$ instead of $k$, and we do not require special assumptions when $v \in A^{\circ} \setminus A^{\bullet}$ and the update is a deletion.

    Up to the point where we prove that the $k$-boundedness is preserved, the proof remains the same (and notice that in the same proof we implicitly establish $\bG_{\mJ} \equiv \bar{\bG}_{\bar{\mJ}}$). It remains to show that \cref{eq:fvsf:1} also holds in this case; that is, $|\bar{x} - \bar{x}'| \leq 4k-2c-2$ implies $|x - x'| \leq 4k-2c$, where $x,\bar{x}$ and $x',\bar{x}'$ are defined as in \cref{clm:fvsg:1} and represent positions in $\bF,\bar{\bF}$ and $\bG, \bar{\bG}$, respectively.
    Because the update occurs in either $\bF$ or $\bG$, but not both, we have either $|x-\bar{x}| \leq 2$ and $|x'-\bar{x}'| = 0$, or $|x-\bar{x}| = 0$ and $|x'-\bar{x}'| \leq 2$. In either case, by the triangle inequality, we have $|x - x'| \leq |\bar{x} - \bar{x}'| + |x'-\bar{x}'| + |x-\bar{x}| \leq 4k-2c$.
\end{claimproof}

\begin{claim}\label{clm:fvsg:2}
        We have $\bar{\Gamma}_v^{2k-c-1} \subseteq \Gamma_v^{2k-c}$ for $v \in P_{\mathrm{old}} \setminus A^{\circ}$.
        Furthermore, in the approximate setting, for $v \in A^{\circ} \setminus A^{\bullet}$ call $\mathsf{id} \in \bar{\Gamma}_v^{2k-c-1}$ \emph{touched} if the update is in $\bG$ and its window satisfies $U(\mJ) \cap \{y, y'\} \neq \emptyset$ before the update or $U(\bar{\mJ}) \cap \{\bar y, \bar y'\} \neq \emptyset$ after it (with $\bar y, \bar y'$ as in the proof of \cref{thm:fvsf}). Every untouched $\mathsf{id} \in \bar{\Gamma}_v^{2k-c-1}$ lies in $\Gamma_v^{2k-c}$ with $U(\mJ) \cap \{y, y'\} = \emptyset$. Over all $v \in A^{\circ} \setminus A^{\bullet}$, there are $\Oh(k \varepsilon^{-6} \log^{6} N)$ touched identifiers, and they can be listed in $\Oh(k \varepsilon^{-6} \log^{7} N)$ time.
 \end{claim}
 \begin{claimproof}
    First, consider the exact setting and let $v \in P_{\mathrm{old}} \setminus A^{\circ}$. Let $\mathsf{id} \in \bar{\Gamma}_v^{2k-c-1}$ and suppose $\mathsf{id} = (\iota_1, \ldots, \iota_{\ell_v}, v, q_1,\ldots, q_{\ell_v})$. Let $x_i$ and $\bar{x}_i$ be the positions of $\iota_i$ before and after the update, respectively. Thus, $\mathsf{id}$ corresponds to $\bar{\mJ} = \langle \fragmentco{\bar{x}_1}{\bar{x}_1+q_1}, \ldots, \fragmentco{\bar{x}_{\ell_v}}{\bar{x}_{\ell_v}+q_{\ell_v}} \rangle \in W^{2k-c-1}(\bar{\mI}_v)$, which satisfies $|\bar{x}_{i}-\bar{a}_{v,i}| \leq 4k-2c-2$ and $|\bar{x}_{i}+q_i-\bar{b}_{v,i}| \leq 4k-2c-2$ for all $i \in \fragment{1}{\ell_v}$.

    Since again we have either $|\bar{a}_{v,i}-a_{v,i}|=0$, $|\bar{b}_{v,i}-b_{v,i}|=0$, and $|\bar{x}_{i}-x_{i}| \leq 2$, or $|\bar{a}_{v,i}-a_{v,i}|\leq 2$, $|\bar{b}_{v,i}-b_{v,i}| \leq 2$, and $|\bar{x}_{i}-x_{i}| = 0$, we conclude that $|x_{i}-a_{v,i}| \leq |\bar{x}_{i}-\bar{a}_{v,i}| + |\bar{x}_{i}-x_{i}| + |\bar{a}_{v,i}-a_{v,i}| \leq 4k-2c$. Similarly, $|{x}_{i}+q_i-{b}_{v,i}| \leq |\bar{x}_{i}+q_i-\bar{b}_{v,i}|+|\bar{x}_{i}-x_{i}|+|\bar{b}_{v,i}-b_{v,i}| \leq 4k-2c$. Therefore, we conclude $\mathsf{id} \in {\Gamma}_v^{2k-c}$. As a side note, since $v \notin A^{\circ}$ and $U(\mJ) \cap \{y,y'\} = \emptyset$ (which follows from \cref{clm:fvsg:0}), no update position intersects $U(\mI_v)$ or $U(\mJ)$. Furthermore, we know that $\bar{x}_i$ also existed before the update (and thus $x_i$ is well-defined).

    In the approximate setting, an update in $\bF$ shifts $\mI_v$ by at most two positions, moves no window, and leaves $Z_v$, the offsets of $z \in Z_v$ in $\mI_v$, and hence $\tilde{\sL}_{v,z}$ unchanged, whereas an update in $\bG$ moves the windows and not $\mI_v$. Let $v \in A^{\circ} \setminus A^{\bullet}$ and let $\mathsf{id} \in \bar{\Gamma}_v^{2k-c-1}$ be untouched, generated after the update by a triple $(z, q, z')$ with $q \in \tilde{\sL}_{v,z}$ (resp.\ by a length $m$) for the threshold $2k-c-1$. The start identifiers of $\mathsf{id}$ and the anchor $z'$ (both occurrences of which lie in $U(\bar{\mJ})$, as $q_1', q_2' \geq 1$) exist before the update with the same offsets, $q$ (resp.\ $m$) is allowed for the threshold $2k-c$ as $\tilde{\sL}^{2k-c-1}(\cdot) \subseteq \tilde{\sL}^{2k-c}(\cdot)$, and the endpoint computation of the exact setting applies; hence the same triple (resp.\ length) generates $\mathsf{id} \in \Gamma_v^{2k-c}$ before the update, and $U(\mJ) \cap \{y, y'\} = \emptyset$ by definition. For $v \in P_{\mathrm{old}} \setminus A^{\circ}$, every identifier is untouched by \cref{clm:fvsg:0}, which gives the first statement in the approximate setting. The touched identifiers are generated by the triples of \cref{lem:relevant_bound}\ref{it:relevant_bound:iv}, applied with $2k-c$ in place of $k$ to $\bG$ and $x \in \{y, y'\}$ and with $2k-c-1$ in place of $k$ to $\bar{\bG}$ and $x \in \{\bar y, \bar y'\}$, and by the mono-interval windows containing $x$, listed by the at most $\min(m, 4k+1)$ starts in $\fragmentoc{x-m}{x}$ per length $m$; the bounds follow from \cref{lem:relevant_bound}\ref{it:relevant_bound:ii}--\ref{it:relevant_bound:iv} and $|A^{\circ}| = \Oh(k \log n)$.
\end{claimproof}

The previous claims help us characterize the terms for which invariants are preserved.

\begin{claim}\label{clm:fvsg:3}
    Let $v \in P_{\mathrm{old}} \setminus A^{\bullet}$, and consider $\mathsf{id} \in \bar{\Gamma}_v^{2k-c-1} \cap \Gamma_v^{2k-c}$. Suppose that the interval $\mJ$ before the update, corresponding to $\mathsf{id}(v,\mJ) = \mathsf{id}$ (resp. the pair $(i, \mJ)$ corresponding to $\mathsf{id}(v,i,\mJ) = \mathsf{id}$), satisfies $U(\mJ) \cap \{y,y'\} = \emptyset$. Moreover, let $\bar{\mJ}$ be the corresponding interval after the update such that $\mathsf{id}(v,\bar{\mJ}) = \mathsf{id}$ (resp. $\mathsf{id}(v,i,\bar{\mJ}) = \mathsf{id}$). Then, if either Invariant~\ref{it:inv:a} or Invariant~\ref{it:inv:b} holds for $\mJ$ (resp. $i,\mJ$) before the update, the same invariant holds for $\bar{\mJ}$ (resp. $i, \bar{\mJ}$) after the update, without requiring any modifications to its corresponding stored data.
\end{claim}
\begin{claimproof}
    We only provide the proof for the exact setting; an almost analogous argument applies to the approximate case.

    If Invariant~\ref{it:inv:a} holds for $v,\mJ$, then combining the fact that $U(\mJ) \cap \{y,y'\} = \emptyset$ with \cref{clm:fvsg:1} tells us that the same value we store for $v,\mJ$ also maintains valid bounds for $v,\bar{\mJ}$ (by \cref{clm:fvsg:1}, the lower bound remains the same, while the upper bound only gets looser). Observe that if Invariant~\ref{it:inv:a} does not hold for $v,\mJ$, then we store no value, and we will consequently store no value for $v,\bar{\mJ}$ (which covers the ``if and only if'' part of the storing invariant we have for such $v,\mJ$).

    Otherwise, suppose Invariant~\ref{it:inv:b} holds for $v,\mJ$. Since $v$ survives up to update $c$ and $v \notin A^{\bullet} \cup P_{\mathrm{new}}$, it follows that $v$ also exists in $\bar{\bF}$. Consequently, $v$ survives up to update $c+1$.
    Moreover, if there exists $\mL \in W^{4k}(\ul{\mI}_v)$ (resp. $\mL \in \widetilde{W}^{4k}(\ul{\mI}_v)$) such that $\bG_{\mJ} \equiv \ul{\bF}_{\mL}$, then by \cref{clm:fvsg:1}, $\bG_{\mJ} \equiv \bar{\bG}_{\bar{\mJ}}$. Thus, for the same $\mL$, we have $\bar{\bG}_{\bar{\mJ}} \equiv \ul{\bF}_{\mL}$.
\end{claimproof}

We observe that for nodes $v \in P_{\mathrm{old}} \setminus A^{\circ}$, no update is required. Indeed, by \cref{clm:fvsg:2}, we know that $\bar{\Gamma}_v^{2k-c-1} \subseteq \Gamma_v^{2k-c}$. Thus, for each $\mathsf{id} \in \bar{\Gamma}_v^{2k-c-1}$, as long as the corresponding interval $\mJ$ satisfies $U(\mJ) \cap \{y,y'\} = \emptyset$, \cref{clm:fvsg:3} ensures that the invariants are preserved by simply leaving the storage as is. The condition $U(\mJ) \cap \{y,y'\} = \emptyset$ follows immediately from \cref{clm:fvsg:0}. Moreover, the number of identifiers $\mathsf{id} \in \bar{\Gamma}_v^{2k-c-1}$ for which Invariant~\ref{it:inv:c} holds also does not increase.

Regarding the nodes in $A^{\circ} \cup P_{\mathrm{new}}$, after locating them via a DFS, we restore the invariant by proceeding bottom-up from the nodes with the smallest height to the largest, in a manner similar to \cref{thm:fvsf}.

Now, consider an arbitrary node $v \in A^{\circ} \cup P_{\mathrm{new}}$ visited during this process. In the exact algorithm, we recompute via \cref{clm:fvsg:recompute} all values $t_{v}(\bar{\mJ})$ for $\bar{\mJ} \in W^{2k-c-1}(\bar{\mI}_v)$, and we store them with their corresponding identifiers in $\bar{\Gamma}_v^{2k-c-1}$ (replacing any previously stored value at the same identifier) such that, after this step, Invariant~\ref{it:inv:a} holds for all identifiers in $\bar{\Gamma}_v^{2k-c-1}$.
In the approximate setting, we only perform this full recomputation when $v \in A^{\bullet} \cup P_{\mathrm{new}}$. Therefore, it remains to specify our procedure for the approximate case when $v \in A^{\circ} \setminus A^{\bullet}$; let us now assume this is the case.

For any such $v$, we list the touched identifiers of \cref{clm:fvsg:2} and establish Invariant~\ref{it:inv:a} for them via \cref{clm:fvsg:recompute}; no value is removed. Every untouched $\mathsf{id} \in \bar{\Gamma}_v^{2k-c-1}$ lies in $\Gamma_v^{2k-c}$ with $U(\mJ) \cap \{y, y'\} = \emptyset$ by \cref{clm:fvsg:2}, so by \cref{clm:fvsg:1} and \cref{clm:fvsg:3} the invariants are preserved where we do not recompute.

\subparagraph*{Initialization at each batch start.}
We just showed how to maintain the invariant for $c+1$ provided that it holds for $c$. It remains to show that it holds when $c = 0$.

\begin{claim}\label{clm:inv:init}
    Suppose that $\ted(\ul{\bF}, \ul{\bG}) \leq 2k$. Then, there are at most $\Oh(k^6 \log n)$ (resp. $\Oh(k^2 \varepsilon^{-6} \log^{6} N)$) many $v \in T(\ul{\bF})$ and $\mJ \in W^{2k}(\ul{\mI}_v)$ (resp. $v \in T(\ul{\bF})$, $i \in \fragment{1}{\ell_v}$, $\mJ \in \widetilde{W}^{2k}(\langle \ul{I}_{v,i}\rangle)$ and $v \in T(\ul{\bF})$, $z \in Z_v$, $\mJ \in \widetilde{W}^{2k}_z(\ul{\mI}_v)$) for which Invariant~\ref{it:inv:b} does not hold, meaning that Invariant~\ref{it:inv:c} holds instead.

    Therefore, setting $\Phi_0 = \gamma \cdot k^6 \log n$ (resp. $\gamma \cdot k^2\varepsilon^{-6} \log^{6} N$) for a sufficiently large constant $\gamma$ ensures that the invariant holds at the beginning.
\end{claim}
\begin{claimproof}
    Let $\mA : \ul{\bG} \onto \ul{\bF}$ be an optimal alignment of cost at most $2k$ (from \cref{clm:TD_ub_diff_ted} follows that we can assume it is $2k$-bounded).
    Recall that $\mS(\ul{\bG})\position{y} = \$$ for $y \notin \fragmentco{0}{2n'}$ and $\mS(\ul{\bF})\position{x} = \$$ for $x \notin \fragmentco{0}{2n}$. The windows $\mJ \in W^{2k}(\ul{\mI}_v)$ and the witnesses $\mL \in W^{4k}(\ul{\mI}_v)$ below may contain such padding positions: as $U(\ul{\mI}_v) \subseteq \fragmentco{0}{2n}$ and $|2n - 2n'| \leq 4k$ (\cref{prop:size_lb}), we have $U(\mJ) \subseteq \fragmentco{-4k}{2n'+8k}$ and $U(\mL) \subseteq \fragmentco{-8k}{2n+8k}$. Define $\mA' \colon \mathbb{Z} \to \mathbb{Z} \cup \{\bot\}$ by
\[
    \mA'(y) \coloneqq
    \begin{cases}
        x & \text{if } (y, x) \in \mA, \\
        y & \text{if } y < 0, \\
        y - 2n' + 2n & \text{if } y \geq 2n', \\
        \bot & \text{otherwise}.
    \end{cases}
\]
Note that $\mA'$ is not an alignment, but it is strictly monotone and satisfies $|y - \mA'(y)| \leq 4k$ whenever $\mA'(y) \neq \bot$.\footnote{The former as $\mA$ is non-crossing and $\mA'(y) < 0 \leq \mA'(y') < 2n \leq \mA'(y'')$ for all $y < 0 \leq y' < 2n' \leq y''$ with $\mA'(y') \neq \bot$; the latter as $\mA$ is $2k$-bounded and $|2n - 2n'| \leq 2 \ted(\ul{\bF}, \ul{\bG}) \leq 4k$.}
    Let $P$ be the set of all $y \in \fragment{0}{2n'}$ such that at least one of the following four conditions holds:
    \begin{enumerate}[(1)]
        \item \label{it:inv:init:1} $\mA'(y) = \bot$;
        \item \label{it:inv:init:2} $\mA'(y) \neq \bot$ and the labels of $\mS(\ul{\bG})\position{y}$ and $\mS(\ul{\bF})\position{\mA'(y)}$ differ;
        \item \label{it:inv:init:3} $\mA'(y), \mA'(y-1) \neq \bot$ and $\mA'(y-1) \neq \mA'(y) - 1$;
        \item \label{it:inv:init:4} $\Other(y) \in P$ by \ref{it:inv:init:1}, \ref{it:inv:init:2}, or \ref{it:inv:init:3} (for $y \in \fragmentco{0}{2n'}$; $2n'$ has no twin).
    \end{enumerate}
    Each $y \in P$ by \ref{it:inv:init:1}, \ref{it:inv:init:2}, or \ref{it:inv:init:3} is charged to half the cost of an edit (for \ref{it:inv:init:3}, to $\fragmentoo{\mA'(y-1)}{\mA'(y)} \subseteq \fragmentco{0}{2n}$, which by monotonicity is non-empty, consists of deleted positions, and is disjoint for distinct $y$), and each $y \in P$ by \ref{it:inv:init:4} to its twin. This means $|P| \leq 2 \cdot 4k = 8k$.

    Consider first the exact setting. For every $v \in T(\ul{\bF})$ and $\mJ \in W^{2k}(\ul{\mI}_v)$ such that $U(\mJ) \cap P = \emptyset$, we claim that there exists $\mL \in W^{4k}(\ul{\mI}_v)$ (whose endpoints may thus deviate from those of $\ul{\mI}_v$ by $2 \cdot 4k = 8k$, as required by Invariant~\ref{it:inv:b}) such that $\ul{\bF}_{\mL} \equiv \ul{\bG}_{\mJ}$. If $\mJ = \langle \fragmentco{a_1'}{b_1'}, \ldots, \fragmentco{a_{\ell_v}'}{b_{\ell_v}'}\rangle$, then for every $i \in \fragment{1}{\ell_v}$ and $y \in \fragmentco{a_i'}{b_i'}$, we have $y \notin P$ and $\mA'(y) \neq \bot$ (by \ref{it:inv:init:1} for $y \in \fragment{0}{2n'}$, by definition otherwise). Assume first that every $J_i$ is non-empty, and set $L_i \coloneqq \fragmentco{\mA'(a_i')}{\mA'(b_i'-1)+1}$. For $y-1, y \in J_i$, we have $\mA'(y-1), \mA'(y) \neq \bot$ and $y \notin P$, hence $\mA'(y-1) = \mA'(y) - 1$ by \ref{it:inv:init:3} (for $y \notin \fragment{0}{2n'}$ by definition); thus $\mA'(a_i' + \hat{y}) = \mA'(a_i') + \hat{y}$ for all $\hat{y} \in \fragmentco{0}{b_i' - a_i'}$, and in particular $|L_i| = |J_i|$ and $\mA'(b_i'-1)+1 = \mA'(a_i') + |J_i|$.

    We claim that
    $\mL = \langle L_1, \ldots, L_{\ell_v} \rangle$
    satisfies the above properties. By the shift identity, $\mB \coloneqq \{(y, \mA'(y)) \mid y \in U(\mJ)\}$ is a bijection between positions in $\mJ$ and $\mL$, and it preserves labels by \ref{it:inv:init:2} and because $\mA'$ maps padding positions to padding positions. Moreover, it preserves the twin relation: for $(y,x) \in \mB$ with $y \in \fragmentco{0}{2n'}$, we have $(y,x) \in \mA$, so either both $\Other(y) \notin U(\mJ)$ and $\Other(x) \notin U(\mL)$, or $(\Other(y),\Other(x)) \in \mB$, by \cref{def:TED_on_biintervals}\ref{def:TED_on_biintervals:iii} and the injectivity of $\mA'$. Thus, $\ul{\bF}_{\mL} \equiv \ul{\bG}_{\mJ}$. (For $\ell_v = 2$, the conditions on the relative position of the two intervals in the definition of $\equiv$ hold as well: since $\mA'$ is strictly monotone, $L_1$ precedes $L_2$ if and only if $J_1$ precedes $J_2$, and likewise with the roles exchanged, and if $J_1 \cap J_2 \neq \emptyset$, then $U(\mJ)$ is a single interval on which $\mA'(y) - y$ is constant, so both components are shifted by the same amount.)

    To see that $\mL \in W^{4k}(\ul{\mI}_v)$, we use the triangle inequality: since $|a_i' - \mA'(a_i')|, |a_i' - a_{v,i}| \leq 4k$, we get $|a_{v,i} - \mA'(a_i')| \leq 8k$, and similarly $|b_{v,i} - (\mA'(b_i'-1)+1)| \leq 8k$.

    Note that the above proof carries over almost as is to the approximate setting (as $|L_i| = |J_i|$ and $\tilde{\sL}^{2k}(\cdot) \subseteq \tilde{\sL}^{4k}(\cdot)$, the length constraints of the $4k$-families are inherited from those of the $2k$-families), where this time we consider $i \in \fragment{1}{\ell_v}$, $\mJ \in \widetilde{W}^{2k}(\langle \ul{I}_{v,i}\rangle)$; or $z \in Z_v$, $\mJ \in \widetilde{W}^{2k}_z(\ul{\mI}_v)$. In these cases, we look for $\mL \in \widetilde{W}^{4k}(\langle \ul{I}_{v,i}\rangle)$ or $\mL \in \widetilde{W}^{4k}_z(\ul{\mI}_v)$, respectively, such that $\ul{\bF}_{\mL} \equiv \ul{\bG}_{\mJ}$ holds. For $\mJ \in \widetilde{W}^{2k}_z(\ul{\mI}_v)$ generated by a node $z' \in \ul{\bG}$, both occurrences of $z'$ lie in $U(\mJ)$ (as $q_1', q_2' \geq 1$) and hence not in $P$, so $z'$ is matched to a node $z''$ of $\ul{\bF}$ $\mA'(\Left(z')) = \Left(z'')$ and $\mA'(\Right(z')) = \Right(z'')$, and the collection $\mL$ constructed above is anchored at $z''$ with the same tuple $q$, i.e., $\mL \in \widetilde{W}^{4k}_z(\ul{\mI}_v)$.

    Note that in the following proof, we assume that both intervals of $\mJ$ are non-empty (otherwise, the projection onto $\mA$ is ill-defined). It is not difficult to see that if the $i$-th interval $J_i$ of $\mJ$ is empty, then for the corresponding interval $L_i$ of $\mL$, we can simply choose any empty $L_i \in W^{4k}(\ul{I}_{v,i})$, e.g., $L_i = \fragmentco{\ul{a}_{v,i}}{\ul{a}_{v,i}}$; note that an empty $J_i \in W^{2k}(\ul{I}_{v,i})$ forces $|\ul{I}_{v,i}| \leq 8k$, so this $L_i$ indeed lies in $W^{4k}(\ul{I}_{v,i})$.

    This allows us to conclude that Invariant~\ref{it:inv:b} holds for all $v \in T(\ul{\bF})$ and $\mJ \in W^{2k}(\ul{\mI}_v)$, except for those where $U(\mJ) \cap P \neq \emptyset$, for which Invariant~\ref{it:inv:c} holds. Since $U(\mJ) \cap P \neq \emptyset$ implies that $\bigcup_{p \in P} \fragment{p-8k}{p+8k} \cap U(\ul{\mI}_v) \neq \emptyset$, we can bound the number of such instances by $\sum_{p \in P, v \in A^{\circ}_{4k}(p)} |W^{2k}(\ul{\mI}_v)|$. Here, $A^{\circ}_{4k}$ is defined as in \cref{lem:relevant_bound}, and the sum by \cref{lem:relevant_bound}\ref{it:relevant_bound:i} evaluates to $\Oh(k \cdot k \log n \cdot k^4) = \Oh(k^6 \log n)$.
    For the approximate setting, we can use \cref{lem:relevant_bound}\ref{it:relevant_bound:ii} and \cref{lem:relevant_bound}\ref{it:relevant_bound:iii} to bound the number of terms for which Invariant~\ref{it:inv:b} does not hold.
\end{claimproof}

\subparagraph*{Runtime in the exact case.}
For the exact case, keeping all black-box data structures up to date (including the delayed updates) costs up to $\Oh(k^7 \log^{2} N)$ amortized time per update; this worst-case bound is achieved by \cref{thm:fvsf}.

To maintain the invariants after each update, we need to bound the overall number of times we use \cref{clm:fvsg:recompute}.
Since we always recompute from scratch, this amounts to $|A^{\circ}|+|P_{\mathrm{new}}|$ times $\Oh(k^4)$ (the number of intervals $\mJ$ in any $W^{k}(\mI_v)$). Now, $|A^{\circ}|$ and $|P_{\mathrm{new}}|$ are $\Oh(k \log n)$ and (on average) $\Oh(\log n)$ by \cref{lem:relevant_bound}\ref{it:relevant_bound:i} and the properties of $T(\bF)$. For each such recomputation, we spend $\Oh((1+\phi)k^2 \log^{3} N)$ time, where $\phi$ is a decrease in potential that never increases.
Since the potential at the beginning is $\Theta(k^6 \log n)$, over $k$ updates, the runtime is $\Oh((k \log n \cdot k^4 + k^6 \log n / k) \cdot k^{2} \log^{3} N) = \Oh(k^7 \log^{4} N)$.

\subparagraph*{Runtime in the approximate case.}
Similar to \cref{thm:fvsf}, the number of times we use \cref{clm:fvsg:recompute} is bounded by $\Oh((|A^{\bullet}| +|P_{\mathrm{new}}|) k\varepsilon^{-5} \log^{5} N) = \Oh(k\varepsilon^{-5} \log^{6} N)$ on average. On the other hand, the number of times we use \cref{clm:fvsg:recompute} for nodes in $A^{\circ} \setminus A^{\bullet}$ is bounded using \cref{lem:relevant_bound}\ref{it:relevant_bound:ii} and \cref{lem:relevant_bound}\ref{it:relevant_bound:iii}. The $\Oh(k \varepsilon^{-6} \log^{6} N)$ touched identifiers of \cref{clm:fvsg:2} are listed in $\Oh(k \varepsilon^{-6} \log^{7} N)$ time.

Overall, we use it at most $\Oh(k\varepsilon^{-6} \log^6 N)$ times.
Similar to before, for each such recomputation, we spend $\Oh((1+\phi) \cdot \varepsilon^{-9} \log^{13} N)$ time, where $\phi$ is a decrease in potential that never increases.
Since the potential at the beginning is $\Theta(k^{2} \varepsilon^{-6} \log^{6} N)$, over $k$ updates, the runtime is $\Oh((k\varepsilon^{-6} \log^6 N + k^{2} \varepsilon^{-6} \log^{6} N/k) \cdot \varepsilon^{-9} \log^{13} N) = \Oh(k \varepsilon^{-15}\log^{19} N)$.
Here, we omitted the other data structures we maintain from the analysis, since their update costs are non-dominant.
\end{proof}

We remark that some logarithmic factors above can be saved by using hash tables instead of balanced binary trees. However, we chose to use the latter to obtain a deterministic algorithm, at least in the exact case. For the approximate case, the spine node sampling would need to be derandomized to achieve the same deterministic guarantee.

Finally, we can prove our two main results on dynamic \TED.

\dynapproxted

\begin{proof}
    We maintain the dynamic algorithm of \cref{thm:fvsg}\eqref{it:fvsg:2} with a smaller $\varepsilon'$ set to $\Theta(\varepsilon/\log N)$, fixed in advance; note that $h_r = \Oh(\log n) = \Oh(\log N)$ for a sufficiently small constant.

    After each update, if $|n' - n| > k$, then we report that $\ted(\bF, \bG) > k$ (which is correct because of \cref{prop:size_lb}). Otherwise, we have $|n' - n| \leq k$ and, most importantly, we have $\mJ \in W^{k}(\mI_r)$, where $r$ is the root of $T(\bF)$, for which $\mI_r = \langle \fragmentco{0}{2n} \rangle$ and $\mJ = \langle \fragmentco{0}{2n'} \rangle$.

    Thus, we can use \cref{lem:round} to obtain a value $t$ from the value stored by \cref{thm:fvsg}\eqref{it:fvsg:2} that satisfies $\ted(\bF, \bG) = \ted(\bF_{\mI_r}, \bG_{\mJ}) \leq t \leq (1+46\varepsilon')^{h_r+1} \ted^{(k)}(\bF_{\mI_r}, \bG_{\mJ}) = (1+46\varepsilon')^{h_r+1} \ted^{(k)}(\bF, \bG)$.

    Since $h_r = \Oh(\log n)$, we have $t \leq (1+46\varepsilon')^{h_r+1} \ted^{(k)}(\bF,\bG) \leq (1+\varepsilon) \ted^{(k)}(\bF,\bG)$ for a sufficiently small hidden constant. If $t > (1+\varepsilon)k$, we report $\ted(\bF,\bG) > k$; this is correct, because $\ted(\bF,\bG) \leq k$ would give $\ted^{(k)}(\bF,\bG) = \ted(\bF,\bG)$ by \cref{clm:TD_ub_diff_ted} and hence $t \leq (1+\varepsilon)k$. Otherwise, we output $t$: if $\ted(\bF,\bG) \leq k$, then $\ted(\bF,\bG) \leq t \leq (1+\varepsilon)\ted(\bF,\bG)$, and if $\ted(\bF,\bG) > k$, then $\ted(\bF,\bG) \leq t \leq (1+\varepsilon)k < (1+\varepsilon)\ted(\bF,\bG)$; in both cases $t$ is a $(1+\varepsilon)$-approximation.

    The running time is dominated by the update time of \cref{thm:fvsg}\eqref{it:fvsg:2}.
\end{proof}

\dynexactted

\begin{proof}
    The proof is identical to that of \cref{thm:dynapproxted}, except that we use the exact counterpart rather than the approximate one throughout. The only minor simplification is that \cref{thm:fvsg}\ref{it:fvsg:1} stores the value $t$ directly, so we do not need to use \cref{lem:round} to obtain it.
\end{proof}

\subsection{Omitted Proofs and Data Structure Constructions}
\label{sec:omitted}

\subsubsection{Proof of \cref{lem:findNN}}

The primary ingredients for \cref{lem:findNN} are an existing data structure for dynamic 2D orthogonal range reporting~\cite{M03} (\cref{lem:dynamic-range-reporting}) and a technique for approximating $L_1$ nearest neighbors via orthogonal range reporting (\cref{lem:red_nn}).

\begin{lemma}[Dynamic 2D Orthogonal Range Reporting~\cite{M03}]
\label{lem:dynamic-range-reporting}
Let $(X, \leq_X)$ and $(Y, \leq_Y)$ be totally ordered sets whose elements are accessed exclusively through comparison oracles. There exists a deterministic data structure that maintains a finite set $S \subseteq X \times Y$ of size $n$ under insertions and deletions, supporting orthogonal range emptiness queries: Given $x_1, x_2 \in X$ and $y_1, y_2 \in Y$, the structure determines whether there exists an $(x, y) \in S$ such that $x_1 \leq_X x \leq_X x_2$ and $y_1 \leq_Y y \leq_Y y_2$ (and if so, reports it). Both updates and queries are supported in time $\mathcal{O}(\mathsf{T_{\mathrm{comp}}}\log n)$ where $\mathsf{T_{\mathrm{comp}}}$ bounds the time for comparing the order between elements in $(X, \leq_X)$ and $(Y, \leq_Y)$.

Crucially, no assumption is made that $X$ or $Y$ are drawn from a fixed universe. The active coordinates may evolve dynamically with insertions and deletions, and the data structure maintains correctness by relying solely on the relative ordering of the currently stored elements. \lipicsEnd
\end{lemma}

\begin{lemma}\label{lem:red_nn}
    Suppose there are ordered sets $X = \{x_1, \ldots, x_n\}$ and $Y = \{y_1, \ldots, y_m\}$, along with a subset $S \subseteq X \times Y$.
    Assume there exists a data structure capable of answering range emptiness queries in $\Oh(\sT)$ time. Specifically, given a query rectangle $\fragment{a}{a'} \times \fragment{b}{b'}$, it verifies whether there exists an element $(x_i, y_j) \in S$ such that $(i,j) \in \fragment{a}{a'} \times \fragment{b}{b'}$ (and if so, reports it).

    Then, for any $0 < \varepsilon \leq 1$, in $\Oh(\sT \varepsilon^{-1} \log(n+m) + \sT \varepsilon^{-2})$ time, we can answer the following approximate query: Given integers $\hi$ and $\hj$ (possibly outside $\fragment{1}{n}$ and $\fragment{1}{m}$) and a range $R = \fragment{a}{a'} \times \fragment{b}{b'}$, find an index pair $(i,j) \in R$ with $(x_i,y_j) \in S$ that satisfies
    \[
        |i-\hi| + |j-\hj| \leq (1+\varepsilon) \big(|i^*-\hi| + |j^*-\hj|\big),
    \]
    where $(i^*, j^*) = \argmin_{(i,j) \in R \mid (x_i,y_j) \in S} \big(|i-\hi| + |j-\hj|\big)$, or report that no such pair exists.
\end{lemma}

\begin{proof}
    Define $|i^*-\hi| + |j^*-\hj| = d^{*}$ and for an integer $d$ and $z \in \fragment{0}{\floor{1/\varepsilon}}$ define $Q_{d,z,\varepsilon}$ as
    \[
        Q_{d,z,\varepsilon} \coloneqq \fragment{-\floor{z \cdot \varepsilon d} +\hi}{\hi+ \floor{z \cdot \varepsilon d}} \times \fragment{-\floor{d - z \cdot \varepsilon d} + \hj}{\hj +\floor{d - z \cdot \varepsilon d}}
    \]
    We also set $Q_{d,\varepsilon} \coloneqq \bigcup_{z} Q_{d,z,\varepsilon}$.
    In the following two claims, we prove that $Q_d$ provides a good approximation to the $L_1$ ball of radius $d$ around $(\hi, \hj)$.

    \begin{claim}\label{claim:red_nn:1}
        For any distance $d \geq d^* / (1-\varepsilon)$, we have $(i^*, j^*) \in Q_{d,\varepsilon}$.
    \end{claim}
    \begin{claimproof}
        Let $u = |i^* - \hi|$ and $v = |j^* - \hj|$, so $u + v = d^*$.
        To show $(i^*, j^*) \in Q_{d,\varepsilon}$, we must prove there exists an integer $z \in \fragment{0}{\floor{1/\varepsilon}}$ such that $u \leq \floor{z \varepsilon d}$ and $v \leq \floor{d - z \varepsilon d}$.
        We choose $z = \lceil u / \varepsilon d \rceil$.

        First, we verify $z$ is valid. Since $u \leq d^* \leq d(1-\varepsilon)$, we have $u / \varepsilon d \leq (1-\varepsilon)/\varepsilon = 1/\varepsilon - 1$. Thus, $z \leq \floor{1/\varepsilon}$.

        Next we prove the bound on $u$. From $z = \ceil{u / \varepsilon d} \geq u /\varepsilon d$ follows $z \varepsilon d \geq u$. Since $u$ is an integer, $\floor{z \varepsilon d} \geq u$.
        Finally, we prove the bound on $v$. From $z = \ceil{u / \varepsilon d} \leq u/\varepsilon d + 1$ follows $z \varepsilon d \leq u + \varepsilon d$, which rearranged gives $d - z \varepsilon d \geq d - u - \varepsilon d = d (1-\varepsilon) - u \geq d^* - u = (u+v)-u = v$. Because $v$ is an integer it follows that $\floor{d - z \varepsilon d} \geq v$.
    \end{claimproof}

    \begin{claim}\label{claim:red_nn:2}
        For any $(i,j) \in Q_{d,\varepsilon}$ we have $|i-\hi|+|j-\hj| \leq d$. In particular, for $d < d^*$, we have  $(i^*, j^*) \notin Q_{d,\varepsilon}$.
    \end{claim}
    \begin{claimproof}
        Consider any $(i,j) \in Q_{d,\varepsilon}$. By the definition of $Q_{d,\varepsilon}$, there is some integer $z \in \fragment{0}{\floor{1/\varepsilon}}$ such that $(i,j) \in Q_{d,z,\varepsilon}$.
        From the boundaries of the rectangle $Q_{d,z,\varepsilon}$, we have:
        \[
            |i - \hi| + |j - \hj| \leq \floor{z \varepsilon d} + \floor{d - z \varepsilon d} \leq z \varepsilon d + d - z \varepsilon d \leq d,
        \]
        which concludes the proof.
    \end{claimproof}

    \subparagraph*{The algorithm.} We first replace $\hi$ (resp. $\hj$) by the nearest index in $\fragment{1}{n}$ (resp. $\fragment{1}{m}$); this shifts $|i-\hi| + |j-\hj|$ by the same constant for all $(i,j) \in \fragment{1}{n} \times \fragment{1}{m}$, so the guarantee is preserved, and afterwards $d^* \leq n+m$. Now, to find the desired approximation we do the following. First, test whether $(x_i, y_j) \in S$ for every pair $(i,j) \in R$ such that $|i - \hi| + |j - \hj| \leq 2/\varepsilon$. We do this by executing a single range emptiness query $\fragment{i}{i} \times \fragment{j}{j}$ for each such $(i,j)$. This allows us to recover $d^*$ exactly when $d^* \leq 2/\varepsilon$.

    Otherwise, when $d^* > 2/\varepsilon$, by \Cref{claim:red_nn:1} and \cref{claim:red_nn:2} that it suffices to find the smallest $d$ such that $Q_{d, \varepsilon/3} \cap R$ contains a pair $(i,j)$ with $(x_i,y_j) \in S$ (a binary search is suffices because $Q_{d,\varepsilon} \subseteq Q_{d+1,\varepsilon}$: as $z \leq \lfloor 1/\varepsilon \rfloor$, both $\lfloor z \varepsilon d \rfloor$ and $\lfloor d - z\varepsilon d \rfloor$ are nondecreasing in $d$), and return the reported point $(i,j)$. Because such a $d$ satisfies
    \[
        d^{*} \leq d \leq \ceil{d^{*} / (1-\varepsilon/3)} \leq d^{*} / (1-\varepsilon/3) + 1 \leq (1+\varepsilon/2) d^{*} + \varepsilon/2 \cdot d^* = (1 + \varepsilon) d^*,
    \]
    we obtain $|i-\hi| + |j-\hj| \leq d \leq (1+\varepsilon) d^{*}$.

    \subparagraph*{Runtime Analysis.}
    We execute at most $\Oh(1/
    \varepsilon^2)$ range emptiness queries to cover the case $d^* \leq 2/\varepsilon$. For the converse case $d^* > 2/\varepsilon$, note that $Q_{d, \varepsilon/3}$ is defined as a union of $\Oh(\varepsilon^{-1})$ rectangles, meaning its intersection with the orthogonal query range $R$ also consists of at most $\Oh(\varepsilon^{-1})$ rectangles. Consequently, evaluating whether $Q_{d, \varepsilon/3} \cap R$ contains such a pair takes $\Oh(\sT \varepsilon^{-1})$ time. Over the $\Oh(\log(n+m))$ steps of the binary search, this yields the claimed running time.
\end{proof}

Combining \cref{lem:dynamic-range-reporting}, \cref{lem:red_nn}, and \cref{lem:uniq_strs} yields the following intermediate result.

\begin{lemma}\label{lem:find_NN_pre}
    Let $\bF$ be a forest that undergoes updates.

    Then, for any $0 < \varepsilon \leq 1$, there is a data structure that supports the following queries in time $\Oh(\varepsilon^{-1} \log^3 N + \varepsilon^{-2} \log^2 N)$: given $R = \fragment{a}{a'} \times \fragment{b}{b'}$ and a query point $(\ell,r) \in \mathbb{Z}^2$ find a node $\hat{z}$ of $\bF$ with $(\Left(\hat{z}),\Right(\hat{z})) \in R$ that satisfies
    \[
        |\ell - \Left(\hat{z})| + |r - \Right(\hat{z})| \leq (1+\varepsilon)(|\ell - \Left(z^*)| + |r - \Right(z^*)|),
    \]
    where $z^* = \arg \min_{z \in \bF \mid (\Left(z),\Right(z)) \in R} |\ell - \Left(z)| + |r - \Right(z)|$, or report that no such node exists.
    Updates are supported in time $\Oh(\log^2 N)$.
\end{lemma}
\begin{proof}
    We again employ uniquely identified strings from \cref{lem:uniq_strs}, associating two unique identifiers, $\iota_z$ and $\iota_z'$, to each node $z \in \bF$ to represent the first and second occurrences of $z$ in $\mS(\bF)$. Consequently, at any point in time while $z$ exists, we maintain the invariants $\mathsf{pos}(\iota_z) = \Left(z)$ and $\mathsf{pos}(\iota_z') = \Right(z)$.

    Thus, the set $I = \bigcup_{z \in \bF} \{\iota_z, \iota_z'\}$ forms a totally ordered set that is implicitly ordered by the mapping $\mathsf{pos}(\cdot)$. Comparing any two elements in $I$ can be performed in $\Oh(\log N)$ time by evaluating this $\mathsf{pos}(\cdot)$ function.

    This formulation allows us to represent all nodes $z \in \bF$ as points within a subset $S \subseteq I \times I$, maintained via the data structure from \cref{lem:dynamic-range-reporting}, such that $z \in \bF$ if and only if $(\iota_z, \iota_z') \in S$.
    The same data structure, allows us to execute the following queries: given $R = \fragment{a}{a'} \times \fragment{b}{b'}$, we can determine whether there exists a node $z \in \bF$ such that $(\Left(z), \Right(z)) \in R$. To do so, we first clip $R$ to $\fragment{0}{2n-1}^2$ (which loses no node; if $R$ becomes empty, we report that no node exists), then retrieve the identifiers $\iota_a = \mathsf{id}(a)$, $\iota_a' = \mathsf{id}(a')$, $\iota_b = \mathsf{id}(b)$, and $\iota_b' = \mathsf{id}(b')$, and subsequently query the data structure using $\iota_a, \iota_a'$ and $\iota_b, \iota_b'$.

    Upon any update to $\bF$, we update the underlying string identifiers and correspondingly insert or remove the associated point from $S$. Finally, the claim follows by an application of \cref{lem:red_nn}.
\end{proof}

Although the following lemma is not used in the proof of \cref{lem:findNN} (it will be used later for \cref{thm:has_match}, we include it here due to its strong similarity to the proof of \cref{lem:find_NN_pre}.

\begin{lemma}\label{lem:findspines}
    Let $\bF$ be a forest that undergoes updates.
    Then, there is a data structure that supports the following queries in time $\Oh(\log^3 N)$: given $R = \fragment{a}{a'} \times \fragment{b}{b'}$ find a node $\hat{z}$ of $\bF$ with $(\Left(\hat{z}),\Right(\hat{z})) \in R$ that minimizes $\Left(\hat{z})$ (or alternatively, maximizes $\Left(\hat{z})$).
    Updates are supported in time $\Oh(\log^2 N)$.
\end{lemma}

\begin{proof}
    We maintain, as in \cref{lem:find_NN_pre}, the identified strings from \cref{lem:uniq_strs}, and define $\iota_z$, $\iota_z'$, and $I = \bigcup_{z \in \bF} \{\iota_z, \iota_z'\}$ similarly. Moreover, we maintain the subset $S \subseteq I \times I$ via the data structure from \cref{lem:dynamic-range-reporting}, such that $z \in \bF$ if and only if $(\iota_z, \iota_z') \in S$.
    This allows us to use queries in $\Oh(\log^2 N)$ time to check if there is at least one node $\hat{z}$ satisfying $(\Left(\hat{z}),\Right(\hat{z})) \in R$. To find the one minimizing $\Left(\hat{z})$, we can perform a binary search by modifying the $\fragment{a}{a'}$ range while leaving the $\fragment{b}{b'}$ range untouched. Similarly, we can also find the one maximizing $\Left(\hat{z})$.
\end{proof}

This brings us to the proof of \cref{lem:findNN}.

\findNN

\begin{proof}
    We maintain the data structure from \cref{lem:find_NN_pre} for $\bG$ (and not $\bF)$ and the tree $T(\bF)$.

    Fix the input to a query, and let $\mJ' = \langle J_1', J_2' \rangle$, where $J_i' \coloneqq \fragmentco{a_i'}{b_i'}$.

    Note that for each node $z'' \in \bG$, the bi-interval $\mJ'' \coloneqq \langle J_1'', J_2'' \rangle$, defined by $J_1'' \coloneqq \fragmentco{\Left(z'')-q_1}{\Left(z'')+q_1'}$ and $J_2'' \coloneqq \fragmentco{\Right(z'')-q_2}{\Right(z'')+q_2'}$, belongs to $\widetilde{W}_{z,q}^{k}(\mI_v)$ if and only if all of the following hold: $|a_{v,1}-\Left(z'')+q_1| \leq 2k$, $|b_{v,1}-\Left(z'')-q_1'| \leq 2k$, $|a_{v,2} - \Right(z'')+q_2| \leq 2k$, and $|b_{v,2}-\Right(z'')-q_2'| \leq 2k$. These constraints are equivalent to $\Left(z'') \in \fragment{a}{a'}$ and $\Right(z'') \in \fragment{b}{b'}$, where:
    \begin{align*}
        &a \coloneqq \max(-2k + a_{v,1} + q_1, -2k + b_{v,1} - q_1'),
        &&a' \coloneqq \min(2k + a_{v,1} + q_1, 2k + b_{v,1} - q_1'), \\
        &b \coloneqq \max(-2k + a_{v,2} + q_2, -2k + b_{v,2} - q_2'),
        &&b' \coloneqq \min(2k + a_{v,2} + q_2, 2k + b_{v,2} - q_2').
    \end{align*}
    Furthermore, we observe that $\mathsf{dist}(\mJ', \mJ'') = |a_1' - (\Left(z'')-q_1)| + |a_2' - (\Right(z'')-q_2)|$.

    This implies that to find the desired $\mJ''$, it suffices to execute the query from \cref{lem:find_NN_pre}, setting the query point $(\ell,r)$ to $(a_1'+q_1, a_2'+q_2)$, and defining the boundaries $a, a', b, b'$ as above.
\end{proof}

\subsubsection{Proof of \cref{lem:relevant_bound}}

\relevant
\begin{proof}
    For any node $v$, let $B_v$ denote its number of children. Recall that $\{\mI_{v_i}\}_{i=1}^{B_v}$ associated with its children $v_1, \ldots, v_{B_v}$ form a partition of $\mI_v$.

    Because the root $r$ of $T(\bF)$ satisfies $\mI_{r} = \langle \fragmentco{0}{2n} \rangle$, it is immediate that the set of nodes $A^{\bullet}(x)$ is either empty or forms a single root-to-leaf path. Hence, $|A^{\bullet}(x)| = \Oh(\log n)$.
    Furthermore, if a node $v \in A^{\circ}_{k}(x)$ has a parent $w$, then $w \in A^{\circ}_{k}(x)$ as well, since $U(\mI_v) \subseteq U(\mI_w)$. Thus, the set of nodes $A^{\circ}_{k}(x)$ is either empty or forms a connected subtree that includes the root of $T(\bF)$. To show that $|A^{\circ}_k(x)| = \Oh(k \log n)$, it suffices to show that at every level of $T(\bF)$, there are $\Oh(k)$ nodes $v$ such that $\fragment{x-2k}{x+2k} \cap U(\mI_v) \neq \emptyset$. This is true because at every level, every position in $\fragment{x-2k}{x+2k}$ can be assigned to at most one node on that level. This concludes the proof of \ref{it:relevant_bound:i}.

    Next, let
    $R_k(x) \coloneqq \{\ (v,i) \mid v \in T(\bF),\ i \in \fragment{1}{\ell_v},\ I_{v,i} \cap \fragment{x-2k}{x+2k} \neq \emptyset \ \}$.
    \begin{claim}\label{clm:relevant_bound}
        Both of the following hold:
        \begin{enumerate}[(a)]
            \item \label{it:clm:relevant_bound:a} Let $v \in T(\bF)$, $i \in \fragment{1}{\ell_v}$, and $J \in W^{k}(I_{v,i})$ with $|J| \leq 2|I_{v,i}|/\varepsilon$. If $x \in J$, or $x \in \mathsf{co}(J, I_{v,i})$ and $|\mathsf{co}(J, I_{v,i})| > 2k$, then $(v,i) \in R_k(x)$.
            \item \label{it:clm:relevant_bound:b} For every level $L$ of $T(\bF)$,
            \[
                |\{(v,i) \in R_k(x) \mid v \in L,\ |I_{v,i}| > k\}| = \Oh(1), \qquad
                |\{(v,i) \in R_k(x) \mid v \in L,\ |I_{v,i}| > \varepsilon k\}| = \Oh(\varepsilon^{-1}),
            \]
            and
            \(
                \sum_{(v,i) \in R_k(x),\ v \in L,\ |I_{v,i}| \leq k} |I_{v,i}| = \Oh(k).
            \)
        \end{enumerate}
    \end{claim}
    \begin{claimproof}
        For \ref{it:clm:relevant_bound:a}, let $I_{v,i} = \fragmentco{a}{b}$ and $J = \fragmentco{a'}{b'}$, so $|a'-a|, |b'-b| \leq 2k$ and $J \subseteq \mathsf{co}(J, I_{v,i}) \subseteq \fragmentco{a-2k}{b+2k}$.
        If $a = b$, then $|J| \leq 2|I_{v,i}|/\varepsilon = 0$, so $x \notin J$ and $|\mathsf{co}(J, I_{v,i})| = |a'-a| \leq 2k$.
        If $a < b$, both hypotheses give $x \in \fragmentco{a-2k}{b+2k}$, so $\max(a, x-2k) \in I_{v,i} \cap \fragment{x-2k}{x+2k}$.
        For \ref{it:clm:relevant_bound:b}, the intervals $I_{v,i}$ with $v \in L$ are pairwise disjoint by \cref{thm:final_decomp_tree}\ref{it:final_tree_2}, so at most one of them contains $x-2k$, at most one contains $x+2k$, and every other $I_{v,i}$ with $(v,i) \in R_k(x)$ satisfies $I_{v,i} \subseteq \fragment{x-2k}{x+2k}$.
        Hence, at most $2 + (4k+1)/k$ such pairs have $|I_{v,i}| > k$, at most $2 + (4k+1)/(\varepsilon k)$ have $|I_{v,i}| > \varepsilon k$, and those with $|I_{v,i}| \leq k$ have total length at most $2k + 4k + 1$.
    \end{claimproof}
    To prove \ref{it:relevant_bound:ii}, note that every $J \in \widetilde{W}^k(I_{v,i})$ satisfies $J \in W^{k}(I_{v,i})$ and $|J| \leq |I_{v,i}|/\varepsilon$, so by \cref{clm:relevant_bound}\ref{it:clm:relevant_bound:a} the sum in \ref{it:relevant_bound:ii} is at most
    \[
        \sum_{(v,i) \in R_k(x)} |\{J \in \widetilde{W}^k(I_{v,i}) \mid x \in J\}| + \sum_{(v,i) \in R_k(x)} |\{J \in \widetilde{W}^k(I_{v,i}) \mid |\mathsf{co}(J, I_{v,i})| > 2k\}|.
    \]
    We bound both sums by $\Oh(k\varepsilon^{-2}\log^2 N)$ separately.
    \begin{itemize}
        \item To bound $\sum_{(v,i) \in R_k(x)}|\{J \in \widetilde{W}^k(I_{v,i}) \mid x \in J\}|$, we partition $R_k(x)$ into $\mathsf{long}_k \coloneqq \{(v,i) \in R_k(x) \mid |I_{v,i}| > k\}$ and $\mathsf{short}_k \coloneqq R_k(x) \setminus \mathsf{long}_k$.
        If $|I_{v,i}|> k$, then we use that $|\widetilde{W}^k (I_{v,i})| \leq \Oh(k\varepsilon^{-1} \log N)$, combined with \cref{clm:relevant_bound}\ref{it:clm:relevant_bound:b}, by which each level contains $\Oh(1)$ pairs $(v,i) \in \mathsf{long}_k$. Therefore we have
        \begin{align*}
            \sum_{(v,i) \in \mathsf{long}_k} |\{J \in \widetilde{W}^k(I_{v,i}) \mid x \in J\}| = \Oh(k\varepsilon^{-1}\log^2 N).
        \end{align*}
        Next, if $|I_{v,i}|\leq k$, we will use the fact that for every allowed length $q \in  \tilde{\sL}^k(|I_{v,i}|)$, there are at most $q$ intervals in $\widetilde{W}^k(I_{v,i})$ of length $q$ that intersect position $x$. Thus we can write
        \begin{align}
             \sum_{(v,i) \in \mathsf{short}_k} |\{J \in \widetilde{W}^k(I_{v,i}) \mid x \in J\}|
            &\leq  \sum_{(v,i) \in \mathsf{short}_k} \sum_{\substack{q \in \tilde{\sL}^k(|I_{v,i}|)\\ q\leq \varepsilon^{-1}|I_{v,i}|}}q\nonumber\\
            &\label{eq:relevant_1}\leq \sum_{(v,i) \in \mathsf{short}_k} \Oh(\varepsilon^{-1}\log N \cdot \varepsilon^{-1}|I_{v,i}|)\\
            &\label{eq:relevant_2}\leq \Oh(k\varepsilon^{-2}\log^2 N)
        \end{align}
       where in \cref{eq:relevant_1} we use the fact that $|J| \leq \varepsilon^{-1}|I_{v,i}|$ for every $J \in \widetilde{W}^k(I_{v,i})$, and \cref{eq:relevant_2} is due to \cref{clm:relevant_bound}\ref{it:clm:relevant_bound:b}, by which the $|I_{v,i}|$ with $(v,i) \in \mathsf{short}_k$ on one level sum to $\Oh(k)$.

    \item To bound $\sum_{(v,i) \in R_k(x)}|\{J \in \widetilde{W}^k(I_{v,i}) \mid |\mathsf{co}(J, I_{v,i})| > 2k\}|$, we partition $R_k(x)$ into $\mathsf{long}_{\varepsilon k} \coloneqq \{(v,i) \in R_k(x) \mid |I_{v,i}| > \varepsilon k\}$ and $\mathsf{short}_{\varepsilon k} \coloneqq R_k(x) \setminus \mathsf{long}_{\varepsilon k}$.

    If $|I_{v,i}| > \varepsilon k$, then there are at most $\Oh(\varepsilon^{-1})$ such pairs $(v,i)$ on one level by \cref{clm:relevant_bound}\ref{it:clm:relevant_bound:b}, so we can afford to count all $\Oh(k\varepsilon^{-1}\log N)$ intervals in $\widetilde{W}^k(I_{v,i})$. This gives us
    \begin{align*}
        \sum_{(v,i) \in \mathsf{long}_{\varepsilon k}}|\{J \in \widetilde{W}^k(I_{v,i}) \mid |\mathsf{co}(J, I_{v,i})| > 2k\}| =  \sum_{(v,i) \in \mathsf{long}_{\varepsilon k}}\Oh(k \varepsilon^{-1}\log N) = \Oh(k\varepsilon^{-2}\log^2 N).
    \end{align*}
    If $|I_{v,i}| \leq \varepsilon k$, then note that $|J| \leq \varepsilon^{-1}|I_{v,i}| \leq k$ for all $J \in \widetilde{W}^k(I_{v,i})$. Let $I_{v,i} = \fragmentco{a_{v,i}}{b_{v,i}}$ and $J = \fragmentco{a'}{b'}$.
    Thus, for $|\mathsf{co}(J, I_{v,i})| > 2k$, we must have $2k - |I_{v,i}| < a_{v,i} - a' \leq 2k$ or $2k - |I_{v,i}| < b' - b_{v,i} \leq 2k$, since $J \supseteq I_{v,i}$ would give $|\mathsf{co}(J, I_{v,i})| = |J| \leq k$. So such a $J$ would have $\Oh(|I_{v,i}|)$ possible starting or ending positions. Therefore, using \cref{clm:relevant_bound}\ref{it:clm:relevant_bound:b}, we bound
    \begin{align*}
        \sum_{(v,i) \in \mathsf{short}_{\varepsilon k}}|\{J \in \widetilde{W}^k(I_{v,i}) \mid |\mathsf{co}(J, I_{v,i})| > 2k\}| \leq \sum_{(v,i) \in \mathsf{short}_{\varepsilon k}} \Oh(|I_{v,i}|\varepsilon^{-1}\log N) \leq \Oh(k\varepsilon^{-1}\log^2 N).
    \end{align*}
    Therefore $\Oh(k\varepsilon^{-2}\log^2 N)$ is a loose upper bound for $\sum_{(v,i) \in R_k(x)}|\{J \in \widetilde{W}^k(I_{v,i}) \mid |\mathsf{co}(J, I_{v,i})| > 2k\}|$.
    \end{itemize}
    Finally, we will prove \ref{it:relevant_bound:iii} and \ref{it:relevant_bound:iv}. Fix $v \in T(\bF)$ with $\ell_v = 2$ and $z \in Z_v$. Every $\mJ = \langle J_1, J_2 \rangle \in \widetilde{W}^{k}_z(\mI_v)$ is determined by a node $z' \in \bG$ and a tuple $q = (q_1, q_1', q_2, q_2') \in \tilde{\sL}_{v,z}$ via $J_1 = \fragmentco{\Left(z') - q_1}{\Left(z') + q_1'}$ and $J_2 = \fragmentco{\Right(z') - q_2}{\Right(z') + q_2'}$, where each coordinate of $q$ ranges over $\Oh(\varepsilon^{-1} \log N)$ values, $q_1, q_1' \leq |I_{v,1}|/\varepsilon$, and $q_2, q_2' \leq |I_{v,2}|/\varepsilon$. In the notation of \ref{it:relevant_bound:iv}, $\mJ = \mJ_{q,z'}$, and every $\mJ \in \widetilde{W}^{k}_z(\mI_v)$ arises from at least one pair $(q, z')$; hence it suffices to count the pairs $(q, z')$ with $\mJ_{q,z'} \in W^{k}(\mI_v)$ that satisfy the condition of \ref{it:relevant_bound:iii}, which proves both items at once. Moreover, $\mJ \in W^{k}(\mI_v)$ confines $\Left(z')$ to at most $4k+1$ positions once $q_1$ is fixed. Since $J_i \in W^{k}(I_{v,i})$ and $|J_i| = q_i + q_i' \leq 2|I_{v,i}|/\varepsilon$, \cref{clm:relevant_bound}\ref{it:clm:relevant_bound:a} shows that $\mJ$ is counted in \ref{it:relevant_bound:iii} only if $x \in J_i$ or $|\mathsf{co}(J_i, I_{v,i})| > 2k$ for some $i$ with $(v,i) \in R_k(x)$. In particular, only $v \in A^{\circ}_k(x)$ contribute, as claimed in \ref{it:relevant_bound:iv}. We count the pairs $(z', q)$ with $x \in J_1$ for $(v,1) \in R_k(x)$; the case $x \in J_2$ is symmetric, and the condition $|\mathsf{co}(J_i, I_{v,i})| > 2k$ for $(v,i) \in R_k(x)$ is handled as in \ref{it:relevant_bound:ii}: for $|I_{v,i}| > \varepsilon k$, there are $\Oh(\varepsilon^{-1})$ such nodes $v$ per level by \cref{clm:relevant_bound}\ref{it:clm:relevant_bound:b} and we count all $\Oh(k \varepsilon^{-4} \log^4 N)$ pairs $(z', q)$, whereas for $|I_{v,i}| \leq \varepsilon k$, where $|J_i| \leq 2|I_{v,i}|/\varepsilon \leq 2k$ excludes $J_i \supseteq I_{v,i}$, the condition confines the start or the end of $J_i$, and hence $\Left(z')$ or $\Right(z')$ for a fixed coordinate of $q$, to $\Oh(|I_{v,i}|)$ positions, giving $\Oh(\varepsilon^{-4} \log^{4} N \cdot |I_{v,i}|)$ pairs.

For fixed $(q_1, q_1')$, the condition $x \in J_1$ leaves at most $\min(q_1 + q_1', 4k+1)$ choices for $z'$, namely $\Left(z')$ lies in $\fragmentoc{x - q_1'}{x + q_1} \cap \fragment{a_{v,1} + q_1 - 2k}{a_{v,1} + q_1 + 2k} \cap \fragment{b_{v,1} - q_1' - 2k}{b_{v,1} - q_1' + 2k}$, an interval that is empty unless $(v,1) \in R_k(x)$ by \cref{clm:relevant_bound}\ref{it:clm:relevant_bound:a}, and $(q_2, q_2')$ contributes a factor of $\Oh(\varepsilon^{-2} \log^2 N)$. If $|I_{v,1}| > k$, there are $\Oh(1)$ such nodes $v$ per level by \cref{clm:relevant_bound}\ref{it:clm:relevant_bound:b}, and we obtain $\Oh(\varepsilon^{-2} \log^2 N \cdot k \cdot \varepsilon^{-2} \log^2 N) = \Oh(k \varepsilon^{-4} \log^4 N)$ pairs for each $(v, z)$. If $|I_{v,1}| \leq k$, summing $q_1 + q_1' \leq 2|I_{v,1}|/\varepsilon$ over the $\Oh(\varepsilon^{-2} \log^2 N)$ pairs $(q_1, q_1')$ gives $\Oh(\varepsilon^{-5} \log^4 N \cdot |I_{v,1}|)$ pairs for each $(v, z)$, and the lengths $|I_{v,1}|$ of these nodes on one level sum to $\Oh(k)$ by \cref{clm:relevant_bound}\ref{it:clm:relevant_bound:b}. Summing over the $\Oh(\varepsilon^{-1} \log N)$ samples $z \in Z_v$ and the $\Oh(\log n)$ levels of $T(\bF)$ yields $\Oh(k \varepsilon^{-6} \log^6 N)$ in all cases, which proves \ref{it:relevant_bound:iii} and the bound of \ref{it:relevant_bound:iv}. For the listing in \ref{it:relevant_bound:iv}, we iterate over $v \in A$, $z \in Z_v$, $q \in \tilde{\sL}_{v,z}$, and $i \in \{1,2\}$, compute the above interval for $\Left(z')$ (for $i = 2$, its analogue $\fragmentoc{x - q_2'}{x + q_2} \cap \fragment{a_{v,2} + q_2 - 2k}{a_{v,2} + q_2 + 2k} \cap \fragment{b_{v,2} - q_2' - 2k}{b_{v,2} - q_2' + 2k}$ for $\Right(z')$) from the positions of the endpoints of $\mI_v$, and, for each position $p$ in it, check whether $p$ is the $i$-th occurrence of a node $z'$ and whether $\mJ_{q,z'} \in W^{k}(\mI_v)$, all in $\Oh(\log N)$ time by \cref{lem:uniq_strs}. This costs $\Oh(\log N)$ per $(v, z, q, i)$, i.e., $\Oh(|A| \varepsilon^{-5} \log^{6} N)$ in total, plus $\Oh(\log N)$ per position, and the number of positions is bounded by the sum just computed.
\end{proof}

\subsubsection{Proof of \cref{thm:has_match}}

Our goal here is to prove \cref{thm:has_match}. To quickly find equivalent subforests, we design fingerprints for forests in a manner similar to the approach of \cite{DGHKS25} (although we do so in a slightly more general setting). The basis for our fingerprints is standard string fingerprints; thus, to apply them, we first need to define the string representations of the trees we operate on.

\begin{definition}\label{def:forest-to-strings}
    Given two forests $\bF$ and $\bG$, we define the strings $F$ and $G$ by taking the strings $\mS(\bF)$ and $\mS(\bG)$, except that for a node $v$ with label $\sigma$, we use $\mathtt{(}^{\sigma}$ for the first occurrence of $v$ and $\mathtt{)}^{\sigma}$ for the second occurrence; padding positions keep the symbol $\$$.
\end{definition}

For mono-intervals, it turns out that string equality directly implies subforest equivalence.

\begin{proposition}\label{prp:matchmono}
    Let $\mI = \langle \fragmentco{a}{a+q}\rangle$ and $\mJ = \langle \fragmentco{a'}{a'+q}\rangle$ be mono-intervals of the same length $q$. Then $\bF_{\mI} \equiv \bG_{\mJ}$ if and only if $F \fragmentco{a}{a+q} = G\fragmentco{a'}{a'+q}$.
\end{proposition}
\begin{proof}
The ``only if'' direction is clear.
For the ``if'' direction, we assume $F \fragmentco{a}{a+q} = G \fragmentco{a'}{a'+q}$.
Since parenthesis pairs never cross (\cref{fct:non_crossing}), within any contiguous substring, the standard stack-based matching (applied independently per label $\sigma$) uniquely pairs each $)^{\sigma}$ with the nearest unmatched $(^{\sigma}$ to its left. Because the two substrings are identical, this matching is exactly the same in both.

We observe that each matched pair of positions $(x_1,x_2)$ corresponds to a single node, and each unmatched position corresponds to a node appearing exactly once in the subforest. Define a mapping $f$ by taking the node of $\bF_{\mI}$ at each matched pair or unmatched position and mapping it to the node of $\bG_{\mJ}$ at that exact same pair or position. This mapping $f$ is a bijection: it is injective because distinct nodes in $\bF_{\mI}$ occupy distinct positions, and it is surjective because every node in $\bG_{\mJ}$ is covered by the shared matching. Furthermore, it preserves labels, since the equality $F\position{a+i} = G\position{a'+i}$ forces $\ell(v) = \ell(f(v))$ at every position $i \in \fragmentco{0}{q}$, and it preserves the occurrence type (opening or closing) by the exact same equality. Hence, $f$ witnesses $\bF_{\mI} \equiv \bG_{\mJ}$.
\end{proof}

For bi-intervals, string equality alone is not sufficient, so we must include additional information. Fortunately, it turns out that comparing only the information related to at most two spine nodes is sufficient to establish equivalence.

\begin{proposition}\label{prp:matchbi}
    Let $\mI = \langle I_1, I_2\rangle$ and $\mJ = \langle J_1, J_2 \rangle$, where $I_i = \fragmentco{a_i}{a_i+q_i}$ and $J_i = \fragmentco{a_i'}{a_i'+q_i}$, be bi-intervals with lengths $q_1$ and $q_2$ such that $a_1 + q_1 \leq a_2$ and $a_1' + q_1 \leq a_2'$ (that is, $I_1$ precedes $I_2$ and $J_1$ precedes $J_2$; the remaining relative positions are treated at the end of the proof). Moreover, let $v$ (resp. $u$) be the node of $\bF$ with the smallest $\Left(v) \in I_1$ (resp. largest $\Left(u) \in I_1$) such that $\Right(v) \in I_2$ (resp. $\Right(u) \in I_2$), and define $v',u'$ similarly for $\bG$ and $J_1,J_2$. Then $\bF_{\mI} \equiv \bG_{\mJ}$ if and only if all of the following hold:
    \begin{enumerate}[(i)]
        \item $F\fragmentco{a_i}{a_i+q_i} = G\fragmentco{a_i'}{a_i'+q_i}$ for all $i \in \fragment{1}{2}$;
        \label{it:prp:matchbi:i}
        \item $v = \bot$ if and only if $v' = \bot$, and if $v \neq \bot$, then $\Left(v)-a_1 = \Left(v')-a_1'$, $\Left(u)-a_1 = \Left(u')-a_1'$, $\Right(v)-a_2 = \Right(v')-a_2'$, and $\Right(u)-a_2 = \Right(u')-a_2'$.
        \label{it:prp:matchbi:ii}
    \end{enumerate}
\end{proposition}

\begin{proof}
    Again, the ``only if'' direction is clear.

    For the ``if'' direction, assume that \ref{it:prp:matchbi:i} and \ref{it:prp:matchbi:ii} hold.
    By \ref{it:prp:matchbi:i} and \cref{prp:matchmono}, applied to each side of the bi-interval, we obtain two bijections, $f_1$ and $f_2$, between the nodes on the corresponding sides of $\bF_{\mI}$ and $\bG_{\mJ}$. Moreover, the nodes within each side that appear only once under each of the bijections occur at the same offsets and carry the same labeled-parenthesis symbols.

    To conclude, it remains to show that $f_1$ and $f_2$ map the spine nodes consistently. If, in \ref{it:prp:matchbi:ii}, we have $v = v' = \bot$, then we know that the nodes appearing only once are all distinct, and the union of the two bijections already yields $\bF_{\mI} \equiv \bG_{\mJ}$; thus, we may assume this is not the case.

    Otherwise, let $u_1, \ldots, u_s$ be the spine nodes of $\bF_\mI$ ordered such that $\Left(u_1) < \cdots < \Left(u_s)$ (hence $\Right(u_1) > \cdots > \Right(u_s)$ by nesting), and let $u_1', \ldots, u_{s'}'$ be those of $\bG_\mJ$ ordered analogously.
    By~\ref{it:prp:matchbi:ii}, we know that $\Left(u_1)-a_1 = \Left(u_1')-a_1'$, $\Right(u_1)-a_2 = \Right(u_1')-a_2'$, $\Left(u_s)-a_1 = \Left(u_{s'}')-a_1'$, and $\Right(u_s)-a_2 = \Right(u_{s'}')-a_2'$.
    Since, by~\ref{it:prp:matchbi:i}, the unmatched openings in $I_1$ and $J_1$ occur at the same offsets with the same labels, and the spine nodes' openings form a contiguous block delimited by the four selected spine nodes (if not, we would have a contradiction with \cref{fct:non_crossing}), we know that $s=s'$ and that all spine nodes are indeed mapped consistently in both $f_1$ and $f_2$.

    The case in which $I_2$ precedes $I_1$ (and hence $J_2$ precedes $J_1$) is symmetric, with the roles of the two intervals exchanged. If the two intervals of $\mI$ intersect, then $\bF_{\mI} \equiv \bG_{\mJ}$ holds if and only if \ref{it:prp:matchbi:i} holds and $a_2 - a_1 = a_2' - a_1'$: the ``only if'' direction is \cref{def:equiv}\ref{def:equiv:iii}, and for the ``if'' direction note that $I_1 \cup I_2$ and $J_1 \cup J_2$ are then intervals of the same length whose strings coincide by \ref{it:prp:matchbi:i}, so \cref{prp:matchmono} applied to them shows that the common shift preserves labels, occurrence types, and the $\Other$ relation on $I_1 \cup I_2$, and hence on both intervals. In all remaining combinations of relative positions, $\bF_{\mI} \not\equiv \bG_{\mJ}$ by definition (for non-empty components; if a component is empty, \cref{def:equiv} imposes no condition on its position, and \ref{it:prp:matchbi:i} for the other component suffices).
\end{proof}

\Cref{prp:matchmono} and \cref{prp:matchbi}, together with state-of-the-art deterministic fingerprints, allow us to prove \cref{thm:has_match}.

\begin{lemma}[Deterministic Dynamic Fingerprints;
  follows from {\cite[Section~8]{KK22}}]
  \label{lem:dynamic-fingerprints}
  Let $S$ be a string of length~$n$ over an integer alphabet
  of size~$\Oh(\poly(n))$ that undergoes at most $Q$ character insertions and deletions.
  There exists a deterministic data structure that supports
  the following operations:
  \begin{description}
      \item[Fingerprint query:] Given positions $x, y$,
      compute a fingerprint $h(S\fragmentco{x}{y})$ of the
      fragment $S\fragmentco{x}{y}$; the fingerprint has
      $\Oh(\log (n + Q))$~bits and satisfies $h(X) = h(Y)$ if and
      only if $X = Y$, for any two strings $X, Y$ that
      arise as fragments of (possibly different versions of)
      the maintained strings.
      \item[Update:] Insert or delete a character in~$S$.
  \end{description}
  Both operations take (worst-case) time $\Oh(\log^{1+o(1)} (n + Q))$. \lipicsEnd
\end{lemma}

\hasmatch

\begin{proof}
    We maintain the data structure from \cref{lem:dynamic-fingerprints} for the concatenation of the strings $F \circ G$, where $F$ and $G$ are defined as in \cref{def:forest-to-strings}. Moreover, we maintain the data structure of \cref{lem:findspines} for both $\bF$ and $\bG$.
    Both data structures start empty, except for the $8k+1$ padding characters $\$$ in front of and after each of $F$ and $G$, inserted at initialization and never deleted (as in the proof of \cref{thm:fvsf}). By the conventions at the beginning of this section, the string $F \circ G$ has length at most $\Oh(N)$, undergoes $Q = \poly(N)$ character edits in total, and its labels are mapped to an integer alphabet of size $\poly(N)$, so every operation of \cref{lem:dynamic-fingerprints} takes $\Oh(\log^{1+o(1)} N)$ time.

    For a mono-interval $\mI = \langle \fragmentco{a}{b} \rangle$ in $\bF$, we define its \emph{fingerprint} as $h(\bF_{\mI}) = h(F\fragmentco{a}{b})$, where the hash $h$ on fragments is taken from \cref{lem:dynamic-fingerprints}. Similarly, for a bi-interval $\mI = \langle \fragmentco{a_1}{b_1}, \fragmentco{a_2}{b_2} \rangle$ in $\bF$ with an empty component, we define its fingerprint as $h(\bF_{\mI}) = (h(F\fragmentco{a_1}{b_1}), h(F\fragmentco{a_2}{b_2}))$, as \cref{def:equiv}\ref{def:equiv:iii} imposes no condition on the position of an empty interval. If both components are non-empty and $b_1 \leq a_2$, let $z$ and $z'$ be the spine nodes in $\bF_{\mI}$ that minimize and maximize $\Left(z)$, respectively. If $z$ and $z'$ do not exist, we define the \emph{fingerprint} of $\mI$ as $h(\bF_{\mI}) = (h(F\fragmentco{a_1}{b_1}), h(F\fragmentco{a_2}{b_2}), \bot, \bot, \bot, \bot)$. Otherwise, we define it as $h(\bF_{\mI}) = (h(F\fragmentco{a_1}{b_1}), h(F\fragmentco{a_2}{b_2}), \Left(z)-a_1, \Right(z)-a_2, \Left(z')-a_1, \Right(z')-a_2)$.
    If instead $b_2 \leq a_1$, we define the fingerprint symmetrically, with the roles of the two intervals exchanged, and if the two intervals intersect, we define it as $(h(F\fragmentco{a_1}{b_1}), h(F\fragmentco{a_2}{b_2}), a_2 - a_1)$. In all cases, we prepend to the fingerprint the relative position of the two intervals (the first precedes the second, the second precedes the first, they intersect, or a component is empty), so that fingerprints of bi-intervals with different relative positions never coincide.
    We define similar fingerprints for the subforests in $\bG$.
    Note that, given the data structures we maintain, the fingerprints for a given interval collection can be computed in time $\Oh(\log^3 N)$, which is dominated by the time required to locate the spine nodes in \cref{lem:findspines}.

    \subparagraph{The Invariant.}
    Now, for Algorithm~\ref{it:has_match:1}, we maintain the invariant that at each node $v \in T(\bF)$, we store the fingerprints $h(\bF_{\mL})$ for all $\mL \in W^{4k}(\mI_v)$ in a balanced binary tree\footnote{We use lexicographic ordering, and we maintain separate trees for mono-intervals and bi-intervals.}.
    More precisely, the tree contains one entry per window $\mL$, keyed by the pair $(h(\bF_{\mL}), \mathsf{id}(v, \mL))$ with $\mathsf{id}(v, \mL)$ as in \cref{thm:fvsf}. Thus, windows with equal fingerprints are kept as separate entries, so that removing the entry of one window (upon an update) does not affect the others, and the entries sharing a fingerprint form a contiguous range of the tree, any element of which serves as a witness. For Algorithm~\ref{it:has_match:2}, we maintain a similar invariant but for a different set of $\mL$: namely, for all $\mL \in \widetilde{W}^{4k}(\langle I_{v,i} \rangle)$ with $i \in \fragment{1}{\ell_v}$, and for all $\mL \in \widetilde{W}^{4k}_z(\mI_v)$ with $z \in Z_v$.

    This allows us to answer queries efficiently: we compute the fingerprint for the query interval $\mJ$ in time $\Oh(\log^3 N)$, and then spend time $\Oh(\log N)$ to check whether this fingerprint is present in the tree stored at $v$ and, if so, to return the window $\mL$ of any entry with this fingerprint (every such window satisfies $\bF_{\mL} \equiv \bG_{\mJ}$). Correctness follows directly from \cref{prp:matchmono} and \cref{prp:matchbi}.

    \subparagraph{Handling Updates.}
    Next, we describe the procedure for handling updates to $\bF$ and $\bG$, and updating the stored fingerprints. Upon an update to $\bG$, we do not need to update any of the stored fingerprints.
    Otherwise, let $y$ and $y'$ denote the indices where the update occurs\footnote{Specifically, if the update is a substitution or deletion in $\bF$ of a node $w \in \bF$, we set $y = \Left(w)$ and $y' = \Right(w)$. Otherwise, if a new node $w$ is inserted into $\bF$, we choose $y$ and $y'$ such that the new bi-order traversal is given by $\mS(\bF)\fragmentco{0}{y} \circ w \circ \mS(\bF)\fragmentco{y}{y'} \circ w \circ \mS(\bF)\fragmentco{y'}{2n}$.}.

    Similarly to \cref{thm:fvsf}, let $P_{\mathrm{new}}$ denote the set of affected nodes in $T(\bar{\bF})$, and let $P_{\mathrm{old}}$ denote the remaining set of nodes in $T(\bar{\bF})$. We also define $A^{\circ}$ as $A^{\circ} \coloneqq (A^{\circ}_{4k}(y) \cup A^{\circ}_{4k}(y')) \cap P_{\mathrm{old}}$.

    Note that for all $v \notin A^{\circ} \cup P_{\mathrm{new}}$, no fingerprints need to be updated because the updates fall too far away (using arguments similar to \cref{thm:fvsf} and \cref{thm:fvsg}). To update nodes $v \in A^{\circ} \cup P_{\mathrm{new}}$, we recompute the fingerprints as follows (the set $A^{\circ}$ can be found via a DFS). For nodes $v \in P_{\mathrm{new}}$, we update all fingerprints; for the other nodes in $A^{\circ}$ with $\{y, y'\} \cap U(\mI_v) \neq \emptyset$ ($\Oh(\log n)$ nodes by \cref{lem:relevant_bound}\ref{it:relevant_bound:i}, where $Z_v$ is redrawn) and, for Algorithm~\ref{it:has_match:1}, at all nodes of $A^{\circ}$, we also update all fingerprints. For Algorithm~\ref{it:has_match:2} at the remaining nodes $v \in A^{\circ}$, we recompute the fingerprints of the windows $\mL$ with $U(\mL) \cap \{y, y'\} \neq \emptyset$ (listed by \cref{lem:relevant_bound}\ref{it:relevant_bound:iv} with $4k$ in place of $k$ and, for mono-intervals, by trying the $\min(|\mL|, 8k+1)$ starts for each length) and of the identifiers entering the stored families, and we delete the identifiers leaving them, where the latter two sets are computed by \cref{clm:fvsf:2} with $4k$ in place of $k$; an entry is looked up under its old fingerprint, which we store with its identifier.

    \subparagraph{Runtime.}
    The number of fingerprint recomputations for Algorithm~\ref{it:has_match:1} is bounded by $\Oh(|P_{\mathrm{new}} \cup A^{\circ}|)$ times $\Oh(k^4)$ (the maximum number of $\mL$ in any $W^{4k}(\mI_v)$). Since the size of the former set $P_{\mathrm{new}}$ is at most $\Oh(\log n)$ on average, and the latter $A^{\circ}$ is at most $\Oh(k \log n)$ by \cref{lem:relevant_bound}\ref{it:relevant_bound:i}, we obtain an $\Oh(k^{5} \log^{4} N)$ update time.
    For Algorithm~\ref{it:has_match:2}, the bounds for the dominant recomputations for nodes in $A^{\circ}$ are given by \cref{lem:relevant_bound}\ref{it:relevant_bound:ii} and \cref{lem:relevant_bound}\ref{it:relevant_bound:iii} (applied with $4k$ in place of $k$), which bound the number of mono-interval and bi-interval windows to be recomputed by $\Oh(k\varepsilon^{-6}\log^{6} N)$ in total, plus the $\Oh(k \varepsilon^{-6} \log^{6} N)$ identifiers entering or leaving the stored families over all nodes of $A^{\circ}$, found in $\Oh(k \varepsilon^{-6} \log^{7} N)$ time by \cref{clm:fvsf:2} with $4k$ in place of $k$ (its auxiliary information is rebuilt together with the fingerprints at the fully recomputed nodes), and all $\Oh(k \varepsilon^{-5} \log^{5} N)$ windows at the $\Oh(\log n)$ fully recomputed nodes amortized per update; as each recomputation takes $\Oh(\log^{3} N)$ time, this results in an update time of $\Oh(k\varepsilon^{-6}\log^{9} N)$.

    The update times of the other maintained data structures are negligible compared to the update time required to maintain the fingerprints.
\end{proof}

%% file: AI_acknowledgement.tex
\section*{Acknowledgments}
The authors used GPT-6 Astra to assist with
debugging proofs and improving language and presentation. All ideas behind the algorithms are the brainchildren of real humans.
The authors take responsibility for all content.

%% file: appendix.tex
\section{A Comparison Between Bi-Intervals, Super-Windows and Contexts}
\label{sec:comp_appendix}

Our definition of bi-interval $\mI = \langle I_{\Left},I_{\Right}\rangle$ generalizes \emph{super-windows} of~\cite{SS22}, defined as subforests $\bF_{\langle I_{\Left},I_{\Right}\rangle}$ with $I_{\Left} = \fragment{\Left(u)}{\Left(u')}$ and $I_{\Right}=\fragment{\Right(u')}{\Right(u)}$ for some nodes $u,u'\in \bF$, and \emph{contexts} of~\cite{DGHKSS22,DGHKS23,KS25,DGHKS25}, which are also defined to be balanced, that is, they cannot partially contain any node. 

Crucially, compared to a super-window and a context, for a bi-interval $\mI = \langle I_{\Left},I_{\Right}\rangle$ the subforest $\bF_{\mI}$ does not need to be balanced. 
Because of this last property, for every subforest $\bF_\mI$ and tree alignment $\mA : \bF \onto \bG$, there is a corresponding subforest $\bG_{\mJ}$ that $\mA$ aligns with $\bF_\mI$. \cref{ex:imbalanced} shows why the image $\bG_{\mJ}$ of a balanced subforest $\bF_\mI$ may need to be imbalanced.

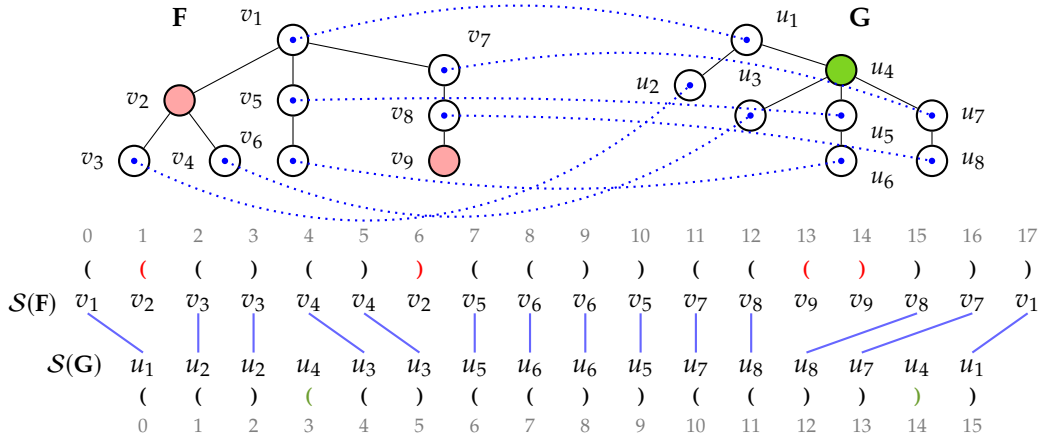
\begin{figure}[t!]
    \centering
    \input{figures/pre/better_forest_merged_TO}
    \caption{Copy of \cref{fig:forest_merged}, repeated here for convenience.}
     \label{fig:forest_merged_rep}
\end{figure}

\begin{example}\label{ex:imbalanced}
In the forest $\bF$ of \cref{fig:forest_merged_rep}, consider a balanced subforest $\smash{\bF_{\langle I_{\Left}, I_{\Right} \rangle}}=\langle v_2 v_3; v_3 v_4 v_4 v_2 \rangle$
with $I_{\Left} = \fragment{1}{2}$ and $I_{\Right} = \fragment{3}{6}$.
With respect to the tree alignment $\mA : \bF \onto \bG$ depicted in \cref{fig:forest_merged_rep}, the corresponding subforest $\bG_{\langle J_{\Left}, J_{\Right} \rangle}$ must contain $u_2$ (shared between $J_{\Left}$ and $J_{\Right}$) and $u_3$ (fully in $J_{\Right}$) since they are matched to $v_3$ and $v_4$, respectively, and it cannot contain any of the nodes $u_1,u_5,u_6,u_7,u_8$ since they are matched to nodes of $\bF$ outside $\smash{\bF_{\mI}}$.
The only possibility is $\smash{\bG_{\langle J_{\Left}, J_{\Right} \rangle}}=\langle u_2; u_2 u_4 u_3 u_3 \rangle$ with $J_{\Left} = \fragment{1}{1}$ and $J_{\Right} = \fragment{2}{5}$.
This subforest is imbalanced because it partially contains $u_4$.
\end{example}

